%% file: thesislowres.tex
\documentclass[12pt,a4paper]{report}

\usepackage{epsf,epsfig}
\usepackage{latexsym,amssymb,amsmath,mathrsfs,amsfonts,psfrag}
\usepackage{setspace,cite} %double spacing for text, single for captions,
\usepackage[hang,small,bf]{caption} 
\usepackage{fancyhdr}
\usepackage{a4}
\usepackage{graphics}

\input{newcommands.tex}

\begin{document}
\newpage

%Puts page numbering of preamble in roman and of main body of thesis in
%arabic. Also defines how chapters and sections are made
\pagenumbering{arabic}
\setcounter{page}{1} \pagestyle{fancy}
\renewcommand{\chaptermark}[1]{
 \markboth{\chaptername \ \thechapter:\,\ #1}{}
}
\renewcommand{\sectionmark}[1]{\markright{\thesection\,\ #1}}

%DEFINES TITLE PAGE, and contains abstract, acknowledgements, etc.

\include{header}

\newpage

%sets up headers for lefthand and righthand pages. To alter, edit
%these lines and the chaptermark/sectionmark lines above
%NOTE some of these lines are redundant, but I haven't had time to
%optimize the size of this file, and anyway, IT WORKS as is.
\addtolength{\headheight}{3pt}
\fancyhead{}
\fancyhead[LE]{\sl\leftmark}
\fancyhead[LO,RE]{\bf\leftmark}
\fancyhead[RO]{\sl\rightmark}
\fancyfoot[C]{\thepage}
\pagenumbering{arabic}

\singlespacing  % you must use setspace.sty to get this. setspace also
                % defines the below two spacing options. It's magic.
%\doublespacing
%\onehalfspacing
\input{chapter1.tex}  %4pages
\input{chapter2.tex}
\input{chapter3.tex}
\input{chapter4sm.tex}
\input{chapter5sm.tex}
\input{chapter6.tex}
\input{end.tex}
\appendix
\input{appendixD.tex}
\input{appendixA.tex}
\input{appendixB.tex}

\input{appendixC.tex}
\singlespacing
\clearpage
\addcontentsline{toc}{chapter}{\protect\numberline{Bibliography}}
\bibliographystyle{utphys}
\bibliography{bib}

\end{document}

%% file: newcommands.tex
\newcommand{\be}{\begin{equation}}
\newcommand{\ee}{\end{equation}}

\newcommand{\bm}[1]{\mbox{\boldmath $#1$}}

\def\bd{\begin{document}}
\def\ed{\end{document}}
\def\nn{\nonumber}
\def\bea{\begin{eqnarray}}
\def\eea{\end{eqnarray}}
\let\bm=\bibitem
\let\la=\label

\newcommand{\EQ}[1]{\begin{equation} #1 \end{equation}}
\newcommand{\AL}[1]{\begin{subequations}\begin{align} #1 \end{align}\end{subequations}}
\newcommand{\SP}[1]{\begin{equation}\begin{split} #1 \end{split}\end{equation}}
\newcommand{\ALAT}[2]{\begin{subequations}\begin{alignat}{#1} #2 \end{alignat}\end{subequations}}
\def\beqa{\begin{eqnarray}}
\def\eeqa{\end{eqnarray}}
\def\beq{\begin{equation}}
\def\eeq{\end{equation}}

\def\hf{{\textstyle \frac{1}{2}}}
\def\rhob{\bar{\rho}}
\def\etab{\bar{\eta}}
\def\abar{\bar a}
\def\sigmabar{\bar\sigma}
\def\etabar{\bar\eta}
\def\zetabar{\bar\zeta}
\def\mubar{\bar\mu}
\def\nubar{\bar\nu}
\def\N{{\cal N}}
\def\sst{\scriptscriptstyle}
\def\thetabar{\bar\theta}
\def\Tr{{\rm Tr}}
\def\one{\mbox{1 \kern-.59em {\rm l}}}
 \def\Nh{\hat{N}}

\newlength{\myVSpace}% the height of the box
\newcommand\xstrut{\raisebox{-.5\myVSpace}% symmetric behaviour,
  {\rule{0pt}{\myVSpace}}%
}
\newcommand{\ncp}{*} %ncp = noncommutative product
\def\star{\ncp}
\def\a{\alpha}      \def\da{{\dot\alpha}}
\def\b{\beta}       \def\db{{\dot\beta}}
\def\c{\gamma}  \def\G{\Gamma}  \def\cdt{\dot\gamma}
\def\d{\delta}  \def\D{\Delta}  \def\ddt{\dot\delta}
\def\e{\epsilon}        \def\vare{\varepsilon}
\def\f{\phi}    \def\F{\Phi}    \def\vvf{\f}
\def\h{\eta}
\def\k{\kappa}
\def\l{\lambda} \def\L{\Lambda}
\def\m{\mu} \def\n{\nu}
\def\o{\omega}
\def\p{\pi} \def\P{\Pi}
\def\r{\rho}
\def\s{\sigma}  \def\S{\Sigma}
\def\t{\tau}
\def\th{\theta} \def\Th{\Theta} \def\vth{\vartheta}
\def\X{\Xeta}
\def\z{\zeta}

\def\cA{{\cal A}} \def\cB{{\cal B}} \def\cC{{\cal C}}
\def\cD{{\cal D}} \def\cE{{\cal E}} \def\cF{{\cal F}}
\def\cG{{\cal G}} \def\cH{{\cal H}} \def\cI{{\cal I}}
\def\cJ{{\cal J}} \def\cK{{\cal K}} \def\cL{{\cal L}}
\def\cM{{\cal M}} \def\cN{{\cal N}} \def\cO{{\cal O}}
\def\cP{{\cal P}} \def\cQ{{\cal Q}} \def\cR{{\cal R}}
\def\cS{{\cal S}} \def\cT{{\cal T}} \def\cU{{\cal U}}
\def\cV{{\cal V}} \def\cW{{\cal W}} \def\cX{{\cal X}}
\def\cY{{\cal Y}} \def\cZ{{\cal Z}}

\def\As {{A \hspace{-6.4pt} \slash}\;}
\def\bs {{b \hspace{-6.4pt} \slash}\;}
\def\Ds {{D \hspace{-6.4pt} \slash}\;}
\def\ds {{\del \hspace{-6.4pt} \slash}\;}
\def\ss {{\s \hspace{-6.4pt} \slash}\;}
\def\ks {{ k \hspace{-6.4pt} \slash}\;}
\def\ps {{p \hspace{-6.4pt} \slash}\;}
\def\pas {{{p_1} \hspace{-6.4pt} \slash}\;}
\def\pbs {{{p_2} \hspace{-6.4pt} \slash}\;}

\def\Fh{\hat{F}}
\def\Vh{\hat{V}}
\def\Xh{\hat{X}}
\def\ah{\hat{a}}
\def\xh{\hat{x}}
\def\yh{\hat{y}}
\def\ph{\hat{p}}
\def\xih{\hat{\xi}}

\def\psit{\tilde{\psi}}
\def\Psit{\tilde{\Psi}}
\def\tht{\tilde{\th}}

\def\At{\tilde{A}}
\def\Qt{\tilde{Q}}
\def\Rt{\tilde{R}}
\def\Nt{\tilde{N}}

\def\at{\tilde{a}}
\def\st{\tilde{s}}
\def\ft{\tilde{f}}
\def\pt{\tilde{p}}
\def\qt{\tilde{q}}
\def\vt{\tilde{v}}
\def\nt{\tilde{n}}

\def\delb{\bar{\partial}}
\def\bz{\bar{z}}
\def\bD{\bar{D}}
\def\bB{\bar{B}}
\def\lqcdmb{\Lambda_{QCD}/m_b}
\def\lud{\lambda_u^{(d)}}
\def\lcd{\lambda_c^{(d)}}
\def\lus{\lambda_u^{(s)}}
\def\lcs{\lambda_c^{(s)}}
\def\lpd{\lambda_p^{(d)}}
\def\lps{\lambda_p^{(s)}}

\def\bk{{\bf k}}
\def\bl{{\bf l}}
\def\bp{{\bf p}}
\def\bq{{\bf q}}
\def\br{{\bf r}}
\def\bx{{\bf x}}
\def\by{{\bf y}}
\def\bR{{\bf R}}
\def\bV{{\bf V}}

\newcommand{\gs}{\hat{\gamma}_{\rm s}^{(0)}}
\newcommand{\svs}{\vbox{\vskip 5mm}}
\newcommand{\gssndr}{\hat{\gamma}_{\rm s}^{(1)}}
\newcommand{\mvs}{\vbox{\vskip 8mm}}
\newcommand{\gem}{\hat{\gamma}_{\rm e}^{(0)}}

%% file: header.tex
\thispagestyle{empty}

% ******* Title page ******* 
% ************************** 

\topskip 1cm

\begin{titlepage}
  \vfill
  \begin{center}
    {\Huge\bf Non-leptonic B-decays in and beyond QCD Factorisation\\}
    \vspace{4.cm}
    \doublespacing
    {\large A thesis presented for the degree of\\ 
      Doctor of Philosophy\\
      by\\}
    \vspace{1.cm}
    {\LARGE\bf Angelique N Talbot}\\
    \vspace{3.cm}
    {\large Institute for Particle Physics Phenomenology\\
      University of Durham\\
      September 2005\\}
    \singlespacing
    \vfill
    \scalebox{0.2}{\includegraphics{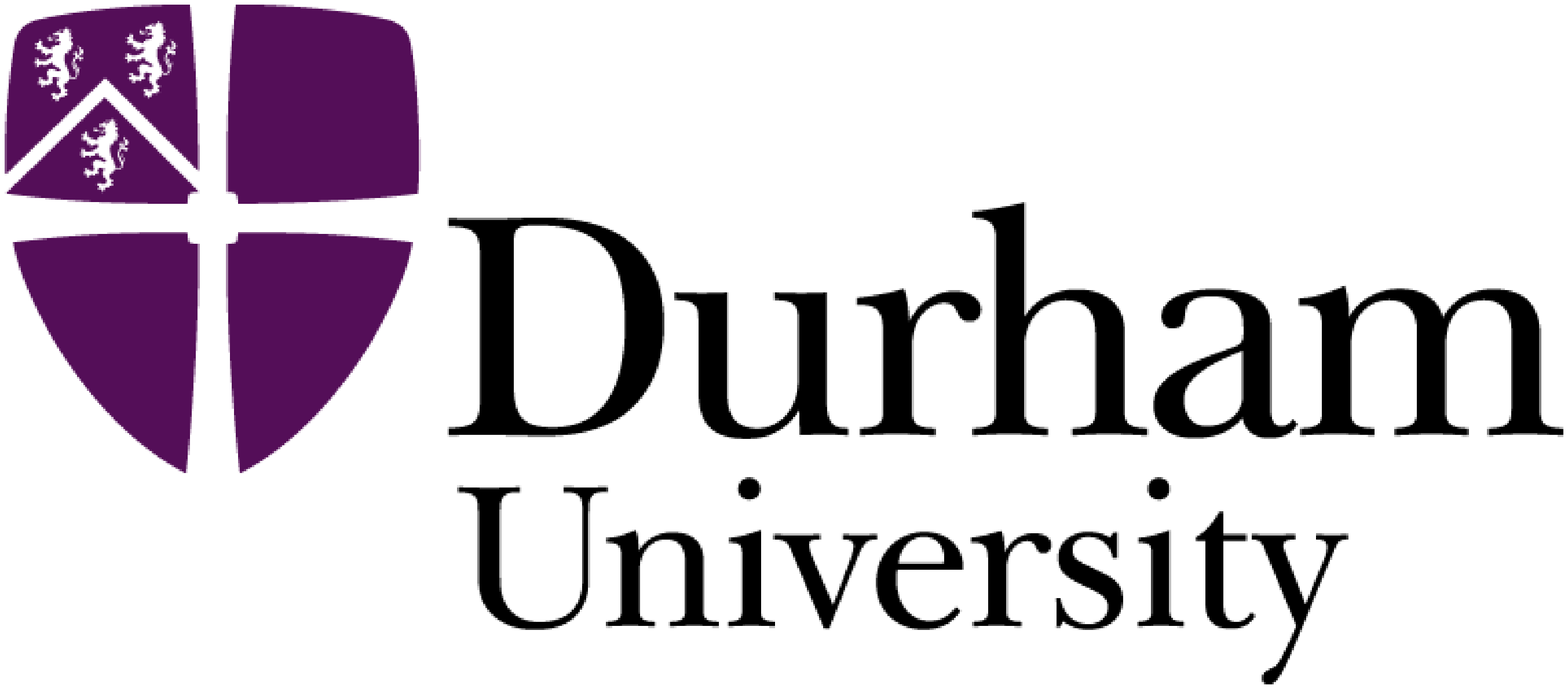}}
  \end{center}
\end{titlepage}

\onehalfspacing

%ABSTRACT
\begin{center}
  {\Huge\bf Abstract\\}
\end{center}
\vspace{2.cm}

This thesis examines the non-leptonic $B$-decays within QCD factorisation and beyond, to challenge the assumptions and limitations of the method.  We analyse the treatment of the distribution amplitudes of light mesons and present a new model described by simple physical parameters.  The leading twist distribution amplitudes of light mesons describe the leading non-perturbative hadronic contributions to exclusive QCD reactions at large energy transfer, for instance electromagnetic form factors.  Importantly, they also enter into the two-body $B$ decay amplitudes described by QCD factorisation.  They cannot be calculated from first principles and are described by models based on a fixed-order conformal expansion, which is not always sufficient in phenomenological applications.  We derive new models that are valid to all orders in the conformal expansion and characterised by a small number of parameters related to experimental observables.  

Motivated by the marginal agreement between the QCD factorisation results with the experimental data, in particular for $B\to\pi\pi$, we scrutinise the incalculable non-factorisable corrections to charmless non-leptonic decays.  We use the available results on $B\to\pi\pi$ to extract information about the size and nature of the required non-factorisable corrections that are needed to reconcile the predictions and data.  We find that the best-fit scenarios do not give reasonable agreement to $2\sigma$ until at least a 40\% non-factorisable contribution is added.  Finally we consider the exclusive $B\to V\gamma$ decays, where we analyse the recently updated experimental data within QCD factorisation and present constraints on generic supersymmetric models using the mass insertion approximation.  

%\doublespacing
\onehalfspacing
\pagenumbering{roman}
\setcounter{page}{1} \pagestyle{plain}

%\addcontentsline{toc}{chapter} 
%                 {\protect\numberline{Abstract\hspace{-76pt}}} 

% ********************************** 
% ******** Acknowledgements  ******** 
% ********************************** 
\chapter*{Acknowledgements}
%\addcontentsline{toc}{chapter} 
%                 {\protect\numberline{Acknowledgements\hspace{-76pt}}} 

First and foremost I would like to thank my supervisor, Patricia Ball for
giving me the opportunity to work with her.  I am grateful for the continued
help and guidance throughout my time here, and if I have gained even a little
of her efficiency and dedication then I am all the more thankful.      

My thanks also go to the others I have had the pleasure to work with,
especially to Martin Gorbahn for always trying to answer even my most stupid
of questions, and to Gareth Jones, to whom I pass the torch to continue our
project as I move on to higher climbs.  A special mention must also go to the
most excellent IPPP C\'{e}ilidh Band 
for two fun years and three successful performances, particularly to Gudi
Moorgat-Pick for her endless enthusiasm even when 800 miles away.

To Tom Birthwright, Paul Brooks, Mark Morley-Fletcher, Richard Whisker and
James Haestier, the best office mates anyone could ever ask for.  I am
thankful for the relaxed atmosphere and willingness to share advice and
adventures, despite the noise and the odd cricket ball.  

Above all others, I have eternal gratitude for the unlimited support and
encouragement from my Mum \& Dad and my boyfriend Michael.  People you can
always talk to about anything is a wonderful blessing, even if far away.  I
have learnt so much from their calm and balanced outlook, that I hope
I close this chapter of my life a better person.  I know at least, there is
nothing that a hug and a glass of Amarone cannot solve.  

This work was supported by a PPARC studentship which is gratefully
acknowledged.      

\chapter*{Declaration}
%\addcontentsline{toc}{chapter} 
%                 {\protect\numberline{Declaration\hspace{-76pt}}} 
I declare that no material presented in this thesis has previously been submitted
for a degree at this or any other university.

The research described in this thesis has been carried out in collaboration
with Dr. Patricia Ball and has been published
as follows:

\begin{itemize}
\item \textit{Models for Light-Cone Meson Distribution Amplitudes},\\
Patricia Ball and Angelique N Talbot\\
JHEP \textbf{0506} (2005) 063 [arXiv:hep-ph/0502115]
\end{itemize}
The copyright of this thesis rests with the author.  No quotation from it should be published without their prior written consent and information derived from it should be acknowledged.

\singlespacing

\tableofcontents

%\listoffigures

\pagestyle{plain}

\input{preface}

%%% Local Variables: 
%%% mode: latex
%%% TeX-master: "thesis"
%%% End: 

%% file: preface.tex
\chapter*{Introduction}
\addcontentsline{toc}{chapter} 
                 {\protect\numberline{Introduction\hspace{-96pt}}}

The quest for comprehension of the physical world around us has dominated
philosophy and science since the beginning of history.  The ancient peoples
looked up to the stars and the Gods for answers.  The philosophers called
upon the elements of nature -- the Ancient Greeks, such as Leucippus,
Democritus or Epicurus were the first to analyse and categorise the nature
of all things.  The belief that matter was composed of the four
fundamental elements: earth, air, fire and water survived for over two thousand
years.  It was they who introduced the first ``elementary particles'', the
indivisible \textit{atomos}.  As mysticism and natural philosophy morphed into
scientific principle, the foundations of modern particle physics were established;
beginning with the physical theories of Newtonian mechanics and gravity to
the work of Thompson and Rutherford which allowed  physics to leave the 19th
century knowing the ``indivisible atoms'' were not in fact, indivisible at all.

The birth of the 20th century and the pioneering study of quantum mechanics
was followed in the 1950s and 60s by the discovery of a bewildering variety and
number of particles, once physicists had discovered
that protons and neutrons were themselves composite particles.  Yet, as with Mendeleev's
periodic table of 1869, these particles could again be categorised and classified
revealing a simple underlying structure in terms of a few elementary
particles.  The ideal to capture the beauty and
complexity of nature in a simple mathematical form culminated in the
formulation of the Standard Model of particle physics.  This is based upon
quantum field theory and gauge principles -- combining the electromagnetic,
weak and strong interactions into one unified theory.  To date the 
Standard Model is the most successful mathematical theory of particle physics ever
created.  Particle accelerators have probed scales down to 
$10^{-16}$cm, and this theory is consistent with virtually all physics down to
this scale -- indeed the Standard Model has passed indirect tests which
probe even shorter distances.  

The Standard Model has 25 elementary particles, so perhaps one may ask ``how
deep the rabbit hole goes''?  The search for another layer of structure
beneath the quarks and leptons has however so far been fruitless.  More
importantly, the Higgs boson which is required to generate mass for all of
the Standard Model particles, has still not been discovered in collider experiments.
There are a number
of other unresolved issues with the Standard Model, that have led to recent conclusions
that it cannot be complete, and is more likely to be an ``effective theory''
of some more encompassing theory at higher energy.  These issues
include some fundamental questions -- such as the large number of arbitrary
parameters that appear in the Standard Model Lagrangian.  Why are there copies
of quarks and leptons in three generations?  Why are the masses split in the
way they are?  Why is the weak scale so different from the expected Planck
scale?  These questions cannot be answered with our current understanding.
On top of all these questions, the problem of including General Relativity
with the Standard Model still remains -- combining gravity and quantum
mechanics produces a non-renormalisable quantum field theory -- this again
suggests the presence of another theory, and hence some new physics at some
higher energy.   

Many ideas have been put forward to solve these problems, the
most popular being supersymmetry, string theory or extra dimensions.  Yet
to date there is no theory that is truly simpler or less arbitrary than the
Standard Model.  More importantly, there is little  
experimental evidence for the existence of the holy grail of ``new
physics'' and to guide the directions of the model builders.  There are
perhaps a handful of discrepancies between the 
theoretical predictions and experimental results, none of which have been
significant enough to claim as new physics.  

Within this thesis I work predominantly within the Standard Model, taking
some brief forays into generic supersymmetric models.  To
summarise the motivation in a line, I say the following:  `before any claims
of new physics can be made, we must be absolutely sure that our theory
predictions are as accurate as we can make them'.  The general aim being to better
understand some elements of the Standard Model -- specifically relating to
the weak decays of $B$-mesons.

Beauty decays are a rich and powerful playground for studying flavour physics
and CP violation -- we can test predictions and constrain Standard Model parameters. From
experimental observables (decay rates, parameters etc.) we can perform
indirect searches for new physics via any measured deviation from the
Standard Model expectation.  Due to the rapid decay of the top quark, the $b$
quark is the heaviest quark to bind into mesons which can be observed.
Theoretical techniques based upon an expansion in the heavy $b$ quark mass
can provide vast simplifications of calculations and model-independent
predictions of observables.   

The theoretical framework within which any
$B$-physics analysis is based is that of effective field theory, where the
operator product expansion and renormalisation group evolution are invaluable
techniques which we exploit.  These encompass the concepts of factorisation
which allow the calculation of decay amplitudes to be separated into
perturbatively calculable short-distance Wilson coefficients, and
long-distance matrix elements.  We need some non-perturbative method such as
QCD sum rules or Lattice QCD to calculate these matrix elements fully.  For
the exclusive two-body decays of the $B$-meson we can make use of the
powerful technique of QCD factorisation which makes use of the
hierarchy between the heavy $b$ quark mass $m_b$, and the intrinsic scale of
QCD, $\Lambda_{\mathrm{QCD}}$.  This gives a factorisation formula for the
evaluation of hadronic matrix elements, and is a further separation of
long-distance contributions from a set of perturbative short-distance
contributions.  The long-distance part must still be calculated via some
non-perturbative technique, but is of much less complexity than the original
matrix element.

QCD factorisation was developed by Beneke, Buchalla, Neubert and Sachrajda (BBNS)
\cite{Beneke:1999br,Beneke:2001ev,Beneke:2003zv} for exclusive decays of the $B$ into
two mesons.  The method is however not without
its uncertainties, nor gaps where calculations cannot be completed
without some model-dependent assumptions.  It is these uncertainties,
in the decays of the $B$-meson into non leptonic final states,
that we discuss in this work.  

The main new result of this work involves examining the uncertainty from one
of the important non-perturbative inputs in the factorisation formulae,
namely the light-cone distribution amplitudes of the light mesons.  We
develop and introduce a set of new models for the distribution amplitude
which improve upon the truncated conformal 
expansion that is widely used in the literature.  We quantitatively take
into account contributions from higher order moments in the conformal
expansion by a resummation to all orders.  We show how these models can be
expressed in terms of a few simple physical parameters that are directly
related to experimental observables.

We then present an analysis of the second major source of uncertainty to QCD
factorisation, namely corrections of order $(\Lambda_{\mathrm{QCD}}/m_b$) to the
leading calculation performed in the heavy quark limit ($m_b\to\infty$), that
cannot be calculated in a model-independent manner.  
Evidence suggests that QCD factorisation may considerably
underestimate these corrections in some decay channels, specifically
$B\to\pi\pi$.  We construct various scenarios of additional non-factorisable
contributions and use the wealth of experimental data currently available
from the $B$ factories and accelerator experiments to quantify how good the
agreement is with the experimental measurements.    

Finally, we present an analysis using recently released (August 2005) results
on the $B\to(\rho,\,\omega)\,\gamma$ decays, based on the extension of QCD factorisation
to  radiative decays $B\to V\gamma$ by Bosch and Buchalla
\cite{Bosch:2001gv}.  We also consider the possibility of new physics in the
$b\to d \gamma$ and $b\to s \gamma$ decays, and use the new experimental
results to constrain contributions from generic minimal supersymmetric models
using the mass 
insertion approximation. \\

The structure of the thesis can be outlined as follows:

We begin with a whistlestop tour of the basic concepts of the Standard Model
and the theoretical techniques required for an analysis in effective field theory;
renormalisation group perturbation theory and a discussion of the $\Delta B =
1$ effective Hamiltonian.  We conclude this introductory chapter with a brief
introduction to two important non-perturbative techniques: QCD sum rules and
Lattice QCD.  We make use of many results from these methods and so it is
important to understand their basis and limitations.  

Chapter \ref{chp:QCDF} provides an
in-depth discussion on the formulation of QCD factorisation.  We
begin by placing it in context with a discussion of naive factorisation, and
then follow with details of the structure of the factorisation formula and
its input.  We introduce the ``non-factorisable corrections'' and the
problems with calculation of 
power-suppressed diagrams in a model-independent manner.  We also include the
isospin decomposition of 
$B\to\pi\pi$ amplitudes which is used extensively in our Chapter 4 analysis, 
and finally conclude with the limitations of the QCD
factorisation framework.

Chapter \ref{chp:LCDA} presents the main result of this work: the development of a new
resummed model of the light-cone distribution amplitude for light mesons.  We
present a detailed discussion of the treatment of the DA, beginning with the
application of conformal symmetry techniques, and the expression of the DA as
a partial wave expansion in conformal spin.  In order to place our new models
in context we give a brief review of the literature and previous constructions
with specific reference to the pion wavefunction, before going on to present
full details of our new models.  These sections are primarily based on the
work presented in \cite{Ball:2005ei}.  We discuss the full implications of
the new models, and give numerical results to show how variation of the size of
contribution from the higher-order moments can affect the branching ratios
and CP asymmetry predictions, using the example of $B\to\pi\pi$.

We then go on in Chapter \ref{chp:NFC}, to present our analysis on the non-factorisable
corrections to the charmless $B$-decays.  We investigate how the extensive
experimental information available can be used to extract information about the size
and nature of non-factorisable corrections that are needed to provide
agreement between the prediction and measurement of $B\to\pi\pi$ branching
fractions and CP asymmetries.  We split the $B\to\pi\pi$ isospin amplitudes
into factorisable and non-factorisable parts -- the former being calculated
via QCD factorisation and the later fitted to the experimental data.  We show
that there is evidence for sizable non-factorisable corrections in the
$B\to\pi\pi$ system, and discuss the application of these results to $B\to\pi
K$.  We also discuss the possibility and likelihood of a charming penguin
contribution to the charmless decays.

Our final analysis is presented in Chapter \ref{chp:ERB}, and discusses the exclusive
radiative decays $B\to(\rho,\omega)\gamma$ and $B\to K^*\gamma$.  We examine
the predictions from QCD factorisation in the context of the recent
experimental results for this system.  These decays are rare decays occurring
with branching fractions of $10^{-5}$ or less, as they occur only at the loop
level within the Standard Model.  They are as such, very sensitive to the
possibility of new physics.  After introducing the basic principles and
motivation for one of the most popular extensions of the Standard Model --
the Minimal Supersymmetric Standard Model (MSSM) -- we discuss how the mass
insertion approximation can be utilised to constrain the parameter space for
new physics in the $b\to (d,s)$ transitions.  We give graphical constraints
on insertions $\delta^d_{13}$ and $\delta^d_{23}$.

Finally, we present our conclusions and outlook for the future in Chapter
\ref{chp:CONC}. 

We consign various technical details to the Appendices: We give the explicit
formulae for the $\Delta B = 1$ Wilson coefficients; the analytic evolution
of the LCDA; a set of useful expressions from QCD factorisation, 
specifically the decay amplitudes for $B\to\pi\pi$ and the annihilation
contributions to the $B\to V\gamma$ decays.  We also present a full summary
of numerical input parameters that have been used in our analyses.

%%% Local Variables: 
%%% mode: latex
%%% TeX-master: "end"
%%% TeX-master: t
%%% End: 

%% file: chapter1.tex
\chapter{Basic concepts of B-physics}
\begin{center}
  \begin{quote}
    \it
    It's a job that's never started that takes longest to finish
  \end{quote}
\end{center}
\vspace{-4mm}
\hfill{\small J.R.R. Tolkien}
\vspace{5mm}

The Standard Model is the cornerstone of particle physics and this chapter introduces
its most pertinent features.  We begin with a brief summary of the structure of
the Standard Model and then introduce the concepts of effective theories that underpin
calculations of B-decays.  Via a discussion on the renormalisation group we
give the full effective Hamiltonian for $\Delta B = 1$ decays.  Finally, we
discuss two important methods of calculating non-perturbative information --
QCD sum rules and Lattice QCD.

\section{The Standard Model}

\begin{figure}[h]
  $$\epsfxsize=0.6\textwidth\epsffile{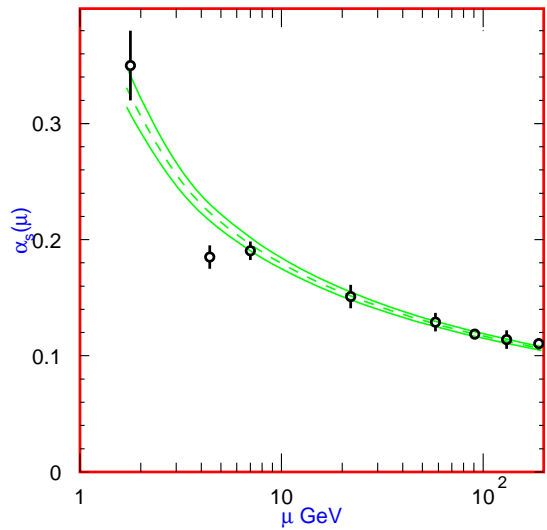}$$
  \vskip-12pt
  \caption[ ]{Running of $\a_s(\mu)$ with energy scale $\mu$ and summary of
  experimental measurements.  Data points are (in increasing $\mu$) from $\tau$
  widths, $\Upsilon$ decays, deep inelastic scattering, JADE data, TRISTAN
  data and $e^+e^-$ event shapes  \cite{Eidelman:2004wy}. } 
  \label{fig:alpha}
\end{figure}

The Standard Model is the most successful and comprehensive theory of particle
interactions to be developed in modern times.  The model conjoins the theories
of strong and electroweak forces into a unified framework based upon gauge
symmetries.  The dynamics can be described by a single fundamental Lagrangian,
constructed of contributions from the three sectors: Quantum Chromodynamics
(QCD), electroweak interactions and the Higgs sector.  The gauge structure of
the Standard Model is summarised as
$$ SU(3)_C\otimes SU(2)_L\otimes U(1)_Y$$
QCD is the theory of the strong interaction that describes the gauge
interactions between quarks and gluons, and was first introduced in the
1970's \cite{Gell-Mann:1964nj,Han:1965pf, Politzer:1973fx,Fritzsch:2002jv,Fritzsch:1973pi}.  This force acts upon ``colour
charge'' and is based upon the gauge group $SU(3)_C$.  The eight generators of
this group represent eight force carriers (the gluons) which communicate the
force between coloured objects.  The quarks carry colour charge and most
importantly so do the gluons due to the \textit{Yang-Mills} or non-Abelian
nature of the QCD gauge theory; this allows the gluons to interact with each
other and is an essential 
ingredient in the \textit{asymptotic freedom} of QCD.  This property allows
the perturbative treatment of the strong interactions at short distances --
where the coupling constant $\a_s$ becomes small.  At
long distances, i.e small energies, the coupling becomes large and there is a
total confinement of the quarks into colourless hadrons.  Figure
\ref{fig:alpha} shows the running of the strong coupling with the energy scale. 

The next sector of the Standard Model is that of the electroweak
interactions, described by the Glashow-Salam-Weinberg model
\cite{Glashow:1961tr,Weinberg:1967tq, Salam}  and which is based on the gauge
group $SU(2)_L\otimes U(1)_Y$.  This theory is a
unified treatment of the weak and electromagnetic interactions, where the
full group is spontaneously broken to QED $U(1)_Q$.  This is the Abelian
theory of pure electromagnetic interactions, which has the photon as its
propagator.  The weak interaction controls all particle decays which do 
not proceed via the dominant strong or electromagnetic interactions.  The
weak decays of the $B$-meson involve a change of flavour of the $b$ quark,
and are controlled by a \textit{charged current interaction}.  We discuss the
dynamics of flavour in detail in the next section.  The spontaneous symmetry
breaking of the electroweak theory occurs via the non-zero
vacuum expectation value of the Higgs field -- a scalar isospin doublet 
\be
   \phi=\left(\begin{array}{c}
               \phi^+ \\
                \phi^0 \\
              \end{array}\right)
\ee
The Higgs field has four degrees of freedom, three of which provide masses to
the electroweak gauge bosons $W^+, W^-, Z^0$.  The remaining degree of
freedom is theorised to manifest itself as a massive scalar boson, but to
date there has been no experimental detection to verify this claim.  Data
from electroweak precision tests suggest the Higgs mass $m_H$, should be light, and a
lower bound from LEP exists of $\sim114$GeV \cite{Barate:2003sz}.  

It would seem that the Higgs is the last missing piece in the particle
puzzle.  However even if the Higgs is found, there are still a number of
issues that need to be addressed:  for example, there is yet to be a consistent method of
incorporating gravity into the model, nor is there an inclusion of masses for
the neutrinos and their recently discovered oscillations.  We also must
address the \textit{hierarchy problem} that 
exists if the Higgs is discovered as light, instead of having a mass $\sim10^{16}$GeV as would be expected from calculation of quantum corrections to $m_H$.

Many different scenarios have been developed to explain these issues and many
more which are not elaborated upon here.  However, before one can study
extensions and revisions of the Standard Model, we must 
understand properly the Standard Model itself!  This thesis is concerned
with the theory behind decays of the $B$-meson, and so we now introduce and
discuss in detail the most 
relevant sector of the Standard Model  -- the dynamics of flavour.

\subsection{The flavour sector}

The fermions appear in the Standard Model in three generations, each
generation identical in all its quantum numbers differing only by the masses
of the particles.  Concerning the electroweak interactions, the quarks and
leptons are spilt into left-handed doublets and a right-handed singlet under
the gauge group $SU(2)_L$.  These contain the leptons:
\begin{eqnarray*}
   \begin{array}{ccc}
   \left(\begin{array}{c} 
               \nu_e \\ 
                 e^- \\ 
         \end{array}\right)_L & 
   \left(\begin{array}{c}
                \nu_\mu \\ 
                  \mu^- \\ 
         \end{array}\right)_L &
    \left(\begin{array}{c}
                 \nu_\tau \\ 
                   \tau^- \\ 
         \end{array}\right)_L \\ 
      e_R & \mu_R & \tau_R 
       \end{array}
\end{eqnarray*}
and the quarks:
\begin{eqnarray*}
   \begin{array}{ccc}
   \left(\begin{array}{c} 
                u\\ 
                d^\prime \\ 
         \end{array}\right)_L &
   \left(\begin{array}{c}
                c \\ 
                s^\prime  \\ 
         \end{array}\right)_L &
    \left(\begin{array}{c}
                t \\ 
                b^\prime \\ 
         \end{array}\right)_L \\
           u_R & c_R & t_R \\
              d_R & s_R & b_R 
      \end{array}
\end{eqnarray*}
The existence and nature of the right-handed neutrino is still under question
and is currently subject to a great amount of theoretical and experimental
probing \cite{Mohapatra:2004vr, Kayser:2005cd}.  This does not affect our
subset of non-leptonic B-decays so we retain the ``standard'', massless neutrino version of the Standard Model.

The electroweak interactions are described by the following Lagrangian, which is made up
of a \textit{charged current} and a \textit{neutral current}.
\begin{eqnarray*}
   \mathcal{L}^{\textrm{EW}}_{\textrm{int}} &=& \mathcal{L}_{CC} +
   \mathcal{L}_{NC} \\
    &=& -\frac{g}{\sqrt{2}}\left[J_\mu^+W^{+\mu} + J_\mu^-W^{-\mu}\right] -
   e\,J^{\textrm{em}}_\mu A^\mu - \frac{g}{\cos{\theta_W}}\left[J^0_\mu Z^\mu   \right]
\end{eqnarray*}
The neutral current part of the Lagrangian is made up of the neutral
electromagnetic and weak currents $J^{\textrm{em}}_\mu$ and $J^0_\mu$ which
are given in terms of the (electric) charge and isospin of the fermions.
We have:
\begin{eqnarray*}
   J^{\textrm{em}}_\mu &=& Q_f\,\bar{f}\gamma_\mu f \\
   J^0_\mu &=& \bar{f} \gamma_\mu \left[\left(I_z^f - 2Q_f\sin^2{\theta_W}\right) -
   I_z^f \gamma_5\right] f
\end{eqnarray*}
summing over all flavours.  The charged current in the quark sector is given
by 
\begin{eqnarray*}
   J^+_\mu =  \big(\bar{u},\,\bar{c},\,\bar{t}\big)_L \gamma_\mu \,V_{\mathrm{CKM}}\,
   \left(\begin{array}{c}
                d \\
               s \\
               b \end{array}\right)_L
\end{eqnarray*}
where the $L$ subscript again represents the left-handed projector
$\tfrac{1}{2}(1-\gamma_5)$ which reflects the vector -- axial-vector ($V-A$)
structure of the weak  
interaction.  $V_{\mathrm{CKM}}$ is the Cabbibo-Kobayashi-Maskawa (CKM)
matrix \cite{Cabibbo:1963yz,Kobayashi:1973fv} and is a 
$3\times3$ unitary mixing matrix which rotates the mass eigenstates $(d,\,s,\,b)$ into
their weak eigenstates $(d^\prime,\,s^\prime,\,b^\prime)$, and allows for
transitions between the quark generations.  The leptonic sector
is described by an analogous mixing matrix which (in the absence of neutrino
masses) is given by the unit matrix.

Symbolically the CKM matrix is written as 
\be
   V_{\mathrm{CKM}} = \left(\begin{array}{ccc}
                    V_{ud} & V_{us} & V_{ub} \\
                    V_{cd} & V_{cs} & V_{cb} \\
                    V_{td} & V_{ts} & V_{tb}  
              \end{array}\right)
\ee
Unitarity ensures that there are no flavour changing neutral currents at
tree level.  The elements $V_{ij}$ are in general complex numbers which are
restricted only by the unitarity condition -- they are free parameters of the
Standard Model and can only be determined by experiment.  In general, an
$n\times n$ unitary matrix is described by $n^2$ parameters.  If this were to
correspond to $n$ quark doublets say, then the phases of each of the $2n$
quark states can be redefined whilst leaving the Lagrangian invariant.
Hence, $V$ should contain $n^2 - (2n-1)$ real parameters.  As an 
orthogonal $n\times n$ matrix can have only $\tfrac{1}{2}n(n-1)$ real
parameters, then we will be left with
$n^2-(2n-1)-\tfrac{1}{2}n(n-1)=\tfrac{1}{2}(n-1)(n-2)$ 
independent residual phases in the quark mixing matrix.  Thus, we see that
$V_{\mathrm{CKM}}$ must be parameterised by three independent angles and one complex
phase; it is this phase which leads
to a non-zero imaginary part of $V$ and which is essential to describing CP
violation in the Standard Model.

The standard parameterisation of $V_{\mathrm{CKM}}$ is written in terms of three angles
$\theta_{ij}$ ($i,j=1,2,3$) and a CP violating phase $\delta$ \cite{Eidelman:2004wy}.
The most useful parameterisation we use in this work is the \textit{Wolfenstein
  parametrisation} \cite{Wolfenstein:1983yz}, where each element is expanded as a power
series in the small parameter $\lambda = |V_{us}| \approx 0.22$.  This reads to
${\cal O}(\lambda^3)$ as 
\begin{equation}
 \label{eq:wolf}
  V_\mathrm{CKM}=\left(
  \begin{array}{ccc} 
     1-\frac{\lambda^2}{2} & \lambda & A\lambda^3 (\rho-i\eta) \\
     -\lambda & 1-\frac{\lambda^2}{2} & A\lambda^2 \\
     A \lambda^3 (1-\rho-i\eta) & -A\lambda^2 & 1
   \end{array}\right)
\end{equation}

We can extend this to higher orders in $\lambda$, and corrections up to order 
$\mathcal{O}(\lambda^5)$ alters only the element $V_{td}$, transforming
$\rho\to\bar\rho=\rho\,(1-\lambda^2/2)$ and
$\eta\to\bar\eta=\eta\,(1-\lambda^2/2)$.  
The latest experimental determination of these parameters is included in a
full summary of input parameters which is given in Appendix \ref{chp:AppC}.

\subsection{Unitarity triangle}

The unitarity of the CKM matrix gives six independent relations, each of
which can be geometrically represented as triangles in an 
Argand diagram.  The relation that is normally employed to represent
unitarity in the B-system is
\be
  \label{eq:unitrel}
   V_{ud}V_{ub}^* + V_{cd}V_{cb}^* + V_{td}V_{tb}^* = 0 
\ee

\begin{figure}[h]
  $$\epsfxsize=0.6\textwidth\epsffile{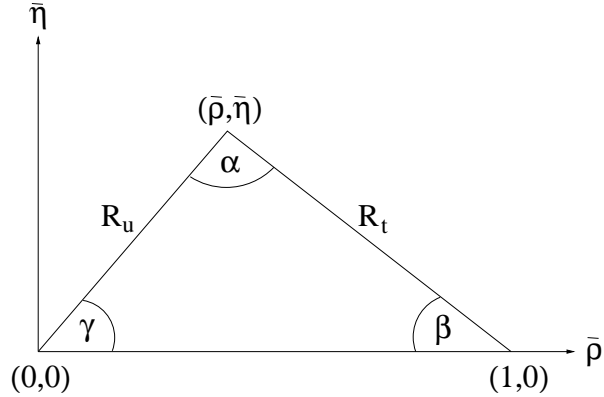}$$
  \vskip-12pt
  \caption[ ]{The unitarity triangle.} 
  \label{fig:tri}
\end{figure}

A phase convention is chosen where $V_{cd}V_{cb}^*$ is real and we rescale
the triangle by
$|V_{cd}V_{cb}^*| = A\lambda^3$.  Graphing this in the complex plane $(\rho,
\eta)$ leads to a the \textit{unitarity triangle} with co-ordinates of the
vertices at (0,0) and (1,0) and the apex at $(\bar\rho, \bar\eta)$ as shown
in Figure \ref{fig:tri}.

The sides are expressed as
\begin{eqnarray}
   \label{eq:RuRb}
      R_u &\equiv& \frac{|V_{ud}V_{ub}^*|}{|V_{cd}V_{cb}^*|} = \sqrt{\bar\rho^2
      + \bar\eta^2} \nonumber\\
 R_t &\equiv& \frac{|V_{td}V_{tb}^*|}{|V_{cd}V_{cb}^*|} = \sqrt{(1-\bar\rho)^2
      + \bar\eta^2}
\end{eqnarray}

\begin{figure}[h!]
  $$\epsfxsize=0.55\textwidth\epsffile{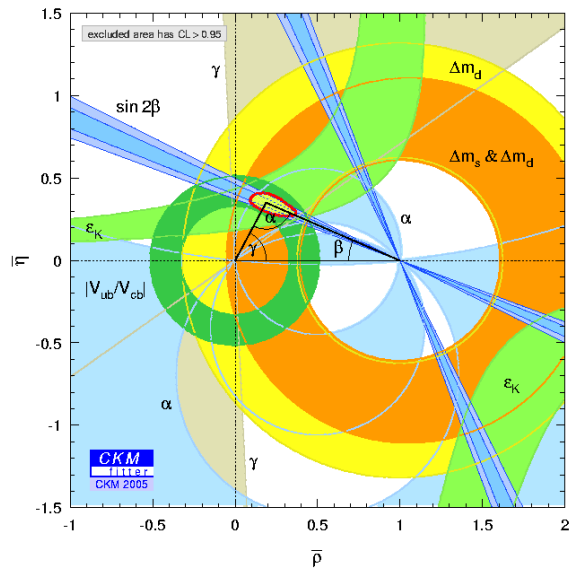}$$
  \vskip-12pt
  \caption[ ]{Constraints on the unitarity triangle from global CKM fit as
  performed by CKM Fitter Group \cite{Charles:2004jd}.} 
  \label{fig:triW05}
\end{figure}

We introduce the shorthand $\lambda_p^{(d)} = V_{pb}V_{pd}^*$ which is used
extensively in phenomenological applications.  The unitarity relation
(\ref{eq:unitrel}) is invariant under phase transformations, which means that
the sides and angles of the triangle remain unchanged with a change of phase
-- hence they are 
physical observables.  These, and the elements of $V_{\mathrm{CKM}}$ have
been subject to as many experimental 
determinations as possible in an attempt to over-constrain the parameters of
the triangle and to test the Standard Model.  The constraints can come from a wide
variety of different decays and parameters: for example $B^0\to J/\psi K_S$ for
$\sin{2\beta}$; $B\to\rho\pi$, $B\to\pi\pi$ for $\alpha$... These constraints
are neatly summarised graphically in Figure \ref{fig:triW05}\footnote{Updated
  results and plots available at: http://ckmfitter.in2p3.fr.}. 

\section{B decays and effective field theory}

\begin{figure}[h]
  $$\epsfxsize=0.7\textwidth\epsffile{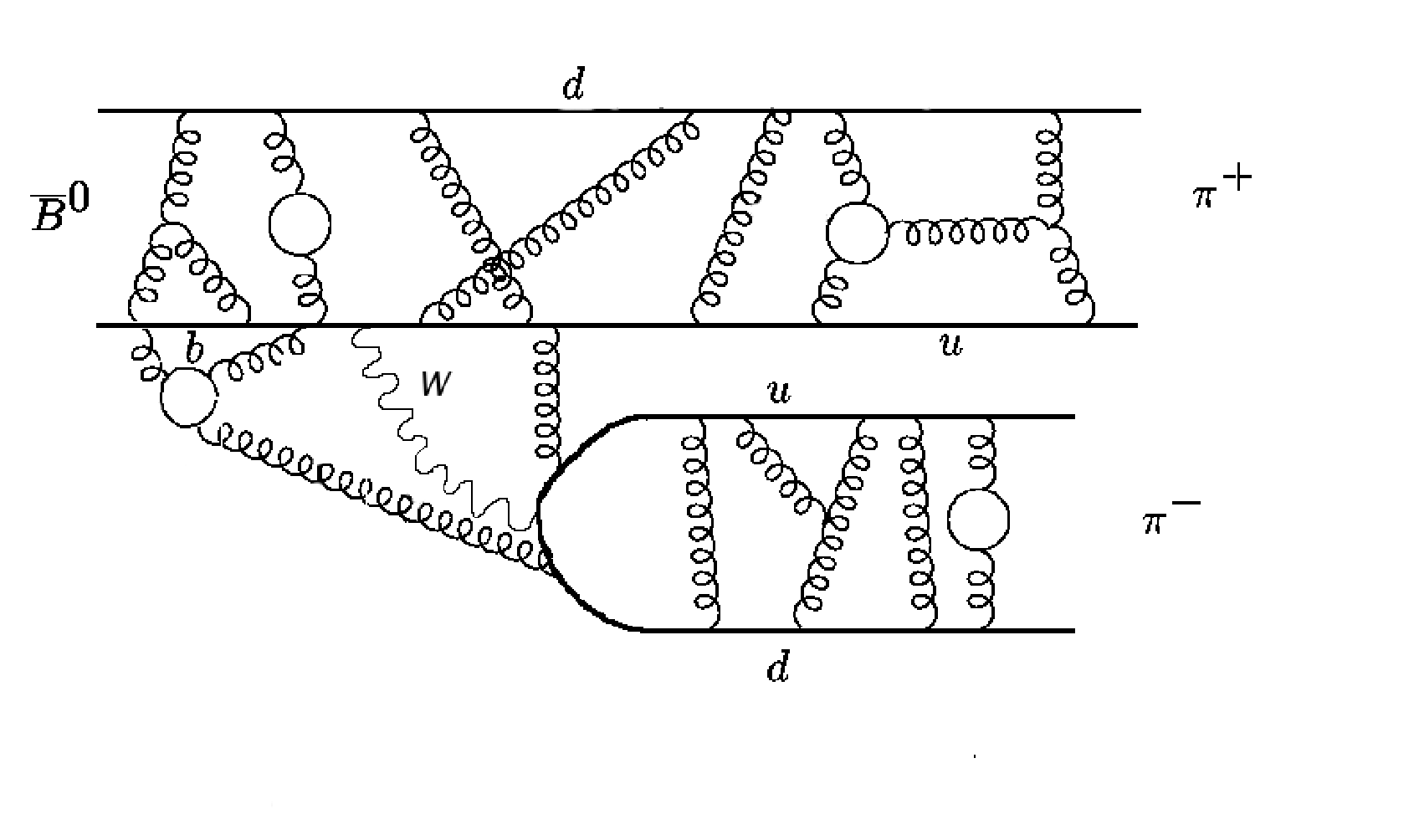}$$
  \vskip-12pt
  \caption[ ]{Illustration of QCD effects in $B\to\pi\pi$ decay.} 
  \label{fig:nonlep}
\end{figure}

There are three types of decays of the $B$-meson, categorised by the final
state decay products.  Firstly there are the leptonic 
decays, such as $B^+\to \ell^+\nu_\ell$ and semi-leptonic decays such as
$B^0\to D^-\ell^+\nu_\ell$.   The decays $B\to\ell^+\nu_\ell + X$, with $X$
as anything, accounts for around 10\% of the total $B$ decay rate.  Finally, and
most important in the context of this work, are the fully hadronic
(non-leptonic) decays.  The complication of calculating weak decays in QCD is
illustrated in Figure \ref{fig:nonlep}.  This demonstrates the non-trivial
interplay between the strong and electroweak forces which determine the
dynamics of the decay.

The decay is characterised by several very different energy scales, from the
mass of the $W$-boson and the heavy quarks to the intrinsic QCD scale and the
masses of the light quarks.  We have the ordering 
 $$ m_t, M_W \gg m_{b,c} \gg \Lambda_{QCD} \gg m_{u,d,s}$$
QCD effects at short distances (i.e. that at higher energies) can be
calculated perturbatively thanks to the asymptotic freedom of the theory.
However, there are unavoidable long-distance dynamics from the confining of
light-quarks into bound states -- the hadrons.  These are characterised by a
typical hadronic scale of $\mu\sim 1$GeV, $\alpha_s(\mu)$ is no longer
small at this scale so we cannot use perturbation theory.  This means that
there are non-perturbative QCD interactions which enter into the calculation
of the decay. 

\begin{figure}[h]
 $$\epsfxsize=0.5\textwidth\epsffile{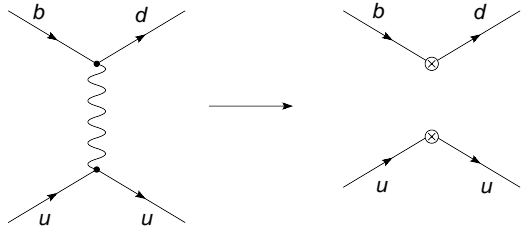} $$ 
  \vskip-12pt
  \caption[ ]{Tree-level diagram for $b\to du\bar{u}$.} 
  \label{fig:eff}
\end{figure}

To systematically disentangle the long and short distance contributions  we
make use of the \textit{operator product expansion} (OPE) \cite{Wilson:1969zs,
  Zimmermann:1972tv,Witten:1977ju}.  The basic idea is that 
any decay amplitude can be expanded schematically in terms of $1/M_W$, since
$M_W$ is much heavier than all of the other relevant momentum scales.  The
products of the quark current operators that interact (via the $W$-- exchange)
are expanded into a series of local operators $Q_i$ multiplied by a
\textit{Wilson coefficient} $C_i$.  These coefficients represent the strength that a
given operator enters into the amplitude.  Schematically we have
\be
   \label{eq:Amps}
   A = C_i\left(M_W/\mu, \alpha_s\right)\cdot\langle Q_i\rangle + \mathcal{O}\left(p^2/M_W^2\right)
\ee
In this way we can define an \textit{effective weak Hamiltonian} to describe
weak interactions at low energies.  In this effective theory the $W$ boson
and the top quark are removed as explicit degrees of freedom -- i.e. they are
``integrated out''.   We describe this as an effective field theory with
$n_f$ ``active'' quarks, i.e. at the scale of $m_b$, we have 5 active
flavours.  As a practical example of this we can consider the
basic (tree-level) $W$-- exchange process of $b\to du\bar{u}$, for which
the OPE gives the amplitude:
\begin{eqnarray*}
   A(b\to du\bar{u})&=& -\frac{G_F}{\sqrt{2}}V_{ud}^*V_{ub}\frac{M_W^2}{k^2 -
   M_W^2}\,\langle Q \rangle \\
    &=& \frac{G_F}{\sqrt{2}}V_{ud}^*V_{ub}\,C\cdot\langle Q\rangle +
   \mathcal{O}\left(\frac{k^2}{M_W^2}\right)
\end{eqnarray*}
with $C=1$ and the local operator
$Q=\left(\bar{d}u\right)_{V-A}\left(\bar{u}b\right)_{V-A}$
This is represented diagrammatically in Figure \ref{fig:eff}.

  The full weak (effective) Hamiltonian, including QCD and electroweak
  corrections has the following structure 
\be
   \label{eq:Heff}
   \mathcal{H}_{eff} = \frac{G_F}{\sqrt{2}}\sum_i V^{\,i}_{\mathrm{CKM}}\,
   C_i(\mu)\, Q_i 
\ee
where the factor $V_{\mathrm{CKM}}^i$ denotes the CKM structure of the
   particular operator.  The decay rate for a two-body non-leptonic decay $B\to M_1M_2$, is then simply 
\be
   \Gamma  = \frac{S}{16\pi
   m_B}|\langle M_1M_2|\mathcal{H}_{\mathrm{eff}}|B\rangle|^2
\ee
with $S=1/2$ if $M_1$ and $M_2$ are identical or $S=1$ otherwise.  The $Q_i$
   denote the relevant local operators which govern the particular decay in
   question; they can be considered as effective point-like vertices.  The
   Wilson coefficients are then seen as  ``coupling constants'' of these
   effective vertices, summarising the    contributions from physics at
   scales higher than $\mu$.  Operators of higher dimensions corresponding to the
   terms of $\mathcal{O}\left(p^2/M_W^2\right)$ can be neglected.   

In short, the OPE gives essentially a factorisation of short and
long distance physics.  The Wilson coefficients contain all the information
about the short distance 
dynamics of the theory, i.e. that at energy scales greater or equal to $\mu$.
They depend intrinsically on the properties of the particles that have been
integrated out of the effective theory, but not on the properties of the
external particles.   The factorisation implies that the coefficients are
entirely independent of the external states, i.e, the $C_i$ are the same for
all amplitudes.      

The long-distance physics, that at
energy scales lower than $\mu$, is parameterised purely by the
process-dependent matrix elements of the local operators.  The
renormalisation scale $\mu$ can be thought of as a ``factorisation scale'' at
which the full contribution splits into the high and low energy parts.

The matrix elements of the local operators are not easily
calculated and as we have discussed, must contain a degree of
non-perturbative information.  Methods of determining these matrix elements using
a systematic formalism are introduced in the next chapter.  The Wilson
coefficients however, can be fully calculated perturbatively by \textit{matching} the full
theory (with the $W$ propagators) onto the effective theory; this ensures that
the effective theory reproduces the corresponding amplitudes in the full
theory.  The steps to compute the $C_i$ are summarised as:
\begin{itemize}
   \item{Compute full amplitude $\mathcal{A}_{full}$ with arbitrary external states}
   \item{Compute the matrix elements $\langle Q_i\rangle$ with same external
   states}
   \item{Extract the $C_i$ using expression (\ref{eq:Amps})}
\end{itemize}

Both ultraviolet and infrared divergences occur in calculation of the
amplitude $\mathcal{A}_{full}$; we discuss the removal of the UV-divergences
in the next section.  In the matching procedure, the IR-divergences are regulated by
setting the momenta of the external quarks to $p^2\neq 0$.  From the
point of view of the effective theory, all of the dependence on $p^2$
(representing the long-distance structure of the full amplitude) is contained
in the matrix elements $\langle Q_i\rangle$.  Hence, the Wilson coefficients
are free from this dependence -- and so independent of the external states.
For convenience, in the matching the external states are chosen to be all
on-shell quarks (or all off-shell), but in general any arbitrary momentum configuration will
work.   

This procedure will give the initial conditions for the Wilson coefficients at the
matching scale, in our case $\mu = M_W$.  We can then use the equations of
\textit{Renormalisation Group Evolution} (RGE) \cite{Christ:1972ms} to find the value at any
desired scale.  

\subsection{Renormalisation group evolution} 

Put simply, renormalisation removes infinities from a theory.  Feynman
diagrams with internal loops often give ultraviolet divergences, as the
virtual particle running through the loop is integrated over all possible
momenta (from zero to infinity).  Renormalisation allows the isolation and
removal of all of these infinities from any physical quantity
\cite{'tHooft:1972fi}.  We relate the
bare (unphysical) parameters with a set of renormalised (physical) parameters
-- such as masses or coupling constants -- and rewrite all of the observables
we need in terms of the new physical quantities.  We can then ``hide'' all of
the divergences in redefinitions of the parameters in the theory Lagrangian.

The procedure of renormalisation introduces a dependence on a dimensionful
parameter known as the \textit{renormalisation scale}, $\mu$.  We can
subsequently obtain the scale dependence of the renormalised parameters from
the $\mu$-independence of the bare ones.   If we choose a set of parameters
at a certain scale $q$ (which gives $g(q), m(q)$ etc.), then the set of all
transformations that relate parameter sets with different values of $q$ is
known as the \textit{renormalisation group}.

An an example, we can consider the coupling constant of QCD: $\a_s(\mu) =
g^2/(4\pi)$.  
The \textit{renormalisation group equation} (RGE) for the running coupling is
given by 
\be
   \mu\frac{dg(\mu)}{d\mu} = \beta(g)
\ee
where the \textit{beta function}, $\beta(g)$, is related to the
renormalisation constant for the coupling.  In QCD this is given
by 
\be
   \beta(g) = -g\left\{\left(\frac{g}{4\pi}\right)^2\beta_0 +
     \left(\frac{g}{4\pi}\right)^4\beta_1 + \ldots \right\}
\ee
with 
\begin{eqnarray*}
   \beta_0 &=& \frac{11}{3}N_c - \frac{2}{3}n_f\\
   \beta_1 &=& \frac{34}{3}N_c^2 - \frac{10}{3}N_cn_f - 2 C_Fn_f
\end{eqnarray*}
$N_c$ is the number of colours, $n_f$ the number of active flavours and $C_F
= \frac{N_c^2 -1}{2N_c}$.
The leading order solution for the coupling $\a_s(\mu)$ is then found via
\be
   \label{eq:runas}
      \a_s(\mu) = \frac{\a_s(\mu_0)}{1+
        \frac{\beta_0}{2\pi}\a_s(\mu_0)\ln{\left(\frac{\mu}{\mu_0}\right)}} 
\ee
This equation can be re-expressed in terms of a mass scale $\Lambda$, which
is the momentum scale at which the coupling becomes strong as $q^2$ is
increased.  We have:
\be
   \a_s(\mu) = \frac{2\pi}{\beta_0\ln{\left(\frac{\mu}{\Lambda}\right)}}
\ee
From (\ref{eq:runas}) we can see that the renormalisation group
equations allow for the resummation of large logarithms, which could
otherwise be problematic.   The large logarithms can spoil the validity of
the perturbative expansion, 
even if the value of $\a_s$ is still small.  The form of the correction terms
to higher orders can be summarised in the following diagram 
\cite{Buchalla:2002pd}, denoting for example, the ``large log'' at scale $\Lambda = M_W$: $L \equiv \ln{\mu/M_W}$. 
\begin{center}
\begin{tabular}{cccccc}
{} & {} & LL & NLL & {}& {} \\
$\alpha_s^1$ & $\to$ & $\a_sL$ & $\a_s$ & {} & {}\\
$\alpha_s^2$ & $\to$ & $\a_s^2L^2$ & $\a_s^2L$ & $\a_s^2$ & {}\\
$\alpha_s^3$ & $\to$ & $\a_s^3L^3$ & $\a_s^3L^2$ & $\a_s^3L$ & $\a_s^3$ \\
{}& {} & $\Downarrow$ &  $\Downarrow$ & {} & {} \\
{}& {} & $\mathcal{O}(1)$ &  $\mathcal{O}(\a_s)$ & {} & {} \\
\end{tabular}
\end{center}
The rows of this table correspond to the expansion in powers of $\a_s$ from
ordinary perturbation theory.  This is no longer true in the presence of the
large logarithms, but is resolved by resumming the terms $(\a_sL)^n$ to
all orders in $n$.  This re-organisation is obtained by solving the RGE
equation.  
Expanding the leading order terms in (\ref{eq:runas}) we have 
\be
   \a_s(\mu) = \a_s(\mu_0)\sum_{m=0}^\infty\left(\frac{\beta_0}{4\pi}\a_s(\mu_0)\ln{\frac{\mu_0^2}{\mu^2}}\right)^m
\ee
This sums logs of the form $\ln{\mu_0^2/\mu^2}$ which can become large
if $\mu_0 \gg \mu$.
We then speak of ``leading logarithmic order'' (LL) and
``next-to-leading logarithmic order'' (NLL), although we carelessly use these
synonymously with the terms LO and NLO.

\subsection{RGE for Wilson coefficients}

As discussed, the Wilson coefficients can be interpreted as effective
couplings for the operators $Q_i$ of the effective Hamiltonian.  These
operators still have to be renormalised (or more specifically the operator
matrix elements $\langle Q_i\rangle^{(0)}$).  This is done using
renormalisation constants for each of the four external quark fields
$Z_q^{1/2}$ and a matrix $Z_{ij}$ which allows operators with equivalent
quantum numbers to mix under renormalisation: we have $\langle
Q_i\rangle^{(0)} = Z^{-2}_qZ_{ij}\langle Q_j\rangle$.  Since the operators
are always accompanied by their corresponding Wilson coefficients the
operator renormalisation is in general entirely equivalent to the renormalisation of their
``coupling constants'' $C_i$.  We can therefore write the RGE for the Wilson
coefficients as 
\be
  \mu\frac{d}{d\mu}C_i(\mu) = \gamma_{ji}(\mu)C_j(\mu)
\ee
where $\gamma_{ij} = \hat\gamma$ is the \textit{anomalous dimension matrix}
for the operators, defined via 
\be
   \gamma_{ij}(\mu) = Z_{ik}^{-1}\frac{dZ_{kj}}{d\,\ln\mu}
\ee
The anomalous dimension matrix (ADM) is itself given in terms of a
perturbative expansion in $\a_s$
\be
   \gamma_{ij} = \left(\frac{\a_s}{4\pi}\right)\gamma_{ij}^{(0)} +
   \left(\frac{\a_s}{4\pi}\right)^2\gamma_{ij}^{(1)} + \mathcal{O}(\a_s^3)
\ee
We can then give the solution for the Wilson coefficients in terms of an
evolution matrix $U_{ij}(\mu,\mu_0)$
$$   C_i(\mu) = U_{ij}(\mu, \mu_0) C_j(\mu_0) $$
The evolution matrix is given generally by
\begin{eqnarray}
   \hat{U}(\mu, \mu_0) = \int_{g(\mu_0)}^{g(\mu)}dg\frac{\hat\gamma^T(g)}{\beta(g)}
\end{eqnarray}
At leading order, this reduces simply to
\begin{eqnarray}
     \hat{U}^{(0)}(\mu, \mu_0) &=&
     \left(\frac{\a_s(\mu_0)}{\a_s(\mu)}\right)^{\frac{\hat\gamma^{(0)T}}{2\beta_0}}
     \nonumber\\
   &=& V
     \left[\left(\frac{\a_s(\mu_0)}{\a_s(\mu)}\right)^{\frac{\pmb{\gamma}^{(0)}}{2\beta_0}}\right]_DV^{-1 }
\end{eqnarray}
 where $V$ is the matrix that diagonalises $\hat\gamma^{(0)T}$ and
 $\pmb{\gamma}^{(0)}$ is the vector (with elements $\gamma_i^{(0)}$)
 containing all of the eigenvalues of the leading order ADM $\hat\gamma^{(0)}$.  
$$ \gamma_D^{(0)}=V^{-1}\gamma^{(0)T}V $$
It is also possible to calculate the exponentiated matrix directly without
diagonalisation.  At the next-to-leading order, we find that the evolution
matrix becomes a 
little more involved, and includes dependence on the NLO ADM
$\hat\gamma^{(1)}$.  The solution is 
\be
   \hat{U}(\mu, \mu_0) = \left[1 +
     \frac{\a_s(\mu)}{4\pi}\hat{J}\right]\,\hat{U}^{(0)}(\mu, \mu_0)\,\left[1-\frac{\a_s(\mu_0)}{4\pi}\hat{J}\right]
\ee
with
\begin{eqnarray*}
    \hat{J} &=& V \hat{S}\, V^{-1}\\
    S_{ij} &=&
    \delta_{ij}\gamma_i^{(0)}\frac{\beta_1}{2\beta_0^2}-\frac{G_{ij}}{2\beta_0+\gamma_i^{(0)}-\gamma_j^{(0)}}
    \\
    \hat{G} &=& V^{-1} \hat\gamma^{(1)T} V
\end{eqnarray*}

In the weak decays of $B$-mesons the matching is performed at the scale
$M_W$, so that both the top-quark and the $W$-boson are integrated out
(removed as explicit degrees of freedom).  This gives matching conditions and
hence values for the Wilson coefficients at this scale.  We can then use the
procedures just outlined to evolve these down to the appropriate scale, e.g $m_b$.  We write 
\be
      C_i(\mu) = \hat{U}(\mu, M_W) C_i(M_W)
\ee
with an expansion of the coefficients given to the same accuracy as the
evolution, i.e. to NLO
\be
  C_i(M_W) = C^{(0)}_i(M_W) + \frac{\a_s(M_W)}{4\pi}C^{(1)}_i(M_W)
\ee
The scale can be any required, e.g $\mu=m_b$ for $B$-decay
amplitudes or $\mu\sim1$GeV for discussions on the wavefunctions of light
mesons.  Care must be taken to include the effects of the flavour thresholds
at lower energies, which can be done simply, by applying the evolution
equations in two stages with the correct number of active quark flavours in
each stage.  

\subsection{The $\Delta B =1$ effective Hamiltonian}\label{sec:Ham} 

We consider a basis for computing non-leptonic $B$ decays with change in
beauty of $\Delta B = 1$.  The expressions below are for those decays with
unchanging strangeness and charm, $\Delta S = \Delta C = 0$, but can easily
be adapted to decays with $\Delta S =1$ by replacing $d\to s$.  

\begin{figure}[h]
  $$\epsfxsize=0.2\textwidth\epsffile{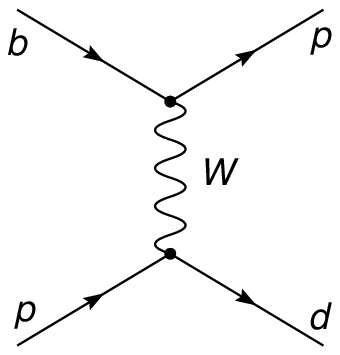} 
  \qquad \epsfxsize=0.2\textwidth\epsffile{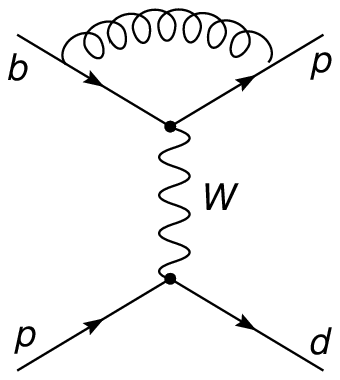} \qquad
  \epsfxsize=0.2\textwidth\epsffile{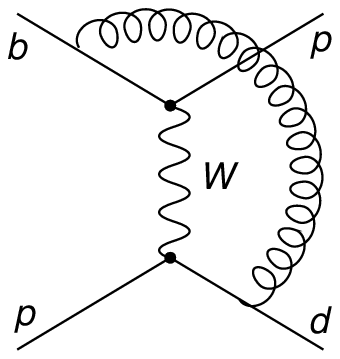} \qquad
  \epsfxsize=0.2\textwidth\epsffile{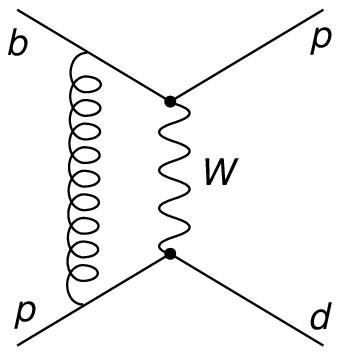}$$ 
   \vskip-12pt
   \caption[ ]{Tree-level exchange and $\mathcal{O}(\a_s)$ corrections; $p = u,c$.} 
   \label{fig:CC}
\end{figure}

We begin with the tree-level process without QCD
corrections, which  is described by one dimension 6 operator.  
When we include QCD corrections, another current-current operator is
generated: these are labelled as $Q_1$ and $Q_2$, although sometimes in the
literature their definitions are interchanged.  The tree-level diagram and
the $\mathcal{O}(\a_s)$ corrections to it are shown in Figure \ref{fig:CC}.

QCD corrections also produce four new gluonic \textit{penguin operators}
$Q_3$ to $Q_6$.  If we include terms from the electroweak sector up to
$\mathcal{O}(\alpha)$ we obtain an additional set of \textit{electroweak
  penguin operators} $Q_7$ to
$Q_{10}$.  These are considered as a next-to-leading order effect due to
their proportionality to the electroweak gauge coupling $\a$.  There are, in
principle, QED corrections to the matrix elements of 
the QCD operators $Q_1\dots Q_6$, but these are suppressed and usually
neglected.  We also have additional terms which contribute in some
$\Delta B = 1$ processes, namely 
the magnetic penguin operator $Q_{7\gamma}$ (which is important for the
radiative decays) and the chromomagnetic penguin $Q_{8g}$.  Examples of all
these diagrams are shown below in Figures
\ref{fig:penguins} and \ref{fig:penguins2}.   
\begin{figure}[h]
  $$\epsfxsize=0.3\textwidth\epsffile{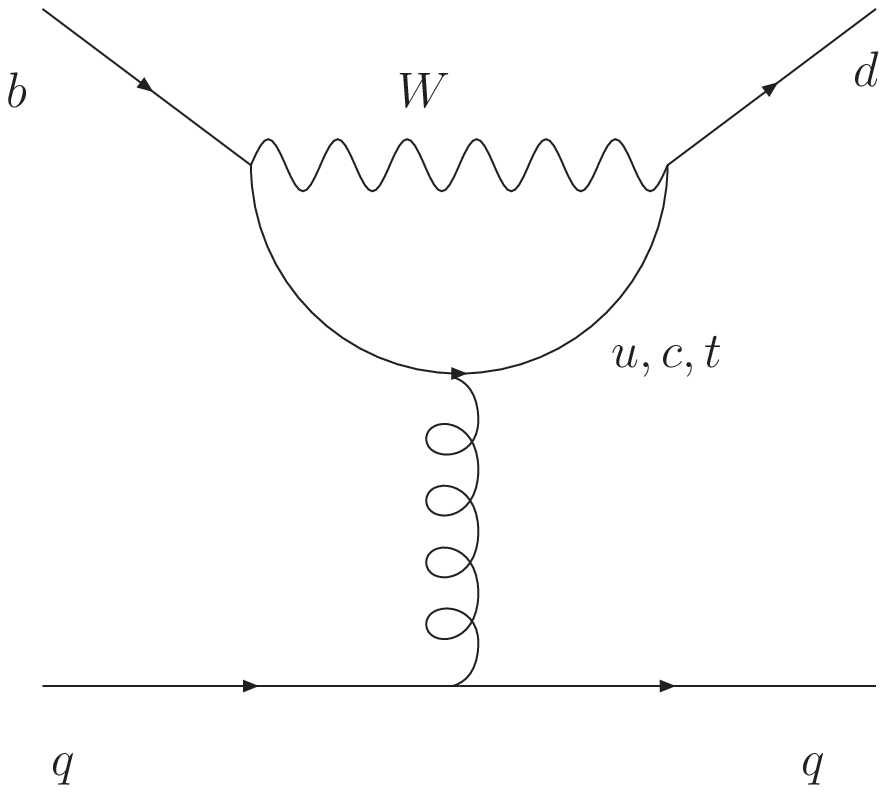} 
  \qquad \epsfxsize=0.3\textwidth\epsffile{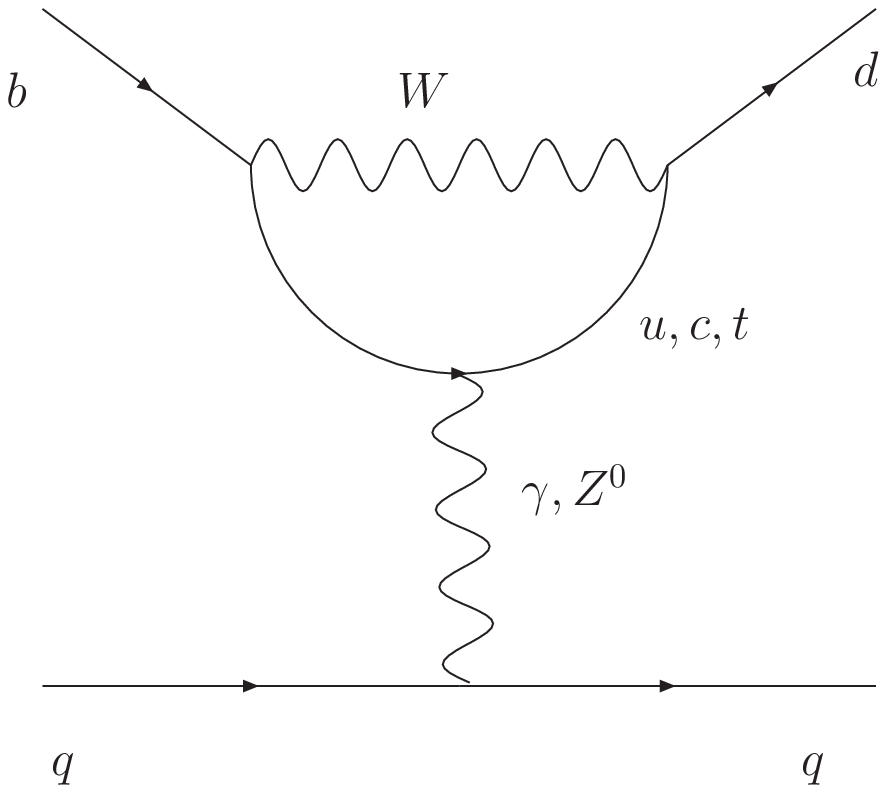} \qquad
  \epsfxsize=0.3\textwidth\epsffile{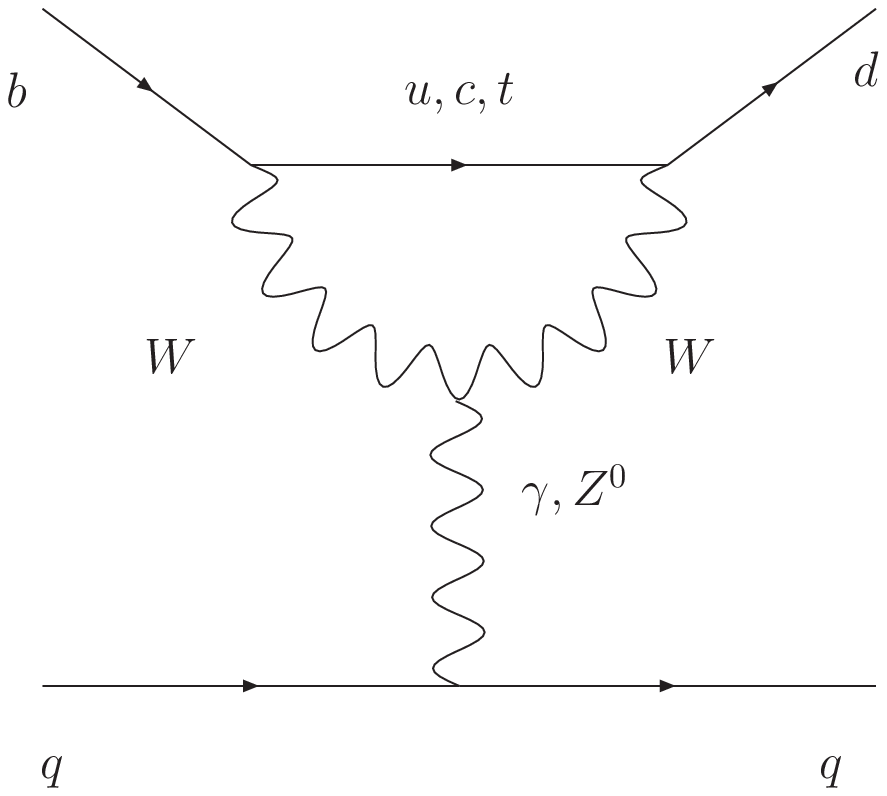}$$ 
   \vskip-12pt
   \caption[ ]{Gluonic and electroweak penguin diagrams.} 
   \label{fig:penguins}
\end{figure}
\begin{figure}[h]
  $$\epsfxsize=0.3\textwidth\epsffile{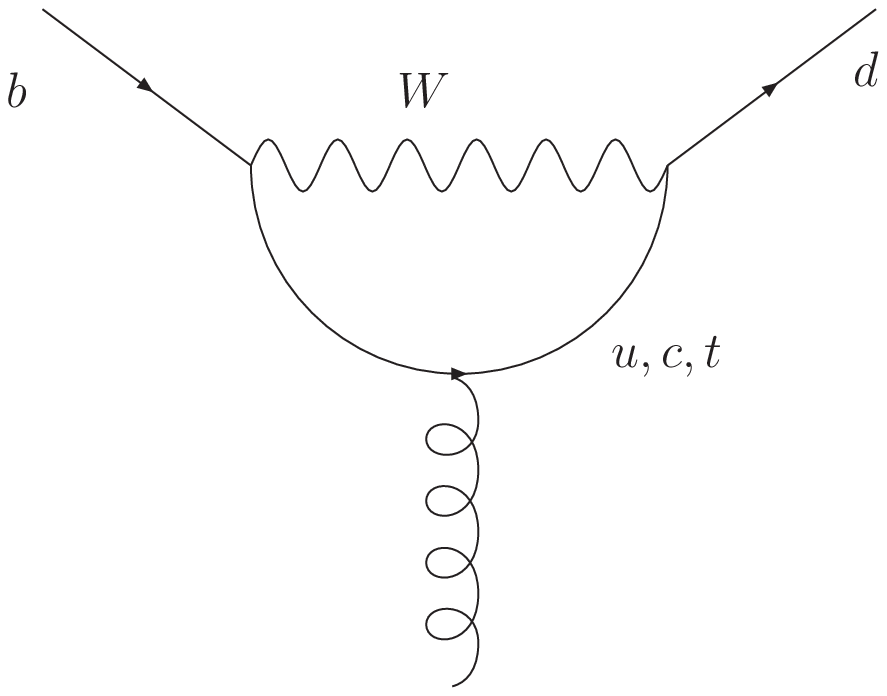} 
  \qquad \epsfxsize=0.3\textwidth\epsffile{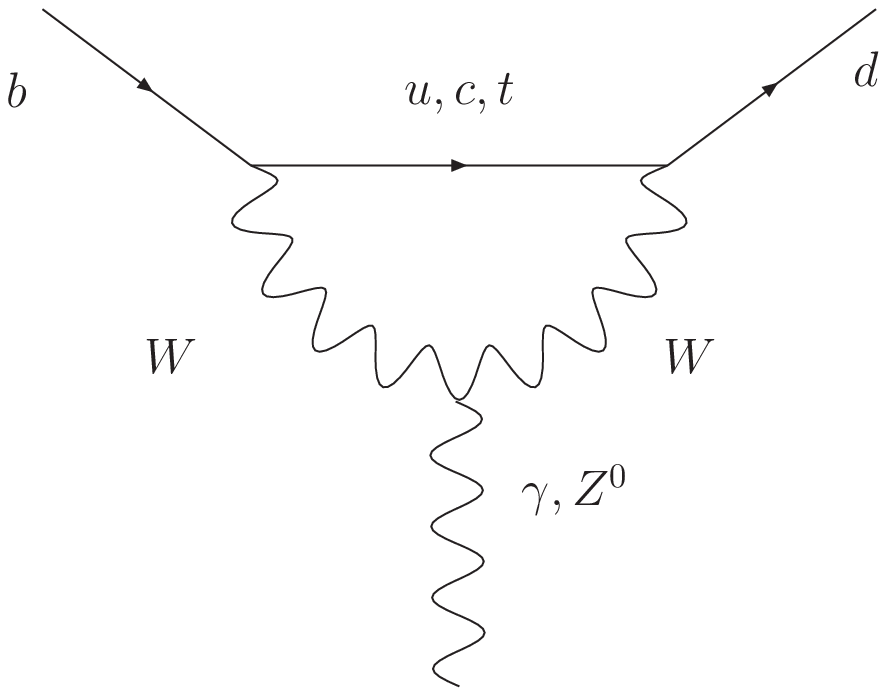}$$
   \vskip-12pt
   \caption[ ]{Magnetic photon and chromomagnetic penguin diagrams.} 
   \label{fig:penguins2}
\end{figure}

This leads us to a $\Delta B=1$ effective Hamiltonian of
\be
   \label{Heff}
      \mathcal{H}_{\mathrm{eff}} = \frac{G_F}{\sqrt{2}}\sum_{p=u,c}\lambda_p
      \left[C_1Q_1^p + C_2Q_2^p + \sum_{i=3,\dots,10} C_iQ_i +
      C_{7\gamma}Q_{7\gamma} + C_{8g}Q_{8g}\right] + \textrm{h.c.}
\ee
with the full operator basis given as 
\begin{eqnarray*}
  Q_1^p &=& \left(\bar{d}_ip_j\right)_{V-A}\left(\bar{p}_jb_i\right)_{V-A}\\
  Q_2^p &=& \left(\bar{d}p\right)_{V-A}\left(\bar{p}b\right)_{V-A}\\
  Q_3 &=&
  \left(\bar{d}b\right)_{V-A}\sum_q\left(\bar{q}q\right)_{V-A}\\
  Q_4 &=&
  \left(\bar{d}_ib_j\right)_{V-A}\sum_q\left(\bar{q}_jq_i\right)_{V-A}\\
  Q_5 &=&
  \left(\bar{d}b\right)_{V-A}\sum_q\left(\bar{q}q\right)_{V+A}\\
 Q_6 &=&  \left(\bar{d}_ib_j\right)_{V-A}\sum_q\left(\bar{q}_jq_i\right)_{V+A}\\
   Q_7 &=&
  \left(\bar{s}b\right)_{V-A}\sum_q\tfrac{3}{2}e_q\left(\bar{q}q\right)_{V+A}\\
  Q_8 &=&
  \left(\bar{s}_ib_j\right)_{V-A}\sum_q\tfrac{3}{2}e_q\left(\bar{q}_jq_i\right)_{V+A}\\
  Q_9 &=&
  \left(\bar{s}b\right)_{V-A}\sum_q\tfrac{3}{2}e_q\left(\bar{q}q\right)_{V-A}\\
  Q_{10} &=&
  \left(\bar{s}_ib_j\right)_{V-A}\sum_q\tfrac{3}{2}e_q\left(\bar{q}_jq_i\right)_{V-A}\\
  Q_{7\gamma} &=&
 \frac{e}{8\pi^2}m_b\,\bar{s}\,\sigma^{\mu\nu}(1+\gamma_5)F_{\mu\nu}\,b\\
  Q_{8g} &=&
  \frac{g}{8\pi^2}m_b\,\bar{s}_i\,\sigma^{\mu\nu}(1+\gamma_5)T^a_{ij}\,b_j\,G^a_{\mu\nu}
\end{eqnarray*}
We use the usual notation that $\left(\bar{q}_1q_2\right)_{V\pm A} =
\bar{q}_1\gamma_\mu(1\pm\gamma_5)q_2$; $i,j$ are colour indicies,
$F_{\mu\nu}$ and $G_{\mu\nu}$ are the photonic and gluonic field strength
tensors respectively.  There is also implied summation over all flavours
$q$.   The matching conditions for the Wilson coefficients and the anomalous
dimension matrices required for scale evolution are given in detail in
Appendix \ref{chp:AppD}.

\section{QCD sum rules}\label{sec:QCDSR}

In later chapters we make use of a number of results from the technique to study hadronic
structure known as \textit{QCD sum rules} \cite{Shifman:1978bx}.  Without any
detailed technical 
discussion, we take a moment here to briefly describe the purpose and form of
the sum rule approach.  They were originally used to calculate simple
characteristics of hadrons, such as masses, but are also applicable to more
complicated parameters such as form factors or hadronic wave functions.  The
sum rule approach is used to calculate non-perturbative effects in QCD.

The hadrons are represented by interpolating quark currents, which are formed
into correlation functions.  These are treated in an operator product
expansion to separate the short and long distance contributions from the
quark-gluon interactions.  The short distance parts are again calculable in
normal perturbative QCD.  The long-distance contributions are
parameterised by universal vacuum condensates or by \textit{light-cone
  distribution amplitudes}.  (See later chapters).  The results of this
calculation are then matched via 
rigorous \textit{dispersion relations} to a sum over the hadronic states.  This gives
us a sum rule which allows us to calculate the observable characteristics of
hadronic ground states.  Together with experimental data this can be used to
determine quark masses and universal non-perturbative parameters.  

There are some limitations in the accuracy of the sum rules which come from
the approximation of the correlation functions.  Additionally there is
uncertainty from the dispersion
integrals, which have complicated and often unknown structure.  The method is
however very successful and all uncertainties can be traced and estimated
well.  

The classical sum rule approach is based on two-point correlators, but this
is not the only approach.  Combining sum rule techniques with a light-cone
expansion gives the very successful technique of \textit{light-cone sum
  rules} \cite{Braun:1988qv,Chernyak:1990ag}.   This procedure
expands the products of currents near the light-cone and involves a partial
resummation of local operators.

\section{Lattice QCD}

Ideally we would like to be able to estimate all of the fundamental
parameters of QCD from first principles.  To do this, we need complete
quantitative control over both the perturbative and non-perturbative
aspects of QCD.  The numerical simulations of \textit{Lattice QCD} in
principle are able to do this -- answering almost any question about QCD and
confinement, from Yang-Mills theories to the running coupling constant to the
topology of the QCD vacuum.  In our work, the relevant parameters that can be
determined via lattice techniques include decay constants
\cite{AliKhan:2001jg,Aoki:2003xb}, form factors 
\cite{Okamoto:2004xg, Shigemitsu:2004ft} and hadronic
matrix elements \cite{Aoki:2003xb}.  As with the results we frequently rely upon from 
QCD sum rules, it is important to understand the importance and limitations
of lattice results that we use; a full review can be found in
\cite{Gupta:1997nd}.

\begin{figure}[h]
  $$\epsfxsize=0.3\textwidth\epsffile{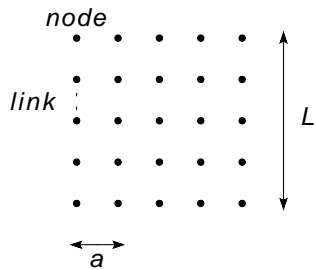} $$ 
   \vskip-12pt
   \caption[ ]{Nodes are separated by a lattice spacing of $a$.  Quark fields
   live on the lattice sites (nodes) and gluons on the links connecting these sites.}  
   \label{fig:lattice}
\end{figure}

QCD can be expressed in Feynman path integrals, which need to be calculated
in the continuous space-time in which we exist.  This is very
difficult but we can make an approximation by discretising the continuous
four-dimensional space-time onto a 4-D lattice as represented in Figure
\ref{fig:lattice}.  The Lagrangian consisting of fields and derivatives of
fields is also discretised by replacing the continuum fields with fields at the lattice
sites, and replacing the derivatives with finite differences of these fields.
The path integrals over different field configurations become
multi-dimensional integrals over the values of the fields at the lattice
sites and on the links.  These can be evaluated numerically, for example using 
Monte Carlo simulation; this is the basic tenet behind lattice QCD.  The problem 
of solving a non-perturbative relativistic quantum field theory in transferred
into a ``simple'' matter of numerical integration.  

Large amounts of computing power are required for simulations, and the
computation time depends on a number of factors -- the overall volume of the
spacetime $L^4$ where $L$ is often taken as $L\sim1-2$fm, and the finiteness
of the grid, $a\sim0.05-0.1$fm.  Ideally the grid spacing should be
small enough that results are within a few percent of the continuum
results.  Errors associated with the discretisation are an important source
of systematic errors in a lattice calculation that need to be controlled.  

The purely gluonic part of QCD can be simply discretised onto the lattice but
complications arise for the quarks due to their fermionic nature.  There are however
a number of different 
methods of including quarks on the lattice such as \textit{Wilson quarks} or
\textit{staggered quarks}, each with its own improvements and disadvantages.
Implementing an algorithm for quarks gives an additional problem as fermionic
variables in  path integrals require the calculation
of a fermionic determinant -- requiring hundreds of matrix inversions for
every vacuum configuration considered.  This calculation amounts to
determining the non-local interaction between the gluons, i.e the computation
of dynamical quark loops.  There is no technical obstruction to performing
this calculation and a number of methods exist for the discretisation of
fermions, however the computation time and difficulty of the
calculation increases by several orders of magnitude!  

This has led to many simulations being performed in the \textit{quenched
  approximation}, where vacuum configurations that include only gluons are
  considered.  This neglects the additional term which gives rise to the sea
  quarks (and hence the ability to produce $q\bar q$ pairs from the vacuum).  
The quenched approximation performs satisfactorily for processes dominated by
valence quarks, but for the converse dynamical quarks are required to
produce correct answers.  Unquenching can have other side effects such as
  allowing processes which require decays into $q\bar q$ pairs e.g. $g\to
  q\bar q \to g$ which will enable the strong coupling to run at the correct
  rate.  This also means that decays such as $\rho\to\pi\pi$ are now visible
  on the lattice, but it will be difficult to determine
  $m_\rho$.   The cost of including the dynamical quarks
increases as we move to lighter and lighter quarks, and it is the lightest
  three $u,\,d,\,s$ that have the most significant effect on most
  quantities. 
   
Heavy quarks, such as the $b$ quark could in principle be treated the same as the
lighter quarks, however in order to gain sufficient accuracy an extremely
fine lattice would be needed.  Instead, exploiting the non-relativistic nature
of the $b$ quark bound states proves much more efficient.  This can be done
using several techniques including \textit{static quarks} -- using the heavy-quark
limit of $m_b=\infty$ and \textit{NRQCD} -- a non-relativistic formulation of
QCD.

%%% Local Variables:
%%% mode: latex
%%% TeX-master: t
%%% End:

%% file: chapter2.tex
\chapter{QCD factorisation}\label{chp:QCDF}
\begin{center}
  \begin{quote}
    \it
   Still round the corner there may wait, \\
   A new road or a secret gate.
  \end{quote}
\end{center}
\vspace{-4mm}
\hfill{\small ``The Lord of the Rings'', J.R.R. Tolkien}
\vspace{5mm}

The concept of factorisation in its various forms is key to many of the
applications of perturbative QCD, including deep inelastic scattering or jet
production in hadron colliders.  Perturbative QCD only describes quarks and
gluons, so in observed, hadronic physics there will always be some
relevant non-perturbative dynamics that must be isolated and dealt
with in a systematic fashion.  In the following sections, we describe the
ideas of ``naive factorisation'' for the hadronic decays of heavy mesons, and
then introduce the ``improved QCD factorisation'' formalism -- a systematic method
for treating exclusive two-body $B$ decays.  There have been a
number of other approaches to try to describe non-leptonic $B$ decays,
although none have been nearly as successful as QCD factorisation.  These
methods include the perturbative QCD (pQCD) methods (which treats the form
factors as perturbatively calculable quantities) or the soft-collinear
effective theory (SCET), which is based upon factorisation of two energy
scales: $m_b^2\gg\Lambda_{\mathrm{QCD}}m_b\gg\Lambda_{\mathrm{QCD}}$.  QCD
factorisation is currently the accepted and most widely used methodology for
treating $B$ decays, and we work exclusively in this framework in this
thesis.  Finally, we conclude this chapter with a discussion of the
limitations of the QCD factorisation method and ask how well predictions
match up with the experimental data.

\section{Naive factorisation}  

A complete understanding of the theoretical framework for non-leptonic $B$
decays is a continuing challenge which is yet to be resolved.  For two-body
decays of $B\to M_1M_2$, the major difficulty involves the evaluation of the
hadronic matrix elements $\langle M_1M_2|Q_i|B\rangle$. %the computation
                                                        %of which requires
                                                        %some assumptions.  

We introduced the concept of factorisation in relation to the  OPE where
we separate the long and short-distance contributions to decay
amplitudes.  Similarly, the hadronic matrix elements for $B$ decays can also be factorised
by disentangling the long and short-distance contributions.  This seems
intuitive for leptonic and semi-leptonic decays, where we can factorise the
amplitude into a leptonic current and the matrix element of a quark current.  Gluons
do not interact with the leptonic current and so the factorisation is exact.
For non-leptonic decays, the full matrix element is separated into a product
of matrix elements of two quark currents; so we must also have
``non-factorisable'' contributions from gluons which can connect these two
currents.  

The first general method proposed is that of \textit{naive factorisation} 
\cite{Ellis:1975hr, Dugan:1990de}.  Here the full matrix element is assumed
to factorise into a product of matrix elements of simple, colour singlet,
bilinear currents.  The first is the transition matrix element between the $B$-meson and
one of the final state 
mesons, and the second is the matrix element of the other final state meson being 
``created'' from the vacuum.  For example
\begin{eqnarray}
\label{eq:naive}
  \langle\pi^+\pi^-|(\bar{u}b)_{V-A}(\bar{d}u)_{V-A}|\bar{B}^0\rangle\longrightarrow\langle\pi^-|(\bar{d}u)_{V-A}|0\rangle\langle\pi^+|(\bar{u}b)_{V-A}|\bar{B}^0\rangle
\end{eqnarray}

These matrix elements can then be expressed in terms of a \textit{form
  factor} $F^{B\to M_1}$ and 
\textit{decay constant} $f_{M_2}$.  The Hamiltonian is re-expressed in terms
of effective coefficients $C_i^{\textrm{eff}}$.  These are
constructed from the Wilson coefficients and all of the scheme and scale
dependent contributions from the hadronic matrix elements.  This ensures the
effective coefficients are free from these dependencies.  The complete matrix
element is then made up of the factorised matrix elements and the
coefficients $a_i$, which are combinations of the effective Wilson
coefficients 
\begin{equation}
  a_i = \left(C_i^\textrm{eff} + \frac{C_{i\pm1}^\textrm{eff}}{N_c}\right) + P_i 
\end{equation}
where $P_i$ involves the QCD and electroweak penguins (but not taking into
account final state flavour) and are zero for $i = 3,\,5$.

A qualitative justification for this approach comes from the concept of \textit{colour transparency} \cite{Bjorken:1988kk,Fakirov:1977ta} which
implies the decoupling of soft (low-energy) gluons from colour-singlet pairs
of quarks, e.g. the final state pion in example (\ref{eq:naive}).  This
occurs as the heavy-quark decays are highly energetic, so the 
hadronisation of the $q\bar{q}$ pair occurs far away from the remaining quarks.

The main source of uncertainty in this approach is the neglect of the
``non-factorisable'' contributions, exacerbated by the assumption that all
final state interactions are absent.  Corrections arise from the exchange of
gluons between the two quark currents, which allows a 
dynamical mechanism (namely re-scattering in the final state) to create a
strong phase between different amplitudes.  The second major issue is that
the cancellation of renormalisation scale dependence in the amplitude is
destroyed, as the scale  independence of the form factor and
decay constants in the factorised matrix element 
is in conflict with the scale dependence of the original matrix
element. Unphysical dependencies are also caused by the Wilson
coefficients at NLL level which develop scheme dependence in addition to
the scale dependence, and for which there is no cancellation from the (scheme
independent) factorised matrix elements.

\section{QCD factorisation}

QCD factorisation was introduced by Beneke, Buchalla, Neubert and Sachrajda
(BBNS) \cite{Beneke:1999br,Beneke:2000ry,Beneke:2001ev}, and is based upon the important simplifications that occur
in the limit where the $b$ quark mass is large as compared with the strong
interaction scale, $m_b \gg \Lambda_{QCD}$.  The factorisation that occurs in
this approach is the separation of the long-distance dynamics (the matrix
elements) and  the short-distance interactions which depend only on the large
scale $m_b$. The short-distance contributions are calculated perturbatively
to order $\alpha_s$, and the long-distance information is encoded in various
process independent non-perturbative parameters, or is obtained directly from
experiment.  However, it is structurally much simpler than the original matrix elements. 

In the limit  $m_b \gg \Lambda_{QCD}$, the underlying
physics involves the decoupling of fast-moving light mesons, produced from point-like
interactions from the weak effective Hamiltonian, from the soft QCD
interactions.  This is a result of the concept of colour transparency as
discussed above, the systematic implementation of which is provided by QCD factorisation.  It
also provides a soft factorisation which allows the matrix elements to
be given at leading order in the $\Lambda_{QCD}/m_b$ expansion.  The full matrix elements can then be represented in the form
\begin{eqnarray}
  \label{QCDF1}
     \langle M_1M_2|Q_i|B\rangle = \langle M_1|j_1|B\rangle\langle
     M_2|j_2|0\rangle\left[1 + \sum_nr_n\alpha_s^n +
     \mathcal{O}\left(\Lambda_{QCD}/m_b\right)\right]
\end{eqnarray}
where $r_n$ denote the radiative corrections in $\alpha_s$, and $j_i$ are
bilinear quark currents.  If the order $\alpha_s$ corrections are neglected,
we see that at leading order in $\Lambda_{QCD}/m_b$ we recover the naive
factorisation results.

In the QCD factorisation approach, the non-factorisable power-suppressed
corrections are in general neglected, with two non-trivial exceptions:  the
hard-scattering spectator interactions and annihilation contributions which are
\textit{chirally enhanced} and cannot be ignored.  In the context of this thesis, it is important to note that these cannot be calculated
within the actual framework of QCD factorisation; they are included in the BBNS
approach via model-dependent assumptions.  This will be discussed in detail
in the following sections.

\subsection{Structure of the QCD factorisation formula}

The QCD factorisation framework is represented by a ``master formula'' for
the matrix element  $\langle M_1M_2|Q_i|B\rangle$, where the final state mesons can
be ``heavy-light''(e.g. $B\to D\pi$) or ``light-light'' (such as $B\to\pi\pi$).
For exclusive non-leptonic decays to two light mesons, we have the following
expression for $\langle M_1M_2|Q_i|B\rangle$, to leading order in the
$\Lambda_{QCD}/m_b$ expansion:
\begin{eqnarray}
  \label{eq:QCDF}
      \langle M_1M_2|Q_i|B\rangle &=& F_j^{B\to
      M_1}(m_2^2)\int_0^1du\,T_{ij}^I(u)\,\Phi_{M_2}(u) + [M_1
      \leftrightarrow M_2] \nonumber\\
         && + \int_0^1 d\xi\, du\,dv\,
      T_i^{II}(\xi,u,v)\,\Phi_B(\xi)\,\Phi_{M_1}(v)\,\Phi_{M_2}(u) 
\end{eqnarray}
$\Phi_M$ denote the \textit{light cone distribution amplitudes} (LCDA)
for the valence quark states.  Both the LCDA $\Phi_M$ and the $B\to M$ form factor, are
much simpler than the original non-leptonic matrix element, and can be
calculated by some non-perturbative technique such as QCD sum rules, on the
lattice, or taken directly from experiment.

The perturbative information is contained in the hard-scattering kernels
$T_{ij}^I(u),\, T_{ik}^I(v)$ and $T_i^{II}(\xi,u,v)$, which are calculable
functions dependent on the light-cone momentum fractions of the constituent
quarks in the light final state mesons ($u,\,v$) and the $B$ meson ($\xi$).
All of the hard-scattering kernels and the LCDA have a factorisation scale
dependence. The expression is split into two categories of
contributions -- ``type I'' and ``type-II''.  The type-I contributions
consist of the leading terms which give the tree level
contributions, and hard vertex and penguin corrections at
$\mathcal{O}(\alpha_s)$.  The type II contributions originate from hard interactions
between the spectator quark and the emitted meson, which enter at
$\mathcal{O}(\alpha_s)$.  If the spectator quark in the interaction can only
form into one of the final state mesons (such as in $B^0\to\pi^+K^-$), then
the second form factor term is absent.  

We can see that from the tree level terms in $T^I$
we reproduce the naive factorisation results, and the convolution integral in
(\ref{eq:QCDF}) will reduce to a meson decay constant.  We also get a large
simplification in the case where one of the mesons is heavy ($B\to H_1M_2$),
as the spectator interactions 
will be power-suppressed in the heavy quark limit and the type II kernel
$T^{II}$ is absent. 

\subsection{Non-perturbative parameters}\label{sec:NPP}

We will now elaborate further on the non-perturbative input into the
factorisation master formula, and introduce the concept of light-cone
distribution amplitudes which will be discussed in detail in Chapter
\ref{chp:LCDA}.

\subsubsection{Light-cone distribution amplitudes for light mesons}

The distribution amplitude $\Phi_M(u, q^2)$ is in essence a probability amplitude for a
meson to be found in a particular state, i.e for finding a valence quark with
a \textit{light-cone} longitudinal momentum fraction $u$ in the meson at a momentum $q^2$,
independent of the process.
The light-cone frame for some vector $p_\mu = (p_0,\, p_1,\, p_2,\, p_3)$,
is a choice of co-ordinates which naturally distinguishes between the transverse and
longitudinal degrees of freedom.  Two of the spatial dimensions (transverse) remain
unchanged and the other (longitudinal) spatial dimension is combined with the
temporal co-ordinate via
\be
   p_\pm = \frac{p_0 \pm p_3}{\sqrt{2}}
\ee
so that $p_\mu = (p_+,\, p_-,\, \vec{p}_\perp)$, and $p^2=0$ for a light-like
vector. The distribution amplitude is defined in terms of the expectation
value of non-local operators, near the light-cone.  For example, considering
the pion distribution amplitude $\phi_\pi$, we express the LCDA  in terms of
a matrix element of a gauge-invariant non-local operator defined between the
vacuum and the pion \cite{Chernyak:1983ej}  
\be
   \label{eq:DAop}
   \langle0|\bar{d}(z)\gamma^\mu\gamma_5[z,0]u(0)|\pi(P)\rangle 
\ee
$[x,y]$ is a path-ordered gauge factor along the straight line
connecting the points $x, y$ in order to preserve gauge invariance and is known as
a \textit{Wilson line}
\be
 \label{eq:wilsonline}
   [x,y] = P exp\left[ig\int_0^1dt\,(x-y)_\mu A^\mu(tx+(1-t)y)\right]
\ee
These matrix elements give a set of integral equations of two and three
particle light-cone distribution amplitudes.  These distribution amplitudes
can be classified in terms of \textit{twist}.  The twist of an operator is
defined as its ``dimension minus spin'', i.e.
\be
   t = d-s
\ee
This is related to power counting of $1/Q$, where $Q$ denotes momentum
transfer, and controls the relative size of contributions from the operator product expansion.  In general, an operator
of dimension $d$ in the OPE has a coefficient function with dimension
(mass)${}^{6-d}$ \cite{Peskin}, corresponding to a suppression factor in the Fourier
transform of the OPE of 
\be
   \left(\frac{1}{Q}\right)^{d-2}
\ee
If the operator has spin $s$, the operator matrix element gains $s$
contributions from the momentum $P^\mu$, so the total contribution is of the
order of
\be
   \left(\frac{2P\cdot q}{Q^2}\right)^s\left(\frac{1}{Q}\right)^{d-s-2}
\ee
with $q^2=-Q^2$.  The leading twist is then $t=2$, allowing us to identify a distribution
amplitude of twist-2 and also of higher twists.  The higher-twist
distribution amplitudes contain contributions from lower twists as well.  For
example the twist-3 DA has contributions from both twist-3 and
twist-2 operators.  In the QCD factorisation formalism, and in Chapter
\ref{chp:LCDA}, where we study models of LCDAs, we consider the simplest case of
the leading-twist distribution amplitudes.  From the matrix element in
(\ref{eq:DAop}) the twist-2 DA is defined as
\be
   \langle0|\bar{d}(z)\gamma^\mu\gamma_5[z,0]u(0)\pi(P)\rangle|_{z^2 = 0} = if_\pi
   P^\mu\int_0^1du\,e^{iu(z\cdot P)}\phi_\pi(u, q^2)
\ee
A detailed discussion of light-cone distribution amplitudes is continued in Chapter \ref{chp:LCDA}.

\subsubsection{Light-cone distribution amplitude for B-meson}

Unlike the distribution amplitudes for the light mesons, there is very
limited knowledge of the parameters determining the $B$-meson wavefunction.  In
the context of QCD factorisation, it is the inverse moment of the
distribution amplitude that is most relevant (see below).  For scales much
larger than $m_b$, $\phi_B$ should tend to a symmetric form, as 
with a light meson distribution amplitude.  However, at or below $m_b$, the
distribution is expected to be very asymmetric in momentum fraction $\xi$, with
$\xi \sim \mathcal{O}(\Lambda_{QCD}/{m_b})$. 

The $B$-meson light-cone distribution amplitude only appears in the type-II
(hard spectator interaction) term in the main QCD factorisation formula.  The
amplitudes for these interactions, at order $\alpha_s$, depend only on the
product of the momenta of the light meson which absorbs the spectator quark,
$p^\prime$, and that of the spectator quark itself, $l$.  Converting to
light-cone co-ordinates we can arrange the momenta so that the hard-spectator
amplitude is dependent only on $l_+$. The wavefunction can then be
integrated over the other momenta $l_\perp$ and $l_-$.  Following the method
of \cite{Beneke:2001ev}, the hard-spectator integrals are re-expressed in terms of
the longitudinal momentum fraction $\xi$, and the $B$-meson  LCDA can be
decomposed (at leading order in $1/m_b$) into two scalar wavefunctions.
These wavefunctions describe the distribution of the longitudinal momentum
fraction $\xi = l_+/p_+$ of the spectator quark inside the meson and are
defined via:
\be
   \langle0|\bar{q}_\alpha(0)b_\beta(z)|\bar{B}(p)\rangle =
   \frac{if_B}{4}[(p \hspace{-6pt}/ +
   m_b)\gamma_5]_{\beta\gamma}\int_0^1d\xi\,e^{-i\xi p_+z_-}\,[\Phi_{B1}(\xi) + n\hspace{-7pt}/_-\, \Phi_{B2}(\xi)]_{\gamma\alpha}
\ee
where $n$ is an arbitrary light-like vector, which is chosen in the direction
of one of the final state momenta, $n_- = (1,0,0,-1)$.  The wavefunctions
are normalised
\be
   \int_0^1 d\xi\,\Phi_{B1}(\xi) = 1 \qquad \int_0^1 d\xi \Phi_{B2}(\xi) = 0 
\ee

In the calculations involving the $B$-meson distribution amplitude, we need only the
first inverse moment of the wavefunction $\Phi_{B1}(\xi)$, parameterised as
\be
   \int_0^1d\xi\,\frac{\Phi_{B1}(\xi)}{\xi} = \frac{m_B}{\lambda_B}
\ee
 
There is little information as to the actual value of the parameter
$\lambda_B$.  There is a known upper bound which implies
$3\lambda_B\leq4\bar\Lambda$ \cite{Korchemsky:1999qb}, where $\bar\Lambda = m_B - m_b$,
corresponding to $\lambda_B\lessapprox600$MeV.  Numerical estimates have been
suggested from a number of different models 
\cite{Grozin:1996pq,Korchemsky:1999qb,Keum:2000ph} which we can combine to 
give an estimate of around $\lambda_B = 350\pm150$MeV.

\subsubsection{Form factor}

Hadronic form factors describe the inner structure of the hadron, and are 
functions of scalar variables arising from the decomposition of some matrix
element.  In the coupling of a particle to a photon, we have an
\textit{electromagnetic form factor}, which is a momentum dependent function
reflecting the charge and magnetic moment distribution, and
hence the internal structure.  In this case we are interested in the
\textit{transition form factor}, which describes the overlap of the $B$ meson
and the decay product (i.e. some pseudoscalar meson) during the actual
decay.  Decays of $B\to\pi$ are fully described by three form factors
$F_+,\,F_0$ and $F_T$, which are defined by \cite{Ball:2004ye} 
\begin{align}
  &\langle\pi(p)|\bar{u}\gamma_\mu b|\bar{B}(p_B)\rangle = F_+(q^2)\left\{(p_B
  + p)_\mu - \frac{m_B^2 - m_\pi^2}{q^2}q_\mu\right\} + \frac{m_B^2 -
  m_\pi^2}{q^2}F_0(q^2)q_\mu \\
  &\langle\pi(p)|\bar{d}\sigma_{\mu\nu}q^\nu(1+\gamma_5)
  b|\bar{B}(p_B)\rangle = i \left\{(p_B+p)_\mu q^2 - q_\mu(m_B^2-m_\pi^2)\right\}\frac{F_T(q^2)}{m_B+m_\pi}
\end{align}
where $q=p_B-p,\, q^2 = m_B^2 - 2m_BE_\pi$.  In QCD factorisation the vector
current arises most often.  We should also note that at the scale $q^2 = 0$
the two form factors coincide, $F_+(0) = F_0(0)$.  The asymptotic scaling behaviour of
the form factors at $q^2=0$ is given as \cite{Ruckl:1997hr}
\be
   F_+(0) = F_0(0) \sim \frac{\Lambda_{QCD}}{m_b}^{3/2}
\ee

The form factors can be calculated using various non-perturbative
techniques, such as QCD sum rules (as we discussed in Section
\ref{sec:QCDSR}), and on the lattice \cite{Okamoto:2004xg,Shigemitsu:2004ft}.  The method of light-cone sum rules \cite{Ball:2004ye} allows the calculation of the
form factor within a controlled approximation, relying on the factorisation
of an unphysical correlation function whose imaginary part is related to the
form factor in question.  This correlation function is that of a weak
current and a current with the same quantum numbers as the $B$-meson,
evaluated between the vacuum and the $\pi$; it is related to the form factor
via 
\begin{eqnarray}
  \Pi_\mu(q, p_B) &=& i \int
  d^4x\,e^{iq.x}\langle\pi(p)|TV_\mu(x)j_B^\dagger(0)|0\rangle \\
 &=& \Pi_+(q^2, p_B^2)(p+p_B)_\mu + \Pi(q^2,p_B^2)q_\mu
\end{eqnarray}
where $j_B = im_b\bar{d}\gamma_5b$.  When $p_B^2\ll m_b^2$, these correlation
functions can be expanded on the light-cone
\begin{eqnarray}
  \Pi^{LC}_\pm(q^2, p_B^2) = \sum_n \int_0^1
  du\,T_{\pm}^{(n)}(u,q^2,p_B^2,\mu)\phi^{(n)}(u,\mu)
\end{eqnarray}
This sum runs over contributions from the pion distribution amplitude in
increasing twist, where $n=2$ is the leading-twist contribution.  The
functions $T_\pm$ are hard-scattering amplitudes, determined by a
perturbative series in $\alpha_s$, known to
$\mathcal{O}(\alpha_s)$ at twist-2 and twist-3.  The form factor can then be
extracted via a light-cone sum rule which depends on the 
\textit{spectral density} $\rho_+^{LC}$ of the correlation $\Pi_+^{LC}$
\begin{eqnarray}
  e^{-m_B^2/M^2}m_B^2f_BF_+^{B\to\pi}(q^2) = \int^{s_0}_{m_b^2}ds\,e^{-s/M^2}\,\rho_+^{LC}(s,q^2)
\end{eqnarray}
where $M^2$ and $s_0$ are specific parameters from the sum rule, related to a
Borel transformation \cite{Ball:2004ye}.

\subsection{Contributions to hard-scattering kernels}

It is useful to discuss qualitatively the diagrams that contribute to the
hard-scattering kernels $T_{ij}^I(u)$ and $T_i^{II}(\xi,u,v)$ at the leading
order and with exchange of one gluon.  We use the decay of
$B^0_d\to\pi^+\pi^-$ as an illustrative example.  The factorisation formula
of equation (\ref{eq:QCDF}) has been proved to one-loop for decays into two
light mesons and to two loops for decays into heavy-light final
states. There is also proof to all orders for $B\to D\pi$
\cite{Bauer:2001cu}.  

\begin{figure}[h]
  $$\epsfxsize=0.2\textwidth\epsffile{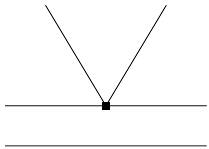}$$
  \vskip-12pt
  \caption[ ]{Leading order contribution to the hard-scattering kernel $T^I$.} 
  \label{fig:factLO}
\end{figure}

\begin{figure}[h]
  $$\epsfxsize=0.2\textwidth\epsffile{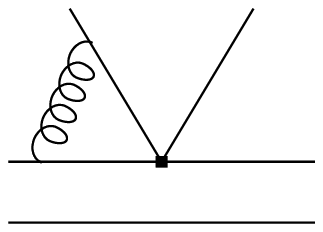} \qquad
  \epsfxsize=0.2\textwidth\epsffile{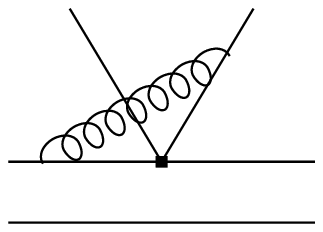}  \qquad
  \epsfxsize=0.2\textwidth\epsffile{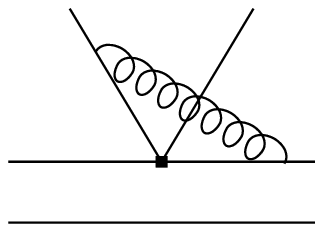}  \qquad
  \epsfxsize=0.2\textwidth\epsffile{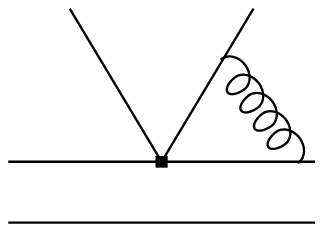}$$ 
  \vskip-12pt
  \caption[ ]{The ``non-factorisable'' vertex corrections.} 
  \label{fig:factvert}
\end{figure}

\begin{figure}[h]
 $$\epsfxsize=0.25\textwidth\epsffile{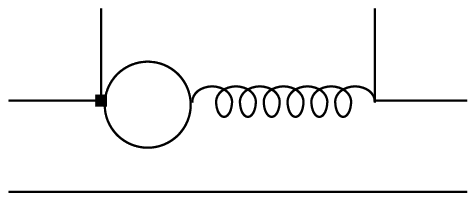} \qquad
  \epsfxsize=0.25\textwidth\epsffile{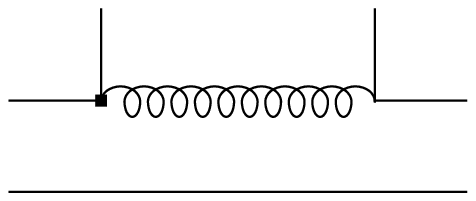} $$  
  \vskip-12pt
  \caption[ ]{Contributions from penguin dipole operator (left) and chromomagnetic
  dipole operator (right).} 
  \label{fig:factpeng}
\end{figure}

The leading order diagram represents the quark level process $b\to
u\bar{u}d$, and there is only one contributing diagram with no hard gluon
interactions, as shown in Figure \ref{fig:factLO}.  
Since the spectator quark is
soft and does not undergo a hard interaction, it is absorbed by the recoiling
meson, and is described by the $B\to\pi$ form factor.  In the heavy-quark
limit, we can represent the scaling of the decay amplitudes as \cite{Beneke:2000ry} 
\begin{equation}
  \mathcal{A}(\bar{B}^0\to\pi^+\pi^-)\sim G_Fm_b^2F^{B\to\pi}(0)f_\pi
\end{equation}
Radiative corrections
to this diagram are suppressed by a power of $\alpha_s$ or
$\Lambda_{QCD}/m_b$, or are already accounted for by the definition of
$F^{B\to\pi}$ or $f_\pi$; gluon exchange diagrams that do not fall into the
above category are those that are ``non-factorisable'' with respect to the
naive factorisation formalism.  The hard-scattering kernel $T^I$ contains
contributions from diagrams of this type, including vertex corrections,
penguin contractions and contributions from the chromomagnetic dipole operator.  These are
illustrated in Figures \ref{fig:factvert} and \ref{fig:factpeng}.

\begin{figure}[h]
  $$\epsfxsize=0.2\textwidth\epsffile{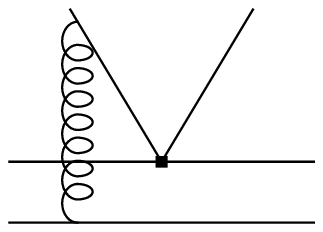} \qquad
  \epsfxsize=0.2\textwidth\epsffile{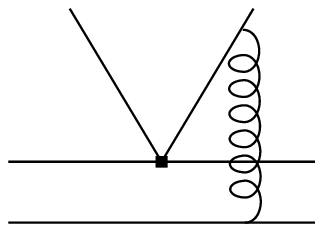} $$ 
  \vskip-12pt
  \caption[ ]{Hard-spectator contributions to kernel $T^{II}$.} 
  \label{fig:facthard}
\end{figure}

The type-II hard-scattering kernel $T^{II}$ contains the hard spectator
interactions, of the type shown in Figure \ref{fig:facthard}.  These would violate
factorisation if there was soft-gluon exchange at leading order, however
they are suppressed because of the endpoint suppression of the light-cone
distribution amplitude for the recoiling $\pi$.   

\section{Basic formulae for charmless B-decays}

It is not necessary to reproduce all of the numerous expressions of which
the QCD factorisation framework is composed, as they are discussed at 
length in the literature.  On the other hand, an exposition of the most
important of these formulae which make up the backbone to the QCD
factorisation method is illuminating, and will be outlined here.  We 
highlight the places where \textit{model-dependent} contributions arise, and
discuss the methods used by BBNS to parameterise them (Section \ref{sec:powersup}).
Our in-depth analysis of non-factorisable contributions to non-leptonic
decays is found in Chapter \ref{chp:NFC}.  The full decay amplitude
expressions as expressed in \cite{Beneke:2001ev} for $B\to\pi\pi$ are
given for completeness in Appendix \ref{chp:AppA}.
 
\subsection{Factorisable contributions}\label{sec:factcont}

We have introduced the non-perturbative input for the factorisation formula
(\ref{eq:QCDF}); what remains is to discuss the perturbative part.  We will
concentrate on the factorisable part -- that which is fully calculable in the
factorisation framework. This is done by translating the effective weak
Hamiltonian into a transition operator so that the matrix element is
expressed, for example, as  
\be
   \langle\pi\pi|\mathcal{H}_{eff}|\bar{B}\rangle = \frac{G_F}{\sqrt{2}}\sum_{p=u,c}\lambda_p\langle\pi\pi|T_p|\bar{B}\rangle
\ee
The transition operator $T_p$ is constructed from operators labelled by the
flavour composition of the final state, corresponding to the operators of
the weak Hamiltonian.  These are multiplied by a QCD factorisation coefficient,
$a_i^p(\pi\pi)$, which contains all of the information from evaluation of diagrams
corresponding to that topology.  

For example, the coefficients $a_1(\pi\pi)$ and $a_2(\pi\pi)$ are related to
the current-current
operators and $a_3(\pi\pi)\ldots a_6(\pi\pi)$ are related to the QCD penguin operators.  The
coefficients $a_7(\pi\pi)\ldots a_{10}(\pi\pi)$ are partially induced by the
electroweak penguin operators $Q_7\ldots Q_{10}$ and are of order $\mathcal{O}(\alpha)$.  These coefficients do not include any 
annihilation topologies -- they are model-dependent, power suppressed
corrections and as such are evaluated separately, as discussed in the next
section. 

The general form of the factorisation coefficients at the next-to-leading
order in $\alpha_s$ is  
\begin{eqnarray}
  \label{eq:coef}
  a_i^p(M_1M_2) &=& \left(C_i + \frac{C_{i\pm1}}{N_c}\right)N_i (M_2)
  \nonumber\\
                &+ &
  \frac{C_{i\pm1}}{N_c}\frac{C_F}{4\pi}\left[\alpha_s(m_b)V_i(M_2) +
  \frac{\alpha_s(\mu_h)4\pi^2}{N_c}H_i(M_1M_2)\right] + P_i^p(M_2) \nonumber \\
\end{eqnarray}
where the upper signs apply for $i$ odd and lower for $i$ even, and the
superscript $p$ again runs over $p = u, c$ (except for $i=1,2$ where there is
no flavour dependence and $p$ is omitted).  The Wilson coefficients $C_i
\equiv C_i(\mu)$ are taken at next-to-leading order in $\alpha_s$. The
quantities that enter these coefficients are split up as follows:

\textbf{Leading order} -- the coefficient $N_i(M_2)$ describes the
normalisation of the relevant LCDA, and is unity in all cases except $i=6,8$.

\textbf{Vertex contributions} -- $V_i(M_2)$, originate from processes as in
Figure \ref{fig:factvert}.  These are calculated via convolution integrals of
the LCDA for 
the meson $M_2$ with a scalar hard-scattering function (this function takes
different forms depending on $i$).   The pion LCDA $\phi_\pi$ is used for
$B\to\pi\pi$ decays, while both $\phi_\pi$ and $\phi_K$ are used in $B\to\pi
K$ decays.  For all $i$ except $i=6,8$, the leading twist distribution
amplitude is used; the $i=6,8$ terms use the twist-3 distribution amplitude,
which is $\Phi^3(x) = 1$ for pseudoscalar mesons.

\textbf{Penguin contributions} -- $P_i(M_2)$, as in Figure \ref{fig:factpeng}.  These are
present at order $\alpha_s$ only for $i=4,6$.  Again, these are convolutions
of a light-cone distribution amplitude with a hard-scattering function, where
the function is dependent on the internal quark mass in the penguin
diagrams.  This enters as $z_p$, where $z_u = 0$ and $z_c = m_c^2/m_b^2$.

\textbf{Hard spectator terms} -- $H_i(M_1M_2)$, as in Figure \ref{fig:facthard}.  These
terms originate from the hard gluon exchange between the final state meson
($M_2$) and the spectator quark, and make up the kernel $T^{II}$.
They are calculated at a lower scale than the other contributions which are
evaluated at $\mu = m_b$.  For these terms we associate a lower ``hard'' scale
$\mu_h\sim(\Lambda_{QCD}m_b)^{1/2}$, where we use $\Lambda_{QCD}\sim\Lambda_h
= 0.5$GeV.  We also evaluate the relevant Wilson coefficients in
(\ref{eq:coef}) at the same scale.  The contributions are again convolution
integrals of the distribution amplitudes of the light, final state mesons
and that of the $B$-meson.  The light LCDA enter both at leading twist, and
in the \textit{chirally enhanced} twist-3 contributions.  Only the
twist-2 contribution is dominated by the hard gluon exchange and so is
calculable, the twist-3 terms having logarithmic endpoint singularities.

The coefficients $a_6(M_1M_2)$ and $a_8(M_1M_2)$ are power suppressed by a
ratio $r_\chi^M$ which is proportional to the QCD quark condensate -- these
are therefore referred to as chirally enhanced corrections. The suppression
ratio is
%% \be
%%    r_\chi^K(\mu) =
%%    \frac{2m_K^2}{\overline{m}_b(\mu)(\overline{m}_q(\mu)+\overline{m}_s(\mu))}  
%% \ee 
%% where $q = u\,(d)$ for charged (neutral) kaons, or correspondingly
\be
   r_\chi^\pi(\mu) =
   \frac{2m_\pi^2}{\overline{m}_b(\mu)(\overline{m}_u(\mu)+\overline{m}_d(\mu))}  
\ee 
for $B\to\pi\pi$ decays.  Although this factor is 
formally suppressed by $\lqcdmb$, it always appears in conjunction with the
coefficients $a_6, a_8$, and can cause a large enhancement of
power-suppressed corrections, as we consider next.

\subsection{Power-suppressed corrections}\label{sec:powersup}

The first main source of non-factorisable corrections to QCD factorisation is
through the chirally-enhanced power corrections that are identified with the
endpoint divergences arising in the hard spectator terms.  These terms are calculated via
\begin{eqnarray}
   H_i(M_1M_2) =
 \frac{f_Bf_\pi}{m_B^2F_0^\pi(0)}\int_0^1\frac{d\xi}{\xi}\Phi_B(\xi)\int_0^1dx\int_0^1dy\left[\frac{\Phi_{M_2}(x)\Phi_{M_1}(y)}{\bar{x}\bar{y}}
 + r_\chi^{M_1}\frac{\Phi_{M_2}(x)\Phi_{m_1}^3(y)}{x\bar{y}}\right] \nonumber\\
\end{eqnarray}
where $\bar{x}\equiv(1-x)$.  This simplifies, for example for $B\to\pi\pi$, to
\begin{eqnarray}
  \label{eq:HS}
  H_i(M_1M_2) &=&
   \frac{f_Bf_\pi}{m_B\lambda_BF_0^\pi(0)}\int_0^1dx\int_0^1 dy
   \left[\frac{\phi_\pi(x)\phi_\pi(y)}{\bar{x}\bar{y}} +
   r_\chi^\pi\frac{\phi_\pi(x)\phi_\pi^3(y)}{x\bar{y}}\right] \nonumber\\
\end{eqnarray}
since the asymptotic twist-3 amplitude $\phi_\pi^3 = 1$ (for all pseudoscalar mesons), and the pion distribution
amplitude is symmetric in exchange $x\leftrightarrow\bar x$.  The first
integral in the bracket in (\ref{eq:HS}) is finite.  However the second,
involving the twist-3 amplitude (where
the explicit power suppression can be seen), is logarithmically divergent.  BBNS
introduce a model-dependent parameterisation of this divergence by defining 
\be
   X_H^M  \equiv \int_0^1\frac{dy}{1-y}\phi_M^3(y) = \int_0^1\frac{dy}{1-y}
\ee
This parameter represents the soft-gluon interactions with the spectator
quark and is expected to be of the order $X_H^M \sim
\ln{\m_b/\Lambda_{QCD}}$.  It is also treated as universal, i.e. the same
value for all cases, so there is no dependence on $M$ or the index $i$ of the
contribution in which it arises.  It is written in the form
\be
   X_H = (1 + \rho_H\,e^{i\varphi_H})\ln{\frac{m_B}{\Lambda_h}}
\ee

\begin{figure}[h]
  $$\epsfxsize=0.2\textwidth\epsffile{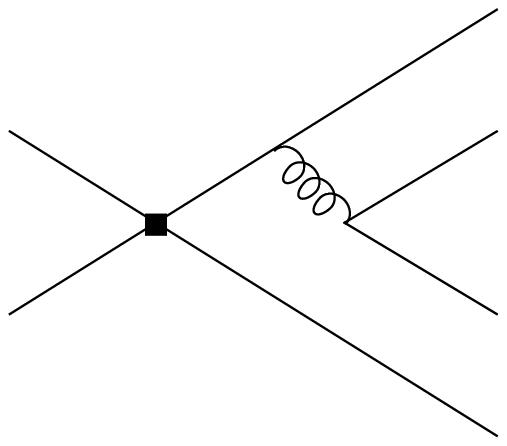} \qquad
  \epsfxsize=0.2\textwidth\epsffile{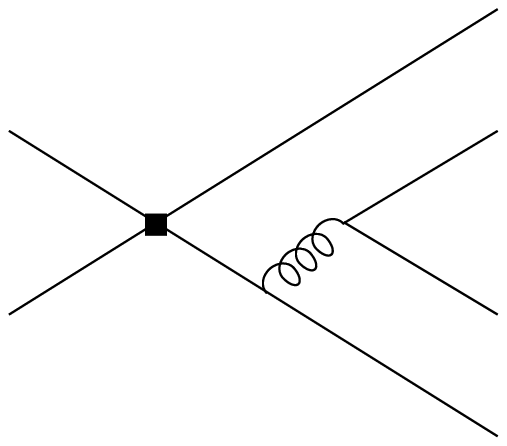} \qquad
  \epsfxsize=0.2\textwidth\epsffile{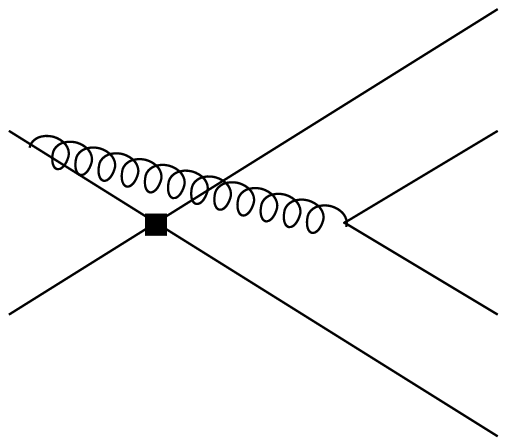} \qquad
  \epsfxsize=0.2\textwidth\epsffile{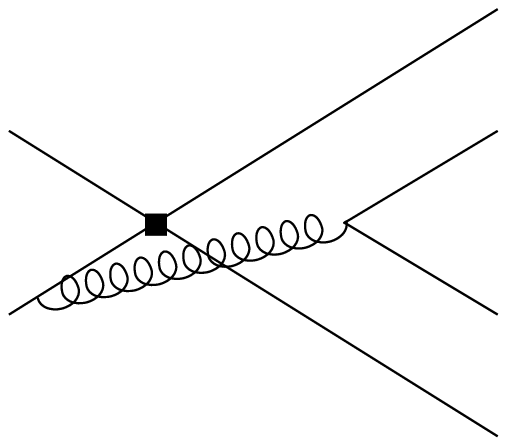} $$ 
  \vskip-12pt
  \caption[ ]{Power suppressed annihilation contributions.} 
  \label{fig:factann}
\end{figure}

The second source of non-factorisable contributions to QCD factorisation is
the evaluation of the annihilation topologies as in Figure \ref{fig:factann}.  

Their
contribution can be numerically significant, especially in
$B\to\pi K$.  The calculational difficulty arises in the appearance of
endpoint singularities even for the leading twist contributions, which cannot
be dealt with in the hard-scattering formalism.  Since there is no consistent
way of treating these contributions, an approximated model ignoring these soft
endpoint divergences is used on top of the QCD factorisation setup in order
to estimate the size of the annihilation contributions.  The contributions are added
at the amplitude level, via a new transition operator $T_p^{ann} $ giving a
new set of coefficients, $b_i(M_1M_2)$.
\be
   \langle\pi\pi|\mathcal{H}_{eff}|\bar{B}\rangle =
   \frac{G_F}{\sqrt{2}}\sum_{p=u,c}\lambda_p\langle\pi\pi|T_p + T_p^{ann}|\bar{B}\rangle
\ee
The divergences that appear are treated in a similar way to the power
corrections for the hard-spectator diagrams, and are written in terms of an
analogous complex parameter $X_A$, introduced via
\be
   \int_0^1\frac{dy}{y} \to X_A \qquad \int_0^1dy\frac{\ln{y}}{y}\to-\frac{1}{2}(X_A)^2
\ee
where again $X_A$ can be simplified under the assumption of universality, so
the value is independent of the identity of the meson or the weak decay
vertex, and is parameterised by 
\be
   X_A = (1 + \rho_A\,e^{i\varphi_A})\ln{\frac{m_B}{\Lambda_h}}
\ee

\subsection{Isospin decompositions for $B\to\pi\pi$}\label{sec:isodec}

For our study of the non-factorisable corrections to charmless B-decays in
Chapter \ref{chp:NFC}, we
decompose the $B\to\pi\pi$ decay amplitudes in terms of their \textit{isospin
  amplitudes}.  The factorisable contributions to these are
calculated within the 
QCD factorisation framework, so it is useful at this point to discuss the decomposition
of the decay amplitudes into their isospin components.   

In $B\to\pi\pi$ decays, the final state ($\pi\pi$) can only have total
isospin of $I=0$ or $I = 2$.  This implies that the three decay amplitudes,
$B^0\to\pi^+\pi^-,\,B\to\pi^0\pi^0$ and $B^+\to\pi^+\pi^0$, have only two
decay paths, with change in total isospin between the initial and final state
of $\Delta I = 1/2$ or $\Delta I = 3/2$.  They must therefore obey a triangle relation
\cite{Gronau:1990ka}
\be
   \sqrt{2}\mathcal{A}(B^+\to\pi^+\pi^0) - \mathcal{A}(B^0\to\pi^+\pi^-) -
   \sqrt{2}\mathcal{A}(B^0\to\pi^0\pi^0) = 0
\ee
and similarly for the CP-conjugate decays.  Assuming isospin invariance in
the matrix elements $\mathcal{A}(B\to\pi\pi) \equiv
\langle\pi\pi|\mathcal{H}_{eff}|B\rangle$, we can decompose the decay
amplitudes into two isospin amplitudes $A_{3/2}$ and $A_{1/2}$ (where the
index corresponds to the change in isospin). Explicitly we have:
\begin{eqnarray}
   \sqrt{2}\mathcal{A}(B^+\to\pi^+\pi^0) &=& 3A_{3/2} \nonumber\\
    \mathcal{A}(B^0\to\pi^+\pi^-) &=& A_{3/2} - A_{1/2} \nonumber\\
    \sqrt{2}\mathcal{A}(B^0\to\pi^0\pi^0) &=& 2A_{3/2} + A_{1/2} 
\end{eqnarray}
with equivalent expressions in terms of $\bar{A}_{i}$ for the
charge-conjugate decays.  We can break the isospin amplitudes down further into
contributions from two weak amplitudes,
\begin{eqnarray}
  A_i = \lambda_u^*A_i^u + \lambda_c^*A_i^c \nonumber \\
  \bar{A}_i = \lambda_uA_i^u + \lambda_cA_i^c
\end{eqnarray}
with $\lambda_q = V_{qb}V_{qd}^*$.  The phase difference between $\lambda_u$
and $\lambda_c$ is related to the angle $\gamma$ of the unitarity triangle:
\be
   \frac{\lambda_u}{\lambda_c} = -\sqrt{\bar{\r}^2 + \bar{\eta}^2}\,e^{i\gamma}
\ee

As was first reported in \cite{Feldmann:2004mg}, it is straightforward to use
the explicit expressions for the decay amplitudes (\ref{eq:piampQCDF}) to obtain
the following expressions for the factorisable part of the isospin
amplitudes:

\begin{eqnarray}\label{eq:A32}
  A_{1/2}^u &=& \frac{A_{\pi\pi}}{6}\left[4a_1 - 2a_2 + 6(a_4^u + r_\chi^\pi
     a_6^u)+ 3(a_7-a_9) + 3(a_{10}^u + r_\chi^\pi a_8^u)\right] \nonumber\\
  A_{1/2}^c &=&  \frac{A_{\pi\pi}}{2}\left[2(a_4^c + r_\chi^\pi
     a_6^c)+ (a_7-a_9) + (a_{10}^c + r_\chi^\pi a_8^c)\right] \nonumber\\
  {}\nonumber\\
  A_{3/2}^u &=& -\frac{A_{\pi\pi}}{6}\left[2a_1 + 2a_2 -3(a_7 - a_9) +
     3(a_{10}^u + r_\chi^\pi a_8^u)\right] \nonumber\\
   A_{3/2}^c &=& -\frac{A_{\pi\pi}}{2}\left[-(a_7 - a_9) +
     (a_{10}^u + r_\chi^\pi a_8^u)\right] 
\end{eqnarray}
with 
\be
   A_{\pi\pi} = i\frac{G_F}{\sqrt{2}}m_B^2F_0^{B\to\pi}(0)f_\pi \qquad
   r_\chi^\pi = \frac{2m_\pi^2}{2m_bm_q}
\ee

\section{Limitations to QCD factorisation}\label{sec:limit}

The predominant problem with the factorisation framework is that it is only
explicitly valid in the heavy quark limit -- the master formula
(\ref{eq:QCDF}) is exact only as $m_b\to\infty$.  However, as we know $m_b$ is
$\sim5$ GeV it is justifiable to examine the validity of 
this limit. The power corrections in the $\Lambda_{\mathrm{QCD}}/m_b$ expansion
are often difficult to calculate, and include non-factorisable corrections
that cannot yet be determined in a model-independent manner.

The expected size of the non-factorisable corrections is
$\mathcal{O}(\lqcdmb)\sim10\%$, yet as we will show in detail in Chapter
\ref{chp:NFC}, the power corrections can be and are often required to be,
considerably larger than this estimate. 

The power-suppressed terms can be large with respect to the leading
order contribution thanks to various different sources of suppression -- such
as small values of Wilson coefficients, CKM factors or colour suppression.  Numerical
enhancement of the power-suppressed terms themselves can also occur, where
the coefficients can be larger than suggested by power counting, e.g., the
chirally enhanced terms in 
coefficients $a_6(\pi\pi)$ or $a_8(\pi\pi)$ discussed in Sections
\ref{sec:factcont} and \ref{sec:powersup}.
     
Comparing the predictions from QCD factorisation to the available
experimental data gives very good agreement in many cases.  For charmless
$B$ decays however, the correspondence between prediction and data gives what has been
called ``The $B\to\pi\pi$ and $B\to\pi K$ puzzles''.  We can consider two
independent ratios of CP-averaged branching fractions:
\begin{equation*}
   R_{+-}^{\pi\pi} =
   2\left[\frac{\mathrm{BR}(B^{\pm}\to\pi^\pm\pi^0)}{\mathrm{BR}(B^0\to\pi^+\pi^-)}\right]\frac{\tau_{B^0}}{\tau_{B^+}} \qquad  R_{00}^{\pi\pi} =
   2\left[\frac{\mathrm{BR}(B^0\to\pi^0\pi^0)}{\mathrm{BR}(B^0\to\pi^+\pi^-)}\right]
\end{equation*}        

The branching ratio for $B\to\pi^+\pi^-$ is unexpectedly small compared to
the prediction, and $B\to\pi^0\pi^0$, unexpectedly large.
$B^\pm\to\pi^\pm\pi^0$ is however in agreement with the theory estimates.
The experimental values, using the HFAG averages, are
$R_{+-}^{\pi\pi}\sim2.2$, $R_{00}^{\pi\pi}\sim0.32$.  Using the default
parameters the QCD factorisation estimates are somewhat different:
$R_{+-}^{\pi\pi}\sim1.25$, $R_{00}^{\pi\pi}\sim0.03$.   Even using the
``favoured scenario'' for the parameters and with enhanced annihilation
contributions we find: $R_{+-}^{\pi\pi}\sim1.8$, $R_{00}^{\pi\pi}\sim0.13$
\cite{Beneke:2003zv}.  An analogous situation exists for $B\to\pi K$ where
moderate discrepancies occur in some branching ratios, such as the ratio of
$B\to\pi^\mp K^\pm$ and  $B^\pm\to\pi^\pm K^0$.

There is a reasonable agreement for the $B\to\pi\pi$ CP asymmetry
predictions from QCD factorisation.  There is however difficulty in making
conclusive statements regarding this due to the discrepancies between the
experimental measurements from {\textsc{BaBar}} and Belle.  This can be seen
graphically in the presentation of our results in Chapter \ref{chp:NFC},
where we clarify and quantify the discrepancies between the predictions and
measurements.       

%%% Local Variables:
%%% mode: latex
%%% TeX-master: t
%%% End:

%% file: chapter3.tex
\chapter{Light-cone meson distribution amplitudes}\label{chp:LCDA}
\begin{center}
 \begin{quote}
    \it
    A light from the shadows shall spring
  \end{quote}
\end{center}
\vspace{-4mm}
\hfill{\small J.R.R. Tolkien}
\vspace{5mm}

This chapter is devoted to the study of distribution amplitudes of light
mesons, where we derive new models characterised by a small number of
parameters directly related to experimental observables -- this is the first
new result of this thesis. We discuss how these new models are constructed
and how they are more descriptive than the current expressions of light meson
distribution amplitudes.  We also show how we can numerically evolve the
new model distribution amplitudes in the scale $\mu$, to exactly reproduce 
the leading order scaling behaviour of the usual approximated DA.  Finally,
within the framework of QCD factorisation, we study the effect of these model
DAs on important 2-body non-leptonic B-decays such as $B\to\pi\pi$. 

\section{General framework}

The basic equation to describe bound states in relativistic quantum field theory is known
as the \textit{Bethe-Salpeter} equation, which was first formulated in the
1950's \cite{Salpeter:1951sz, Gell-Mann:1951rw}.  The leading twist distribution
amplitude of a meson is related to this wave function by integrating out the dependence on
the transverse momentum $p_\perp$,
\be
   \phi(u)\sim\int_{p_\perp^2<\mu^2} d^2p_\perp\,\phi(u,p_\perp)
\ee
The light-cone distribution amplitudes were originally introduced in the
context of hadronic electromagnetic form factors
and the pseudoscalar-photon transition form factor
\cite{Chernyak:1977as,Chernyak:1980dj,Efremov:1979qk,Lepage:1979zb, Lepage:1980fj},
but have attracted interest in $B$ physics 
due to their appearance in QCD sum rules on the light-cone
\cite{Ball:2004rg,Colangelo:2000dp,Khodjamirian:2001bj,Belyaev:1993wp,Ball:1997rj,Ball:1998tj,Ball:1998kk,Khodjamirian:2000ds,Ball:2001fp,Ball:2004ye}, 
form factors \cite{Beneke:2000wa,Bauer:2002aj},
and their use in factorisation for $B$ decay amplitudes \cite{Beneke:1999br,Beneke:2001ev}. 

Using the formalism introduced in Section \ref{sec:NPP}, we can define the
distribution amplitude for a general pseudoscalar meson P in terms of the
matrix element of a non-local operator near the light-cone.  We have
 \begin{eqnarray}
 \langle 0 | \bar q_1(x)\gamma_\mu\gamma_5 [x,-x]
 q_2(-x)|P\rangle & = & i f_P p_\mu \int_0^1 du \, e^{i\xi px} 
 \left[ \phi_P(u) +
   \frac{1}{4}\, m_P^2 x^2 {\mathbb A}_P(u) + O(x^4)\right]\nonumber\\
 &&{} + \frac{i}{2}\,
   f_P m_P^2\,
   \frac{1}{px}\, x_\mu \int_0^1 du \, e^{i\xi px}\,
 \left[{\mathbb B}_P(u)+O(x^2)\right]\hspace*{0.7cm}\nonumber\\
 \label{eq:T2}
 \end{eqnarray}
 where $\xi = 2u-1$ and $[x,y]$ is the Wilson line defined in
 (\ref{eq:wilsonline}).  In equation (\ref{eq:T2}),  $\phi_P$ is the leading
 twist-2 distribution  amplitude, whereas $\mathbb
 A_P$ and $\mathbb B_P$ contain contributions from higher-twist
 operators. The corresponding definitions of vector meson DAs can be found in
\cite{Ball:1998sk,Ball:1998ff}.

The theory of meson DAs is well understood
\cite{Chernyak:1983ej,Braun:1989iv,Ball:1998ff,Ball:1998sk,Ball:1998je}, 
and suggests their parameterisation in terms of a \textit{partial wave
expansion} in conformal spin (see Section \ref{sec:conformal}), allowed due to
the conformal invariance of QCD.  This expansion is in terms of
contributions with different conformal spin, which do not mix with each other
under a change of scale.  This is true to leading  logarithmic accuracy, but
is no longer  the case at higher order, as the underlying symmetry is anomalous. 
The leading-twist distribution amplitude $\phi(u)$, for both pseudoscalar
and vector mesons has a conformal expansion given in terms of \textit{Gegenbauer polynomials} $C_n^{3/2}$,
   \begin{equation}\label{eq:confexp}
   \phi(u,\mu^2)=6 u (1-u) 
    \sum\limits_{n=0}^\infty a_{n}(\mu^2)  C^{3/2}_{n}(2u-1)
 \end{equation}
The Gegenbauer polynomials are generalisations of the associated Legendre
polynomials; the first few are
\begin{eqnarray*}
  C_0^{3/2}(x) &=& 1 \\
  C_1^{3/2}(x) &=& 3x \\
  C_2^{3/2}(x) &=& \frac{3}{2}(-1 + 5x^2) 
\end{eqnarray*}

The coefficients $a_n$ in (\ref{eq:confexp}) are Gegenbauer moments, and renormalise
multiplicatively to leading logarithmic accuracy
 \begin{equation}
   \label{eq:scaling}
     a_n(Q^2) = a_n(\mu^2)
   \left(\frac{\alpha_s(Q^2)}{\alpha_s(\mu^2)}\right)^{
 (\gamma_0^{(n)}-\gamma_0^{(0)})/(2\beta_0)}  
    \end{equation}
where $\beta_0=11 - (2/3) n_f$.  For pseudoscalar and longitudinally
polarised vector mesons, the one-loop anomalous dimension $\gamma_0^{(n)}$,
given by \cite{Gross:1974cs} 
   \begin{eqnarray}
   \gamma_0^{(n)} &=& 8 C_F
   \left(\psi(n+2)+\gamma_E -\frac{3}{4} - \frac{1}{2(n+1)(n+2)}\right)
   \label{eq:1loopandim}
   \end{eqnarray}
and for transversely polarised vector mesons 
   \begin{eqnarray}
   \gamma_0^{(n)} &=& 8 C_F
   \left(\psi(n+2)+\gamma_E -\frac{3}{4}\right)
   \end{eqnarray}
 For $\pi$, $\rho$, $\omega$, $\eta$, $\eta'$ and $\phi$, G-parity
 ensures that $a_{\rm odd} = 0$ and that the DA is symmetric under
 $u\leftrightarrow 1-u$, whereas for $K$ and $K^*$ the nonzero values
 of $a_{\rm odd}$ induce an antisymmetric component of the
 DA. $a_0\equiv 1$ is fixed by normalisation
 $$\int_0^1 \phi(u,\mu^2) = 1$$
 and all the other $a_n$ are intrinsically nonperturbative
 quantities. As they do not mix under renormalisation to leading order accuracy,
 equation (\ref{eq:confexp}) is well suited to construct models for $\phi$:
 truncating the series after the first few terms yields a parameterisation of the DA that is ``stable'' under a change of
 scale, except for the numerical values of $a_n$. Despite there being
 no small expansion parameter, such a {\em 
 truncated conformal expansion} is often a meaningful approximation to the full distribution amplitude.  An advantage
 of this  expansion is that the contributions of higher conformal
spins to the convolution integrals involving the LCDA (and therefore also the
physical amplitudes) are suppressed by the highly
oscillating behaviour of the partial waves. This suggests a construction of
models for the DAs based on a truncated conformal expansion, where only the first few waves are
included.  

We now move on to discuss the construction of the distribution amplitude via
the conformal expansion using the conformal invariance of QCD.

\section{Conformal symmetry}\label{sec:conformal}
\subsection{The conformal group}

The \textit{conformal group} is defined as the group of general co-ordinate transformations of
4-D Minkowski  space that conserve the interval $ds^2=g_{\mu\nu}dx^\mu
dx^\nu$ up to a change of scale -- i.e. transformations that preserve
angles and leave the  light-cone invariant. Transformations of this type are: 
scale  transformations ($x^\mu\to\lambda x^\mu$), inversions ($x^\mu\to
x^\mu/x^2$),  translations ($x^\mu\to x^\mu + c$) and Lorentz rotations.  The conformal group is the 
maximal extension of the Poincar\'{e} group  that leaves the light cone invariant \cite{Braun:2003rp}.  
In four dimensions the full conformal group has 15 generators, which are denoted by 
\begin{center}
   \begin{tabular}{cc}
      $\rm{\bf{P}}_\mu$ \qquad &  4 translations \\
      $\rm{\bf{M}_{\mu\nu}}$ \qquad & 6 Lorentz rotations \\
      $\textbf{D}$  \qquad & dilatation\\
      $\textbf{K}_\mu$ \qquad  & 4 special conformal transformations\\ 
   \end{tabular}
\end{center}

An important subgroup of the full conformal group is the \textit{collinear subgroup}, 
which (as we will see later) is the most relevant in the study of QCD.  Denoted 
$SL(2, \mathbb{R})$, this subgroup is
made up of a special case of the \textit{special conformal transformation} 
\begin{equation}
\label{eq:SCtrans}
   x^\mu\to x^{\mu\prime}=\frac{x^\mu+a^\mu x^2}{1+2a\cdot x+a^2x^2}
\end{equation}
for $a^\mu$ light-like; this can be reduced to 
\begin{equation}
\label{eq:SCll}
   x_-\to x^{\prime}_-=\frac{x_-}{1+2ax_-}
\end{equation}
which maps the light-ray onto itself in the $x_-$ direction.  Transformations of this type together with 
the dilations and translations along the $x_-$ direction make up the collinear subgroup.

Transformations of $SL(2, \mathbb{R})$ are governed by four generators
$\rm{\bf{P}}_+$, $\rm{\bf{M_{-+}}}$, $\textbf{D}$ and $\textbf{K}_-$.  These are most often presented in the linear combinations \cite{Braun:1999te}:
\bea
\label{eq:gen}
   \rm{\bf{L}}_+&=& \rm{\bf{L}}_1+i \rm{\bf{L}}_2=-i \rm{\bf{P}}_+\nonumber\\
   \rm{\bf{L}}_-&=& \rm{\bf{L}}_1 - i \rm{\bf{L}}_2=\tfrac{i}{2}\rm{\bf{K}}_-\nonumber\\
   \rm{\bf{L}}_0&=&\tfrac{i}{2}(\rm{\bf{D}}+\rm{\bf{M}}_{-+})\nonumber\\
   \rm{\bf{E}}&=&\tfrac{i}{2}(\rm{\bf{D}}-\rm{\bf{M}}_{-+})
\eea
so that the algebra of $SL(2, \mathbb{R})$ is written 
\be
   [\rm{\bf{L}}_0, \rm{\bf{L}}_\mp]=\mp\rm{\bf{L}}_\mp \qquad
   [\rm{\bf{L}}_-,  \rm{\bf{L}}_+]=-2\rm{\bf{L}}_0
\ee
These generators act on fundamental fields $\Phi(x)$ which are equivalent to primary fields
in conformal field theory.  If we consider the parton model of hadron states,
   the hadron is replaced by partons moving collinearly in some direction $\vec{n}_\mu$.  We can then consider only the fields confined to the light-cone 
$$ \Phi(x)\to\Phi(\a n)$$
with $\a$ a real  number.  Additionally, we choose the field $\Phi$ as an eigenstate of
the spin operator  so that is has a fixed spin projection $s$ in $+$direction.
%See p8 of hep-ph/0306057 for def of conformal algebra in terms of differential operators!!!
The action of the generators can be described in terms of differential
operators acting on the  primary field. $\phi(\a)\equiv\phi(\a n)$
\bea
\left[\rm{\bf{L}}_+, \Phi(\a)\right] &=& -\partial_\a\Phi(\a)\equiv L_+\Phi(\a), \nonumber\\
\left[\rm{\bf{L}}_-, \Phi(\a)\right] &=& (\a^2\partial_\a + 2j\a)\Phi(\a) \equiv L_-\Phi(\a), \nonumber\\
\left[\rm{\bf{L}}_0, \Phi(\a)\right] &=& (\a\partial_\a+j)\Phi(\a)\equiv L_0\Phi(\a) \nonumber\\
\left[\rm{\bf{E}}, \Phi(\a)\right] &=& \frac{1}{2}(l-s)\Phi(\a)
\eea
The generator $\rm{\bf{E}}$ measures the \textit{collinear twist} of the field, defined as the dimension minus spin projection in $+$ direction.  The field $\Phi(\a x)$, with a fixed spin projection on the lightcone ($x^2=0$) is an eigenstate of the quadratic Casimir operator $L^2$, defined by
\be
   L^2\equiv L_0^2+L_1^2+L_2^2=L_0^2+L_-L_+ \qquad \left[L^2, L_i\right]=0
\ee
so that
\be
L^2\Phi(\a) = \sum_i\left[\rm{\bf{L}}_i, \left[\rm{\bf{L}}_i, \Phi(\a)\right]\right]=j(j-1)\Phi(\a)
\ee
Here $j=\frac{1}{2}(l+s)$ and is referred to as the \textit{conformal spin},
where $l$ is the canonical mass dimension of the field and $s$ the (Lorentz)
spin.  Consequently, we see that the field $\Phi(\a)$ is transformed according to representations of the collinear conformal group $SL(2, \mathbb{R})$, specified by the conformal spin $j$.

\subsection{Conformal symmetry in QCD}

QCD is conformally invariant at leading order, but is broken at next-to-leading-order by the
inclusion of quantum corrections.  However, the conformal spin is seen as a
good quantum number in hard-processes up to small corrections of order
$\a_s^2$.  The non-local operators of QCD can be written in terms of \textit{conformal operators}.   
We can construct a complete basis of local operators on  $SL(2, \mathbb{R})$ from $\Phi(0)$ 
by applying the ``raising'' operator $\rm{\bf{L}}_+$ $k$-times.
\begin{eqnarray*}
   \mathcal{O}_0 &=& \Phi(0) \\
    \mathcal{O}_k &=& \left[\rm{\bf{L}}_+,\ldots,\left[\rm{\bf{L}}_+,\left[\rm{\bf{L}}_+,\Phi(0)\right]\right]\right] = \left(-\partial_+\right)^k\Phi(\a)|_{\a=0} 
\end{eqnarray*}
The primary fields $\Phi(\a)$ can be formally expressed as a Taylor series 
expansion of local conformal operators  
\be
   \Phi(\a ) = \sum_{k=0}^\infty\frac{(-\a)^k}{k!}O_k
\ee
This is a simple example of a \textit{conformal tower}.  It is possible to 
construct conformal towers  for general operators built of a number of primary 
fields and derivatives \cite{Braun:1999te}. We can define a local conformal 
operator $\mathbb{O}_n$ that transforms under the collinear conformal
subgroup.  This is equivalent to demanding the operator satisfies  
\begin{eqnarray*}
   \left[\rm{\bf{L}}^2, \mathbb{O}_n\right] &=& j(j-1)\,\mathbb{O}_n \\
   \left[\rm{\bf{L}}_0, \mathbb{O}_n\right] &=& j\,\mathbb{O}_n  \\
   \left[\rm{\bf{L}}_-, \mathbb{O}_n\right] &=& 0 
\end{eqnarray*}
The tower of operators can be built by repeatedly applying the raising operator, as 
above
\begin{eqnarray*}
   \mathbb{O}_{n,n} &=& \mathbb{O}_n \\
    \mathbb{O}_{n,n+k} &=& \underbrace{\left[\rm{\bf{L}}_+,\ldots,\left[\rm{\bf{L}}_+,\left[\rm{\bf{L}}_+\right.\right.\right.}_k,\left.\left.\left.\mathbb{O}_n\right]\right]\right] = \left(-\partial_+\right)^k\mathbb{O}_n 
\end{eqnarray*}

It is possible to construct explicitly a general local conformal operator as
\cite{Braun:2003rp} 
\be
   \mathbb{O}_n^{j_1,j_2}(x) = \left[\Phi_{j_1}(x)\,P_n^{(2j_1-1,2j_2-1)}\left(\frac{\overrightarrow\partial_+-\overleftarrow\partial_+}{\overrightarrow\partial_+ + \overleftarrow\partial_+}\right)\Phi_{j_2}(x)\right]
\ee
where $P_n^{a,b}(x)$ are Jacobi polynomials and the superscripts indicate the conformal spins of the constituent fields.

Returning to QCD, we can extract different conformal operators related to the 
different spin and dimension of the constituent quark or gluon fields.  
For a quark $\sim$ anti-quark operator at the leading twist we have 
the relevant local conformal operator $\mathbb{O}_n^{1,1}$, given by
\be
  \label{eq:Conf}
 \mathbb{O}_n^{1,1}(x) = (i\partial_+)^n\left[\bar\psi(x)\,\gamma_+\,C_n^{3/2}\left(\overleftrightarrow{D}_+/\partial_+\right)\,\psi(x)\right]
\ee
where the Gegenbauer polynomials are a special case of the Jacobi polynomials 
$P_n^{1,1} = C_n^{3/2}$.  

\subsection{Conformal partial wave expansion} 

The aim of the partial wave expansion is to make full use of the underlying
symmetry of the theory in order to simplify the dynamics of the problem.  In
the case of QCD, we make use of the conformal invariance to create a partial
wave expansion in conformal spin. The conformal expansion of the LCDA is
analogous to the partial wave expansion for a spherically symmetric potential
in quantum mechanics.  Here, the rotational symmetry allows the separation of
angular and radial degrees of freedom.  The angular dependence is encoded in
the spherical harmonics $Y_l^m(\theta, \phi)$ which form an irreducible
representation of $O(3)$, and the radial information is governed by an
evolution equation, namely the 1-d Schr\"{o}dinger equation in $R(r)$.

In QCD, the partial-wave expansion decomposes the distribution amplitude into
longitudinal and transverse degrees of freedom.  The dependence on the
longitudinal momentum is expressed in terms of orthogonal polynomials which
form irreducible representations of $SL(2, \mathbb{R})$.  For a general
two-body wavefunction, these polynomials are the Jacobi polynomials.  For
conformal spins $j_1=j_2$, we obtain the Gegenbauer polynomials, defined by
the generating function 
\be
\label{eq:GPgenfun}
   \frac{1}{(1-2xt+t^2)^\lambda}=\sum_{n=0}^\infty C_n^{(\lambda)}(x)t^n
\ee
A multiple-particle state that is built of primary fields can be expanded in
terms of irreducible representations of $SL(2, \mathbb{R})$ with increasing
conformal spin.  The lowest possible conformal spin for the multi-particle
state equals the sum of spins of the
constituents, and its wavefunction is simply a product of non-degenerate one particle states.  This
lowest state defines what is known as the \textit{asymptotic} distribution
amplitude, as it is reached at the formal limit of $q^2\to\infty$. This state
is expressed, in the general case, as
\be
   \phi_{as}(u_1, u_2\dots u_m) =
   \frac{\Gamma(2j_1+\dots+2j_m)}{\Gamma(2j_1)\dots\Gamma(2j_m)}
   u_1^{2j_1-1}u_2^{2j_2-1}\dots u_m^{2j_m-1}
\ee
where $u_i$ is the momentum fraction of the $i^{th}$ constituent, and the
normalisation is chosen so that $\int[du_i]\phi_{as}(u_i)=1$.

\subsubsection{Light-cone meson distribution amplitudes}

The distribution amplitude for the pion $\phi_\pi(u)$ is an expansion over 
an infinite series of Gegenbauer polynomials, with multiplicatively
renormalisable co-efficients 
\be
\label{eq:phipi}
   \phi_\pi(u, \mu^2)=6u(1-u)\sum_{n=0}^\infty a_n(\mu^2)C_n^{3/2}(2u-1)
\ee
The Gegenbauer polynomials form an orthonormal set on the interval $0<u<1$
with weight function $u(1-u)$ and satisfy the orthogonality condition
\be
   \int_0^1du\,u(1-u)\,C_n^{3/2}(2u-1)C_m^{3/2}(2u-1) = \delta_{mn}N_n
\ee
with
\be
 N_n = \frac{(n+1)(n+2)}{4(2n+3)}
\ee
The renormalisation group equation for $\phi_\pi(u, \mu^2)$ is known as 
the \textit{Efremov-Radyushkin-Brodsky-Lepage} (ER-BL) evolution equation \cite{Lepage:1980fj, 
Efremov:1979qk}
\be
\label{eq:ER-BL}
   \mu^2\frac{d}{d\mu^2}\phi_\pi(u, \mu^2)=\int_0^1dvV(u, v, \a_s)\phi_\pi(v, \mu^2)
\ee
The evolution kernel $V(u, v, \a_s)$ has been calculated perturbatively 
to next-to-leading order in $\a_s$ \cite{Dittes:1983dy, Mikhailov:1984ii}
\be
   V(u, v, \a_s) = \frac{\a_s}{2\pi}V_0(u,v) + \frac{\a_s^2}{4\pi^2}V_1(u,v)
\ee

The pion distribution amplitude can be expressed as matrix elements of 
renormalised local operators
\be   \langle0|\bar{d}(0)\gamma_+\gamma_5(i\overleftrightarrow{D}_+)^nu(0)|
\pi^+(p)\rangle=if_\pi(p_+)^{n+1}\int_0^1du(2u-1)^n\phi_\pi(u,\mu^2)
\ee
so the relevant conformal operators (\ref{eq:Conf}) are then \cite{Braun:2003rp}
\be
\label{eq:relops}
O_n(x)=(i\partial_+)^n\left[\bar{d}(x)\gamma_+\gamma_5C_n^{3/2}\left
(\overleftrightarrow{D}_+/\partial_+\right)u(x)\right]
\ee
Due to the flavour structure of this operator, there is no mixing with
operators involving two-gluon fields,  nor with operators of three or more
fields, as these appear at higher twist.  Thus the operators of
(\ref{eq:relops})  do not mix at the leading order, and are multiplicatively 
renormalised. 
The evolution equation (\ref{eq:ER-BL}) can then be solved by a conformal spin expansion
\be
\phi_\pi(u, \mu^2) = \sum_{n=0}^\infty\frac{u(1-u)}{N_n}C_n^{3/2}(2u-1)\langle\langle O_n(\mu^2)\rangle\rangle
\ee
where
\be
\label{eq:Rop}
   \langle\langle O_n(\mu^2)\rangle\rangle = \int_0^1du\,C_n^{3/2}(2u-1)\,\phi_\pi(u, \mu^2)
\ee
The sum runs over even values of $n$ to ensure symmetry of the pion DA, due to
charge conjugation invariance: $\phi_\pi(u) = \phi_\pi(1-u)$.  We can
construct the Gegenbauer moments, $a_n(\mu^2)$, using the reduced matrix
elements of the conformal operators given in (\ref{eq:Rop}), such that 
\bea
   \label{eq:an}
      a_n(\mu^2)&=&\frac{2(2n+3)}{3(n+1)(n+2)}\langle\langle O_n(\mu^2)\rangle\rangle \nonumber\\
                &=&\frac{2(2n+3)}{3(n+1)(n+2)}\int_0^1du\,C_n^{3/2}(2u-1)\,\phi_\pi(u, \mu^2)
\eea

\section{Non perturbative input}\label{sec:nonpert}

In general, not much is known about the amplitudes of these
partial waves and hence the numerical values of the moments $a_n$.  There is
some information available on the lowest moments of the $\pi$ DA, but much
less for the other mesons.  For the $\pi$ and $\eta$ mesons, there is some experimental
information available on the $\pi(\eta)\gamma\gamma^*$ transition form factor \cite{CLEO:1},
which can be used to extract the values of the first two amplitudes.  The
$\gamma\gamma^*\to\pi$ form factor factorises into a partonic hard scattering
amplitude and a soft hadronic matrix element parameterised by the pion DA
$\phi_\pi$.  The leading twist expression for $F_{\pi\gamma}$ is given by
\cite{Musatov:1997pu,Brodsky:1997dh}:
\begin{eqnarray}
   F_{\pi\gamma}(q^2) &=& \frac{4}{\sqrt{2}}\int_0^1dx\,\phi_\pi(x,
   q^2)\,T^H_{\gamma\to\pi}(x,q^2) \nonumber\\
    &=&  \frac{4}{\sqrt{2}q^2}\int_0^1dx\,\frac{\phi_\pi(x,
   q^2)}{x}\,\left(1+\mathcal{O}(\alpha_s)\right)
\end{eqnarray}
 using $\phi_\pi(x) = \phi_\pi(1-x)$ and since the amplitude for the
 subprocess $\gamma\gamma^*\to\pi$ (labelled $T^H_{\gamma\to\pi}$) is 
\be
   T^H_{\gamma\to\pi}(x,q^2) = \frac{1}{(1-x)q^2}\left(1+\mathcal{O}(\alpha_s)\right)\nonumber
\ee 
This means that the $\gamma-\pi$
transition form factor approximately probes the sum of the Gegenbauer moments
via
\be
   \label{eq:delta1}
   \int_0^1 dx\,\frac{\phi_\pi(x)}{x} = 1 + \sum_na_n
\ee
By making some assumption (such as truncating this series after $n=4$) it
is possible to extract information about the moments.

On the theory side, there exist a few lattice calculations for the second
moment of the $\pi$ DA; some quite dated \cite{Martinelli:1987si,DeGrand:1987vy,Daniel:1990ah},
and a recent retry \cite{DelDebbio:1999mq,DelDebbio:2002mq}.  Unfortunately, these results are
still preliminary and cannot yet be used in phenomenological
applications. Other theoretical calculations have been done using QCD sum
rules for both pseudoscalar and
vector mesons; these are reviewed below.

\section{Previous constructions of $\phi_\pi$}\label{sec:premods}
 
This section briefly reviews and consolidates the previous calculations (and
constructions) of the pion DA and its moments $a_n$.

\begin{figure}[h]
  $$\epsfxsize=0.75\textwidth\epsffile{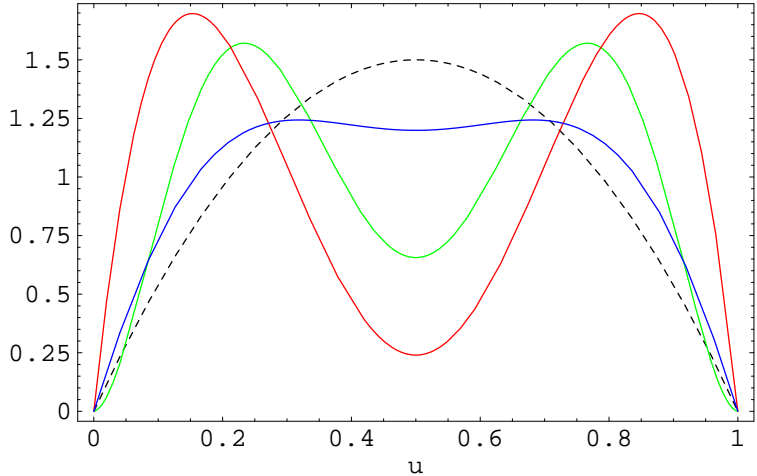}$$
  \vskip-12pt
  \caption[ ]{Model pion distribution amplitudes for comparison between
  asymptotic (dotted), BZ (blue), BMS (red) and CZ (green), all taken as scale $q^2 =
  1$ GeV$^2$.} 
  \label{fig:phimods}
\end{figure}

The earliest calculation of one of the moments of the pion DA was that of
$a_2$ by Chernyak and Zhitnitsky (CZ) \cite{Chernyak:1983ej}.  They developed
a model of the pion 
DA based on local QCD sum rules whereby $\phi_\pi$ is expanded in terms of
local operators from which the $a_n$ can be extracted.  The form of the DA is
strongly  peaked at the endpoints, due to the approximation of the vacuum
quark  distribution by a
delta function and derivatives of, the non-perturbative contributions to the
sum rule forcing the DA to be end-point concentrated.  %%  This sum rule for the
%% wave function $\phi_\pi^{\,CZ}$ reads:
%% \begin{eqnarray}
%%    f_\pi^2\,\phi_\pi^{\,CZ}(u) &=& \frac{3M^2}{2\pi^2}u(1-u) +
%%    \frac{\alpha_s\langle GG\rangle}{24\pi M^2}\left[\delta(u) +
%%    \delta(1-u)\right] \nonumber\\
%%    && + \frac{8}{81}\frac{\pi\alpha_s\langle
%%    \bar{q}{q}\rangle^2}{M^4}\left\{11\left[\delta(u) + \delta(1-u)\right] +
%%    2\left[\delta^\prime(u) + \delta^\prime(1-u)\right]\right\} \nonumber\\
%% \end{eqnarray}
%% with $M^2$ as the Borel parameter characterising the exponential suppression
%% of contributions from the higher states.  
The DA is shown graphically later in this section in Figure
\ref{fig:phimods}, and is described by a model wavefunction of
\be
   \phi_\pi^{\,CZ}(u) = 30u(1-u)(1-2u)^2
\ee
which corresponds to the numerical result
\be
  a_2(0.5\textrm{GeV}) = 2/3  
\ee
The drawback of extracting the $a_n$
in this way is that the expansion of an
intrinsically non-local quantity, namely the pion DA, in terms of contributions from
the local operators causes an increased sensitivity to the non-perturbative
effects.  The coefficients of the condensates in the sum rule increase with
$n$ and can in fact dominate over the perturbative contributions for a large
enough value of $n$.  This implies that this method should be reliable for at least
$n=2$ but not for large $n$.  They also obtained the result 
\be
   \phi\left(1/2, 1/2\,\textrm{GeV}\right)  = 0
\ee
neglecting all $a_{n\ge 4}$.  This however is an artifact of the neglect of
higher order moments. 

This method was improved by Braun and Filyanov \cite{Braun:1988qv}, who
found that $\phi(1/2)$ was required to be non-zero; they found a
constraint on  $\phi(1/2)$ at the scale of 1 GeV of
\be
  \phi\left(1/2, 1\,\textrm{GeV}\right) = 1.2\pm 0.3 = \frac{3}{2} - \frac{9}{4}a_2(1) + \frac{45}{16}a_4(1)
\ee
They also re-determined the first two moments using the CZ procedure:
\be
  a_2(1\textrm{GeV}) = 0.44  \qquad a_4(1\textrm{GeV}) = 0.25 \nonumber
\ee

Other calculational methods have also been used to determine the moments of
the DA.  For example, the most recent calculation is from Ball and
Zwicky \cite{Ball:2004rg,Ball:2004ye} using light-cone
sum rules, who have determined the first two moments as
\be
  a_2(1\textrm{GeV}) = 0.115  \qquad a_4(1\textrm{GeV}) = -0.015 \nonumber
\ee

Another alternative method is to use sum rules with \textit{non-local condensates},
as developed in  
\cite{Bakulev:2001pa, Bakulev:2004cu,
  Mikhailov:1991pt,Mikhailov:1988nz,Mikhailov:1986be}.  This method
constructs the DA by connecting the dynamic properties of the pion with the
QCD vacuum structure.  This allows for the fact that quarks and gluons can
travel through the QCD vacuum with non-zero momentum $k_q$, which implies a
non-zero average virtuality, $\langle k_q^2\rangle = \lambda_q^2$, using the
ratio 
\be
   \lambda_q^2 = \frac{\langle\bar{q}\sigma g G g\rangle}{2\bar{q}q}
\ee
The most recent calculations using this technique was by Bakulev, Mikhailov and Stefanis
(BMS) \cite{Bakulev:2004cu}, who used a virtuality  of $\lambda_q^2 =
(0.4\pm0.1)\textrm{GeV}^2$ 
and extracted the values of $a_2$ to $a_{10}$ from the non-local condensate
sum rules.  The non-local quark condensate represents a partial resummation
of the operator product expansion to all orders, for the vacuum expectation
value of the (non-local) operator $\langle\bar{q}(0)E[0, z]q(z)\rangle$.
This is expressed in terms of analytic functions $F_{S,V}$, whose
derivatives are related to condensates of the corresponding dimension 
\begin{eqnarray}
  \label{eq:NLC}
  M(z)\equiv\langle\bar{q}(0)E[0,
  z]q(z)\rangle=\frac{\langle\bar{q}(0)q(0)\rangle}{4}\left[F_S(z^2) - \frac{i\hat{z}}{4}F_V(z^2)\right]
\end{eqnarray}
$E[0, z]$ is an appropriate Wilson line operator, taken in an appropriate
gauge so that the path-ordered exponential equals unity.  The lowest order
condensates are found via standard QCD sum rules \cite{Harrison:1998yr}, and are
\be
   Q^3 = \langle\bar{q}q\rangle \qquad Q^5 =
   ig_s\langle\bar{q}G^{\mu\nu}\sigma_{\mu\nu}q\rangle \qquad Q^6 =
   \langle (\bar{q}q)^2\rangle
\ee
The functions $F_{S,V}$ are given in terms of a virtuality distribution which
describes the distribution of the vacuum fields in terms of $\lambda_q^2$
\be
   F_{S,V}(z^2) = \int_0^\infty e^{\a z^2/4}f_{S,V}(\a)d\a
\ee
where
\be
   \int_0^\infty f_{S,V}(\a)\,d\a = \Bigg\{\begin{array}{cc} 1 & \textrm{S}\\ 0
     & \qquad \textrm{V and chiral limit}\\ \end{array}
\ee
In the BMS approach, this distribution is modelled by performing an expansion
where the large average virtuality (compared to the relevant hadronic scale)
is included in only the first term of the series, instead of the usual
expansion of the non-local condensate in terms of local condensates $\langle\bar{q}q\rangle$,
$\langle\bar{q}D^2q\rangle\dots$.  This corresponds to taking the form of the
virtuality distribution as $f_S(\a)=\delta(\a-\lambda_q^2/2)$ and $f_V \sim
\a_sQ^3\delta^\prime(\a-\lambda_q^2/2)$ \cite{Mikhailov:1986be}.

The pion distribution amplitude can then be connected directly to the non-local
condensates by means of a sum rule
\begin{eqnarray}
  \label{eq:NLCSR}
  f_\pi^2\phi_\pi^{\,BMS}(x)&=&\int_0^{s_0}\rho^{pert}(x,
  s)e^{-s/M^2}ds+\frac{\a_s\langle GG\rangle}{24\pi M^2}\Phi_G(x,
  M^2) \nonumber\\
  && \qquad +\frac{16\pi\a_s\langle\bar{q}q\rangle^2}{81M^4}\sum_{i=S, V,
  T_j}\Phi_i(x, M^2)
\end{eqnarray}
The $s$ and $M$ are Borel parameters and the $\Phi_i$ are local condensates,
with $i$ running over all scalar, vector and tensor types.  Using this sum
rule provides values for the first ten Gegenbauer moments of the pion
distribution amplitude; for $n>4$, the $a_n$ were found to be $\sim10^{-3}$
and so were neglected
in their calculations.  The non-zero moments were found to be
\be
    a_2(1.16\textrm{GeV}) = 0.19 \qquad  a_4(1.16\textrm{GeV}) = -0.13
\ee

Finally, the extraction of the moments $a_2$ and $a_4$ from the 
photon-pion transition form factor (as discussed above), has been developed and
refined by a succession of authors
\cite{Khodjamirian:1997tk,Schmedding:1999ap,Bakulev:2002uc,Bakulev:2003cs},
culminating in the constraint 
\be 
    a_2(1\textrm{GeV}) + a_4(1\textrm{GeV}) = 0.1\pm0.1 
\ee

The moments of the DA which are obtained in all of these models are
summarised in the table below, where the moments are all normalised to a scale of
$q^2 = 1$ GeV$^2$
 \begin{center}
    \begin{tabular}{|c|c|c|}\hline
    \textit{Model} \hspace{20pt} & $a_2$ \hspace{0pt}  & $a_4$ \\ \hline
    AS\hspace{20pt} & 0 & 0  \\ 
    CZ\hspace{20pt} & 0.56 & 0  \\ 
    BF\hspace{20pt} & 0.44 & 0.25  \\ 
    BMS\hspace{20pt} & 0.20 & -0.14  \\ 
    BZ\hspace{20pt} & 0.115& -0.015\\\hline  
  \end{tabular} \newline 

\end{center}

The models of the DAs are shown for comparison in
Figure \ref{fig:phimods}.\\

\section{New models for LCDA}\label{sec:newmodels}

Here we discuss the main new result of this chapter, where we introduce new
models for the leading-twist distribution amplitudes of the $\pi$ and the $K$
based on the fall-off behaviour of the $n$th Gegenbauer moment of these
amplitudes, $a_n$ in $n$, which is assumed to be power-like
\cite{Ball:2005ei}.  The models are
formulated in terms of a few parameters, notably the first inverse moment of
the DA (which for the $\pi$ is directly related to experimental data) and the strength of the fall-off of the $a_n$, and can be summed to
all orders in the conformal expansion, parameterising the full DA at a
certain energy scale.  Although we focus on the pseudoscalar mesons $\pi$ and
$K$, the models we propose are equally well applicable to vector mesons
$\rho,\, \omega,\, K^*$ and $\phi$.  

\subsection{Motivation}

 For many processes involving distribution amplitudes it is usually argued
 that a truncated conformal expansion would be sufficient for the
 calculation of physical amplitudes so long as the perturbative
 scattering amplitude is ``smooth'' --- the reason being the highly
 oscillatory behaviour of higher order Gegenbauer polynomials.
 In order to quantify this statement, consider the simplest case of one meson
 in the initial or final state, so that the convolution integral reads
 \begin{equation}
 I = \int_0^1 du\,\phi(u) T(u)
 \end{equation}
 where $T$ is the perturbative scattering amplitude.

This can be separated into three cases:
 \begin{itemize}
 \item[(i)] $T$ is nonsingular for $u\in[0,1]$,
 \item[(ii)] $T$ has an integrable singularity at one of the endpoints,
 \item[(iii)] $T$ contains a non-integrable singularity at one of the
   endpoints.
 \end{itemize}
 As a typical example for case (i), consider $T(u) = \sqrt{u}$, which yields
 $$\int_0^1 du\,\phi(u) \sqrt{u} = \sum_{n=0}^\infty \frac{(-1)^n
   36(n+1)(n+2)}{(2n-1)(2n+1)(2n+3)(2n+5)(2n+7)} \, a_n$$
 This result implies a strong fall-off $\sim 1/n^3$ of the coefficients of
   higher Gegenbauer moments $a_n$: assuming $a_i\equiv 1$ for all $i$, already
   the first three terms in the sum account for 98.8\% of the full
   amplitude. In reality the convergence will be even better as all
   existing evidence points at $|a_n|\ll 1$ for $n\geq 1$. 

 As an example for case (ii) consider $T(u) = \ln u$, giving 
 $$\int_0^1 du\,\phi(u) \ln(u) = -\frac{5}{6}\, a_0 + \sum\limits_{n=1}^\infty
 \frac{(-1)^{n-1}}{n(n+3)}\, 3 a_n$$
 The singularity at $u=0$ evidently worsens the convergence of the
 series; again assuming $a_i\equiv 1$, the first three terms now
 overshoot the true result by 35\%. In order to approximate the full
 amplitude to within 5\% one now has to include nine terms, but the
 convergence will again be better in practise, thanks to the fall-off
 of $a_n$ with $n$.

 Case (iii) is more complicated and  depends on the
 asymptotic behaviour of the $a_n$. For $T(u)=1/u$, for instance, one obtains
 \begin{equation}
   \label{eq:nn}
    \int_0^1 du\,\phi(u)\,\frac{1}{u} = 3 \sum_{n=0}^\infty (-1)^n a_n
 \end{equation}
Here the amplitude is finite only if the $a_n$ fall off sufficiently
 fast in $n$.
 For stronger endpoint divergences the coefficients multiplying the $a_n$
 start to grow in $n$ and for $T\sim 1/u^2$ the integral diverges, even
 for the asymptotic DA, which would indicate a breakdown of
 factorisation for that process.

This discussion suggests that models of $\phi$, based on a conformal
 expansion truncated after the first few terms, can be used appropriately for
 cases (i) and (ii), but are less reliable for case (iii). Convolutions with
 $T\sim 1/u$ are very relevant both in hard perturbative QCD
 e.g. $\gamma\gamma^*\to\pi$ (Section \ref{sec:nonpert}), and in decays such
 $B\to\pi\pi$. The $a_n$ can have different weighting in the different
 convoluted amplitudes, and have the highest impact for the convolutions of type
 (iii).  Instead of basing a parameterisation of $\phi$ on $a_2$
 and $a_4$ (taking all $a_{n>4}=0$), it is not unreasonable to use the case in
 (\ref{eq:nn}), where all $a_n$ enter with the highest possible weight factor. This is the
 basic idea behind our models of leading-twist DAs.

\subsection{Model construction}

In beginning to construct a model of the (leading-twist) pion DA we consider
 the two properties that any viable model must fulfil.  Firstly, the large-$q^2$
 behaviour must be satisfied for all light-meson DAs, in that the
 distribution amplitude must tend to the asymptotic form $\phi_\pi(u,
 \mu^2\to\infty) = 6u(1-u)$.  Also, in order for QCD factorisation in B decays
 to make sense the first inverse moment, $\int_0^1 du\,\phi_\pi(u)/u$, which
 as we have seen is related to the $\pi\gamma\gamma^*$ transition form
 factor, must exist.  Both these conditions are fulfilled for models
 based on a truncated conformal expansion, which has in addition the
 prediction that for $u\to 0,1$, $\phi\sim u(1-u)$, independently of
 the factorisation scale.   This prediction is in
 general, not fulfilled for models that are not truncated at fixed order
 in the conformal expansion, as shown below. 

As was introduced in equation (\ref{eq:delta1}), the first inverse moment of
the pion DA $\phi_\pi$ is related to the sum of all the Gegenbauer moments.
We define this as our first ``important parameter'', $\Delta$, via:
\be
   \label{eq:delta}
   \int_0^1du\frac{\phi(u, \mu^2)}{3u}\equiv\Delta(\mu^2) = 1 +
   \sum_{n=1}^\infty(-1)^na_n(\mu^2)  
\ee

%%  Available experimental data for the $\pi$, as summarised in
%%  Ref.~\cite{Stefanis1},  
%%  point at a value of $\Delta$ around $1.1$
%%  at the scale $\mu\approx1.2\,$GeV, which implies that the infinite sum
%%  be convergent. Hence, even at the comparatively low scale 1.2$\,$GeV the
%%  $a_{2n}$ must fall off fast enough in $n$. 

If this sum is to be convergent then the $a_{2n}$ must fall-off in $n$
sufficiently fast (recalling that for the $\pi$, all odd-numbered $a_n$
vanish due to G-parity).  If we assume an asymptotic equal-sign behaviour for
the moments with large $n$, then the slowest possible fall-off is power-like
$a_{2n}\sim 1/n^p$, with $p$ slightly larger than 1.  When the
distribution amplitude is defined with a power-like fall-off it is possible to
explicitly sum over the Gegenbauer moments.  This is done using the
generating function for the Gegenbauer polynomials:
\be
  \label{eq:genfun}
  f(\xi,t) = \frac{1}{(1-2 \xi t + t^2)^{3/2}} = \sum_{n=0}^\infty
  C_n^{3/2}(\xi)\, t^n
\ee

To show this, we begin from the full conformal expansion for the pion distribution
amplitude, suppressing notation for the $\mu^2$-dependence for compactness
and defining $\bar{u}\equiv (1-u)$.  We define the moments with a power-like
fall-off as
\be
   a_n = \frac{1}{(n/b + 1)^a}
\ee
for $n$ even.  This gives us an expression for the DA (for even $n$) of 
\begin{eqnarray*}
   \phi^+(u) &=& 6u\bar{u}\,\sum_{n=0}^\infty C_n^{\,3/2}(2u-1)\,a_n \nonumber\\
   &=&
   6u\bar{u}\,\sum_{n=0}^\infty C_n^{\,3/2}(2u-1)\,\left[\frac{1}{(n/b + 1)^a}\right]
\end{eqnarray*}
where $\phi^+$ denotes the same-sign behaviour of the Gegenbauer moments.  We can quite easily extend this across all values of $n$, giving
\begin{eqnarray*}
   \phi^+(u) = 3u\bar{u}\,\sum_{n=0}^\infty C_n^{\,3/2}(2u-1)\,\big(1+(-1)^n\big)\left(\frac{b}{b+n}\right)^a
\end{eqnarray*}
We can express the $(a, b)$ dependent part of this equation in terms of an
analytic integral equation 
\be
   \left(\frac{b}{b+n}\right)^a =
   \frac{1}{\Gamma(a)}\int_0^1\,dt\,(-\ln{t})^{a-1}\,t^{n/b} \nonumber
\ee
giving
\be
   \phi^+(u) = \frac{3u\bar{u}}{\Gamma(a)}\int_0^1\,dt\,(-\ln{t})^{a-1}\,\sum_{n=0}^\infty
   C_n^{\,3/2}(2u-1)\,\big(1+(-1)^n\big)\,t^{n/b} \nonumber
\ee
The sum can be split into two parts
\be
   \sum_{n=0}^\infty C_n^{\,3/2}(2u-1)\,t^{n/b} + \sum_{n=0}^\infty C_n^{\,3/2}(2u-1)\,(-t)^{n/b}\nonumber
\ee
which can be directly related to the generating function given in
equation (\ref{eq:genfun}), and gives a final expression for the DA of 
 \begin{equation}\label{eq:eqx} 
 \tilde\phi_{a,b}^+(u) = \frac{3u \bar u }{\Gamma(a)}\,
  \int_0^1 dt (-\ln t)^{a-1}\, \bigg( f\left(2u-1,t^{1/b}\right)
  + f\left(2u-1,-t^{1/b}\right)\bigg)
  \end{equation}

Using an analogous method we can obtain a re-summed DA with alternating-sign
behaviour of the Gegenbauer moments,
  $$a_{n} = \frac{(-1)^{n/2}}{(n/b+1)^a} \quad \mbox{for $n$ even}$$
of
\begin{equation}\label{eq:eqy} 
  \tilde\phi_{a,b}^-(u) = \frac{3u \bar u }{\Gamma(a)}\,
   \int_0^1 dt (-\ln t)^{a-1}\, \bigg(f\left(2u-1,it^{1/b}\right)
  + f\left(2u-1,-it^{1/b}\right)\bigg)
  \end{equation}
From equation (\ref{eq:delta}) we can find the corresponding values of
$\Delta$ 
\be
   \Delta_{a,b}^+ = (b/2)^a \zeta(a,b/2) \qquad \Delta_{a,b}^- = (b/4)^a (\zeta(a,b/4) -
\zeta(a,1/2+b/4))
\ee
 where $\zeta(a,s)= \sum_{k=0}^\infty 1/(k+s)^a$ is a generalisation of the Riemann
 zeta function known as the Hurwitz zeta function.  In order to obtain
 models for arbitrary values of $\Delta$ we split off the asymptotic DA and
 write
\begin{eqnarray}\label{eq:model}
  \phi^\pm_{a,b}(\Delta) &=& 6 u \bar u + \frac{\Delta-1}{\Delta_{a,b}^\pm -1}
  \left(\tilde{\phi}^\pm_{a,b}(u) - 6 u \bar u\right) 
  \end{eqnarray}
valid for $a\ge1$ and $b>0$.  This equation implies that the asymptotic DA is
recovered for $\Delta =1 $ and also from $\phi^+_{a,b}$ in the limit
$a\to1$.  Examples of these DAs for various values of $a,
b$ are shown in Figures \ref{fig:mods} and \ref{fig:mods2} in the next section. 

The fall off of the $a_n$ in inverse powers of $n$ gives a compact,
closed expression for the DA using equation (\ref{eq:model}).  This sort of
fall-off behaviour is in fact intrinsic to QCD, which we can see if we
consider the behaviour of the 
moments $a_n(\mu^2)$ under a change of scale.  Taking the models defined
at the hadronic scale $\mu\sim1.2$GeV, we know the $a_n$ scale with $\mu$
according to equation (\ref{eq:scaling}).  For large $n$ we have
  $$\gamma_0^{(n)}\stackrel{n\to\infty}{\approx} 8 C_F \ln n + O(1)
  $$
  and 
  $$ a_n(Q^2) \stackrel{n\to\infty}{\approx} \frac{1}{n^{4 C_F/\beta_0\,
      \ln (1/L)}} \,a_n(\mu^2)$$
  with $L = \alpha_s(Q^2)/\alpha_s(\mu^2)$. This shows that the leading-order
      scaling induces a power-like fall-off of the $a_n$, at least for large
      $n$.  Another consequence of this is that as $Q^2\to\infty$ ($L\to 0$)
      the suppression of the higher order moments is power-like so that
  $$\tilde{\phi}^{\pm}(Q^2\to\infty) = \tilde{\phi}^\pm(a\to\infty) = 6 u
      (1-u)$$
 Hence the DAs $\phi_{a,b}^\pm$ defined in (\ref{eq:model}) approach the
      asymptotic DA in this limit, so both of the necessary conditions for
      the DA construction are satisfied. 
\subsection{Properties of model DAs}

 Once the parameters $\Delta,\,a$ and
$b$ are specified at a certain reference scale, the model DA can be evolved
to a different scale via the evolution equation (\ref{eq:ER-BL}), using the
method which we describe in Section \ref{sec:numev} -- which shows how the evolution
equation can be solved numerically to leading logarithmic accuracy.  The
procedure for the full analytic evolution of a DA expressed as a conformal
expansion is included for completeness in Appendix \ref{chp:AppB}.  

%  The change of the
%% model DAs $\phi^\pm_{3,3}(1.2)$ is illustrated in
%% Fig.~\ref{fig:extra}, for an evolution from $\mu=1.2\,$GeV to
%% $\mu=m_b=4.8\,$GeV. 
%% \FIGURE{
%%  $$\epsfxsize=0.45\textwidth\epsffile{fig0a.eps}\qquad
%%  \epsfxsize=0.45\textwidth\epsffile{fig0b.eps}$$
%%  \caption[]{Evolution of $\phi^+_{3,3}(1.2)$ (left) and
%%  $\phi^-_{3,3}(1.2)$ (right) from $\mu=1.2\,$GeV (solid curves) 
%% to $\mu=4.8\,$GeV (dashed curves). 
%%  For comparison we also show the asymptotic DA
%%  (dash-dotted curves).}\label{fig:extra}}

   After evolution to higher scales, the DA is no longer described by the
   function $\phi^\pm_{a,b}(\Delta)$ with a simple suitably chosen set of
   $\Delta,\,a$ and $b$.  This lack of ``form-invariance'' under changes of
   scale is not however disadvantageous to our model.  We determine the
  value of $\Delta$ at some low scale around $\mu = 1.2$GeV  (chosen
   according to the analysis of \cite{Bakulev:2004cu }) and ideally the other
   parameters $a,b$ could then be fixed from some experimental or theoretical
   determination.  There is an analogy to this elsewhere in QCD, for the
   \textit{parton distribution function}, which has a 
   parameterisation fixed at a low scale (typically $\sim2$ GeV), but has parameters
   which are fitted to experimental data obtained at a large 
   variety of scales \cite{Martin:1987vw,Martin:1988aj,Harriman:1990hi}. 

\begin{figure}[h!]
  $$\epsfysize=0.3\textwidth\epsffile{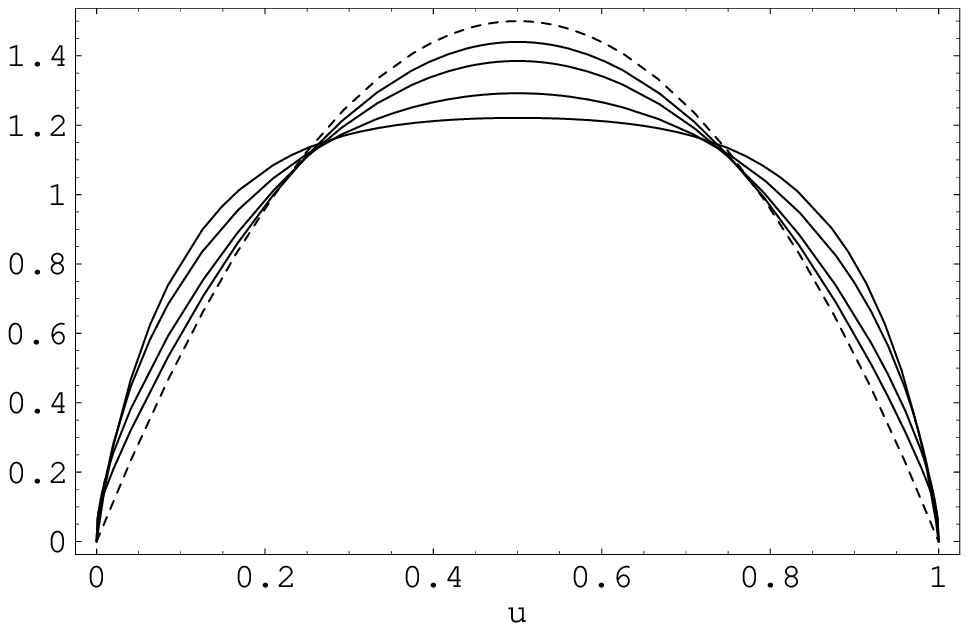}\qquad
  \epsfysize=0.3\textwidth\epsffile{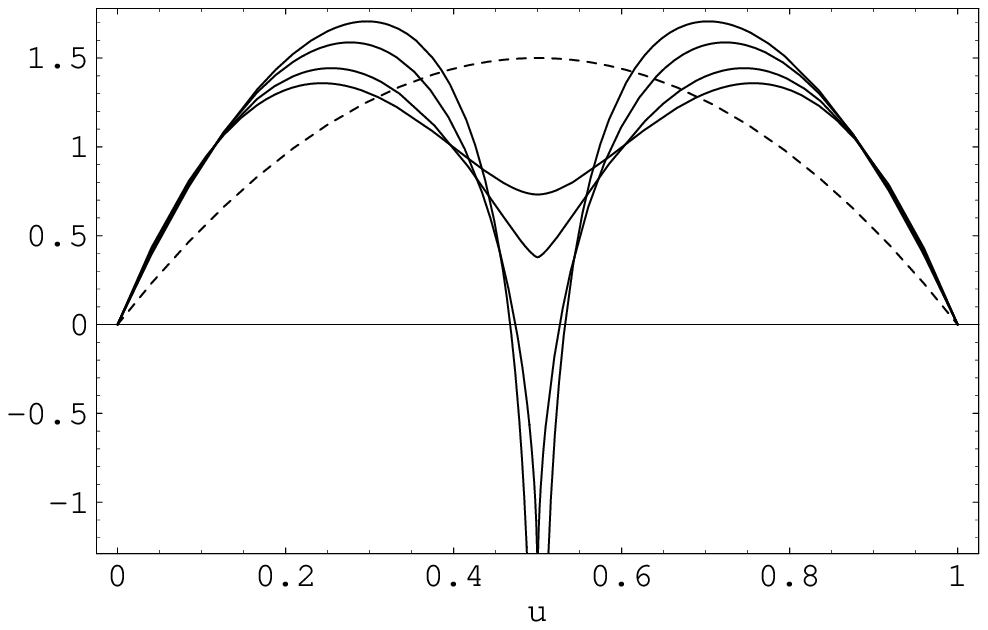}$$
  \caption[]{Left: Examples for model DAs $\phi_{a,b}^+$ as 
 functions of $u$, for
    $a=1.5,2,3,4$, and constant $b=2$ and $\Delta=1.2$ 
  (solid curves). For $a\to 1$, $\phi^+_{a,b}$  approaches the asymptotic
    DA. Right: the same for $\phi_{a,2}^-$.  The asymptotic DA is also
    shown for comparison (dashed curve).}\label{fig:mods}
\end{figure}

\begin{figure}[h!]
  $$\epsfysize=0.3\textwidth\epsffile{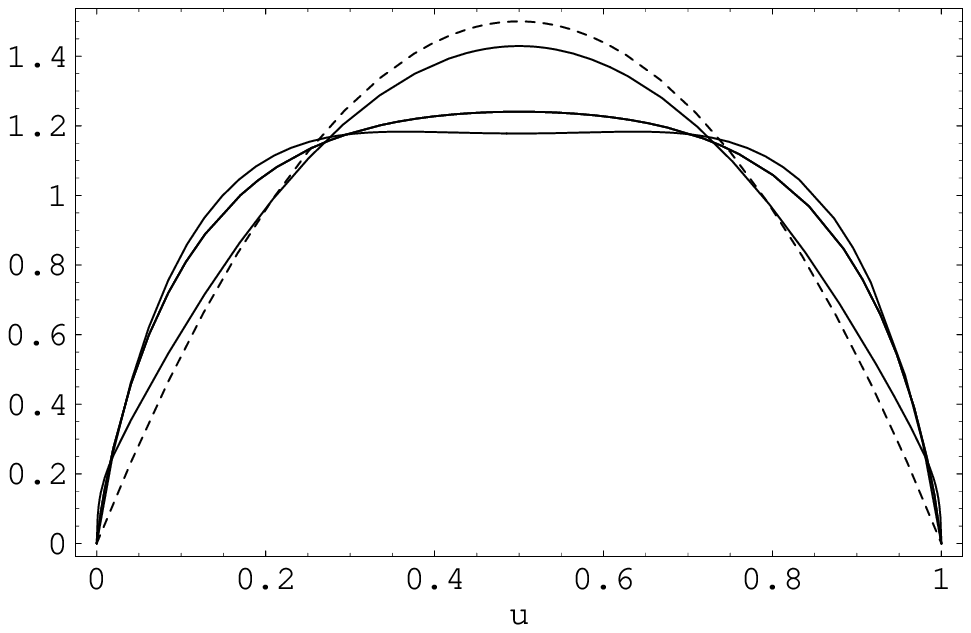}\qquad
  \epsfysize=0.3\textwidth\epsffile{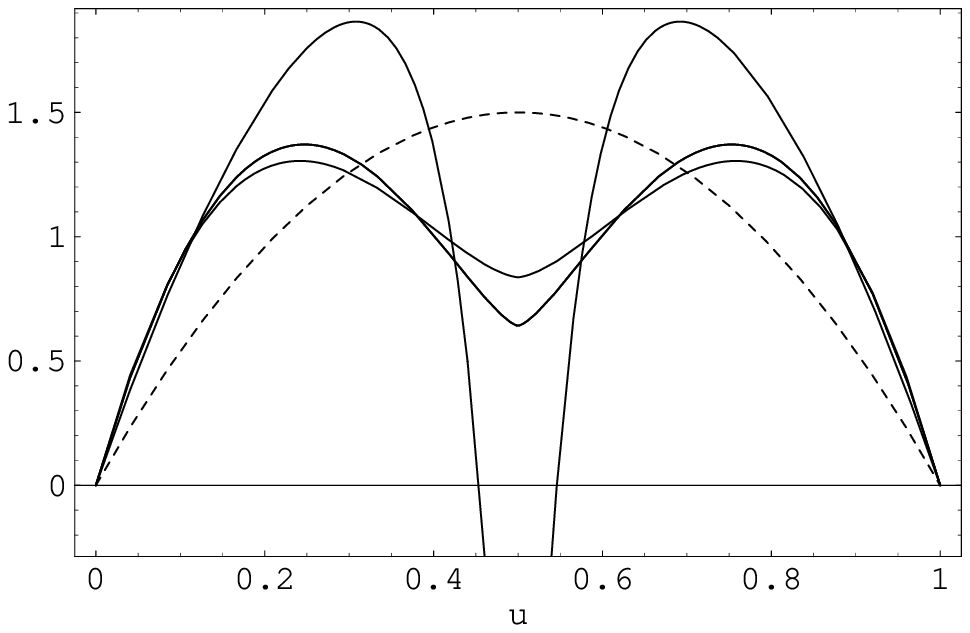}$$
  \caption[]{Left: Examples for model DAs $\phi_{a,b}^+$ as 
 functions of $u$, for
    $b=0.1,1,10$, and constant $a=3$ and $\Delta=1.2$ 
  (solid curves).  Right: the same for $\phi_{3,b}^-$.  The asymptotic DA is also
    shown for comparison (dashed curve). }
   \label{fig:mods2}
\end{figure}

Figures \ref{fig:mods} and \ref{fig:mods2} show that the two models $\phi^+$ and $\phi^-$ have quite
a dissimilar functional dependence on the momentum fraction $u$, notably
that $\phi^-$ becomes non-analytic at $u=1/2$ for values of $a\le3$.  The
spike at $u=1/2$ characteristic for $\phi^-$ causes these models to
significantly deviate from the asymptotic distribution amplitude even for
$\Delta $ close to 1.   

What we can also discern from the form of $\phi_{a,b}$ is that it is much more
sensitive to variation in the fall-off parameter $a$, than $b$.  Since the experimental data available are too
scarce to constrain all of the parameters of the model we fix $b\equiv2$.  Using the set of model DAs $\phi_{a,b}$ we can
find the possible values of the lowest non-zero Gegenbauer moments $a_2$ and
$a_4$.  For $a = 1$, $\phi$ approaches the asymptotic DA with $a_2 = a_4 = 0$;
at $a\to\infty$ the models approach the standard NLO fixed order conformal
expansion with $a_2 = \Delta -1$ and $a_4 =0$.  

As discussed above, for models based on the truncated conformal expansion, 
the endpoint behaviour of the DA (as $u\to0,1$) is assumed to follow 
$\phi\sim u(1-u)$.  This is no longer true for our new models for all 
possible values of $a$.  For example, for the model $\phi^+$, we have
\begin{eqnarray*}
   \phi^+ &\sim& \sqrt{u(1-u)} \qquad \hspace{44.5pt}\textrm{for a = 2} \\
   \phi^+ &\sim& u(1-u)\ln{u(1-u)} \qquad \textrm{for a = 3} \\
   \phi^+ &\sim& u(1-u) \qquad \hspace{57pt}\textrm{for a $>3$}
\end{eqnarray*}

We see that the linear endpoint behaviour is only obtained for models 
with higher values of $a$.  This is not however a drawback to the model.  
The original argument in the literature for this endpoint behaviour was put 
forward in \cite{Chernyak:1983ej} from the calculation of DA moments via QCD sum rules:
\be
   \langle\xi^n\rangle = \int_0^1du\,\xi^n\phi(u)\stackrel{n\to\infty}{\sim}\frac{1}{n^2} \nonumber
\ee
This is a result of a leading order calculation of the perturbative 
contributions to the QCD sum rule.  The NLO result \cite{Ball:2003sc, Ball:1996tb} does not have the
same dependence, instead following  
$$\phi\sim u(1-u)\ln^2{\left(u/(1-u)\right)}$$
which is equivalent to $a_n\sim(1/n^3)$.  The large-$n$ behaviour of
the non-perturbative terms cannot be determined from the sum rules,
but there is no reason why it should not follow the behaviour of NLO
perturbation theory or some other scaling.  

\subsection{Constraints on model parameters}

The asymptotic distribution amplitude is recovered for all models with
$\Delta =1$ and with $\phi^+$ in the limit of
$a\to1$.  We must impose the constraint that
$a\ge1$, otherwise the models $\phi_{a,b}^\pm$ will not vanish at the
endpoints $u=0,\,1$.  In order to constrain our parameters, we can now
introduce the experimental restrictions on our parameter set $\Delta, a, b$.
As discussed above, we fix $b\equiv 2$ and we also fix our reference
scale as $\mu = 1.2$ GeV.  The available experimental data for the $\pi$, as
summarised in \cite{Bakulev:2004cu}, points to a value of $\Delta$ around $1.1$
at the scale $\mu\approx1.2\,$GeV.

If we require $a_2$ to be positive, which as seen in Section \ref{sec:premods}
is a reasonable conclusion from all of the previous determinations,
then we require $\Delta\ge1$.  We can infer an upper bound from experimental
data or by imposing the requirement that $a_2\le0.2$ -- the upper bound as
found by the most recent analyses.  This corresponding bound on $\Delta$ is
$\Delta\le1.2$ for $\phi^+$, and is smaller again for $\phi^-$.  The final
constraint we must consider is the allowed range of $\phi_\pi(1/2)$ from
light-cone sum rule analyses, which give \cite{Braun:1988qv}:
    $$0.9\leq \phi_\pi (1/2,1\,{\rm GeV}) \leq 1.5$$
 For $\Delta>1$, $\phi(1/2)$ is always smaller than 1.5, so only the lower
 bound is relevant. The upper bound on $\Delta^\pm$ for various values of a
 (as implied by the lower bound $\phi_\pi(1.2)\ge0.9$) is 

\begin{center}
 \begin{tabular}{|l|llllll|}
 \hline
 $a$ & 2 & 3 & 4 & 5 & 6 & $\infty$\\\hline
 $\Delta^+_{\rm max}$ & 2.04 & 1.58 & 1.43 & 1.36 & 1.33 & 1.27\\
 $\Delta^-_{\rm max}$ & 1.04 & 1.11 & 1.16 & 1.19 & 1.22 & 1.27\\\hline
 \end{tabular}\newline
\end{center}

The constraint on $\Delta^+$ is weaker than those discussed above, but for
$\phi^-$ the minimum value of $\phi_\pi(1/2)$ poses a nontrivial constraint
on $\Delta$.  Concentrating on the general characteristics of the
distribution amplitudes in order to retain a generality for the
other psuedoscalar and vector mesons, we do not refine these constraints
further using specific data for the pion.  It is therefore possible to draw
up a set of general constraints on the (symmetric) part of our model DAs:
\begin{itemize}
  \item $1\leq \Delta\leq 1.2$ with $0\leq a_2\leq 0.2$ for
      $\phi^+_{a,2}$: this is based on the observation \cite{Chernyak:1983ej} that
      DAs of mesons with higher mass tend to become narrower, 
  \item $1\leq \Delta \leq{\rm Min}(1.2,\Delta^-_{\rm max})$  for $\phi^-_{a,2}$,
      with $\Delta^-_{\rm max}$ given in the table above,
  \item $b=2$, lacking further data.
\end{itemize}

\subsection{Extension to $\phi_K$}

As mentioned before, we have to distinguish between those mesons for which the
$a_{\rm odd}$ vanish due to G-parity, such as the $\pi$, and the strange
mesons, for which the odd moments are in general non-zero and induce an
antisymmetric part into the DA. 

\begin{figure}[h]
  $$\epsfysize=0.3\textwidth\epsffile{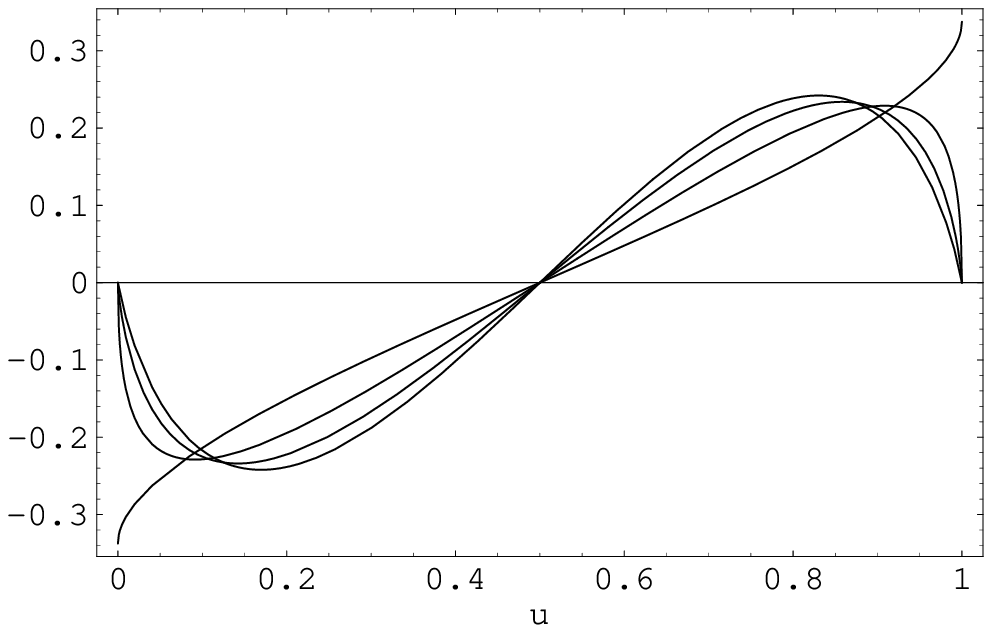}\qquad
  \epsfysize=0.3\textwidth\epsffile{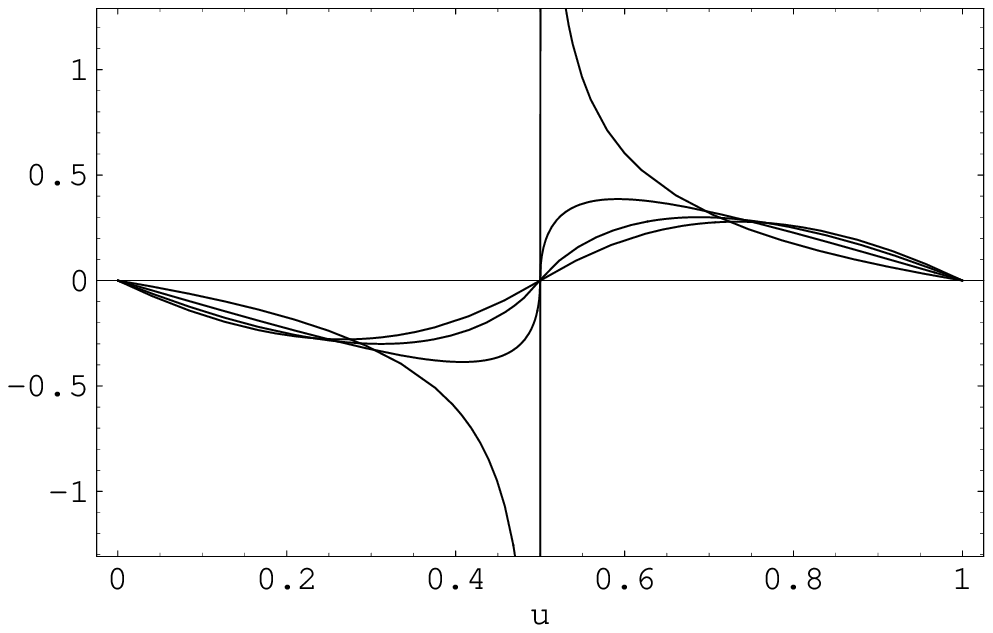}$$
  \caption[]{Models for the antisymmetric contributions to the twist-2
    DA for $a_1 = 0.15$. $\psi_c$ is shown as function of $u$ for
    $c\in\{1,2,3,4\}$.  Left: $\psi_c^+$, Right: 
  $\psi_c^-$.  Like
    the symmetric models $\phi_a^-$, $\psi_c^-$ 
  is non-analytic at $u=1/2$ for $c\leq 3$.}\label{fig:modsx}
\end{figure}

We can now consider the extension of our new model DAs to the leading-twist
DA of the Kaon, $\phi_K$.  This differs from $\phi_\pi$ by the contribution
of odd Gegenbauer moments.  We can construct a model for this antisymmetric
part in an analogous way as before, where we introduce $\Delta^{\rm asym}, c, d$ in
place of $\Delta, a, b$.  Since there is almost no information at all about
the form of the antisymmetric DAs we set $d\equiv 2$ immediately.  We have
 \begin{eqnarray}
 \tilde\psi_c^+(u) &= &\frac{3u \bar u }{\Gamma(c)}\, 
 \int_0^1 dt (-\ln t)^{c-1}\, \left( f(2u-1,\sqrt{t})
 - f(2u-1,-\sqrt{t})\right)\nonumber\\
 \tilde\psi_c^-(u) &= &\frac{3u \bar u }{i\Gamma(c)}\, 
 \int_0^1 dt (-\ln t)^{c-1}\, \left( f(2u-1,i\sqrt{t})
 - f(2u-1,-i\sqrt{t})\right) \nonumber\\
 \end{eqnarray}
The models give a value of the first odd moment as $a_1=(2/3)^c$.  If we wish
to redefine these expressions to account for arbitrary $a_1$, we can write  
\begin{equation}
  \psi_c^{\pm} = a_1 (3/2)^c \tilde\psi_c^\pm(u)
\end{equation}
An example of such a model is shown in Figure \ref{fig:modsx}. 
The scale evolution proceeds as for the symmetric part.  There is very little
information about the $a_1$ moment for the Kaon -- not
even the sign is known reliably \cite{Ball:2004ye}.  From a
heuristic standpoint it is expected $a_1\ge0$.  This is because the DA is expected
to be skewed towards larger values of $u$, since $u$ denotes the momentum of
the $s$-quark (the heavier quark) in the meson.  The first calculation using 
QCD sum rules was from Chernyak and Zhitnitsky \cite{Chernyak:1983ej}, who
confirmed this intuition with a central value of
\be
   a_1^K(1\,\textrm{GeV}) = 0.17 
\ee
This result was countered by the work of \cite{Ball:2003sc}, who claimed to
find a sign error in the CZ method as well as adding higher order radiative
corrections; this gave a value of $a_1$ of
\be
   a_1^K(1\,\textrm{GeV}) = -0.18 
\ee
It was reported in \cite{Ball:2004ye} that this re-analysis of the CZ sum
rule calculation gave a problem with the $q^2$ dependence of the form
factors.  In light of these thoughts we use the original result from
\cite{Chernyak:1983ej} and use the evolved value of $a_1 = 0.15$ at
$1.2$GeV.  We can then combine the asymmetric and symmetric parts of the
wavefunction to give a total DA for the $K$.  We can write
\be
   \Delta^{\textrm{tot},\pm} = \Delta + \Delta^{\textrm{asym},\pm}\nonumber
\ee
where $\Delta$ is the sum of the even-valued moments from the symmetric
contributions to the DA, and we defined $\Delta^{\textrm{asym},\pm}$
\begin{eqnarray*}
   \Delta^{\textrm{asym},+} &=& \int_0^1du\frac{\psi_c^+(u)}{3u} =
   -a_1(3/2)^c\zeta(c, 3/2) \\
 \Delta^{\textrm{asym},-} &=& \int_0^1du\frac{\psi_c^-(u)}{3u} =
   -a_1(3/4)^c\left[\zeta(c, 3/4)-\zeta(c,5/4)\right] \\
\end{eqnarray*}

\section{Numerical evolution of model DA}\label{sec:numev}

 Using the evolution equation for the meson DA 
 \be
    \mu^2\frac{\partial}{\partial\mu^2}\phi(u, \mu^2) = \int_0^1dv\,V(u,v,\mu^2)\phi(v,\mu^2)\label{eq:EK}
 \ee
 the kernel is given, to leading order in $\alpha_s$, as 
 $$V(x,y,\mu) = C_F\,\frac{\alpha_s(\mu)}{2\pi}\, V_0(x,y) + O(\alpha_s^2)$$
 with, for pseudoscalar mesons,
 \begin{eqnarray*}
 V_0(x,y) & = & V_{BL}(x,y) - \delta(x-y) \int_0^1 dz V_{BL}(z,y)\\
 V_{BL}(x,y) & = & \frac{1-x}{1-y}\left( 1 + \frac{1}{x-y}\right)
 \Theta(x-y) + \frac{x}{y} \left(1 + \frac{1}{y-x}\right) \Theta(y-x)
 \end{eqnarray*}

\begin{figure}[h]
  $$\epsfysize=0.4\textwidth\epsffile{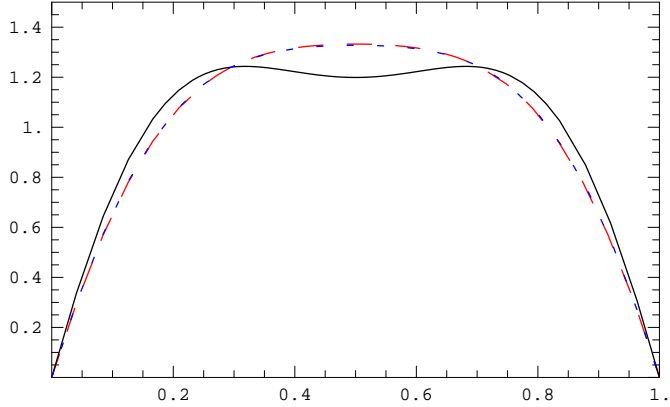}$$
  \caption[]{Numeric (dashed red line) and analytic evolution
    (dotted blue line) of sample DA from 1.2GeV to 5.3GeV; un-evolved DA
    shown for comparison.}\label{fig:numevol} 
\end{figure}

\begin{figure}[h]
  $$\epsfysize=0.3\textwidth\epsffile{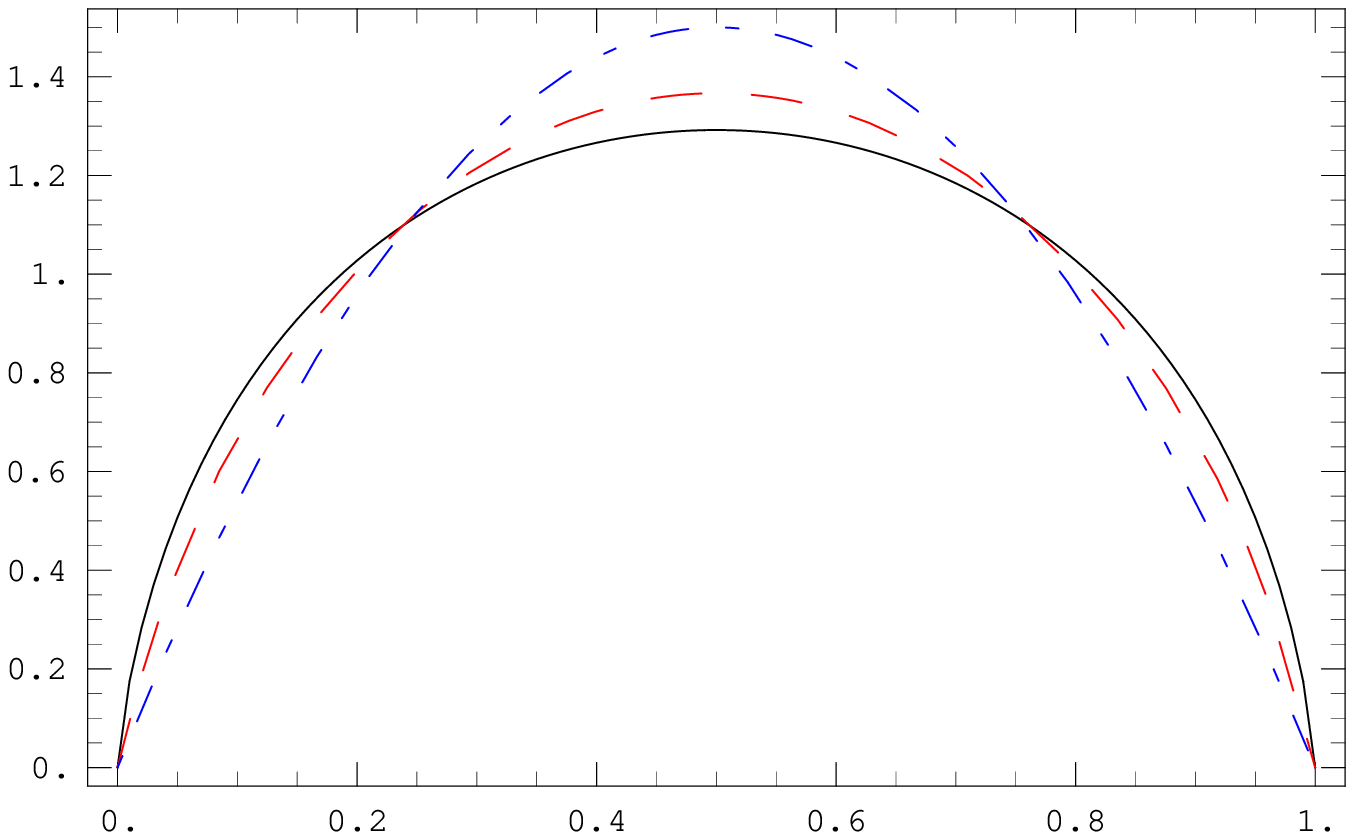}\qquad
  \epsfysize=0.3\textwidth\epsffile{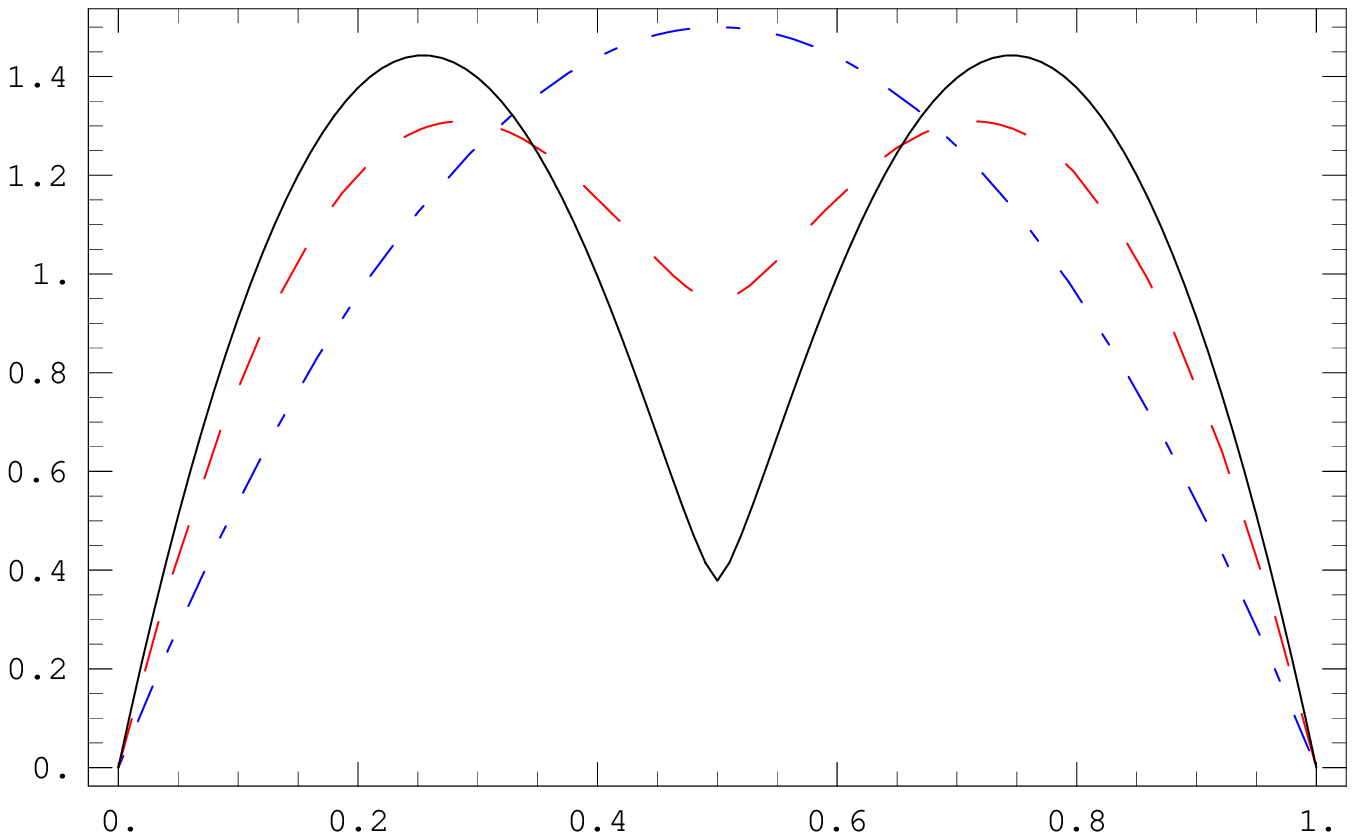}$$
  \caption[]{Evolution of model DAs. Left: $\phi_{3,3}^+(1.2)$ and Right:
  $\phi^-_{3,3}(1.2)$ from scale $\mu = 1.2\textrm{GeV}$ (solid curves) to
  $\mu = 4.8\textrm{GeV}$ (dashed red line).  We also show the asymptotic
  curve for comparison (dot-dash blue line).}\label{fig:evolution}
\end{figure}

Equation (\ref{eq:EK}) is not well suited to the numerical evolution of
$\phi$, as it requires the use of an explicit formula for $\alpha_s$, which
is itself the leading order solution of a renormalisation group equation. In
order to exactly reproduce the correct LO scaling behaviour of the Gegenbauer
moments numerically, 
we can rewrite (\ref{eq:EK}) in terms of a differential equation in
$\alpha_s$.  To LO this is written as 
 \begin{equation}
 -\frac{\beta_0}{2C_F}\,\alpha_s \,\frac{\partial}{\partial
  \alpha_s}\,\phi(x,\alpha_s) = \int_0^1 dy\,V_0(x,y)\phi(y,\alpha_s)
 \end{equation}
which can be solved iteratively using Euler's method:
 $$\phi(x,\alpha_s-\Delta \alpha_s) = \phi(x,\alpha_s) +
 \frac{2C_F}{\beta_0}\,\frac{\Delta\alpha_s}{\alpha_s} \,\int_0^1 dy
 V_0(x,y) \phi(y,\alpha_s)$$
We have checked that for $\Delta\alpha_s=0.01$ and with 21 mesh
points in $x$ we correctly reproduce the known scaling behaviour of
the truncated conformal expansion to within one per mille.  We can show this
 for example, using the standard conformal expansion truncated at $a_4$
\begin{equation*}
   \phi(u, \a_s) = 6u(1-u)\left[1 + a_2(\a_s) C^{3/2}_2(2u-1) + a_4(\a_s) C^{3/2}_4(2u-1)\right]
\end{equation*}
 using the values for the first two moments from Ball and Zwicky
 \cite{Ball:2004ye} given at $\mu=1.2\textrm{GeV} \Rightarrow \a_s^{LO} = 0.6$ as
 $a_2(0.6) = 0.115,\, a_4(0.6) =-0.015$.  Figure \ref{fig:numevol} shows the
 results for evolving this wavefunction up to a scale of 5.3 $(\a_s^{LO} =
 0.26)$ using the numerical method, and for
 the full analytical evolution using the formulae in Appendix \ref{chp:AppB}.
 As we can see clearly, both methods give equivalent results.
The change of our model DAs with the scale $\mu$ is shown for an example model
($\phi^\pm_{3,3}$) in Figure \ref{fig:evolution}.
\newpage
 \section{Application to $B\to\pi\pi$} 

It is at this point that we can consider the effects of the leading-twist DA
on the non-leptonic $B$ decays calculated within QCD factorisation.  As shown
in Chapter \ref{chp:QCDF}, the matrix elements determining the $B\to\pi\pi$
decay amplitudes are written as convolution integrals involving the leading
twist DA.  We discussed in Section \ref{sec:limit} how the recent
experimental data from the non-leptonic $B$ decays (specifically $B\to\pi\pi$)
imply the failure of QCD factorisation to explain the branching
ratios and CP asymmetry measurements.  
This section investigates the effect of a non-standard DA on the predictions
of QCD factorisation.

In this study, we use the full QCD factorisation formulae for the
factorisable contributions (including the calculated $1/m_b$ corrections) and
also include the model-dependent corrections for the hard-scattering terms
(parameterised by the complex number $X_H$) which we include at the ``default
level'' of $X_H = \ln{(m_\pi/\Lambda_h)}\sim2.4$ \cite{Beneke:2001ev}.  We do
not however include the non-factorisable weak annihilation terms.  We use the
input parameters as quoted in \cite{Beneke:2003zv} except for that of the
leading-twist $\phi_\pi$, which we replace by our model parameter and the
$B\to\pi$ transition form factor which we take from light-cone sum rules
as found in \cite{Ball:2004ye}.

We begin by considering the time-dependent CP asymmetry in $B\to\pi^+\pi^-$,
which is defined as
\begin{eqnarray}
  A_{CP}^{\pi\pi}&=&\frac{\Gamma(\bar B_0\to\pi^+\pi^-)-
\Gamma(B_0\to\pi^+\pi^-)}
  {\Gamma(\bar B_0\to\pi^+\pi^-)+\Gamma(B_0\to\pi^+\pi^-)} =
  S_{\pi\pi} \sin{(\Delta m t)} + C_{\pi\pi} \cos{(\Delta m t)}\nonumber\\
\end{eqnarray}
where the mixing-induced and direct asymmetries ($S_{\pi\pi}$ and
$C_{\pi\pi}$ respectively) depend on the unitarity triangle angles $\beta$ and $\gamma$ via
\begin{equation}\label{eq:SC}
  S_{\pi\pi} =\frac{2\,\textrm{Im}\lambda_{\pi\pi}}{1+|\lambda_{\pi\pi}|^2}
   \qquad 
 C_{\pi\pi} =\frac{1-|\lambda_{\pi\pi}|^2}{1+|\lambda_{\pi\pi}|^2}
\end{equation} 
with
\begin{equation}
\lambda_{\pi\pi}=e^{-2i\beta}\frac{e^{-i\gamma}+P/T}{e^{i\gamma}+P/T}
\end{equation}

In practise, we calculate the CP asymmetry via its parameterisation in Wolfenstein
parameters, so that the CKM factors read
\be
   e^{\pm i\gamma} = \frac{\rhob \pm i\etab}{\sqrt{\rhob^2 + \etab^2}} \qquad
   e^{-2i\beta} = \frac{(1-\rhob)^2 - \etab^2 - 2i\etab(1-\rho)}{(1-\rhob)^2
    + \etab^2}
\ee
Although the penguin-to-tree ratio $P/T$ is highly suppressed, it is
not negligible and can be expressed via QCD factorisation in terms of CKM
phases and pure strong interaction parameters.
\begin{eqnarray}
    \frac{P}{T} &=&  \frac{re^{i\phi}}{\sqrt{\rhob^2 + \etab^2}} \nonumber\\
  &=&\left(\frac{1}{\sqrt{\rhob^2 +\etab^2}}\right)
    \frac{\left(a_4^c(\pi\pi) + r_\chi^\pi a_6^c(\pi\pi)\right) +
      \left(a_{10}^c(\pi\pi) + r_\chi^\pi a_8^c(\pi\pi)\right)}{a_1(\pi\pi) +
     \left(a_4^u(\pi\pi) + r_\chi^\pi a_6^u(\pi\pi)\right) +
    \left(a_{10}^u(\pi\pi) + r_\chi^\pi a_{8}^{u}(\pi\pi)\right)} \nonumber\\
\end{eqnarray}
$P/T$ is then given by a ratio of polynomials in the Gegenbauer polynomials
$a_n$ with complex coefficients. This enables us to write the time-dependent CP
asymmetry as 
\be
   S_{\pi\pi} = \frac{2\etab\left[\rhob^2 + \eta^2 - r^2 - \rhob(1-r^2) +
       (\rhob^2 + \etab^2 - 1)\,r\cos{\phi}\right]}{\left[(1-\rhob^2) +
       \etab^2\right]\left[\rhob^2 + \etab^2 + 2r\rhob\cos{\phi}\right]}
\ee

In the limit where  $P/T$ is zero this reduces to
$S_{\pi\pi}=\sin2\alpha$. 

\begin{figure}[h]
  $$\epsfysize=0.3\textwidth\epsffile{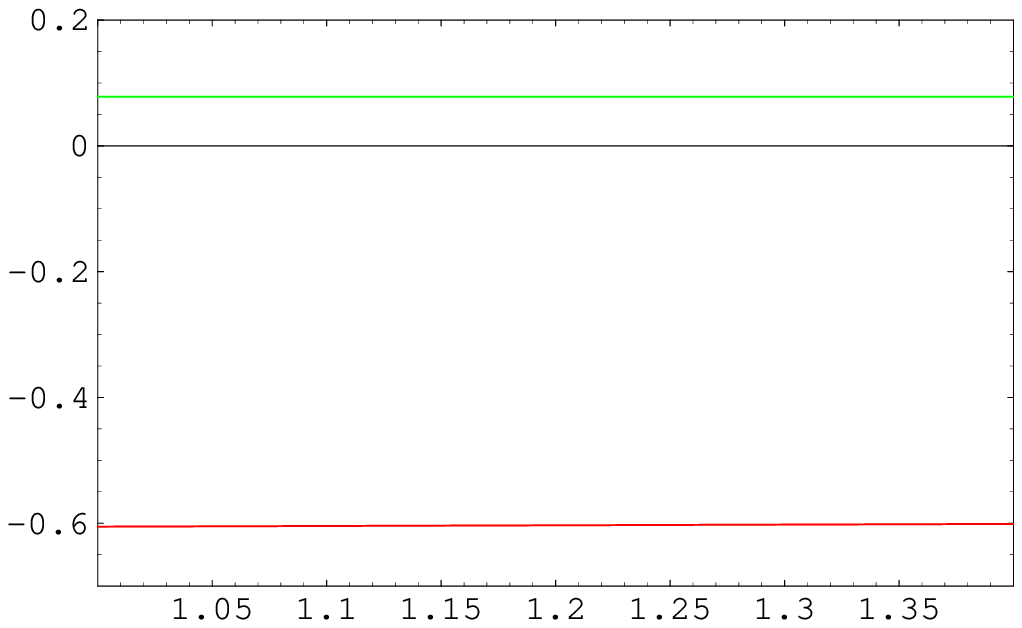}\qquad
  \epsfysize=0.3\textwidth\epsffile{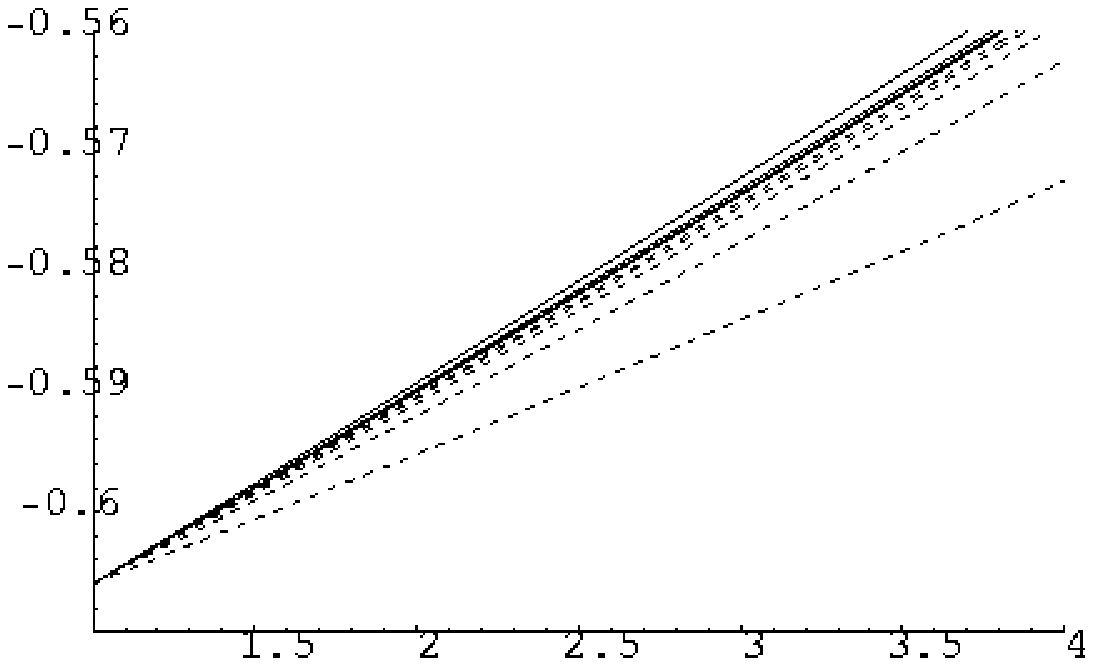}$$
  \caption[]{CP asymmetries in $B\to\pi^+\pi^-$ decays in QCD factorisation.
         Left: In physical region of $\Delta$, $C_{+-}$
  (green) and $S_{+-}$ (red).  Right: $S_{+-}$ as a function of $\Delta$ for
  $a=2,3\dots6$, shown for models with same-sign fall-off (dashed) and
  alternating fall-off (solid lines).  Curves converge as $a$ increases. }\label{fig:CPpipi}
\end{figure}

The variation of the CP asymmetry in terms of $\Delta$ is shown in
Figure \ref{fig:CPpipi}.  The left-hand figure shows the variation within the
physical region for $\Delta$, and shows that both $S_{+-}$ and the direct
asymmetry $C_{+-}$ are 
largely independent of $\Delta$.  The right-hand figure shows the level of
enhancement required to approach a 10\% change in the value of the
asymmetry, and that there is an increased effect for larger value of $a$.
The effect is only significant for unrealistically large values of $\Delta$.
The current experimental results \cite{Aubert:2005av,Abe:2005dz}
\begin{eqnarray*}
  S_{\pi\pi} &=& -0.30\pm0.17\pm0.03 \hspace{10pt}(\textsc{BaBar}) \\
 S_{\pi\pi} &=& -0.67\pm0.16\pm0.06  \hspace{10pt}(\textrm{Belle})
\end{eqnarray*}
can only be accommodated using very extreme values of $\Delta$. For example,
taking a model with same-sign fall-off, to reproduce the {\textsc{BaBar}} result with
$a=2$ would require a value of $\Delta>10$ to be within the $1\sigma$ band
and $\Delta=20$ to approach the central value.  This also produces unphysical values
for the Gegenbauer moments, $a_2(2.2$ GeV$)=7.7$ and $a_4(2.2$
GeV$)=0.5$.  Obtaining a value of $S_{\pi\pi}=-0.67$ is not possible 
for values of $\Delta\geq1$, which is required to keep $a_2$ positive. 

\begin{figure}[h]
  $$\epsfysize=0.3\textwidth\epsffile{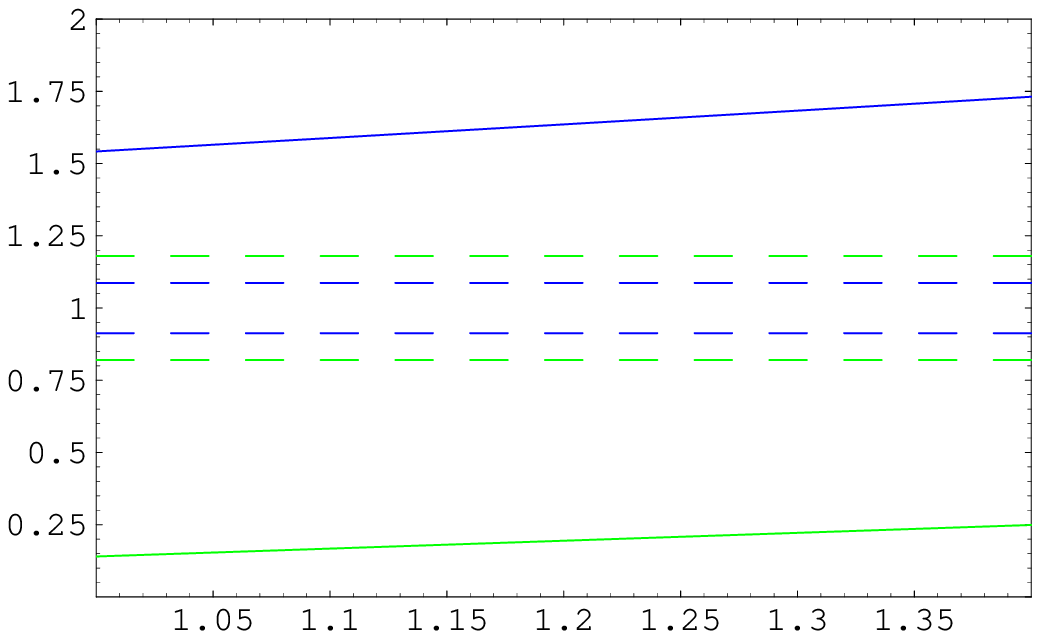}\qquad
  \epsfysize=0.3\textwidth\epsffile{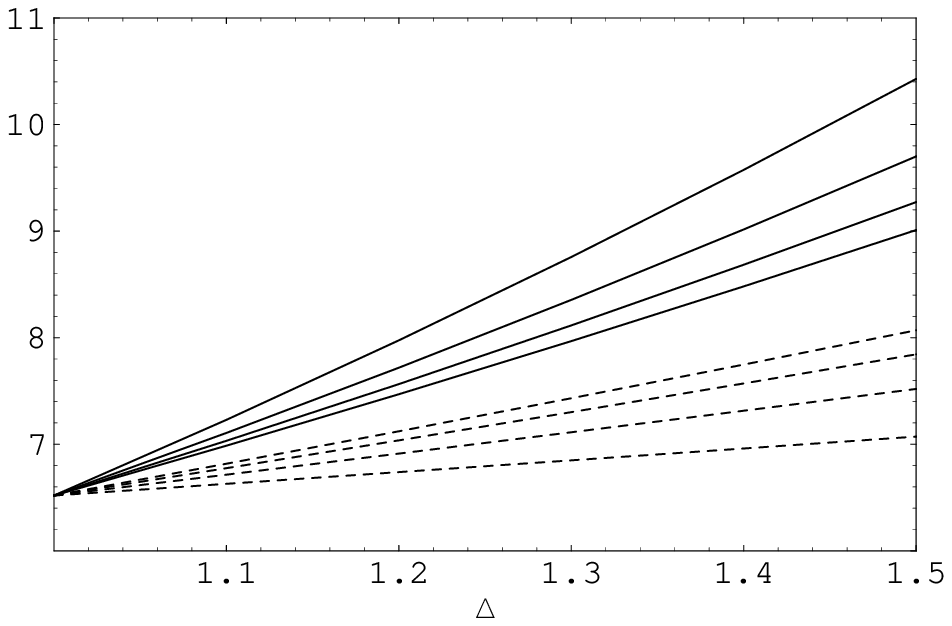}$$
  \caption[]{Left: Ratio of CP-averaged branching ratios for $B\to\pi\pi$
  decays as compared with experimental result as a function of
  $\Delta$, for model DA $\phi^+_{2,2}$; $B^0\to\pi^+\pi^-$ (blue) and
  $B^0\to\pi^0\pi^0$ (green).  Right: Example of $\Gamma(B\to\pi^+\pi^-)$
  (units $10^{-6}$) as
  a function of $\Delta$ for $a=2,3\dots 5$ for same-sign fall-off (dashed
  lines) and alternating (solid lines).  Curves converge for increasing $a$.}\label{fig:BRpipi}
\end{figure}

We now move on to study the effect of the model DAs on the $B\to\pi\pi$
branching fractions.  We find that the effect is significantly more
pronounced on the branching ratios than for the CP asymmetries, as there is
no longer the near 
cancellation of terms that occurs in the calculation of $A_{CP}$. 
Considering first the decay of $B\to\pi^+\pi^-$, the central value from QCD
factorisation (for the asymptotic DA) is found to be 
\begin{equation}
{\rm BR}(B\to\pi^+\pi^-)=
5.5\times10^{-6}|0.25e^{i\cdot15^\circ} + e^{-i\gamma}|^2
\end{equation}
where $\gamma= 60^\circ\pm7^\circ$ \cite{Bona:2005vz}, and the
explicit dependence of the branching ratio on the Gegenbauer moments is again a
polynomial in $a_n$.  Figure \ref{fig:BRpipi} shows the variation of the
branching ratios 
for the model DAs compared with the current experimental data.
These graphs show that large increases away from the asymptotic value are
possible by increasing $\Delta$ within its physical range.  However,
we see that in the case of $\pi^0\pi^0$ (where unlike $\pi^+\pi^-$, the enhancement is toward the
experimental value) it is still not possible to reach the $1\sigma$ error
bound within the allowed range of $\Delta$. 

This study does not aim to be a full error analysis and the graphs
presented here are illustrative of the discrepancies that exist in the
$\pi\pi$ system for QCD factorisation predictions. These results show that
the lack of agreement between the predictions and the experimental results is
not a result of the uncertainties from the leading-twist distribution
amplitude, not even with our model DA with parameters well outside the allowed
range.  An alternative explanation of these discrepancies, 
namely the over-neglect of suppressed non-factorisable corrections, is
discussed next in Chapter \ref{chp:NFC}.

%%% Local Variables:
%%% mode: latex
%%% TeX-master: t
%%% End:

%% file: chapter4sm.tex
\chapter{Non-factorisable corrections to charmless B-decays}\label{chp:NFC}
\begin{center}
  \begin{quote}
    \it
    It does not do to leave a live dragon out of your calculations 
  \end{quote}
\end{center}
\vspace{-4mm}
\hfill{\small ``The Hobbit'', J.R.R. Tolkien}
\vspace{5mm}

This chapter is devoted to a study of the non-factorisable corrections to
non-leptonic charmless decays of the $B$-meson.  This subset of decays is a
crucial testing ground for QCD factorisation (QCDF).  As we have access to
increasingly accurate measurements in the flavour sector, it is essential to
exploit the data to better understand the limitations and potential
``pitfalls'' with QCDF.  Indeed, before we enter the era of LHC physics we
ideally need to have complete grasp of the theoretical framework of $B$
decays, as it is expected this sector will provide complementary,
indirect evidence for new physics searches.  As we discussed in Chapter
\ref{chp:QCDF}, QCDF is only marginally consistent with the data on charmless
B-decays, and investigation is warranted to study the source of this
discrepancy.     

It is known that the factorisable contributions to exclusive $B$ decays can
be estimated using the formalism of QCD factorisation which is exact in the
heavy quark limit, $m_b\to\infty$.  Contributions that cannot 
be treated in this framework, i.e. the ``non-factorisable'' corrections, vanish in
this limit, and are in general treated as unknown hadronic parameters.  

Our study specifically scrutinises the $B\to\pi\pi$ system in the context of
these unknown non-factorisable corrections.  A previous analysis of non-factorisable
corrections was presented in \cite{Feldmann:2004mg}, but we consider a different
approach to the problem, and specifically quantify our comparison of the
theory -- data agreement.  We begin by removing the model
dependence from the QCDF predictions and including ``generic''
non-factorisable (NF) contributions which aim to replicate both the known and
unknown NF corrections.  Our aim is to determine the general size and nature of
the NF corrections and test what exactly is required to reconcile the
predictions with the experimental data.  

Our study analyses three separate scenarios for the inclusion of
non-factorisable effects; in each scenario the sizes and phases of the NF 
contributions are varied.  We also test a scenario which includes an enhanced
\textit{``charming penguin''} contribution, which we discuss in the following
sections.  We analyse the level of NF
corrections required to bring each of the branching fractions and CP asymmetries
within the $B\to\pi\pi$ system in line with their experimental measurements.
Finally, we give a short discussion on the implication of our results on the
predictions for $B\to\pi K$, on which analysis is currently underway \cite{BJT05}.
 
\section{Non-factorisable effects in $B\to\pi\pi$}

\begin{figure}[h!]
    $$\epsfxsize=0.6\textwidth\epsffile{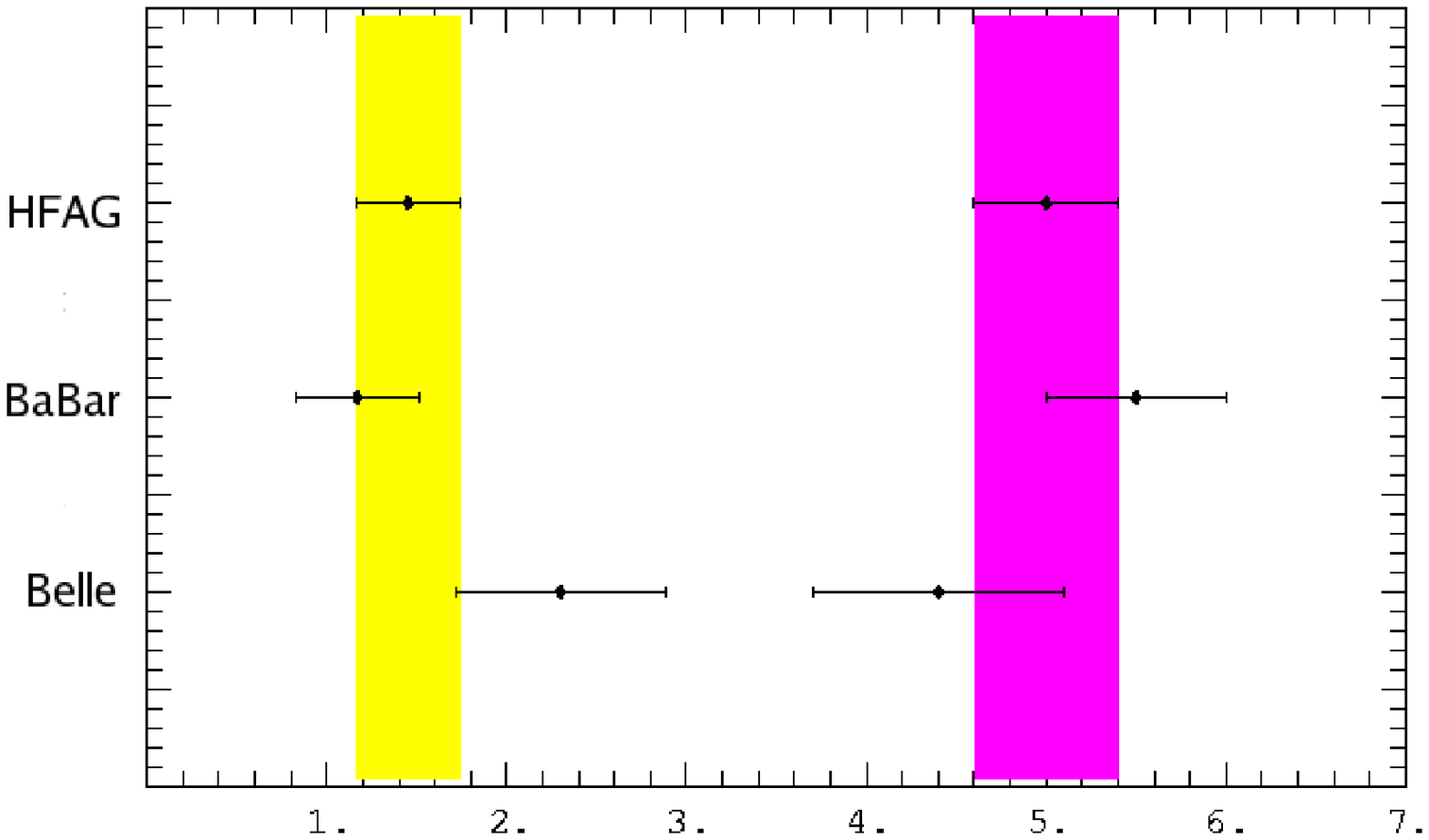}$$
 \vskip-12pt
  \caption[ ]{Plot of experimental results for $B\to\pi\pi$ branching
 fractions in units of $10^{-6}$: From left to right  
 $B\to\pi^0\pi^0$ (Yellow), and  $B\to\pi^+\pi^-$ (Magenta).  HFAG error
 bounds are shown for comparison: $\mathrm{BR}(B\to\pi^0\pi^0)=(1.45\pm0.29)\times10^{-6}$ and
 $\mathrm{BR}(B\to\pi^+\pi^-)=(5.0\pm0.4)\times10^{-6}$.}
   \label{fig:BR}
  $$\epsfxsize=0.4\textwidth\epsffile{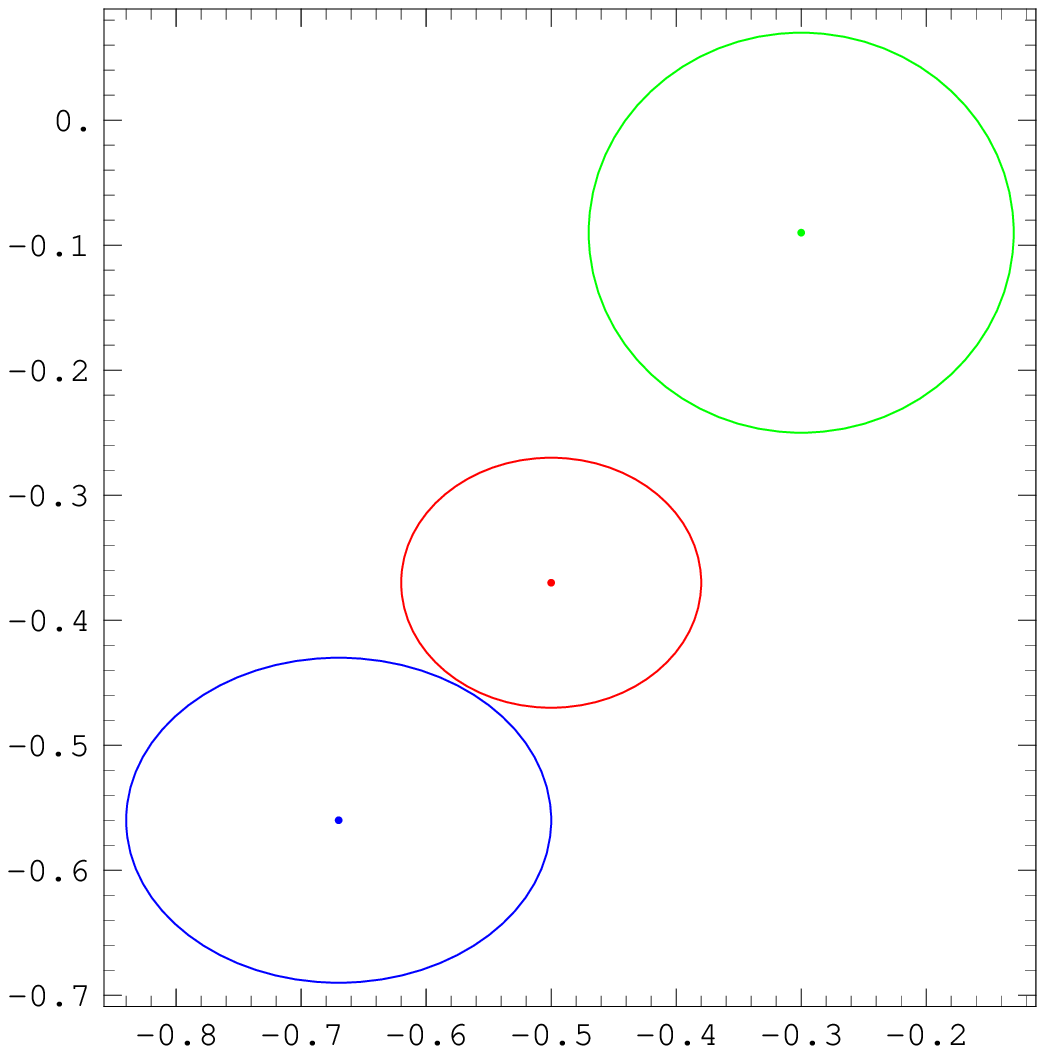}$$
 \vskip-12pt
  \caption[ ]{Plot of experimental results for $B\to\pi\pi$ CP asymmetries
 from HFAG (Red), Belle (Blue) and \textsc{BaBar} (Green).  The horizontal
 axis shows $S_{\pi\pi}$ and the vertical axis $C_{\pi\pi}$.  }
   \label{fig:CPexp}
\end{figure}

The continued improvement of experimental data on the charmless $B$ decays,
most notably for $B\to\pi\pi$, has culminated in a large enough number
 of results (with sufficient accuracy) to fully constrain all the QCD
 parameters relevant for describing this system.  There are now six available
 measurements for the three $B\to\pi\pi$ channels, for the branching ratios
 $\mathrm{BR}(B\to\pi^+\pi^-)$ \cite{Chao:2003ue,Aubert:2005ni}, 
 $\mathrm{BR}(B\to\pi^+\pi^0)$ \cite{Aubert:2004aq, Chao:2003ue} and  
$\mathrm{BR}(B\to\pi^0\pi^0)$ \cite{Aubert:2004aq,Abe:2004mp}, and for the
three CP asymmetry 
measurements $S_{+-},\, 
C_{+-}$ \cite{Aubert:2005av,Abe:2005dz} and $C_{00}$ \cite{Aubert:2004aq,Abe:2004mp}.  The CP asymmetries are defined in
equation (\ref{eq:SC}).    
We use the most recent data (as of September 2005) from 
\textsc{BaBar} and Belle, as well as the combined results from the Heavy
Flavour Averaging Group
(HFAG)\footnote{http://www.slac.stanford.edu/xorg/hfag/index.html}.   Although
this system is well known experimentally, 
there are still some discrepancies between the two $B$-factory experiments,
most notably in the measurement of 
 $B\to\pi^0\pi^0$, and increasingly in the 
measurement of $B\to\pi^+\pi^-$.  These are shown
for illustration in Figure \ref{fig:BR} and the CP asymmetries for
$\pi^+\pi^-$ in Figure \ref{fig:CPexp}. 
 
As discussed in Section \ref{sec:limit}, there are discrepancies between
these results and the QCDF predictions, leading to the coined
``$B\to\pi\pi$ problem''.  In Chapter \ref{chp:LCDA} we examined the effect
of the uncertainty in the hadronic input parameters -- specifically that of
the light-cone distribution amplitude of the pion -- on the $B\to\pi\pi$
observables.  We determined that the resummed models can significantly affect the
branching ratios and can account for both moderate and large
deviations from the asymptotic DA.  However, these models cannot cause
enough enhancement to bring agreement between theory and data, within the
physical range of the model input parameters.  It is logical therefore, to
examine the next large source of uncertainty in the 
theoretical predictions:  non-factorisable corrections.  

The large quantity
of experimental data makes this system ideal to test our 
hypothesis that non-factorisable contributions can have a large effect on
predictions of charmless $B$ decays.  Annihilation contributions to
$B\to\pi\p$ are expected to be small in QCDF, and this is supported by
attempts of direct calculation of these topologies via QCD sum rules
\cite{Khodjamirian:2005wn}.  If this is the case, then the enhanced NF
contributions would be predominantly from hard-gluon exchange, as parameterised by the
non-perturbative parameter $X_H$.  This is supported to some degree by the
results of factorisation fits to charmless data \cite{Cottingham:2005es}.
This study fits both $X_H$ and $X_A$ to all the experimental data on
charmless decays and finds best-fit values for $X_H$ well outside the
expected range.  In our analysis we do not distinguish between the different
sources of NF correction, instead categorising them according to their
contribution to the $B\to\pi\pi$ isospin amplitudes.  Using the decomposition
presented in Section \ref{sec:isodec}, we find that additional contributions
will be required mainly for the $\Delta I=1/2$ amplitude.  

%% In order to test the required levels of non-factorisable corrections that are
%% needed to reconcile with the experimental results, we can use the fully
%% model-independent predictions from QCD factorisation -- without including the
%% (model-dependent) phenomenological parameters $X_H$ and $X_A$.  This enables us to include
%% unconstrained corrections to each decay amplitude.  We also include the
%% possibility of enhancement from charming penguin contributions, which we
%% discuss in the next section.

\section{Charming penguins}

\begin{figure}[h]
    $$\epsfxsize=0.6\textwidth\epsffile{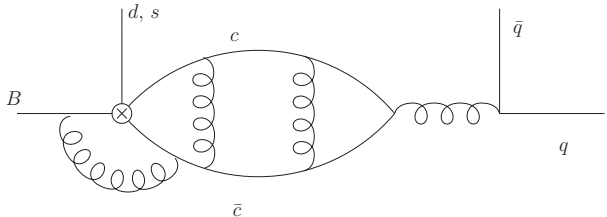}$$
 \vskip-12pt
  \caption[ ]{Example of a charming penguin.}
   \label{fig:BRchp}
\end{figure}

The \textit{charming penguin} is a non-perturbative
$\mathcal{O}\left(\Lambda_{QCD}/m_b\right)$ correction from enhanced penguin
diagrams containing charm loops (Figure \ref{fig:BRchp}).  It was first introduced by Ciuchini et
al. in 1997 \cite{Ciuchini:1997hb, Ciuchini:1997rj}, who added long-distance contributions to decay
amplitudes to improve agreement of fits to experiment.  The charming penguin
originates from a non-perturbative penguin contraction of the leading
operators in the effective Hamiltonian, namely those with $\mathcal{O}(1)$
Wilson coefficients $Q_1$ and $Q_2$, i.e., when the $c$ and $\bar c$
annihilate.  It is expected to give large contributions in some decay
channels, notably $B\to K^+\pi^-$ and $B^+\to K^0\pi^+$, where the
factorised amplitudes are colour or Cabbibo suppressed with respect to the penguins.

\begin{figure}[h]
 $$\epsfxsize=0.4\textwidth\epsffile{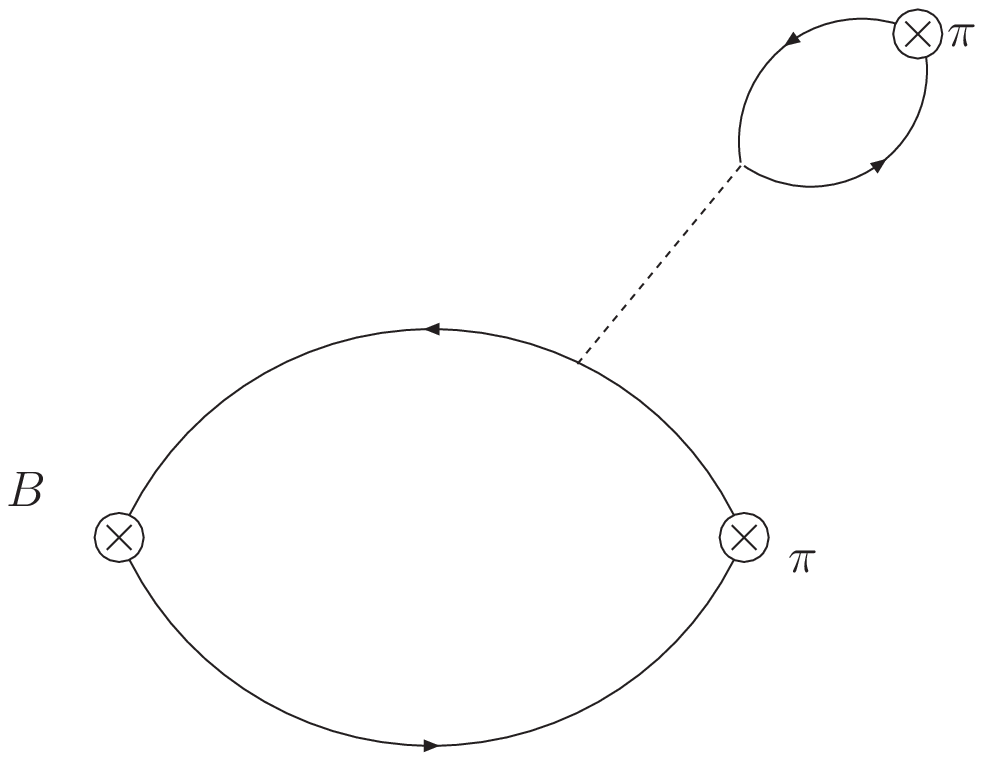} \qquad
  \epsfxsize=0.4\textwidth\epsffile{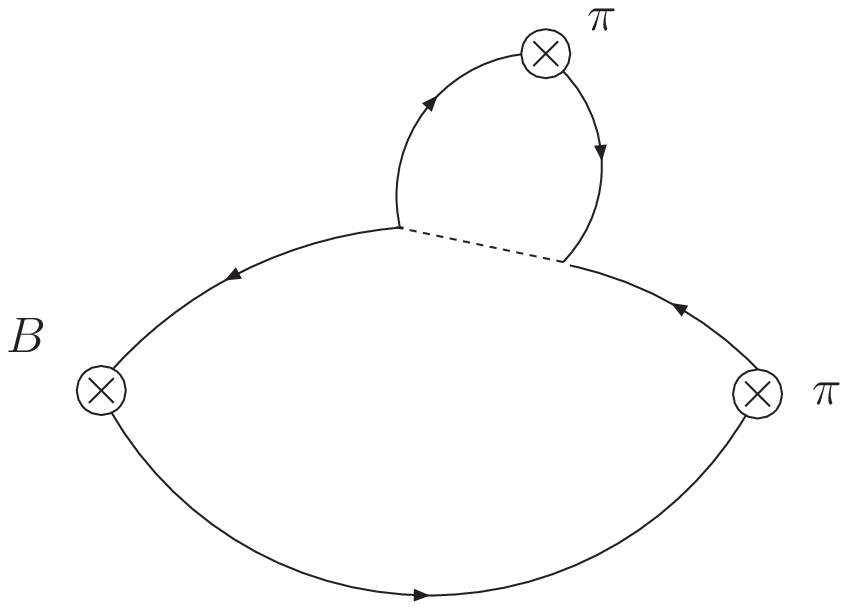} $$  
 \vskip-12pt
  \caption[ ]{Emission diagrams.  Left: colour allowed (DE) topology; Right:
  colour suppressed (CE) topology.  The dashed line represents the
  four-fermion operator.}
   \label{fig:DECE}
\end{figure}

The evidence and support for a charming penguin contribution has fluctuated
since their introduction in \cite{Ciuchini:1997hb}.   The original
parameterisation was in terms of a diagrammatic deconstruction into
renormalisation group invariant quantities or topologies \cite{Buras:1998ra}.  For example
the leading (tree) contributions are described by emission topologies $E_1$
and $E_2$ which are scheme independent combinations of Wilson coefficients
and matrix elements related to the current-current operators $Q_1$ and
$Q_2$, as shown in Figure \ref{fig:DECE}. 

\begin{figure}[h]
    $$\epsfxsize=0.4\textwidth\epsffile{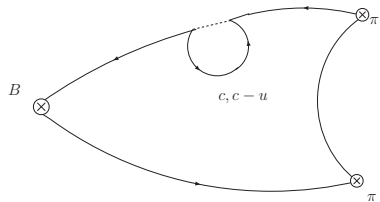}$$
 \vskip-12pt
  \caption[ ]{Example of charming and GIM penguin topologies (CP).}
   \label{fig:chptop}
\end{figure}

The charm and \textit{GIM penguin} followed a topology of the type in
Figure \ref{fig:chptop}.  The charming penguin has only the charm
quark in the loop; the GIM penguins are those where the diagrams always appear in the
combination of $u - c$.    

This description was later updated to include further diagrams (annihilation and penguin
contractions) in addition to the original diagram, to give two
phenomenological parameters $\widetilde{P}_1$ and
$\widetilde{P}_1^{\mathrm{GIM}}$ for the charming and GIM penguins
respectively.  Fits of the charming penguins to $K\pi$ and $\pi\pi$ decays gave promising results
\cite{Ciuchini:2001gv}.  However, when QCD factorisation
was introduced and the non-factorisable corrections were shown to be
calculable in perturbation theory, the charming penguin fell out of favour.
It should be noted that since the charm penguins are expected to have a size
equivalent to a $\Lambda_{\mathrm{QCD}}/m_b$ correction they are not in
fact in conflict with the QCDF results in the $m_b\to\infty$
limit.  If we call into question the size of the power-suppressed corrections
to the $\Lambda_{\mathrm{QCD}}/m_b$ expansion and expect  -- and in fact
demand -- some of the non-factorisable contributions to be large, the charming
penguin can provide significant additional contributions at this order. 

The current QCD factorisation predictions (specifically for the
$B\to\pi\pi$ and $B\to\pi K$ decays) do not satisfactorily explain the
current experimental data.  Both the $B\to\pi\pi$ and $B\to\pi K$
decays may require significant levels of power-suppressed corrections.  It is
therefore quite possible that the charming 
penguin can provide additional enhancement, enough to bring the predictions into
agreement with the current experimental results.

Charming penguins are not Cabbibo enhanced in $B\to\pi\pi$ decays, unlike
the case for $B\to\pi K$.  It is naively expected that, along with the
other $\mathcal{O}\left(\Lambda_{\mathrm{QCD}}/m_b\right)$ corrections, the
charming penguin contributions should be small.  However, as we show in our
study, the predictions for the $B\to\pi\pi$ branching fractions
cannot be brought into agreement with the experimental data without
significant non-factorisable corrections.  Since the charming penguin
is of the same order as the
$\mathcal{O}\left(\Lambda_{\mathrm{QCD}}/m_b\right)$ effects, there is no reason
why a sizable contribution should not exist in these decays.  
             
The charming penguin parameter $\widetilde P_1$, and the GIM-penguin
parameter $\widetilde P_1^{\mathrm{GIM}}$ were expressed as a complex
number fitted to experiment by Ciuchini et al., and added to the factorisable amplitudes via
\begin{equation*}
   P_1 = a_4^c A_{\pi\pi} + \widetilde P_1 \qquad P_1^{\mathrm{GIM}} = (a_4^c
   - a_4^u) A_{\pi\pi} + \widetilde P_1^{\mathrm{GIM}}
\end{equation*}
where the notation is from QCD factorisation, as found in Appendix
\ref{chp:AppA}.   

There is some disagreement over the inclusion of charming
penguins within the factorisation framework.  The  parameter
$\widetilde{P}_1$ contains not only the charming penguin contribution 
but additionally the annihilation and penguin contractions of the penguin
operators.  Ciuchini et al. suggest that all chirally suppressed terms should
be dropped and replaced with this term (these correspond to dropping the coefficients
$a_6^p$ from QCD factorisation).  The proponents of QCD factorisation
however, argue that the ``charming penguin'' contribution is completely
described by the penguin annihilation contributions and non-perturbative
corrections to the coefficient $a_4^c\,$, i.e those contained in $a_6^c$.  They
argue that a large non-perturbative penguin enhancement is implausible,
and that any modification required to bring theory predictions into agreement
with data should be attributed to weak annihilation rather than charming
penguins \cite{Beneke:2001ev}.  

As our study does not include any of the model-dependent calculations of the
$\mathcal{O}(\Lambda_{\mathrm{QCD}}/m_b)$ terms suggested by BBNS, to 
obtain a charming penguin enhancement we need additional 
contributions.  We do not replace all of the chirally-suppressed terms as
suggested by Ciuchini et al., and instead take the charm penguin to be 
in addition to the QCDF penguin diagrams (i.e. keep
the calculable part of $a_6^c$, but remove any dependence on terms
proportional to $X_H$).  We can then allow for the possibility of the
charming penguin explicitly as an enhancement of the
coefficient $\a_4^c = a_4^c + r_\chi a_6^c$.  
                       
\section{$B\to\pi\pi$ Analysis}

We use the expansion of the $B\to\pi\pi$ decay amplitudes in terms of the two isospin
amplitudes $A_{1/2}$ and $A_{3/2}$.  We construct three different
scenarios for inclusion of non-factorisable corrections and enhancements on top of the
factorisable amplitude as calculated within QCD factorisation.  We include only
the (fully) factorisable parts from QCDF, i.e. we do not include
the hard-scattering or annihilation contributions incorporated in the
model-dependent parameters $X_H$ and $X_A$ (Section \ref{sec:powersup}).  NF
corrections are then added according to the prescriptions of our three
scenarios, so we can determine the size and 
nature of corrections needed to reconcile with the experimental data.  

In each scenario we allow the size and phases of the NF contributions to each
isospin amplitude to vary independently.  The full set of $B\to\pi\pi$
branching ratios and CP asymmetries are calculated for each of the
combinations of NF contributions we consider.  We take a defined set of input 
parameters which are also varied within their allowed ranges in order to further
optimise the agreement between theory and experiment.  Our input parameters
are a combination of CKM phases, the $B\to\pi$ form factor and
$\Delta$ from the parameterisation of the pion wavefunction using the models
developed in Chapter \ref{chp:LCDA}.

\begin{center}               
 \begin{tabular}{|l|l|}
 \hline
 {} & \textrm{Input ranges} \\\hline
 $\Delta$ & 1.1$\leq\Delta\leq$ 1.3\\
 $F^{B\to\pi}$ & 0.26 $\pm$ 0.04 \\
 $R_u$ & 0.399 $\pm$ 0.08 \\
 $\gamma$ & 60.3$^\circ \pm$ 6.8$^\circ$ \\\hline
 \end{tabular}\newline
\end{center}

We consider primarily the data from the HFAG which averages over all of the
current data available for each channel. 
However, we consider the data from {\textsc{BaBar}} and Belle separately
where it is required, if the discrepancy between the two experiments is
significant.  Each of our scenarios is constructed with a
``base'' of the fully calculated factorisable part of the $B\to\pi\pi$ isospin
amplitudes, as found using from QCDF.  Different
combinations of non-factorisable corrections and charming penguins are then
added.  

We add a percentage non-factorisable contribution to each of the isospin amplitudes
individually.  The amplitude $A_{3/2}$ is tree dominated with a tiny
contribution from electroweak penguins; the expressions for the isospin
amplitudes given in 
equation (\ref{eq:A32}) show that the amplitude $A_{3/2}^c$ contains only 
contributions from electroweak penguins.  We can therefore safely
assume that there will be no significant non-factorisable contribution to
this particular amplitude.  This is supported by the good agreement between the
QCD factorisation prediction and the measurement for the decay $B\to\pi^+\pi^0$.  

The non-factorisable corrections are added then as
\begin{eqnarray}\label{eq:isononfac}
   A_{3/2} &=& A_{3/2}^F + \lambda_u^*\,
   |A_{3/2}^{u\,F}|\,N_{3/2}\,e^{i\,d_{3/2}}\nonumber\\
    A_{1/2} &=& A_{1/2}^F + \lambda_u^*\,
   |A_{1/2}^{u\,F}|\,N^u_{1/2}\,e^{i\,d_u} + \lambda_c^*\,
   |A_{1/2}^{c\,F}|\,N^c_{1/2}\,e^{i\,d_c} 
\end{eqnarray}
where the notation is intuitive: the superscript $F$ denotes the pure
factorisable amplitudes, the $N_{\Delta I}\,e^{i\,d_{\Delta I}}$ is the size of the
non-factorisable contribution ($0\leq N_{\Delta I}\leq 1$), with arbitrary
phase $d_{\Delta I}$.
The level of non-factorisable corrections is expected to be of the order of
$\bar\Lambda/m_b\sim10\%$, where $\bar\Lambda$ measures the kinetic energy of
the $b$ quark within the $B$ meson.  As a conservative estimate, we allow
corrections of up to 20\% as our ``normal'' expected level, i.e taking each
$N\leq0.2$.  As an indicator for the enhanced corrections we allow double
this estimate with up to 40\% non-factorisable corrections.  Anything beyond
this limit would be approaching the point where the factorisation scheme
itself would begin to break down.  The QCD factorisation scheme is based upon the
calculation of order $\a_s$ corrections, so any terms suppressed by $1/m_b$
should not be much larger than $\a_sm_b$, otherwise it would not make sense to include
small calculable terms and to neglect large incalculable terms.

We also consider the possibility of additional contributions from charming
penguin diagrams, which are not accounted for in the calculation of the
factorisable amplitudes.  The charming penguin contribution arises purely in
the isospin amplitude $A_{1/2}^c$, and as such, can be parameterised as an
enhancement to the factorisation coefficient $\alpha_4^c$
\begin{eqnarray*}
   \alpha_4^c = a_4^c + r_\chi a_6^c\rightarrow\alpha_4^{c \hspace{3pt} eff}
   = R\, \alpha_4^c 
\end{eqnarray*}
where $R$ is some complex factor $R =|R|e^{i\varphi}$ with $|R| > 1 $.  
This will have an effect on both the decay amplitudes for $BR(B\to\pi^+\pi^-)$ and
$BR(B\to\pi^0\pi^0)$, but there is no effect on $BR(B\to\pi^+\pi^0)$ which
has no contribution from $A_{1/2}^c$.

The scenarios are summarised as follows:  
\begin{itemize}
   \item{Scenario I: QCD factorisation + expected level of non-factorisable
   corrections at  $\leq$ 20\%  }
   \item{Scenario II: QCD factorisation + enhanced non-factorisable
   corrections at $\leq$ 40\%  }
   \item{Scenario III: QCD factorisation + expected non-factorisable
   corrections at $\leq$ 20\%
   and charming penguin contribution}
\end{itemize}

In order to quantify the ``quality'' of each of these scenarios we construct
a $\chi^2$ statistic, where the form factor $F_0^{B\to\pi}(0)\equiv F^\pi$ and the phases
of the non-factorisable contributions (NF) are left as fitting parameters.  The
absolute size of the NF 
correction and the CKM parameters $R_u$ and $\gamma$ are taken as fixed, and
the fitting is performed separately over differing values of the pion
distribution amplitude parameter $\Delta$.  We define
\begin{eqnarray*}
  \chi^2 = \sum_{i=1}^6 \frac{[y_i - x_i]^2}{\sigma_i^2}
\end{eqnarray*}
 where $y_i$ are the theory predictions, and $x_i$ are the data values with
 their associated error $\sigma_i$, calculated by combining statistical and
 systematic errors in quadrature.  The index $i$ runs over the 6 available
 experimental measurements of the CP-averaged branching ratios and CP
 asymmetries:
 $\{\Gamma^{+-},\,\Gamma^{00},\,\Gamma^{+0},\,S_{+-},\,C_{+-},\,C_{00}\}$.  We
 then numerically minimise the $\chi^2$ while scanning over the three phases and
 applying the constraint $F^\pi < 0.3$.  We perform this separately
 for the {\textsc{BaBar}}, 
 Belle and HFAG averaged data sets.
The goodness-of-fit then follows the $\chi^2$ probability distribution with 2
degrees of freedom (equal to the number of measurements minus the number of
fitted parameters).  We express the goodness-of-fit as $\chi^2/n_d$, where the
closer the value to 1, the better the fit.  %% ; the probability distribution is
%% illustrated in figure \ref{fig:chi2}.   
%% \begin{figure}[h!]
%%     $$\epsfxsize=0.5\textwidth\epsffile{figures/Fig4/chi2.eps}$$
%%  \vskip-12pt
%%   \caption[ ]{Probability distribution $P(\chi^2, r)$ for $r$ degrees of
%%  freedom; $r=1$ is the red curve and $r$ increases up the spectrum
%%  approaching a Gaussian for $n>20$.}
%%    \label{fig:chi2}
%% \end{figure}

\section{Results}

We now present and discuss the results of the analysis for the $B\to\pi\pi$ system for
each of the specified scenarios in turn.

\subsection{Scenario I}

In this scenario we calculate the CP-averaged branching ratios and CP asymmetries for
$B\to\pi\pi$ decays with the expected level of non-factorisable corrections
($\leq$ 20\%).  We find that it is very difficult to reconcile the
predictions for all of the observables with their
experimental measurements within the experimental error bounds.  We varied all of
the input parameters within their allowed ranges and found a set of optimum
inputs (shown in Table \ref{tab:optinp}) which gave the best agreement
between theory and experiment; the results were approximately the
same for all data sets.  The two CKM parameters optimised at their input
values but the DA parameter $\Delta$ required its maximum allowed 
contribution to best optimise the agreement.  For the form factor we used the
best-fit value as calculated from the minimisation of the $\chi^2$ function
on the full set of $B\to\pi\pi$  data.  \\

\begin{table}[h]
   \begin{center}
     \begin{tabular}{|c|c|c|c|} \hline
     $\Delta$ & $F^\pi$ & $R_u$ & $\gamma$ \\ \hline
     1.3 & 0.22 & 0.399 & $60.3^\circ$ \\ \hline
   \end{tabular}
   \caption{Optimum input parameter; $R_u,\,\gamma$ are the default central
     values ($R_u = \sqrt{\bar{\rho}^2 + \bar{\eta}^2}$) for 20\%
     non-factorisable corrections.}
   \label{tab:optinp}
 \end{center}
\end{table}

\begin{figure}[h]
  $$\epsfxsize=0.4\textwidth\epsffile{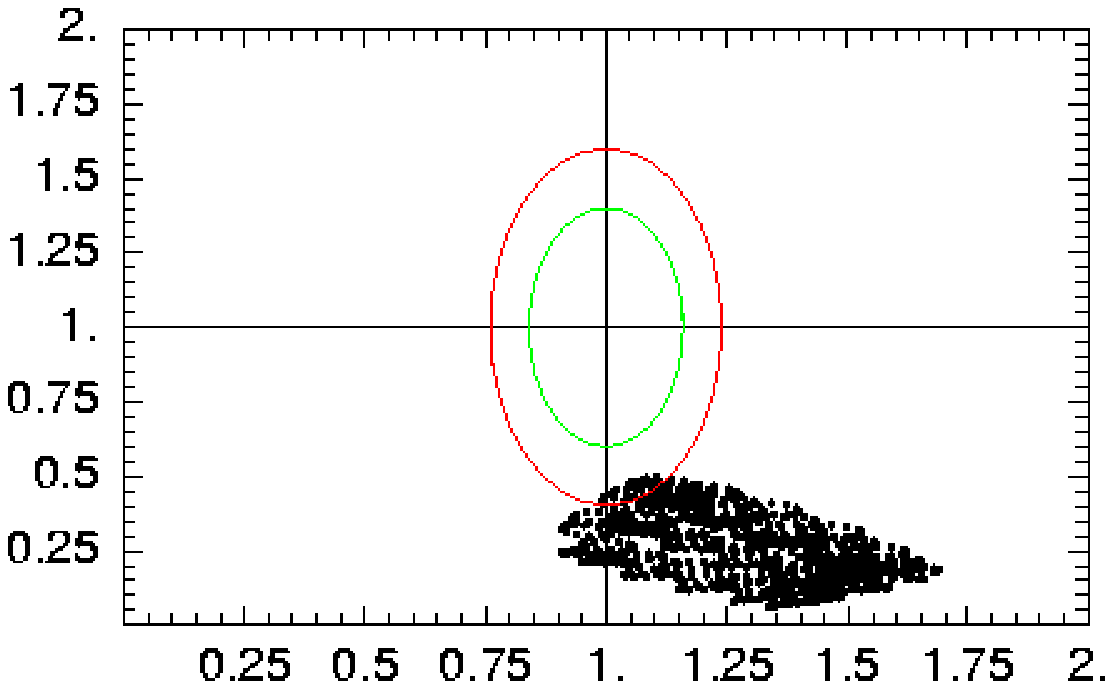}\qquad
  \epsfxsize=0.4\textwidth\epsffile{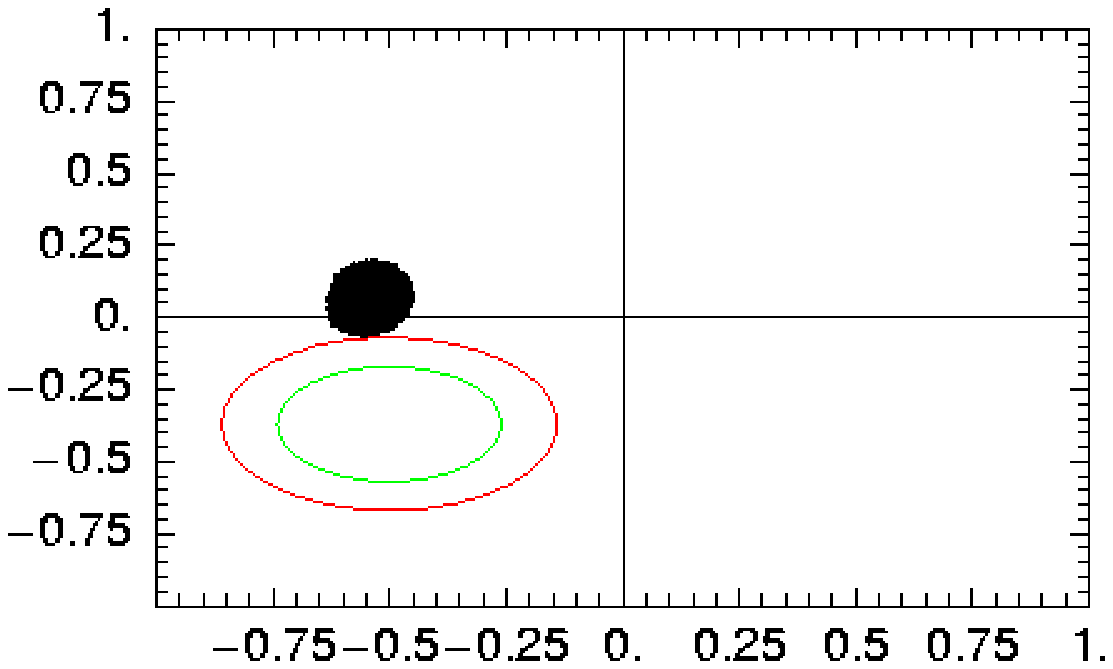}$$
  \vskip-12pt
  \caption[]{Theory/Experimental ratio using HFAG data for $\Gamma^{+-}$
   against $\Gamma^{00}$ (left) and CP asymmetry $S_{\pi\pi}$/$C_{\pi\pi}$
   (right).  Error ellipses shown at $2\sigma$ and $3\sigma$.}  
  \label{fig:HFAG20}
  $$\epsfxsize=0.4\textwidth\epsffile{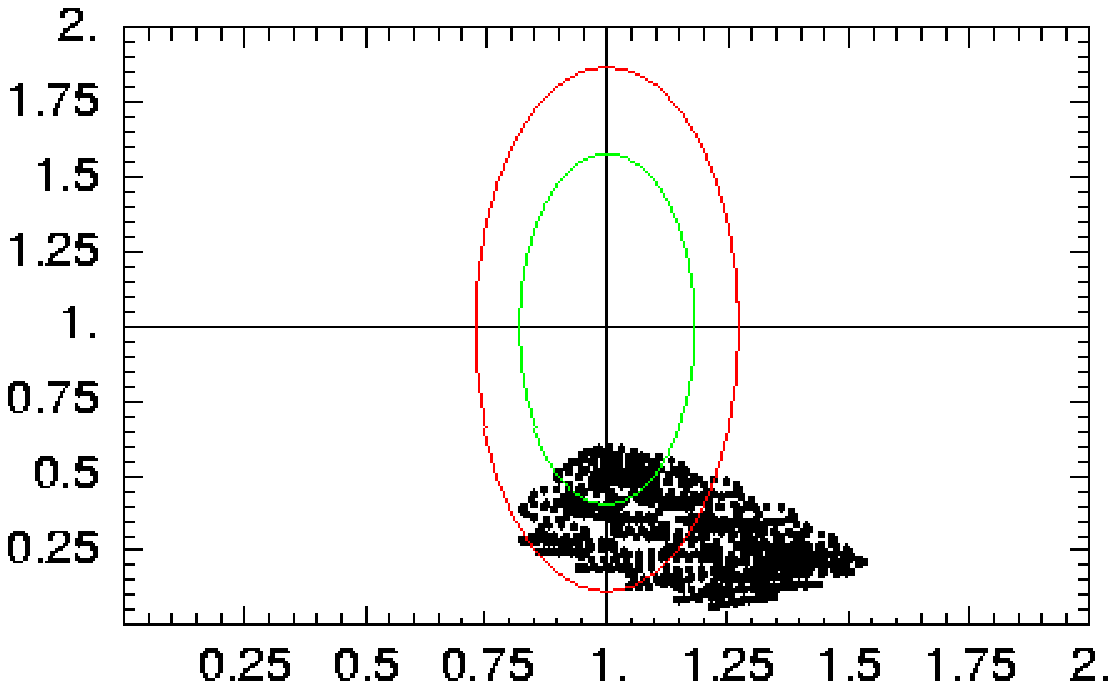}\qquad
  \epsfxsize=0.4\textwidth\epsffile{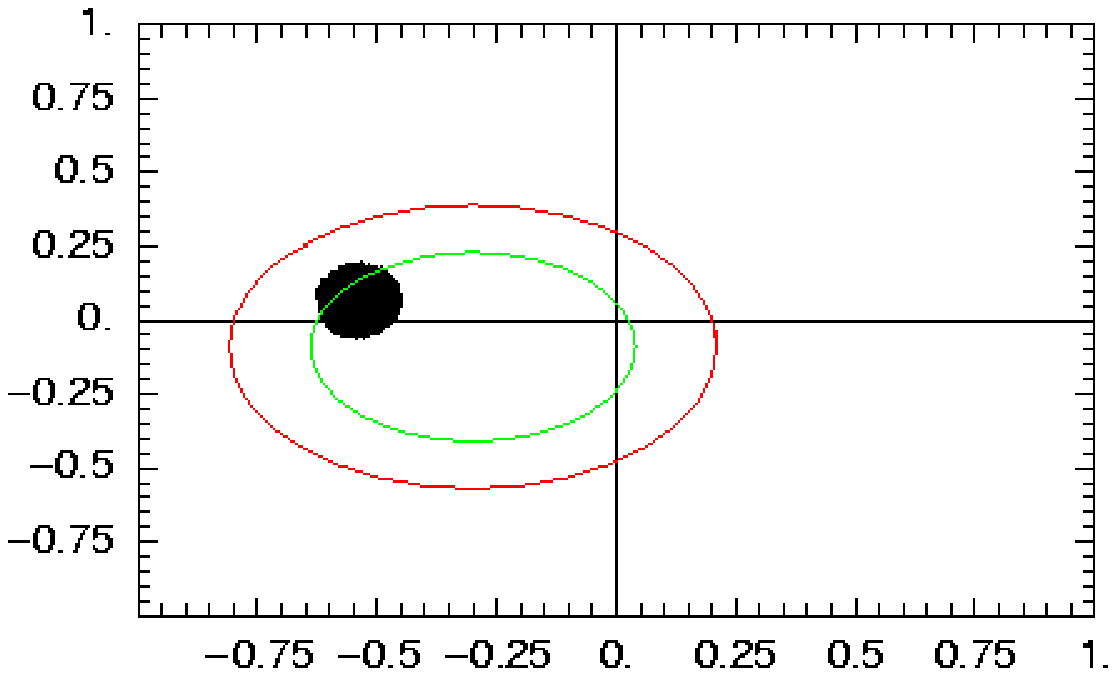}$$
  \vskip-12pt
  \caption[]{As Figure \ref{fig:HFAG20} using only \textsc{BaBar} data set.}  
  \label{fig:BaBar20}
  $$\epsfxsize=0.4\textwidth\epsffile{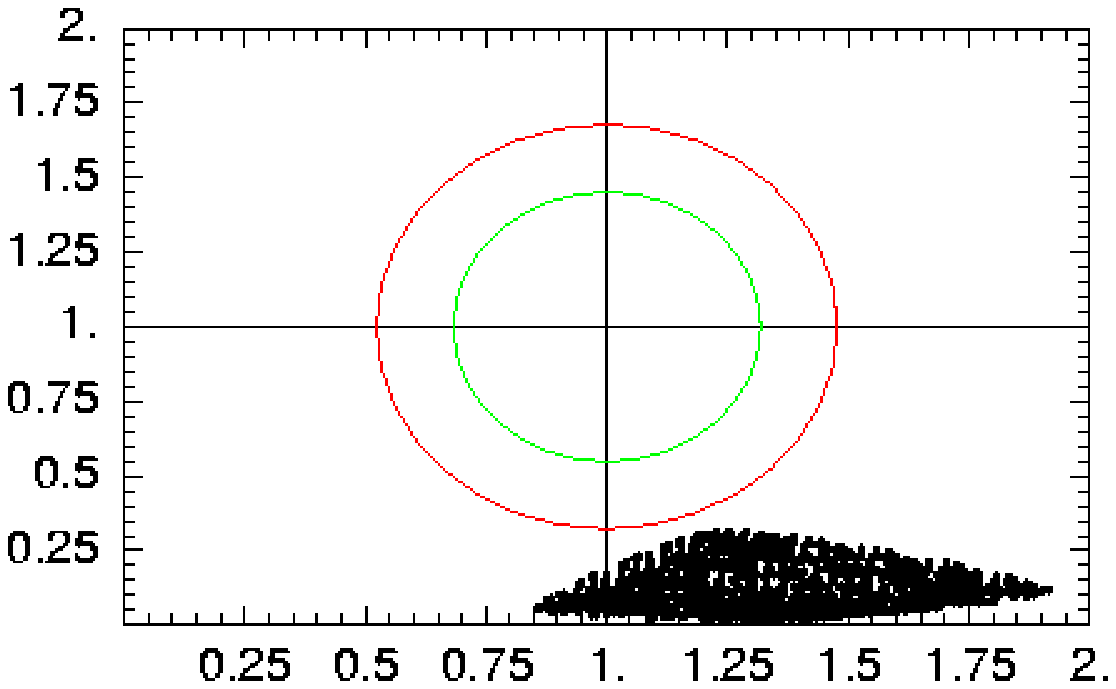}\qquad
  \epsfxsize=0.4\textwidth\epsffile{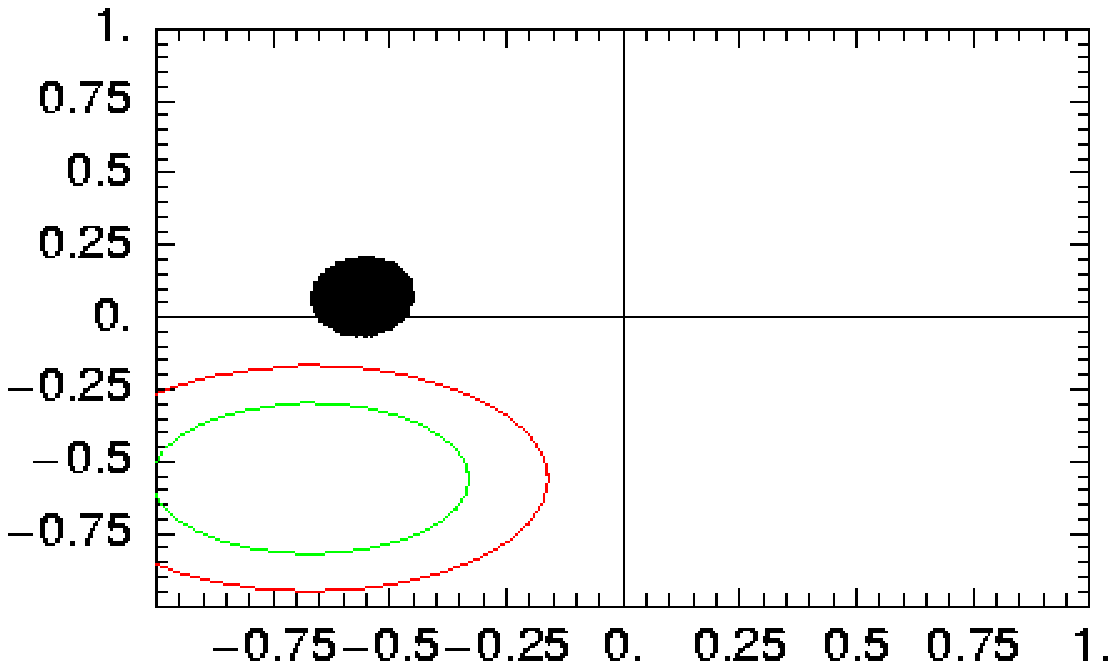}$$
  \vskip-12pt
  \caption[]{As Figure \ref{fig:HFAG20} using Belle data set.}  
  \label{fig:Belle20}
\end{figure}

 Since the optimum scenario is clearly obtained for the maximum possible NF correction,
 we specified the size of the contributions as $N=0.2$ for each amplitude and scanned
 over all possible phase combinations.  We then calculated  
 the ratio of the theory over experimental results for the branching
 fractions and the CP asymmetries.  The most 
 illustrative results are of $B\to\pi^+\pi^-$ against
 $B\to\pi^0\pi^0$ and for $S_{\pi\pi}$ against $C_{\pi\pi}$.  Considering
 first the branching ratios, the value of $\textrm{BR}(B\to\pi^+\pi^0)$ is quite well
 reproduced, but the  $\pi^+\pi^-$ and $\pi^0\pi^0$ branching fractions show
 considerable 
 discrepancy.  This is illustrated for the HFAG result in Figure
 \ref{fig:HFAG20} and also separately for the data sets from \textsc{BaBar}
 and Belle in Figure \ref{fig:BaBar20} and  Figure \ref{fig:Belle20}
 respectively, highlighting the large impact of the differing measurements of
 $\textrm{BR}(B\to\pi^+\pi^-)$.  

 The ellipses shown in Figures \ref{fig:HFAG20} to \ref{fig:Belle20}, are the
 $2\sigma$ and $3\sigma$ error ellipses.   These
 are calculated from the experimental results under the assumption of no
 correlation between the branching ratios, and using the measured
 correlations for the CP asymmetries \cite{Aubert:2005av,Abe:2005dz}.  These
 curves correspond to 
 confidence regions (confidence level across 2 degrees of freedom) of
 $\sim84\%$ and $\sim95\%$ respectively.  

 These figures clearly show that with a 20\% non-factorisable correction, it
 is not possible to reproduce the experimental data within the $2\sigma$ error
 bound, and not even to 3$\sigma$ using the Belle data alone.   
Using the HFAG data and maximal contribution of $\Delta = 1.3$, the $\chi^2$
 minimisation yields a 
 value of $F^\pi= 0.22$ with 
 $\chi^2/r$ of 4.41/2. For the {\textsc{BaBar}} data alone we also obtain 
 $F^\pi= 0.22$ but with a much improved fit and $\chi^2/r =  0.850/2$.

\begin{figure}[h]
   $$\epsfxsize=0.5\textwidth\epsffile{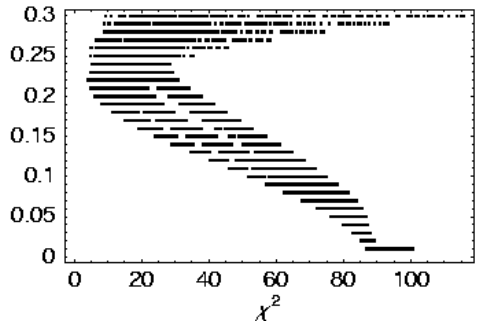}$$
   \vskip-12pt
   \caption[]{Dependence of minimum $\chi^2$ on $F^\pi$ for $\Delta = 1$
   and 20\% NF correction.}   
  \label{fig:chidtn}
\end{figure}

We observe that in increasing the parameter $\Delta$
 from the asymptotic value ($\Delta = 1$) to its physically allowed maximum
 ($\Delta = 1.3$) we find a steadily decreasing $\chi^2$ value, and
 correspondingly increasing best-fit value of $F^\pi$.  The
 distribution of the $\chi^2$ values as a function of the form factor is
 plotted for 20\% NF corrections, for the full range of possible phases in
 Figure \ref{fig:chidtn}.  This shows clearly the minimum $\chi^2$ in agreement
 with the numerical minimisation requiring $F^\pi\sim0.22$.

\subsection{Scenario II}

\begin{figure}[h]
   $$\epsfxsize=0.5\textwidth\epsffile{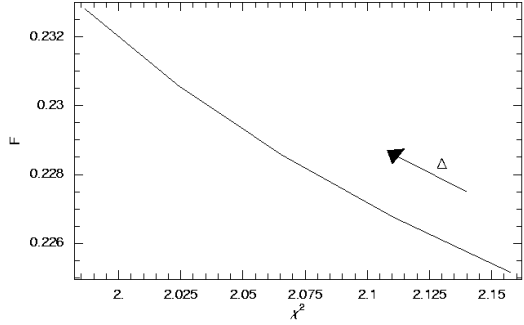}$$
   \vskip-12pt
   \caption[]{Dependence of minimum $\chi^2$ on $F^\pi$ with increasing
   $\Delta$ for 40\% NF correction.} 
    \label{fig:chi40FF} 
\end{figure}

 In a direct extension of Scenario I, we repeat our study with a further
 enhanced non-factorisable contribution beyond that of the expected levels.
 We again use the factorisable isospin amplitudes calculated within QCDF, but
 now add up to 40\% non-factorisable corrections.  As may be
 expected, this reproduces the experimental data quite well.  We find the 
 optimum parameter set to be similar to that from Scenario I; the agreement
 is not greatly improved from changing the CKM input, but it is improved by
 the maximisation of $\Delta$ and use of the best-fit form factor.  For a 40\% NF
 correction, the $\chi^2$  minimisation gives a value of $F^\pi=
 0.23$ at $\chi^2/r = 2.0/2$ when using the maximum correction of $\Delta =
 1.3$.  As in Scenario I, we see that the $\chi^2$ improves as we increase
 $\Delta$, and the best-fit form factor 
 increases correspondingly, around the value of $F^\pi\sim0.23$.  This is
 illustrated in Figure \ref{fig:chi40FF}.

\begin{figure}[h!]
  $$\epsfxsize=0.45\textwidth\epsffile{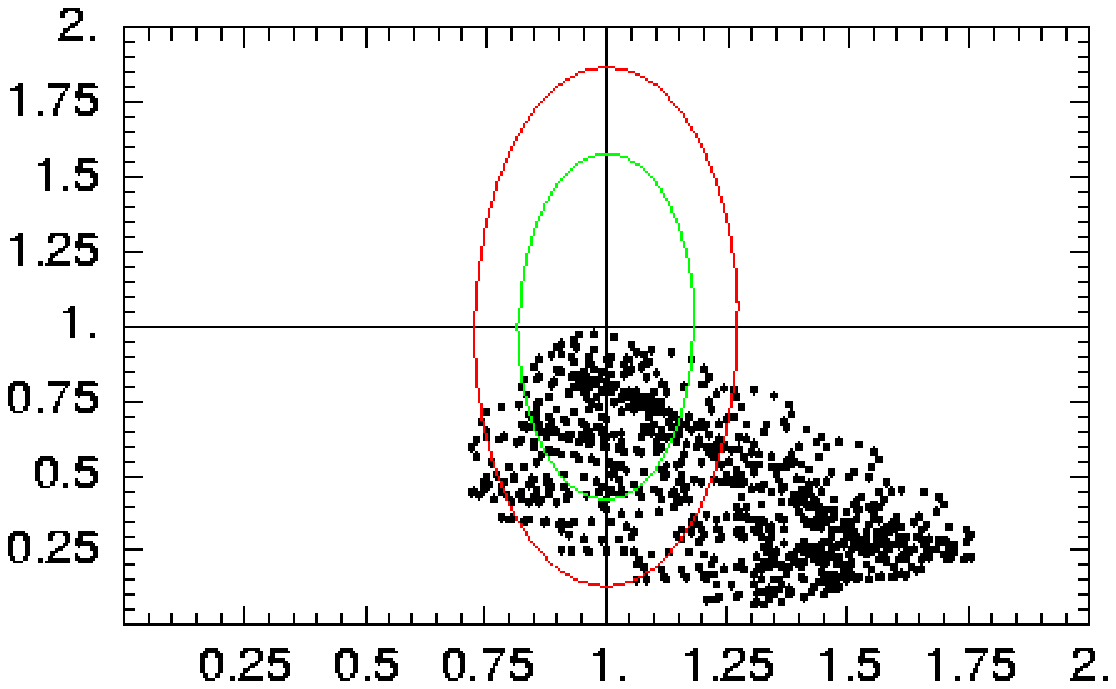} \qquad 
   \epsfxsize=0.45\textwidth\epsffile{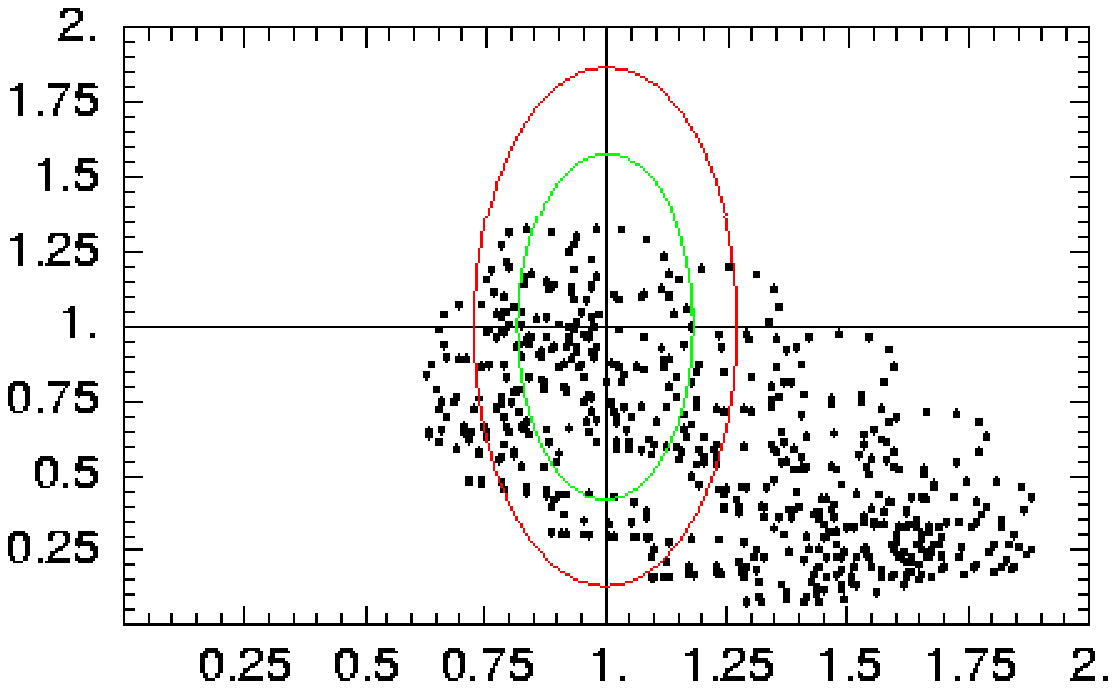}$$
 \vskip-12pt
 \caption[]{The $BR(B\to\pi^+\pi^-),\,BR(B\to\pi^0\pi^0)$ agreement for the
   best-fit case for 30\% NF correction (left) and 40\% (right), both using
  {\textsc{BaBar}} data.} 
  \label{fig:chi40} 
\end{figure}

 We find that there is a range of allowed values of from $0.21\leq
 F^\pi\leq0.23$ which give good agreement with the data. We illustrate the
 best possible agreement (for both a 30\% and 40\% non-factorisable correction)
 using the \textsc{BaBar} data set in Figure \ref{fig:chi40}.

\begin{figure}[h!]
  $$\epsfxsize=0.45\textwidth\epsffile{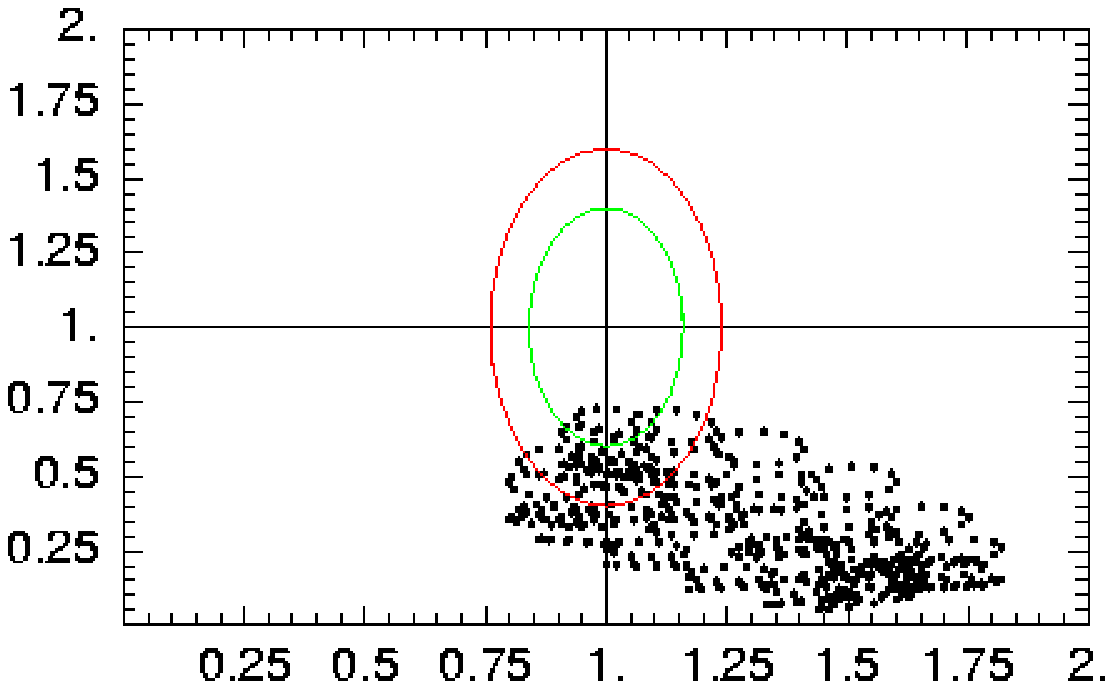}\qquad
 \epsfxsize=0.45\textwidth\epsffile{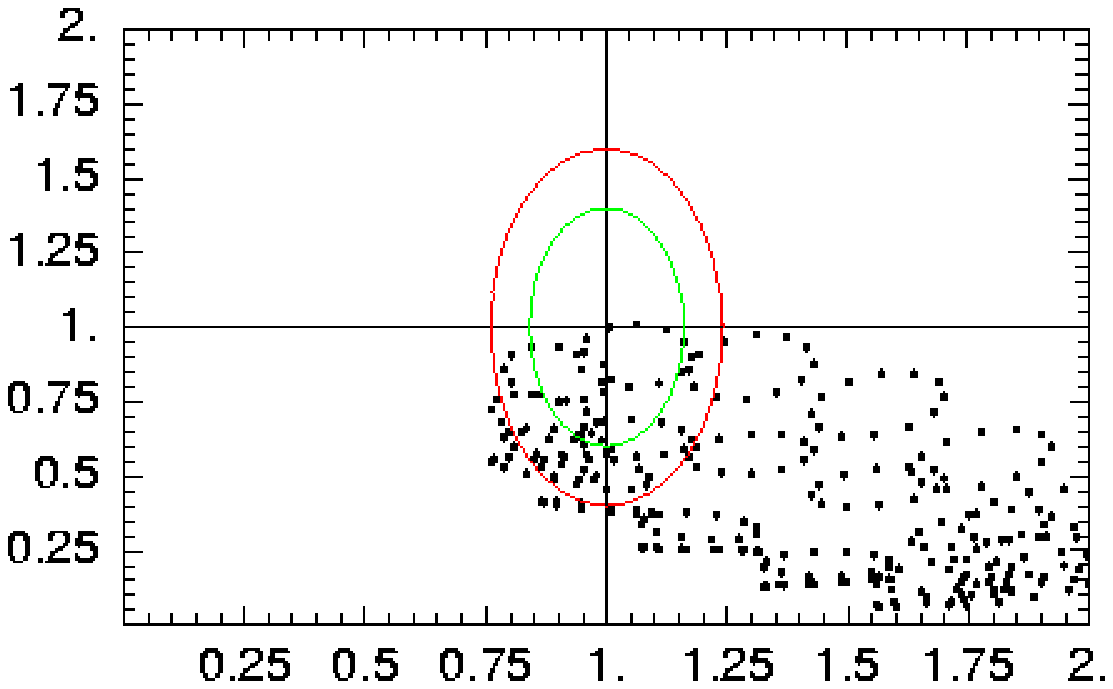}$$
 \vskip-12pt
 \caption[]{The $BR(B\to\pi^+\pi^-),\,BR(B\to\pi^0\pi^0)$ agreement for the
 best-fit case for 30\% NF correction (left) and 40\% (right), using HFAG data.}  
  \label{fig:chiHFAG}
\end{figure}

The results using the HFAG data are shown in Figure \ref{fig:chiHFAG}.  In this
case the experimental results are further away from
the theory predictions, primarily due to the Belle measurement of
$\textrm{BR}(B\to\pi^0\pi^0)$.  We see that the enhanced levels of NF
corrections of around 30\% can bring the theory values close to the
experimental averages, at least to within the $3\sigma$ error, although
larger corrections are needed to approach $2\sigma$ agreement.  The optimum for
 $\Delta$ is again found to be maximal, $\Delta = 1.3$, and for the form factor 
 $F^\pi = 0.22$ and $F^\pi = 0.23$ for the 30\% and 40\%
 cases respectively.

\subsection{Scenario III}

\begin{figure}[h!]
 $$\epsfxsize=0.45\textwidth\epsffile{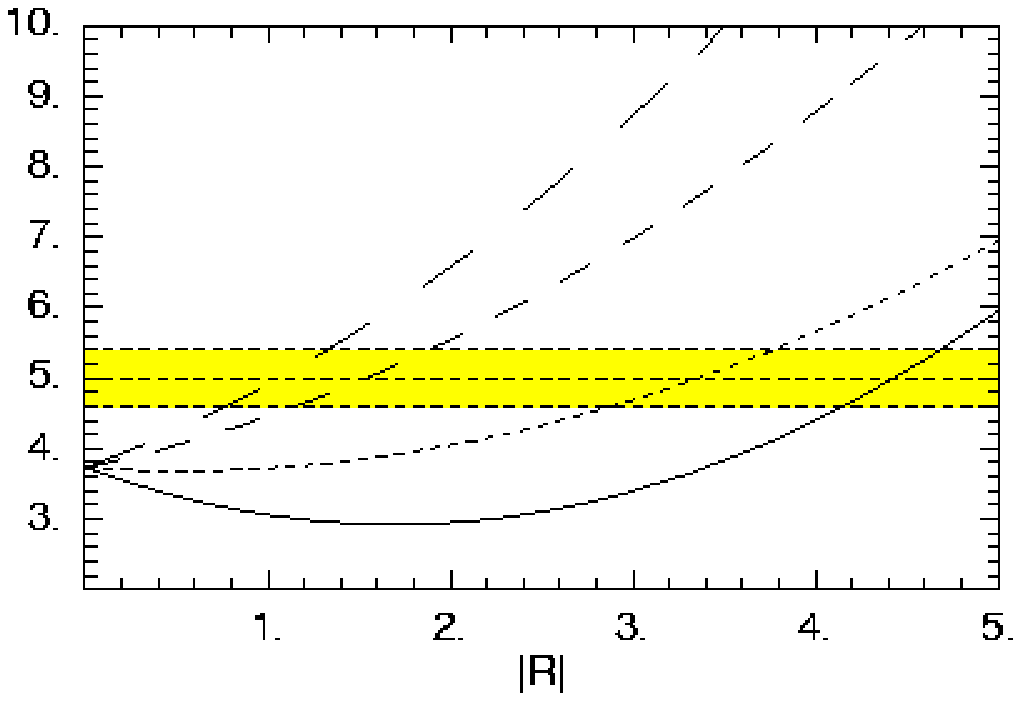}\qquad 
 \epsfxsize=0.45\textwidth\epsffile{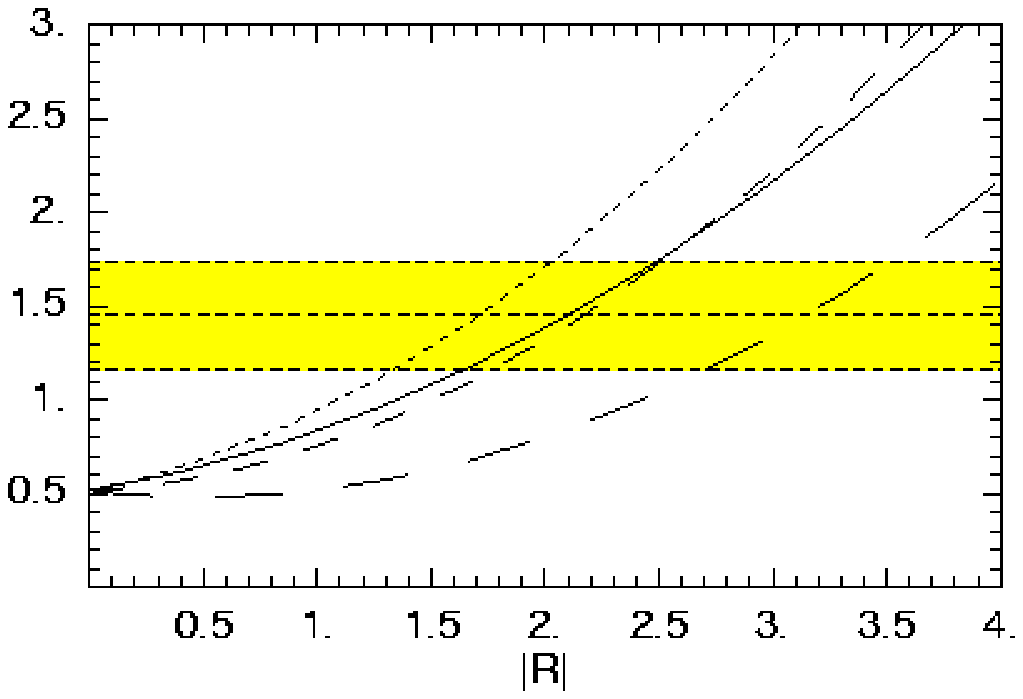}$$
 \vskip-12pt
 \caption[]{Dependence of $BR(B\to\pi^+\pi^-)$ (left) and $BR(B\to\pi^0\pi^0)$
   (right) on charming penguin parameter $R$, in units of $10^{-6}$; curves
   shown for constant 
   $\varphi$ of $0$ (long-dashed), $\pi/4$ (short-dashed), $\pi/2$ (dotted)
   and $\pi$ (solid).  The experimental result and $1\sigma$ error from HFAG
   is shown for comparison.}   
  \label{fig:chp1}
\end{figure}

Our third scenario involves adding contributions from charming penguins in
addition to the expected non-factorisable correction ($\leq$ 20\%) to the
QCDF base.  As discussed above, we expect this to impact on
the values of the branching ratios for $B\to\pi^0\pi^0$ and $B\to\pi^+\pi^-$.
For illustration, Figure \ref{fig:chp1} shows the possible enhancement of these
branching ratios for various sizes of charming penguin contribution.  

This is presented for the expected 20\% non-factorisable correction and our
``best-fit'' set of parameters ($\Delta$, $F^\pi$ and NF phases for each
branching ratio), as found in the analysis with no charming 
penguin contribution (Scenario I).

 The figure shows that it should indeed be possible to improve the
 agreement between the theory and experimental results for both branching
 ratios simultaneously, with a level of 20\% non-factorisable corrections.  
 
 To better gauge and quantify the level of enhancement that is needed, we
 again compare the 
 agreement of the ratio of $B\to\pi^+\pi^-$ and $B\to\pi^0\pi^0$ branching
 fractions, and the $B\to\pi^+\pi^-$ CP asymmetry. The results are shown
 below in Figure \ref{fig:chpHFAG} and  Figure \ref{fig:chpHFAGcp} for sample 
 values of $|R| = 1, 1.5, 2$, scanning over possible phases
 $0\leq\varphi\leq2\pi$ and using the combined HFAG dataset.  
\begin{figure}[h]
 $$\epsfxsize=0.55\textwidth\epsffile{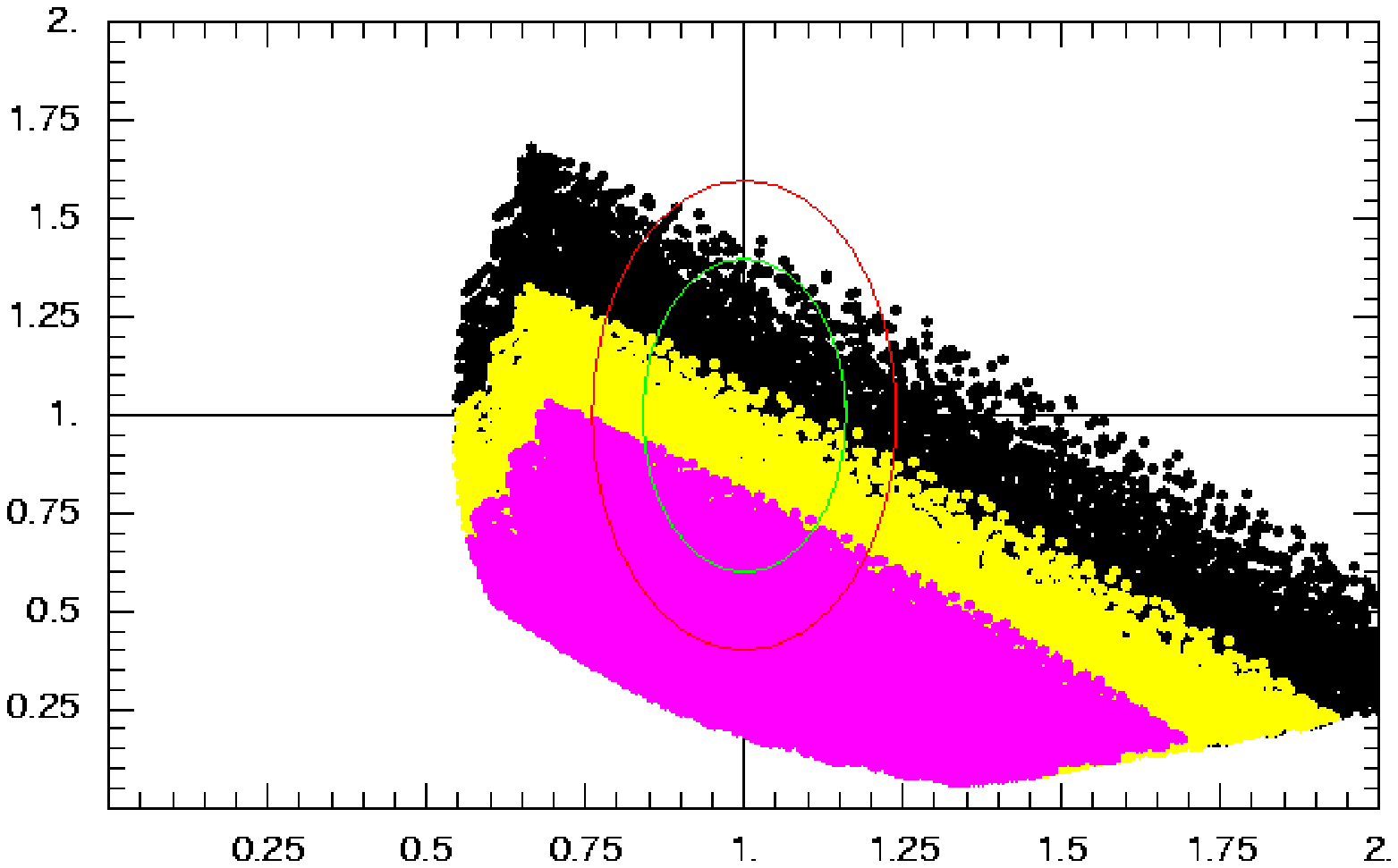}$$
 \vskip-12pt
 \caption[]{Theory/Experimental ratio using HFAG data for $\Gamma^{+-}$
   against $\Gamma^{00}$ with charming penguin contribution of $|R|=1$
   (magenta), $|R|=1.5$ (yellow) and $|R| = 2$ (Black), scanning over all
   possible phases $\varphi$.}  
 \label{fig:chpHFAG}
 $$\epsfxsize=0.55\textwidth\epsffile{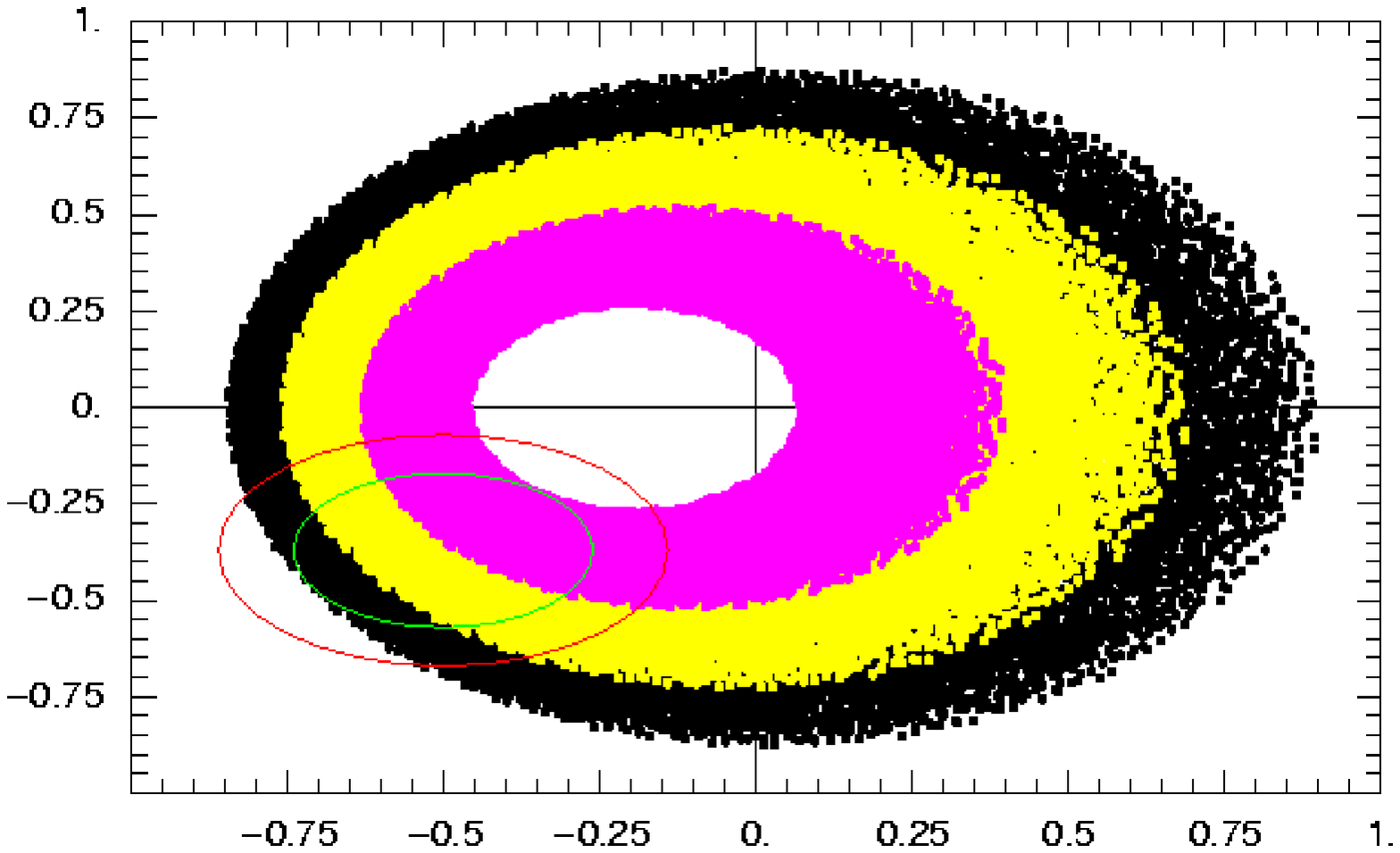}$$
 \vskip-12pt
 \caption[]{As above for CP asymmetry $S_{\pi\pi}$/$C_{\pi\pi}$.}  
 \label{fig:chpHFAGcp}
\end{figure}

 Similarly to our previous scenarios, we find that the input parameters optimise for
 the maximal contribution $\Delta=1.3$ and with $F^\pi = 0.22$.  
 The figures show that considerable enhancement is possible, even for small values of
 $|R|$, and that it is possible to reproduce the data very well.  It is clear
 however that there must be restrictions on the size and the phase of the
 parameter $R$, in order not to overshoot the experimental bounds --
 especially in the case of the CP asymmetry.  Any contribution of magnitude
 $|R|>2$ would be difficult to reconcile with the data.   

As can be seen in the figures, using any magnitude of enhancement between
$|R|=1$ and $|R|=2$ can reproduce the data extremely well for both the branching
ratios and for the CP asymmetry; however, the required phase does not agree
for both.  The best-fit for the branching ratios requires
$\varphi\sim120^\circ$, and for the CP asymmetries $\varphi\sim300^\circ$.
It is the value of $C_{\pi\pi}$, that dictates the 
 best-fit value for the CP plot, as it is very sensitive to the phase of the
 additional charming penguin contributions.  Using our $\chi^2$ minimisation
 technique we found the best-fit to all six experimental results, requiring a
 charming penguin contribution of $R=2e^{i260^\circ}$, resulting in a
 (minimum) $\chi^2 = 0.44$; this is shown below in
 Figure \ref{fig:chpBaBarbest}.    \\

\begin{figure}[h]
   $$\epsfxsize=0.4\textwidth\epsffile{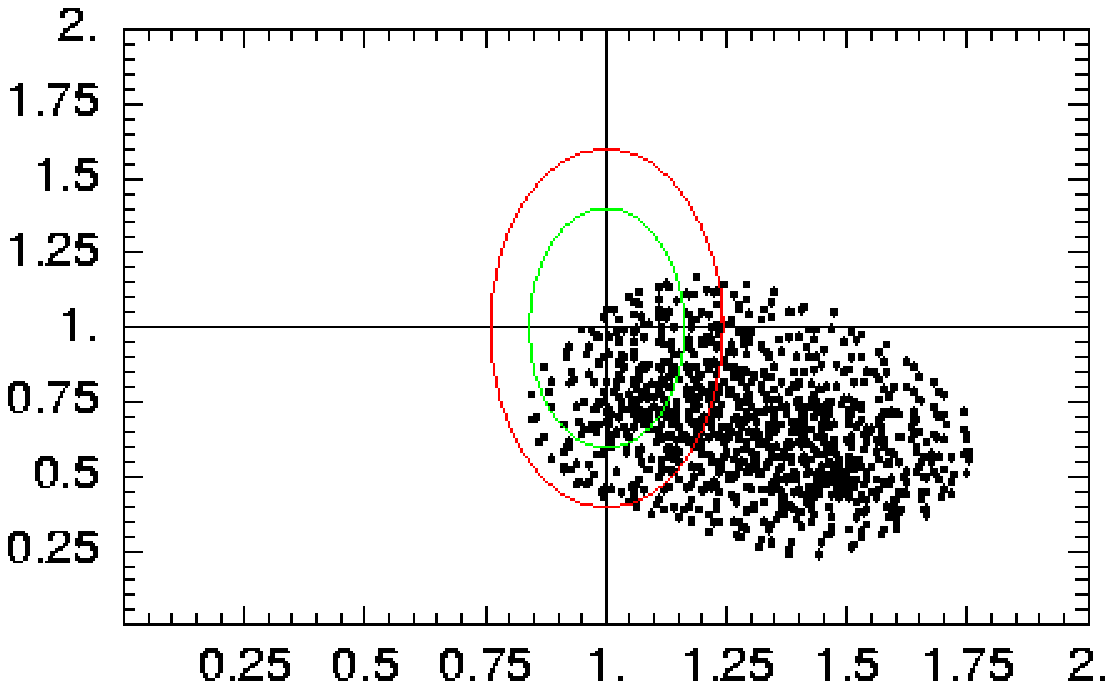}\qquad
 \epsfxsize=0.4\textwidth\epsffile{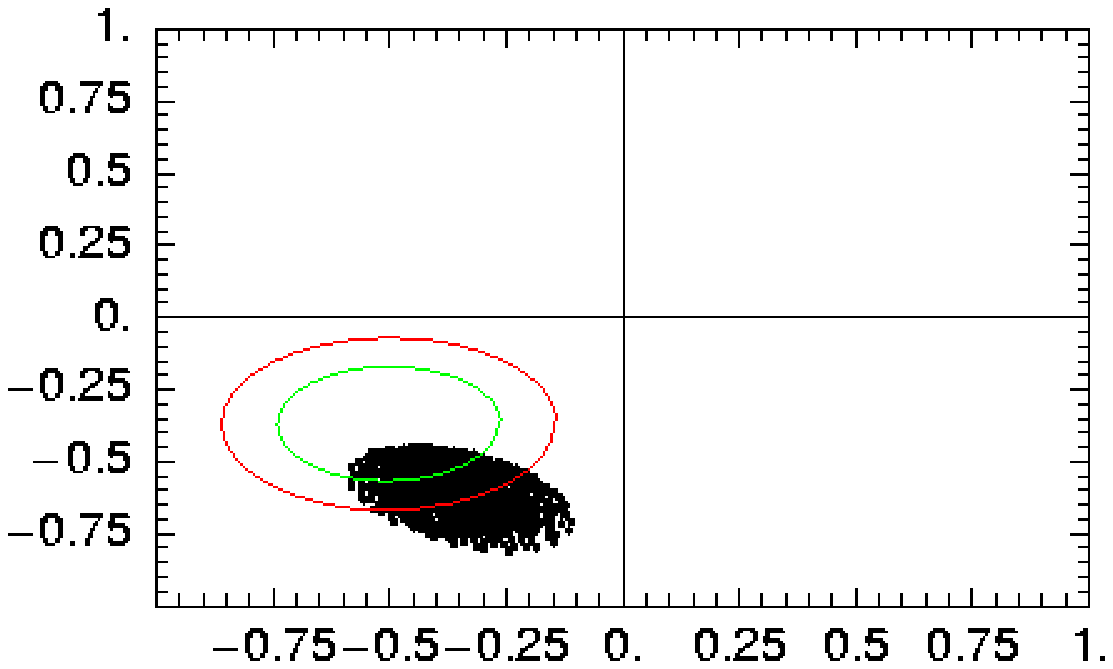}$$
 \vskip-12pt
 \caption[]{Best fit agreement from $\chi^2$ minimisation to full data set
 with charming penguin 
   contribution of $R=2e^{i260^\circ}$, using HFAG data with $\Delta=1.3$,
   $F^\pi=0.22$.}   
 \label{fig:chpBaBarbest} 
\end{figure}

\section{Discussion and comments}

In this chapter we have presented an analysis of the non-factorisable
corrections to $B\to\pi\pi$ and shown how the experimental data can be used
to guide and constrain the levels of contribution needed to gain agreement
between data and theory.  Before discussing the full implications of our
results, it is useful to compare the size of the model-dependent QCD factorisation
parameterisation of the non-factorisable effects to our generic
non-factorisable contributions.  Consider first the corrections arising
from the hard-spectator diagrams, where the divergences from the chirally
enhanced twist-3 contributions to kernel $T_i^{II}$ are parameterised by
complex parameter $X_H$:
\begin{eqnarray*}
   H_i(\pi\pi) = \frac{N_i f_B f_\pi}{m_B\lambda_B F_0^{\pi}(0)}\Big[ 3r_\chi^\pi
   \Delta X_H\Big]
\end{eqnarray*}
with $N_i = 0$ (i = 6, 8), -1 (i = 5, 7) or 1 otherwise.  These contributions
enter into each of the factorisation coefficients $a_i^{II}(\pi\pi)$.  We can
extract the contributions proportional to $X_H$ in both of the isospin
amplitudes $A_{1/2}$ and $A_{3/2}$ and obtain
\begin{eqnarray*}
   (A^p_{1/2})_{HS} &=&
   B_{\pi\pi}\frac{m_B}{\lambda_B}\frac{C_F\pi\alpha_s(\mu_h)}{2N_c^2}\,\mathbb{C}_{1/2}\,\,r_\chi^\pi\Delta
   X_H \nonumber\\
 (A^p_{3/2})_{HS} &=&
  - B_{\pi\pi}\frac{m_B}{\lambda_B}\frac{C_F\pi\alpha_s(\mu_h)}{2N_c^2}\,\mathbb{C}_{3/2}\,\,r_\chi^\pi\Delta
   X_H \nonumber  
\end{eqnarray*}  
with $B_{\pi\pi} = i \frac{G_F}{\sqrt{2}}f_Bf_\pi^2$, 
and the $\mathbb{C}_{\Delta I}$ are combinations of Wilson coefficients given as
\begin{eqnarray*}
     \mathbb{C}_{1/2} &=& \delta_{pu}(4C_2-2C_1)+6C_3 -3C_8+3C_9-3C_{10}\\
     \mathbb{C}_{3/2} &=& \delta_{pu}(2C_2+2C_1)+3C_8+3C_9+3C_{10}
\end{eqnarray*}  
Comparing this contribution to the expression for our generic
non-factorisable correction (\ref{eq:isononfac}), we can estimate
what size of correction corresponds to the ``default'' estimate of
$X_H\sim\ln(m_b/\Lambda_h)\sim2.4$ \cite{Beneke:2001ev}.  For
$\Delta = 1$ and equal size contributions from $N_{1/2}^{\,u}$ and $N_{1/2}^{\,c}$,
we estimate that using the BBNS model for $X_H$ equates to a non-factorisable
correction of $\sim8\%$. 

We can perform a similar decomposition for the other main source of
non-factorisable corrections -- the annihilation contributions -- parameterised
by the parameter $X_A$ in the BBNS model.  There is no correction to the $A_{3/2}$
isospin amplitude (and correspondingly no annihilation contribution to the 
$B\to\pi^+\pi^0$ decay), and the correction to $A_{1/2}$ can be expressed
\begin{equation*}
   (A_{1/2}^p)_{ann} = B_{\pi\pi}\frac{2C_F}{N_c^2}\left[A_1^i\mathbb{C}_1 + A_3^f\mathbb{C}_3\right]
\end{equation*}
with $\mathbb{C}_i$ again combinations of Wilson coefficients
\begin{eqnarray*}
\mathbb{C}_1 &=& \delta_{pu}C_1 + 
C_3+2C_4+2C_6-\tfrac{1}{2}C_9+\tfrac{1}{2}C_{10}+\tfrac{1}{2}C_8 \\
\mathbb{C}_3 &=& C_5+N_cC_6-\tfrac{1}{2}C_7-\tfrac{1}{2}N_cC_8
\end{eqnarray*}
and the amplitudes $A_1^i$ and $A_3^f$, calculated with asymptotic
distribution amplitudes and assuming SU(3) flavour symmetry, are 
\begin{eqnarray*}
   A_1^i &=& 2\pi\alpha_s\left[9(X_A-4+\frac{\pi^2}{3})+r_\chi^\pi
   X_A^2\right]\\
   A_3^f &=& 12\pi\alpha_s\,r_\chi^\pi(2X_A^2 - X_A)
\end{eqnarray*}  
We can see the explicit power-suppression of these terms
relative to the leading gluon exchange contributions via the fact that the
weak annihilation terms are proportional to $f_B$ rather than
$f_B\,m_B/\lambda_B$.  By comparing these expressions with the isospin
expansions of (\ref{eq:isononfac}) and using the default value of
$X_A\sim\ln(m_b/\Lambda_h)\sim2.4$, we obtain a corresponding size for the generic
non-factorisable correction of $\sim5\%$.

From these estimates it becomes very clear how this size of contribution
cannot reproduce the experimental data.  This would correspond to an overall
NF contribution less than the moderate (20\%) correction we tested in
Scenario I, which in itself could not reconcile the predictions with the
experimental data.

It was the marginal agreement of the QCDF predictions with the $B\to\pi\pi$
data that originally motivated us to test the consequences of enhancing the 
non-factorisable corrections to a moderate (20\%) and extreme (40\%) level.
Our results show clearly that even for extremal values of our input
parameters, the best-fit scenarios do not provide reasonable agreement
between theory and experiment until at least 30\% contributions are used.  In
the case of the combined HFAG data, a 40\% contribution is required to obtain
a reasonable $2\sigma$ agreement.  This also highlights the problems with the
discrepancies between the {\textsc{BaBar}} and Belle measurements.

Our final scenario incorporated the possibility of the charming penguin as an
additional non-factorisable effect.  We found that significant enhancement
of the branching ratios is possible, but that a considerable contribution
with $|R|>2$ is not supported by the data.  There are also a number of questions
which must be addressed if we are to consider this contribution as a
realistic contender for a significant, neglected $1/m_b$ correction.
Firstly, the required best-fit phase of the charming penguin parameter $R$
does not naturally agree for both the branching ratios and the CP
asymmetries simultaneously.  The best-fit agreement (shown in Figure
\ref{fig:chpBaBarbest}) is however reasonable, although yet more NF
corrections would be needed to gain a $1\sigma$ agreement with the data.    

Secondly, we must ensure that the size of the contribution suggested by our
results is appropriate with respect to the leading QCD
factorisation terms.  To check this, we can determine the relative size of
the charming penguin contribution to the relevant isospin amplitude,
$A_{1/2}$, compared with the factorisable contribution.  
For the maximum allowed charming penguin contribution $|R|=2$, the ratio of
the charming penguin contribution to the QCDF contribution is
$\sim0.3$, and for a moderate contribution of $|R|=1.5$, we find a ratio of
$\sim0.15$.  We can compare this to the relative size of our generic
non-factorisable corrections, for example with equal phases and
$|N_{1/2}^{\,u,\,c\,}|=0.2$, which provides a correction 15\% the size of the purely
factorisable terms.   

\begin{figure}[h!]
   $$\epsfxsize=0.4\textwidth\epsffile{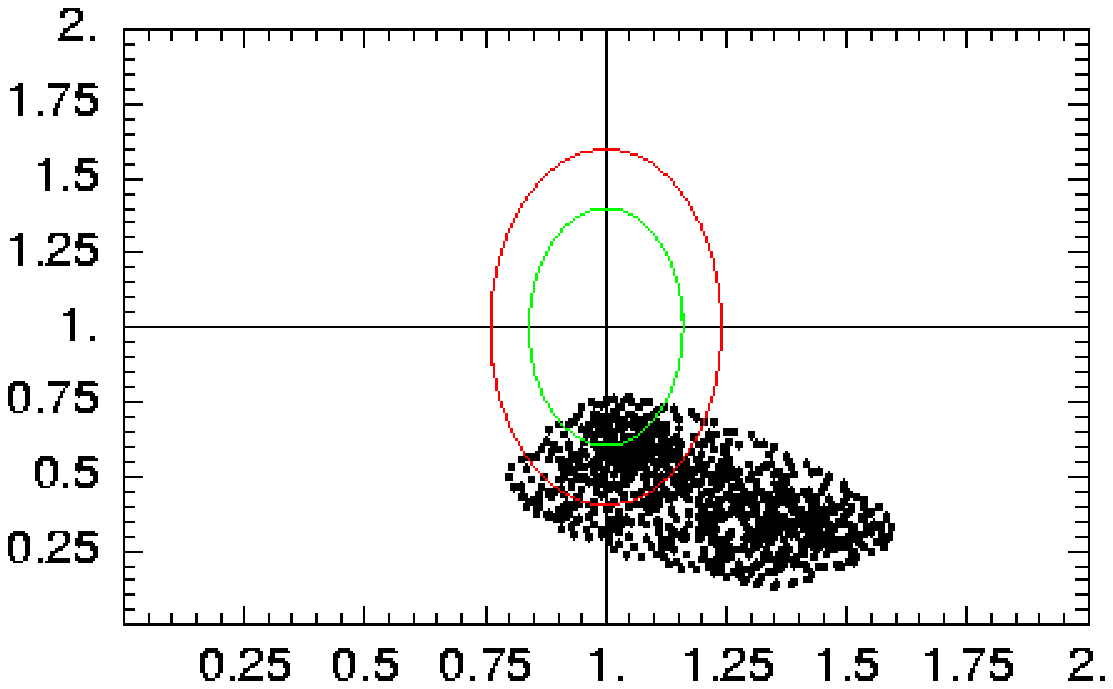}\qquad
 \epsfxsize=0.4\textwidth\epsffile{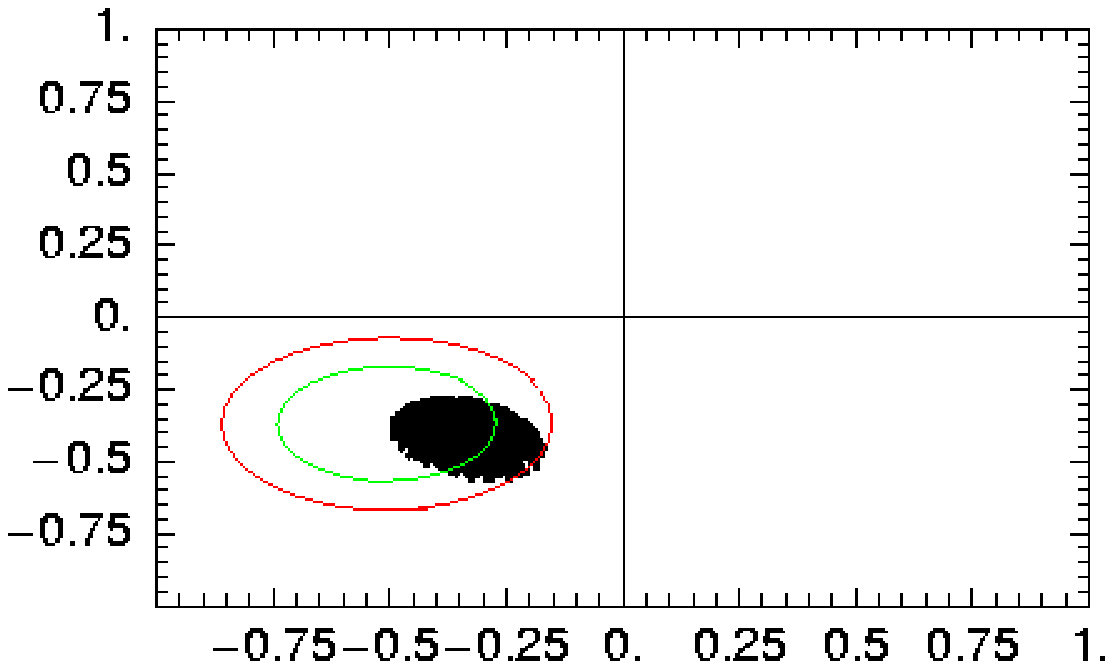}$$
 \vskip-12pt
 \caption[]{Best fit agreement from $\chi^2$ minimisation to full data set
 with charming penguin 
   contribution of $R=1.2e^{i269^\circ}$, using HFAG data with $\Delta=1.3$,
   $F^\pi=0.22$.}   
 \label{fig:chpHFAGbest} 
\end{figure}

It would seem that the enhancement from the charming penguin must be
approached with caution, as even a moderate increase, $|R|=1.5$, gives an
overall contribution equal to the size of the non-factorisable corrections
coming from all of the annihilation and hard-spectator topologies.  This
suggests the implementation of the charming penguins in this way can provide
a relative size contribution too large for them to be realistic additional NF
contributions.  We should then increase the constraint on the value of $R$,
restricting the maximum contribution to be $|R|\sim1.2$.  The theoretical
agreement to the $B\to\pi\pi$ observables would still receive significant
improvement from the additional contributions, although not enough to bring 
both the branching ratios and CP asymmetries into $2\sigma$ agreement using 
only a 20\% generic NF contribution.  The best-fit agreement to the HFAG data
with this constraint is shown in Figure \ref{fig:chpHFAGbest}.

Finally we can consider the application of this work to other charmless
decays, as the techniques and results that we determined for $B\to\pi\pi$ can 
be applied to $B\to\pi K$ \cite{BJT05}.  An isospin decomposition
can be performed in an  
analogous way to $B\to\pi\pi$; there are now six independent isospin amplitudes 
\begin{equation*}
   A^p_{3/2, 1} \qquad A^p_{1/2, 1} \qquad A^p_{1/2, 0}
\end{equation*}
with $p=u,c$.  The factorisable contribution can again be determined from
QCDF, and the non-factorisable contributions defined as
corrections to the above amplitudes.  These corrections can be related to the
non-factorisable corrections to $B\to\pi\pi$ via $SU(3)$ flavour symmetry. 

It is possible in principle to use the best-fit data found from the
$B\to\pi\pi$ case for $\Delta(\pi)$ and $F_0^{B\to\pi}$.  The separate $\pi K$ 
input parameters, $\Delta(K)$ and $a_1^K$ (from the $\phi_K$
distribution amplitude), $F_0^{B\to K}$, and the CKM factors $R_u$ and
$\gamma$, can be varied within their allowed ranges.

There is experimental information available for four $\pi K$ branching
ratios ($\pi^0 K^+, \pi^+ K^0, \pi^0 K^0, \pi^- K^+$), four direct CP
asymmetries $C_{+0}, C_{0+}, C_{00}$ and $C_{-+}$ and one indirect CP
asymmetry for $B\to\pi^0 K^0$.  Using this information, the level of
non-factorisable corrections needed to bring the predictions for the $\pi K$
system into agreement with the data can be tested and constrained.   The
possibility of charming penguins also has more significance in this system,
as they are expected to have a significant contribution to the decays $B\to
K^+\pi^-$ and $B^+\to K^0\pi^+$.

%%% Local Variables:
%%% mode: latex
%%% TeX-master: t
%%% End:

%% file: chapter5sm.tex
\chapter{Exclusive radiative B-decays}\label{chp:ERB}
\begin{center}
  \begin{quote}
    \it
    It is wisdom to recognise necessity when all other courses have been
    weighed, though as folly it may appear to those who cling to false hope...
  \end{quote}
\end{center}
\vspace{-4mm}
\hfill{\small ``The Lord of the Rings'', J.R.R. Tolkien}
\vspace{5mm}

Radiative penguin transitions such as $b\to s\gamma$ or $b\to d\gamma$ are
examples of flavour changing neutral currents.  They are rare decays, arising
only at 1-loop in the Standard Model and are a valuable
test of predictions of flavour physics, enabling us to test the detailed
structure of this sector at the level of radiative corrections.

There has been much work and speculation regarding the
phenomenology of FCNC processes, as they are ideal candidates for indirect
searches for new physics phenomena e.g. as reviewed in
\cite{Krizan:2005dw,Neubert:2002ku}.  The process $b\to s\gamma$ allows 
a large parameter space for new physics and has undergone  
scrutiny by a number of authors
\cite{Ciuchini:2002uv,Borzumati:1999qt,Bertolini:1987pk,Kagan:1998bh,Besmer:2001cj};
this process is seen as ideal to constrain the
possible new physics effects, assisted by available experimental data on
both the inclusive process $B\to X_s\gamma$ and the exclusive
channel $B\to K^*\gamma$.  The $b\to d\gamma$ process is
suppressed with respect to $b\to s\gamma$ by the CKM factor $\lambda_t= V_{tb}V_{td}^*$, and
the exclusive modes have 
previously had only experimental upper bounds from the $B$ factories.  In June of
this year however, Belle announced new measurements of $B\to\rho\gamma$
and $B\to\omega\gamma$ \cite{Abe:2005rj}.  These new data for $b\to d$ decays
and updated results for the $b\to s$ processes present an ideal opportunity
to perform an analysis of the radiative $B$ decays, allowing us to both test the
predictions of the Standard Model and examine the sensitivity of $b\to
(d,\,s)\gamma$ to new physics.

In this chapter, we consider the exclusive channels
$B\to(\rho,\,\omega)\gamma$ and $B\to K^*\gamma$ within the Standard Model,
based on the formulation of $B\to V\gamma$ decays in QCD factorisation from
Bosch and Buchalla 
\cite{Bosch:2001gv}.  We construct ratios of these branching fractions in
order to limit the theoretical uncertainties in the predictions.  We also
examine the effect of the uncertainty from the distribution amplitudes of the
vector mesons -- the $\rho,\,\omega$ or $K^*$ -- by using our new resummed DA
model as developed in Chapter \ref{chp:LCDA}.  By testing the branching fractions
against our DA parameter $\Delta$, we can determine how 
much of an effect the LCDA has upon the final predictions for these decay
channels.  Using the ratio of the $B\to(\rho,\,\omega)\gamma$ and $B\to K^*\gamma$
branching fractions we show how the CKM ratio $|V_{td}/V_{ts}|$ can be
extracted, and obtain an estimate of this quantity using the QCDF
predictions.  
  
We then go on to motivate an analysis of $B\to V\gamma$ decays within generic
supersymmetric models with the so-called \textit{mass insertion
  approximation} \cite{Hall:1985dx}.  We then determine the constraints on such a generic MSSM
model for both $b\to d\gamma$ and $b\to s\gamma$ transitions, using all the
available experimental data from the $B$ factory experiments.  

\section{$B\to V\gamma$ in QCD factorisation}\label{sec:BVgexp}

A model-independent framework for the analysis of the radiative decays has
been presented in \cite{Bosch:2001gv}; we summarise the important points and
calculational formulae here.  The method presents an expansion in the
heavy-quark limit, giving expressions valid at leading order in a
$\Lambda_{\mathrm{QCD}}/m_b$ expansion.  There are contributions 
from power-suppressed annihilation topologies, particularly relevant for the
$B\to\rho\gamma$ decays, where they are numerically enhanced by large Wilson
coefficients.  We implement these corrections following the method of
\cite{Bosch:2002bw}.  

\begin{figure}[h]
 $$\epsfxsize=0.4\textwidth\epsffile{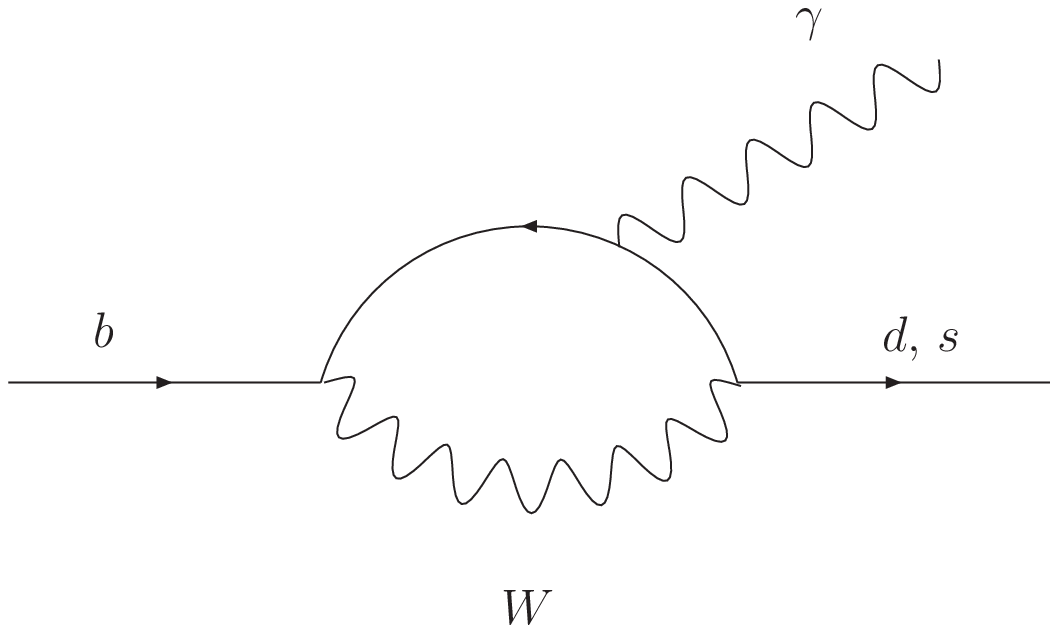} \qquad
  \epsfxsize=0.37\textwidth\epsffile{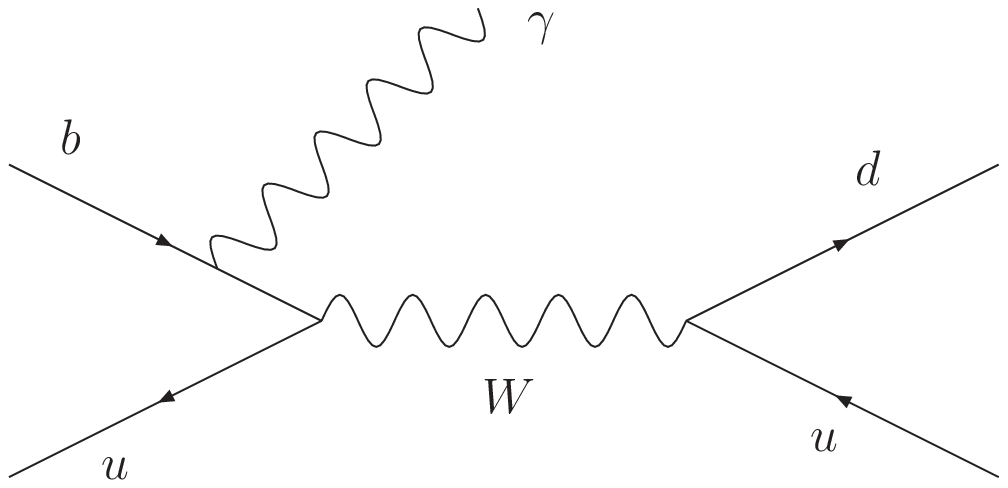} $$  
 \vskip-12pt
  \caption[ ]{Loop diagram for $b\to d\gamma$ and the annihilation diagram
  for $B^-\to\rho^-\gamma$.}
   \label{fig:BVg}
\end{figure}

The process, shown in Figure \ref{fig:BVg}, is dominated by the
electromagnetic penguin operator $Q_{7\gamma}$.  The annihilation
diagram shown is important in $B\to\rho\gamma$, specifically for
the charged decays.

The decay amplitudes are expressed following a similar procedure as in QCDF
for decays into two mesons, i.e. in terms of Wilson
coefficients and hadronic matrix elements, which are then calculated using
a factorisation formula.  The matrix elements for the radiative processes are
found using a factorisation formula valid up to corrections of order
$(\Lambda_{\mathrm{QCD}}/m_b)$ 
\be
  \label{eq:factBV}
   \langle V\gamma(\epsilon)|Q_i|\bar B\rangle = \left[F^{B\to V}(0)T_i^I + \int_0^1\,d\xi\,dv\,T_i^{II}(\xi,v)\,\Phi_B(\xi)\Phi_V(v)\right]\cdot\epsilon
\ee
 where $\epsilon$ is the photon polarisation four-vector.  For the matrix
 element of the operator $Q_7$, the formula (\ref{eq:factBV}) becomes
 trivial:  The kernel $T_i^I$ is a purely kinematic function and the spectator term is
 absent, so the only dependence on non-perturbative input parameters that remain is
 from the $B\to V$ transition form factor.  At the leading order, this is the
 only contribution to the $B\to V\gamma$ decay.  At the next-to-leading order
 we gain additional contributions from the other $\Delta B=1$ operators,
 dominated by $Q_1$ and $Q_8$.      

The matrix elements of the additional operators can be expressed in terms of
the calculable matrix element for $\langle Q_7\rangle$.  There are two
contributions, one from the hard-scattering kernel $T_i^I$ for vertex
corrections, and the other from $T_i^{II}$ which includes the hard-spectator
corrections.  These are given by 
\begin{eqnarray*}
  \langle Q_{1,8}\rangle^I&=&\frac{\a_s(m_b)C_F}{4\pi}\langle Q_7\rangle G_{1,8}\\
  \langle Q_{1,8}\rangle^{II}&=&\frac{\a_s(\mu_h)C_F}{4\pi}\langle Q_7\rangle H_{1,8} 
\end{eqnarray*}

The type-I contributions depend on $G_i(s_p)$, which are simple functions of
$s_p=m_p^2/m_b^2$ (for $p=u,c$) given explicitly in \cite{Bosch:2001gv}.  The
spectator contributions are incorporated into the functions $H_i$, and are
more interesting than the type-I functions:   
\begin{eqnarray*}
  \label{eq:H}
  H_1(s_p)&=&-\frac{2\pi^2}{3N_c}\frac{f_Bf_{V}^\perp}{F^{V}m_B^2}\int_0^1d\xi\frac{\Phi_{B
  }(\xi)}{\xi}\int_0^1dv\,h(1-v, s_p)\,\Phi_\perp(v) \\
  {} \\
 H_8&=&\frac{4\pi^2}{3N_c}\frac{f_Bf_{V}^\perp}{F^{V}m_B^2}\int_0^1d\xi\frac{\Phi_{B
  }(\xi)}{\xi}\int_0^1dv\,\frac{\Phi_\perp(v)}{v}
\end{eqnarray*}
The  hard-scattering function $h(u, z)$ is given as 
\begin{eqnarray}
  \label{eq:h}
  h(u, z)=\frac{4z}{u^2}\left[Li_2\left(\frac{2}{1-\sqrt{\frac{u-4z}{u}}}\right)+Li_2\left(\frac{2}{1+\sqrt{\frac{u-4z}{u}}}\right)\right]-\frac{2}{u}
\end{eqnarray}
and is real for values of $u\leq 4z$, and complex for $u>4z$.  For $z\to0$
(for example the up-quark contribution $z_u\sim 0$), the 
function is regular and becomes
\begin{equation*}
   h(u, 0) = -\frac{2}{u}
\end{equation*}
The distribution amplitude for the $B$-meson is parameterised as before using
the first inverse moment
\begin{equation*}
   \int_0^1\,d\xi\frac{\Phi_B(\xi)}{\xi} = \frac{m_B}{\lambda_B}
\end{equation*}

These type-II contributions also include a dependence on the light-cone
distribution amplitude of the vector meson of the $B\to V\gamma$ decay,
specifically, for the \textit{transversely polarised vector mesons},
$\phi_\perp$.  The vector meson is described at leading twist, by two
distribution amplitudes $\phi_\parallel$ and $\phi_\perp$ corresponding to
longitudinally and transversely polarised mesons.  The vector meson in $B\to
V \gamma$ decays is to leading power transversely polarised, and hence can be
described by $\phi_\perp$, as any contributions from $\phi_\parallel$ are
suppressed.  
 
As with the pseudoscalar mesons, the distribution amplitude $\phi_\perp$ can
be treated using a conformal expansion that is often truncated at the moment
$a_2$.  In our analysis, we wish to examine the effects of a non-standard DA
on the radiative decays, so we use our resummed model DA in place of this
truncated expansion.  For definiteness we use the model
$\phi^+_{3,3}(\Delta)$, where $\Delta=1$ reduces to the asymptotic DA.  This
model applies to the mesons symmetric in the fractional momentum $u$,
i.e. $\pi$, $\rho$, $\omega$.  Where we consider the $K^*$ channels, there is
a non-zero anti-symmetric part to the meson wavefunction which we can also
model using our resummed DA models.  This is constructed in an analogous way
to the symmetric DA, as discussed in Chapter \ref{chp:LCDA}.  We use the
model $\psi^+_4(a_1)$, which allows for the inclusion of an arbitrary value of
the first anti-symmetric Gegenbauer moment $a_1^K$.  We take the value of
$a_1^K$ from our discussion on $\phi_K$ unless otherwise stated.  The value
of $\Delta$ for the $K^*$ wavefunction is the sum of the symmetric and
antisymmetric parts, i.e. 
\begin{equation*}
  \Delta^{\textrm{tot},+}=\Delta + \Delta^{\textrm{asym},+}
\end{equation*}
with $\Delta^{\textrm{asym},+} = -a_1(3/2)^c\zeta(c,\,3/2)$.  Hereafter, we
will refer to the total $\Delta^{\textrm{tot}}$ as $\Delta$ and imply the 
inclusion of the anti-symmetric part.

Finally we combine all of the contributions to the $B\to V\gamma$ decay into
a factorisation coefficient, written as 
\cite{Bosch:2001gv}
\begin{eqnarray}
   \label{eq:a7y}
   a^p_7(V\gamma) = C_{7\gamma}^{\mathrm{NLO}} &+& \frac{\a_s(m_b) C_F}{4\pi} \left[
   C_1^{(0)}(m_b)\,G_1(s_p) + C_8^{(0)\,\mathrm{eff}}(m_b)\,G_8\right] \nonumber\\
           &+& \frac{\a_s(\mu_h) C_F}{4\pi} \left[
   C_1^{(0)}(\mu_h)\,H_1(s_p) + C_8^{(0)\,\mathrm{eff}}(\mu_h)\,H_8\right]
\end{eqnarray}
The Wilson coefficients $C_{1}^{(0)}$ and $C_{8}^{(0)\,\mathrm{eff}}$ as indicated,
are taken at leading order.  

The decay amplitudes are then expressed as
follows, keeping both the charm and up-quark contribution 
\be
   A(B\to V\gamma) = \frac{G_F}{\sqrt{2}}\left[\sum_p\lambda_p^{(q)}
     a_7^p(V\gamma)\right] \langle V\gamma|Q_7|\bar B\rangle
\ee
giving the final $B\to V\gamma$ branching ratios as
\be
   \label{eq:BVgBR}
   \mathrm{BR}(\bar B\to V\gamma) = \tau_B\frac{G_F\a_e
     m_B^3m_b^2}{32\pi^4}\left(1-\frac{m_V^2}{m_B^3}\right)^3
   \left\lvert\sum_p\lambda_p^{(q)}a_7^p(V\gamma)\right\rvert^2
   c_{\,V}^2\left\lvert F^V\right\rvert^2
\ee
where $q=s$ for $V=K^*$ and $q=d$ for $V=\rho$; $c_V = 1$ for $K^*,\,
\rho^-$ and $c_V=1/\sqrt{2}$ for $\rho^0$.  The CP-conjugate branching ratios
are obtained by replacing $\lambda_p^{(q)}\to\lambda_p^{(q)*}$; in our analysis we
     take all branching ratios as CP-averaged.  

The annihilation contributions can be included in the factorisation
coefficient $a_7^p$ via
\begin{equation*}
   a_7^p \longrightarrow a_7^p + a^p_{ann}
\end{equation*}
The dominant annihilation contributions are suppressed by one power of
$\Lambda_{\mathrm{QCD}}/m_b$, but can still be calculated 
within QCD factorisation, (a proof to $\mathcal{O}(\a_s)$ can be found in
\cite{Bosch:2002bw}).  These can give a sizable contribution to the radiative
decays, such as $B\to\rho\gamma$, where the contributions originating from
operators $Q_1$ and $Q_2$ are enhanced by large Wilson coefficients $C_1$ and $C_2$.  
These have small impact on the $B\to K^*$ decays due to CKM suppression,
however this decay can receive a large contribution from the operator $Q_6$
\cite{Kagan:2001zk}.  The annihilation contributions are also sensitive to
the flavour of the light quark of the $B$-meson, and so differentiate
between decays of the $B^0\,(\bar {B^0})$ and the $B^\pm$, and additionally
are dependent on the distribution amplitude $\phi_\perp$.  Since this is
clearly important to our analysis, we describe the contributions $a_{ann}^p$ in more
detail in Appendix \ref{chp:AppA}.  

There is no indication for large power corrections beyond these
calculable annihilation terms \cite{Beneke:2004dp,Bosch:2001gv}.

\section{Analysis and comparison with Belle data}

For channels that have low statistics such as the rare radiative processes,
the experimental measurements often combine different individual channels in
order to improve the significance, and allow a signal to be discovered.
Belle is to date the only experiment to have released an actual measurement
of the $b\to d\gamma$ processes \cite{Abe:2005rj}, where they combine the data
from three channels 
$\rho^0\gamma,\,\rho^+\gamma,\,\omega\gamma$.  The average is defined as
\cite{Ali:2004hn}
\begin{equation*}
   \mathrm{BR}\left[B\to(\rho,\omega)\gamma\right]=\frac{1}{2}\left\{\mathrm{BR} 
   (B^+\to\rho^+\gamma) +
   \frac{\tau_{B^+}}{\tau_{B^0}}\left[\mathrm{BR}(B^0\to\rho^0\gamma) +
   \mathrm{BR}(B^0\to\omega\gamma)\right]\right\} 
\end{equation*}
We can also define an isospin-averaged branching ratio $\mathrm{BR}(B\to
K^*\gamma)$ which is used frequently
\begin{equation}\label{eq:Kstarav}
   \mathrm{BR}(B\to K^*\gamma)=\frac{1}{2}\left\{\mathrm{BR} 
   (B^+\to K^{*+}\gamma) +
   \frac{\tau_{B^+}}{\tau_{B^0}}\mathrm{BR}(B^0\to K^{*0}\gamma)\right\} 
\end{equation}
We
 can predict the branching fractions or CP asymmetries for the $B\to
V\gamma$ decays directly from these averages using the QCD factorisation
 formulae of Section \ref{sec:BVgexp}.  We can also find the ratio of these
 averaged branching fractions which is fitted directly from the experimental data.
 We define, using the CP-averaged branching fractions \cite{Ali:2004hn,Bosch:2004nd} 
\begin{eqnarray}
\label{eq:THratio}
   R\left[(\rho,\omega)\gamma/K^*\gamma\right] &=&
  \frac{\mathrm{BR}\left[B\to(\rho,\omega)\gamma\right]}{\mathrm{BR}(B\to
  K^*\gamma)} \nonumber\\
   &=&
  \left\lvert\frac{V_{td}}{V_{ts}}\right\rvert^2\frac{(1-m_{(\rho,
  \omega)}^2/m_B^2)^3}{(1-m_{K^*}^2/m_B^2)^3}\,\zeta^2\,c_\rho^2\left\lvert\frac{a_7^c((\rho,\omega
  )\gamma)}{a_7^c(K^*\gamma)}\right\rvert^2[1+\Delta R ]   \nonumber\\
\end{eqnarray}
$\zeta$ denotes the form factor ratio $T_1^\rho(0)/T_1^{K^*}(0)$, which characterises   
the $SU(3)$ breaking in the transition form factors.  The form factor
$T_1(0)$ is one of seven independent form factors which describe all $B\to V$
decays, and is the only one relevant for the radiative $B\to V\gamma$ decay.
It is defined in conjunction with two other form factors $T_{2,3}$ via
\cite{Ball:2004rg}
\begin{eqnarray*}
   && \hspace{-50pt}\frac{1}{c_V}\langle V(p)|\bar{q}\sigma_{\mu\nu}q^\nu(1+\gamma_5)b|B(p_B)\rangle =
   i\epsilon_{\mu\nu\rho\sigma}\epsilon^{*\nu}p^\rho_B\,p^\sigma\,2T_1(q^2)\\
 && \qquad \qquad +\,
   T_2(q^2)\left\{e_\mu^*(m_B^2-m_V^2)-e^*q)(p_B+p)_\mu\right\}\\
 && \qquad \qquad +\, T_3(q^2)(e^*q)\left\{q_\mu-\frac{q^2}{m_B^2-m_V^2}(p_B+p)_\mu\right\}
\end{eqnarray*}

The parameter $\zeta$ has been estimated using several methods, the most recent value is from
Ball and Zwicky using light-cone sum rules \cite{BZ05} who find $\zeta^{-1}=1.25\pm0.18$.   
%% Lattice QCD can no longer provide a reliable result for this
%% quantity as the $K^*$ meson is not visible on the lattice -- this is an
%% unfortunate side effect of the un-quenching of lattice calculations, which has
%% been able to provide a light mass to the pion, hence to allow the $K^*$
%% decay

$\Delta R$ is given within QCD factorisation by \cite{Bosch:2004nd}
\begin{eqnarray}
\label{eq:deltar}
   \Delta R = 2\mathrm{Re}\,\delta a\,\left(\frac{R_u^2 -
   R_u\,\cos{\gamma}}{1-2R_u\,\cos{\gamma}+R_u^2} \right) 
\end{eqnarray}
written in terms of $R_u=\sqrt{\bar\rho^2+\bar\eta^2}$ and the
CKM angle $\gamma$.  $\delta a$ is given by 
\be
\label{eq:dela}
   \delta a = \frac{a_7^u((\rho,\omega)\gamma)-
     a_7^c((\rho,\omega)\gamma)}{a_7^c((\rho,\omega)\gamma)}
\ee 
where the factors $a_7$ implicitly include the annihilation contributions,
     which contribute only at order $\Lambda_{\mathrm{QCD}}/m_b$.  If the
     annihilation contributions to $\delta a$ were neglected then
     $\delta=\mathcal{O}(\a_s)$.  To obtain expression (\ref{eq:dela})
we make use of the identity:
\be
   \label{eq:idKs}
   \lambda_c\,a_7^c + \lambda_u\,a_7^u =
   -\lambda_t\,a_7^c\left(1-\frac{\lambda_u}{\lambda_t}\frac{a_7^u-a_7^c}{a_7^c}\right) 
\ee
where the second term in the bracket in (\ref{eq:idKs}) is neglected for $B\to
K^*\gamma$ as it provides a contribution of less than 1\%.  This results in
the correction $\delta a$ depending only on the factorisation coefficients
   from $(\rho,\omega)\gamma$.  

The advantage of calculating the ratio of branching
fractions in this way is to reduce the error from theoretical
uncertainties.  Firstly, from the theory parameters that cancel in
taking the ratio,  including parameters such as $m_b$ which have a 
sizable associated uncertainty.  Secondly, the quantity $\Delta R$ is a small
correction with central values close to zero, primarily due to the smallness
of the CKM factor $f$, multiplying $\mathrm{Re}(\delta a)$.  The ranges of
$\gamma$ and $R_u$ (as suggested in Chapter \ref{chp:NFC}) imply $-0.1\leq
f\leq0.03$.  The contribution to $\Delta R$ specifically from the neutral decays is also
small due to an accidental cancellation between the $\mathcal{O}(\a_s)$ and
annihilation effects in $\delta a$.  The two main sources of uncertainty
remaining in the expression $R_{\mathrm{th}}$ are from $\zeta$ and $\Delta R$.  

We use our prediction for $R_{\mathrm{th}}$ in combination with the experimental 
results to constrain the possibility of new physics in the radiative decays,
or in its absence, to determine the CKM factor $|V_{td}/V_{ts}|$.  We can
use $R_{\mathrm{th}}$ in the following ways -- either to extract the value of
$|V_{td}/V_{ts}|$ assuming $\Delta R$ is fully known, or to extract
information about $\Delta R$ using the best-fitted value for the CKM ratio
from another source.  

We begin by analysing the individual branching ratios for $B\to\rho\gamma$
and $B\to K^*\gamma$ before going on to study the ratio $R_{\mathrm{th}}$.
We summarise the experimental 
 results for the radiative decays in Table \ref{tab:expVg}
 \cite{Aubert:2004te, Nakao:2004th,Eidelman:2004wy,Abe:2005rj}. 
\begin{table}[h]
   \begin{center}
     \begin{tabular}{|lccc|} \hline
     Decay channel & \textsc{BaBar} & Belle & Average \\ \hline
     $B_{\mathrm{exp}}(B\to K^*\gamma)$  & $40.6\pm2.6$ & $43.0\pm2.5$ &
     $42.0\pm1.7$ \\ 
     $B_{\mathrm{exp}}(B^0\to K^{*0}\gamma)$  & $39.2\pm2.0\pm2.4$ &
     $40.1\pm2.1\pm1.7$ & $40.1\pm2.0$  \\ 
     $B_{\mathrm{exp}}(B^+\to K^{*+}\gamma)$  & $38.7\pm2.8\pm2.6$ &
     $42.5\pm3.1\pm2.4$ & $40.3\pm2.6 $ \\ 
     $B_{\mathrm{exp}}(B\to (\rho,\,\omega)\gamma)$  & $<1.9$ &
     $1.34^{+0.34}_{-0.31}{}^{+0.14}_{-0.10}$ & {} \\ 
     $R_{\mathrm{exp}}\left[(\rho\,\omega)\gamma/K^*\gamma\right]$  & {} &
     $0.032\pm0.008^{+0.003}_{-0.002}$ & {} \\ \hline 
   \end{tabular}
   \caption{The CP-averaged branching fractions for exclusive $B\to
     K^*\gamma$ and $B\to\rho\gamma$, in units
     of $10^{-6}$ (with exception of $R_{exp}$) and using the combined branching fractions 
     defined in the text.}
   \label{tab:expVg}
 \end{center}
\end{table}

%{BRCP[rhoom, 1] 10^6, BRCP[rho0, 1] 10^6, BRCP[om, 1] 10^6,   BRCP[rhom, 1] 10^6}
%{2.15383, 0.998235, 0.99602, 2.18474}

\subsection{$B\to(\rho,\omega)\gamma$}

Using the formulae collected above we can obtain a theoretical estimate for
the combined and individual branching ratios for $B\to\rho\gamma$.  With a
conservative estimate of the 
uncertainty on our values, and using the asymptotic
distribution amplitude of the rho and omega mesons ($\Delta(\rho)=\Delta(\omega)=1$), we find
$\mathrm{BR}(B^\pm\to\rho^\pm\gamma) = (1.52\pm0.45)\times10^{-6}$ and
$\mathrm{BR}(B^0\to\rho^0\gamma) = (0.72\pm0.22)\times10^{-6}$. We also find, for the
combined weighted averages 
\begin{eqnarray*}
   \mathrm{BR}(B\to(\rho,\,\omega)\gamma) = (1.60\pm0.48)\times10^{-6}
\end{eqnarray*}
which is above, but within the error bounds of the current experimental
measurements.  We take the values of the form factors as $F^\rho
=0.267\pm0.021$ and $F^\omega=0.242\pm0.022$ \cite{Ball:2004rg}.  The error on
the branching ratio is estimated by taking into 
account the dominant uncertainties on the input parameters:
$F^\rho,\,F^\omega,\,\lambda_B,\, m_b,\,m_c$, $f_B$ and the CKM inputs.  Adding these in
quadrature leads to an error estimate of $\mathcal{O}(30\%)$ for the
$B\to\rho\gamma$ channels.

The distribution amplitude for the rho meson required for the
calculation of the hard-spectator and annihilation contributions to the
process is that for the transversely polarised mesons $\phi_\perp(u, \mu)$.
The leading twist DA (both the transverse and longitudinal)  
is often expressed as a 
truncated conformal expansion, where there is a prediction for $a_2^\perp$
from QCD sum rules \cite{Ball:1998ff,Ball:1998sk} of $a_2^{\,\perp}(\rho,
1\textrm{GeV}) = 0.2\pm0.1$. 
In this analysis we replace the conformal expression of $\phi_\perp(\rho)$
with our resummed DA model.  The suggested region for $\Delta$ was examined
in \cite{Ball:2004rg}, and is based on the indication from all available
calculations that $a_2>0$.  This demands, as for $\phi(\pi)$, that $\Delta>1$.
Using the model DA with $a=3$ suggests $\Delta(\rho) = \Delta(\omega) =
1.15\pm0.10$. In light of this we use a ``physical range'' for $\Delta$ of
$1\leq\Delta\leq1.4$.   
 
\begin{figure}[h]
 $$\epsfxsize=0.6\textwidth\epsffile{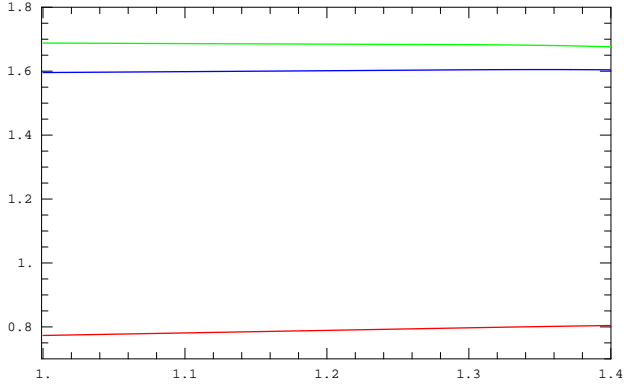} $$  
 \vskip-12pt
  \caption[ ]{Branching fractions for $B\to\rho\gamma$ decays from QCD
 factorisation using our model DA $\phi^+_{3,3}$, in units of $10^{-6}$ and in
  possible physical 
  region of $\Delta$: $B^+\to\rho^+\gamma$ (green), $B\to\rho^0\gamma$ (red)
  and the combined $B\to(\rho,\,\omega)\gamma$ (blue).}
   \label{fig:BVgdel}
\end{figure}

We test the effect of our non-standard DA by plotting the dependence of the $b\to
d\gamma$ branching ratios on the parameter $\Delta$, as is shown in Figure
\ref{fig:BVgdel}.   

This figure shows that within this range the dependence on $\Delta$ appears
relatively small.  There is a downward trend in the $\rho^+\gamma$ case and
conversely an upward trend in the $\rho^0\gamma$ case.  These conspire to
cancel in the combined branching fraction, leaving 
$B\to(\rho,\,\omega)\gamma$ free of any significant dependence on $\Delta$ and of 
uncertainty from the distribution amplitude.   The actual change in the value
of the branching ratio over the full (physical) range of $\Delta$ is of the
order of 3\% for $(\rho^0\gamma)$ and 1\% for the combined ratio.  
 
%%Tee[M1_] := (BRCP[M1, 1.4, 1, 0] - BRCP[M1, 1, 1, 0])/(BRCP[M1, 1, 1, 0])
%%{Tee[rhoom], Tee[rho0], Tee[rhom]}
%%{0.014208, 0.0335316, -0.00456878}

\subsection{$B\to K^*\gamma$}

The theoretical estimate for the $B\to K^*\gamma$ decays can be found in an
analogous way to $B\to \rho\gamma$, using the definition of the weighted average from equation
(\ref{eq:Kstarav}).  We obtain an estimate for the $K^*\gamma$ branching ratio
as $\mathrm{BR}(B\to K^*\gamma) = (5.49\pm1.64)\times10^{-5}$, which is somewhat
higher than, but within the error bounds of, the current experimental
measurement.  The considerable decrease between this value and that
presented in \cite{Bosch:2001gv} is predominantly due to the updated value of the form
factor $F^{K^*} = 0.333\pm0.028$ \cite{Ball:2004rg} from its previous value
of  $F^{K^*} = 0.38\pm0.06$ \cite{Ball:1998kk}.  

\begin{figure}[h!]
 $$\epsfxsize=0.6\textwidth\epsffile{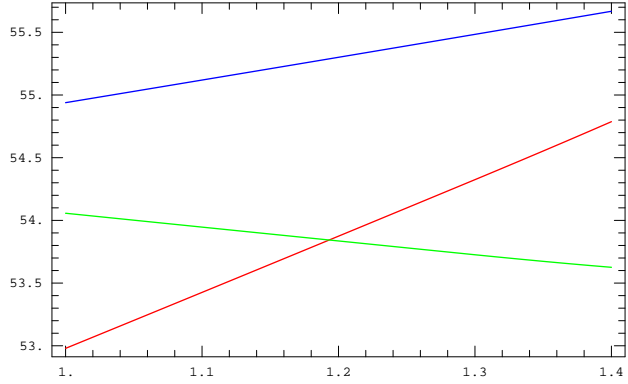} $$  
 \vskip-12pt
  \caption[ ]{Branching fractions (in units of $10^{-6}$) for $B\to K^*\gamma$ decays in QCD
 Factorisation using our model DA with antisymmetric part $\psi_4$, in
  the possible physical
  region of $\Delta$ and constant $a_1=0.13$; $B^+\to K^{*+}\gamma$ (green),
  $B\to K^{*0}\gamma$ (red) and the combined $B\to K^*\gamma$ (blue).}
   \label{fig:BVgKdel}
\end{figure}

We begin by considering the effect of the $K^*$ distribution
amplitude, by testing the dependence of the branching fractions on our
non-standard DA.  Unlike the $\rho$, we have the additional complication for
the $K^*$ meson of an anti-symmetric part to the distribution amplitude.  The 
moments for the truncated conformal expression are largely unknown, although
there have been predictions from QCD sum rules \cite{Ball:2004rg,Ball:2003sc}.  This
gives us:
\begin{equation*}
    a_1^\perp(K^*, 1\textrm{GeV}) = 0.10\pm0.07 \qquad a_2^\perp(K^*,
1\textrm{GeV}) = 0.13\pm0.08
\end{equation*}
We can examine the dependence on both the symmetric part of the $K^*$ DA and
on the leading anti-symmetric moment $a_1^K$.  The value of the symmetric
contribution to $\Delta$ for the $K^*$ is estimated at $\Delta(K^*) =
1.12\pm0.10$ -- we consider the wider range of $1\leq\Delta(K^*)\leq1.4$, in
order to compare directly with the physical range used in the $\rho\gamma$
case.  The dependence of the $B\to K^*\gamma$ 
branching ratios on $\Delta$ with a constant value of $a_1^K=0.13$ is
summarised in Figure \ref{fig:BVgKdel}.  
Although this graph suggests that the $K^*$ branching ratios have a larger
dependence that the $B\to\rho\gamma$ decays, the percentage change across the
full range of $\Delta$ is in fact the same for both sets of decays.  As
before, we have a 1\% 
change for the combined $K^*$ decay and a 3\% change for $K^{*0}$.  
%Teek[M1_] := (BRCPk[M1, 1.4, 0.15, 1, 0] - BRCPk[M1, 1, 0.15, 1, 0])/(BRCPk[
%        M1, 1, 0.15, 1, 0])
%{Teek[Kav], Teek[Ks0], Teek[Ksm]}
%{0.0116507, 0.0283178, -0.00557561}
This effect suggests that the discrepancies between the experimental
values and the theory predictions can 
not be attributed to the uncertainty from the DA -- increasing the
contribution of the higher order moments of the DA by increasing $\Delta$
serves only to increase the prediction of the averaged $K^*$ branching ratio.    

\begin{figure}[h]
 $$\epsfxsize=0.4\textwidth\epsffile{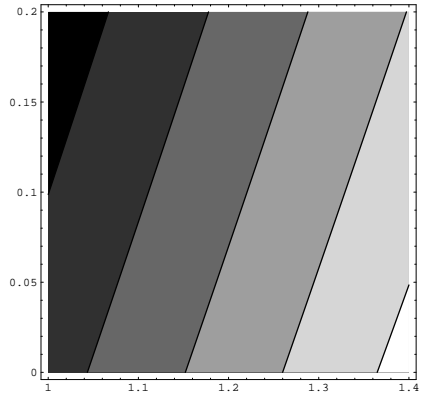} $$  
 \vskip-12pt
  \caption[ ]{Density plot of $\mathrm{BR}(B\to K^*\gamma)$ (with increasing
 values indicated by lighter shading) for variation in $\Delta$ (horizontal
  axis) and $a_1$ (vertical axis).  Contours shown at $5.5,5.52\ldots5.58\times10^{-5}$.}
   \label{fig:BVgKdela1}
\end{figure}

Figure \ref{fig:BVgKdela1} shows the dependence of the $B\to K^*$ branching
ratio on both $\Delta$ and $a_1$, and shows clearly that a higher value of
the moment $a_1$ will lead to lower values of the branching ratio for all
values of $\Delta(K^*)$, although the change is not considerable.  

\subsection{$R\left[(\rho,\omega)\gamma/K^*\gamma\right]$ and the extraction
  of $\lvert V_{td}/V_{ts}\rvert$}

This section proceeds to build on the results from the individual
branching fractions, by calculating the ratio
$R_{\mathrm{th}}\left[(\rho,\omega)\gamma/K^*\gamma\right]$ as introduced
above.  As discussed, this is important for the
extraction of information on
the hadronic quantity $\Delta R$ and the value of the CKM ratio
$\lvert V_{td}/V_{ts}\rvert$.  

\begin{figure}[h]
 $$\epsfxsize=0.6\textwidth\epsffile{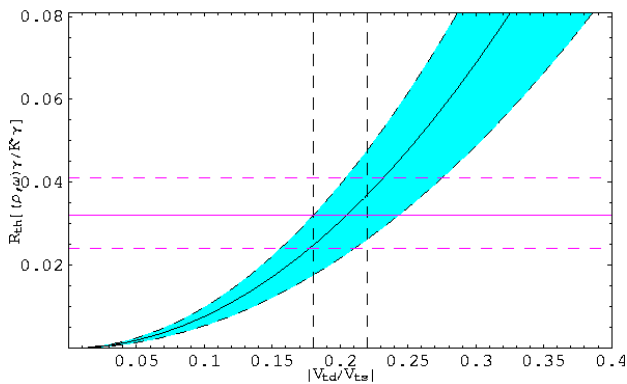} $$  
 \vskip-12pt
  \caption[ ]{Plot of
 $R_{\mathrm{th}}\left[(\rho,\omega)\gamma/K^*\gamma\right]$ (within quoted
 $\pm1\sigma$ errors) against  $\lvert
 V_{td}/V_{ts}\rvert$ with current Belle experimental result and $1\sigma$
 errors shown (horizontal band in magenta) and the SM best-fit for $\lvert
 V_{td}/V_{ts}\rvert$ (vertical band).}
   \label{fig:VTDS}
\end{figure}

We begin by using equation (\ref{eq:THratio}), and the calculation of
$\Delta R$ as in equations (\ref{eq:deltar}) and (\ref{eq:dela}) to find the
ratio of the $(\rho, \omega)\gamma$ and 
$K^*\gamma$ branching ratios.  We use the asymptotic distribution amplitudes for 
$\phi_\perp(\rho)$, $\phi_\perp(\omega)$ and $\phi_\perp(K^*)$, and obtain an
estimate of $R_{\mathrm{th}}$ as   
\begin{equation*}
   R_{\mathrm{th}} =
   \frac{\mathrm{BR}(B\to(\rho,\omega)\gamma)}{\mathrm{BR}(B\to K^*\gamma)} =
   0.030\pm0.008
\end{equation*}
which is in rather good agreement with the experimental measurement. 
The theoretical uncertainty remaining in $R_{\mathrm{th}}$ is dominated by
the form factor ratio $\zeta$, and the value of the quantity $\Delta R$.
We can use the ratio to extract a value of $\lvert
V_{td}/V_{ts}\rvert$.  Using $\zeta^{-1} = 1.25$ and our estimate of $\Delta R$
from QCD factorisation as $\Delta R = 0.006$, we  
obtain:
\begin{equation*}
   \lvert V_{td}/V_{ts}\rvert = 0.195\pm0.051
\end{equation*}
which is well within the SM range of $\lvert V_{td}/V_{ts}\rvert = 0.197\pm
0.013$ \cite{UTfit}.
We plot the dependence of $R_{\mathrm{th}}$ against  $\lvert
V_{td}/V_{ts}\rvert$ in Figure \ref{fig:VTDS}.

The constraints on the unitarity fit are continually improving, so we are
approaching the stage of having consistent measurements for many of the CKM
parameters.  We can then in principle use the measured (fitted) value for $\lvert
V_{td}/V_{ts}\rvert$ in order to examine the parametric uncertainties
entering into $R_{\mathrm{th}}$.  

We expect the dependence on the distribution amplitudes that enter into
$R_{\mathrm{th}}$ to be small, even though there is dependence on
$\Delta(\rho)$, $\Delta(K^*)$ and $a_1^{K^*}$.  We approximate by taking 
$\Delta(\omega) = \Delta(\rho)$.  Increasing the contribution from
$\Delta(\rho)$ acts to increase the ratio, and increasing $\Delta(K^*)$ acts
to decrease it.  Since these effects are of the same magnitude, the overall
effect is negligible.  

\begin{figure}[h!]
 $$\epsfxsize=0.6\textwidth\epsffile{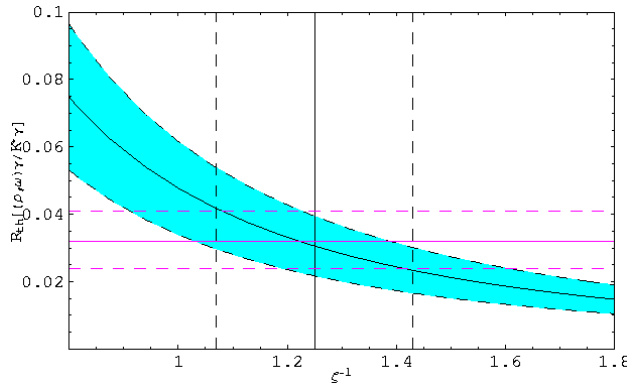} $$  
 \vskip-12pt
  \caption[ ]{Plot of
 $R_{\mathrm{th}}\left[(\rho,\omega)\gamma/K^*\gamma\right]$ (within quoted
 $\pm1\sigma$ errors) against form factor ratio $\zeta^{-1} =
 T_1^{K^*}(0)/T_1^\rho(0)$ (vertical band) with current Belle experimental result
 and $1\sigma$  errors shown (horizontal band in magenta).}
   \label{fig:Zeta}
\end{figure}

The dependence of $R_{\mathrm{th}}$ on the form factor ratio is much greater
and is illustrated 
in Figure \ref{fig:Zeta}.  This shows the sensitivity to changes in the value
of $\zeta^{-1}$ especially any movement towards lower values.  

Our prediction for
$R_{\mathrm{th}}\left[(\rho,\omega)\gamma/K^*\gamma\right]$ is close to the 
central value determined experimentally, showing clearly how well the uncertainties
can be reduced by calculating the combined ratio.  The parametric uncertainty
in the individual branching ratios is dominated by the dependence on the form
factors which enter quadratically in the expressions for $\mathrm{BR}(B\to
V\gamma)$.  The ratio $R_{\mathrm{th}}$ reduces the sensitivities to 
$F^\rho$ and $F^{K^*}$ by limiting the impact of the long-distance hadronic
physics to the SU(3) breaking in the form factor ratio $\zeta$.  The other
main uncertainty is from the QCDF prediction for $\Delta R$; this is largely
due to $\lambda_B$, as it is this that determines the strength of the weak
annihilation contributions.  This is especially relevant for the charged
modes $B\to\rho^\pm\gamma$.   

\begin{figure}[h]
 $$\epsfxsize=0.6\textwidth\epsffile{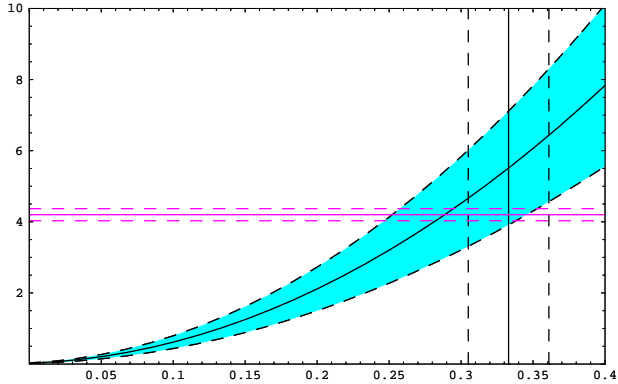} $$  
 \vskip-12pt
  \caption[ ]{Plot of
 $\mathrm{BR}(B\to K^*\gamma)$ in units of $10^-5$ (and within conservative
 estimated errors) against 
 form factor $F^{K^*}(0)$ (vertical band) with current experimental
 average (horizontal band in magenta).}
   \label{fig:KsF}
\end{figure}

It is worth questioning however, why the individual branching ratios have
high central values with respect to their experimental determinations, most
notably, the $B\to K^*\gamma$ branching ratio.  From the
calculation of inclusive $B\to X_s\gamma$, which is in good agreement with
the data, it is justifiable to conclude that the short distance physics is
compatible with the Standard Model expectations \cite{Beneke:2004dp}.  This 
suggests two lines of reasoning: firstly, that the uncertainties which are
removed when calculating the ratio $R_{\mathrm{th}}$ have a considerable
effect on the branching ratio.  We illustrate the sensitivity to the form
factor $F^{K^*}$ in the $K^*\gamma$ branching ratio in Figure
\ref{fig:KsF}.  This shows how the experimental measurement leans
towards a lower value for this result and how the reduction to the
updated value from \cite{Ball:2004rg} leads to such a significant drop in the
theory estimate of $\textrm{BR}(B\to K^*\gamma)$.       

A second line of reasoning is to consider if there are any other factors
contributing to the over-estimation of the branching ratios, which conspire to cancel in the
combined ratio $R_{\mathrm{th}}$.  This could include errors from the
calculation of $\Delta R$, large power corrections
which have been neglected, or new physics in the underlying
$b\to (d,s)\gamma$ transitions.    

We now go on to consider the possibility of new physics in the $B\to V\gamma$
decays.  The errors on the
experimental data are reducing to a level where it will be possible to impose
considerable constraints on new physics models.  As we discussed in the
opening motivation of this chapter, the radiative decays (especially those 
involving the $b\to s$ transition) have long been seen as an excellent 
candidate for searches for new physics.  We proceed to perform an analysis in
a generic minimal supersymmetric model as an example of the most popular extension of the
Standard Model.  We show how effective constraints can be imposed on the parameter
space for possible new physics effects in the radiative decays using the
bounds from the experimental results. 
\newpage

\section{Introduction and motivation for supersymmetry}

Supersymmetry (SUSY) is the most popular low-energy extension of the standard
model (see for example \cite{Drees:1996ca,Martin:1997ns}).  Beyond the elegance of the symmetry itself, which unifies fundamental
fermions and scalars, supersymmetric models have many virtues which can
alleviate some of the issues within the Standard Model.  

The main benefit
that motivates many SUSY models is the resolution of the ``hierarchy
problem''.  In the Standard Model (SM) there are a number of corrections which lead to infinite contributions
to a particular process -- these are removed by renormalisation by
introducing some cut-off energy $\Lambda$.  Quadratic divergences appear in
the contribution of heavy fermions to the Higgs self energy.  These can be
again be removed by renormalisation, but large corrections ($\sim m_f^2$)
remain.  Hence fine-tuning is required to ensure the small Higgs mass
required in order to produce a vacuum expectation value of the correct size
to give the observed $W$ and $Z$ boson masses.  If we are envisioning that
our theory contains the large scales of grand unification, namely the Planck
scale ($m_{PL}\sim10^{19}$GeV), then we have to ask how we can arrange for
the Higgs mass term (and the corresponding electroweak scale) to be so much
smaller then the underlying mass scale $m_{PL}$. Supersymmetry requires that
each scalar particle has a fermionic ``superpartner'' (and vice versa) of the same
mass so that all the quadratic divergences of the scalar mass terms will
automatically vanish.  These cancellations occur at every order in
perturbation theory protecting the light scalar masses.

Supersymmetry is also a natural bridge to incorporate gravity, as
implementing SUSY as a local gauge symmetry demands local co-ordinate
covariance i.e linking to general relativity.  Supersymmetry is also an
integral requirement at some energy scale in most string theories
\cite{GSW:1987}.

If SUSY were to be manifest then all particles and their
superpartners would be mass degenerate.  Since to date, no state which can be
identified as a superpartner has been observed, then we must have broken
supersymmetry at our low energy scales.  The mechanism for the SUSY breaking
is unknown, but as we discuss below it has a large impact on phenomenology.  We
can give rough bounds for the masses of the superpartners; all of the masses
must be over 100GeV (otherwise they should have been 
observed at LEP), but at least some masses have to be less than 1 TeV,
in order to resolve the hierarchy problem. 

\subsection{Low energy MSSM}

The Minimal Supersymmetric Standard Model (MSSM) is the simplest and most popular SUSY
extension to the SM.  The particle content is a set of supermultiplets,
constructed from the SM states and their  
superpartners.   The superpartners of the left and
right handed quarks are denoted as left and
right handed squarks even though the squarks are scalar and have no chiral
structure.  In summary we have the following: 
  \begin{center} 
     \begin{tabular}{|lcccc|} \hline
     {} & {} & spin 0 & spin $\tfrac{1}{2}$ & $\left(SU(3)_c,\,SU(2),\,U(1)_Y\right)$ \\ \hline
     (s)quarks  & Q & $(\tilde{u}_L,\,\tilde{d}_L)$ & $\left({u}_L,\,{d}_L\right)$ 
    & $(3,2,\tfrac{1}{6})$\\  
{}  & U & $\tilde{u}_R^*$ & $u_R^+$  & $(\bar{3},1,-\tfrac{2}{3})$\\  
{}  & D & $\tilde{d}_R^*$ & $d_R^+$  & $(\bar{3},1,\tfrac{1}{3})$\\  \hline
     (s)leptons  & L & $(\tilde{\nu},\,\tilde{e}_L)$ & $\left({\nu},\,{e}_L\right)$ 
    & $(1,2,-\tfrac{1}{2})$\\  
     {}  & E & $\tilde{e}_R^*$ & $e_R^+$ 
    & $(1,1,1)$\\  \hline
 higgs(inos)  & $H_u$ & $(h_u^+, h_u^0)$ & $(\tilde{h}_u^+, \, \tilde{h}_u^0)$
     & $(1,2,\tfrac{1}{2})$\\   
 {}  & $H_u$ & $(h_d^0, h_d^-)$ & $(\tilde{h}_d^0, \, \tilde{h}_d^-)$
     & $(1,2,-\tfrac{1}{2})$\\ \hline
   \end{tabular}
\end{center}
The interactions among the scalar particles are governed by a
superpotential $W$, where the dimensionless numbers {$\mathbf{y_i}$} contain the
Yukawa couplings of the SM fermion fields.  
\be
   W = U\mathbf{y_u}QH_u + D\mathbf{y_d}QH_d + E\mathbf{y_e}LH_d + \mu H_uH_d
\ee 

The gauge bosons of the Standard Model also get a fermionic superpartner
generically called a gaugino.  The \textit{gluino} is the superpartner of the
gluon, and hence carries colour charge. The remaining gauginos corresponding
to superpartners of the W, Z and photon, combined with the higgsinos of the
same quantum numbers all mix together to give four new physical states -- two
charged and two neutral fermions labelled \textit{charginos} and
\textit{neutralinos} respectively.  

Since SUSY cannot be manifest, we need to find a mechanism to break the
symmetry.  It is in principle possible to achieve spontaneous symmetry
breaking of global SUSY, however it is very difficult to
build a realistic model.  The problems with the different spontaneous SUSY
breaking scenarios include, for example, violation of mass sum rules between
quarks and squarks or the need for additional
field content beyond the MSSM.   It is instead expected that SUSY is \textit{softly}
broken by dimensionful parameters such as squark mass terms.  These explicit
breaking terms are put in by hand, and can be thought of as originating from
dynamics beyond the MSSM operating at some very high energy scale and 
``communicated'' to the low energy sector via some (unknown) mechanism.  The
soft breaking Lagrangian contains a host of mass terms for scalar particles
and gauginos, bi- and tri-linear scalar interactions, giving a total of 105 new
physical parameters in its most general form.  In general scenarios there is
no reason why these parameters should be flavour-blind, but often in
particular models many simplifying assumptions are made about the structure
of the flavour sector.  We consider the implications of this in the next
section. 

\subsection{Implications for the flavour sector} 

Given the most general form of supersymmetry breaking as we have discussed
above, there are many new parameters and contributions in the flavour
sector which could impact on the phenomenology of flavour changing and CP
violating processes.  There is potential for many supersymmetric
contributions to these processes, allowing them to arise at rates much higher
than experimentally observed.  This is however coupled with the very stringent
constraints on any new physics, for example, from the fitting of the unitarity
triangle -- which combines all available experimental data on flavour and CP violation.
The success with which this fit determines the CKM parameters to be in
agreement with the Standard Model expectations shows where new large
sources of flavour changing neutral currents or CP violation are not favoured
in a possible new physics model.  

There are a large number of different possible scenarios for SUSY 
breaking \cite{Chung:2003fi}, including the simplifying assumption of
flavour universality at the high energy scale.  This universality removes all
of the terms which could lead to ``dangerous'' FCNC and CP-violating effects
which are already excluded by experiment.  For example, in the minimal
supergravity (mSUGRA) model, which imposes gravity-mediated SUSY breaking, the canonical scenario
has universal scalar masses
\begin{equation*}
   m_{\tilde{q}_{ij}}^2 = \delta_{ij}m_0^2
\end{equation*}
However, even in this picture evolving the dynamics from the high energy (``GUT'' scale) to the
electroweak scale induces flavour non-universality due to the contribution
from the Yukawa couplings \cite{BS:2000}, although they are
expected to be small. 

Alternatively, in other SUSY breaking scenarios, it is possible for the dynamics
of the hidden sector to be communicated to the low energy theory in a flavour
specific way -- even 
before radiative corrections or the RGE evolution are included.  This will 
allow flavour changing neutral currents which are in general not
GIM-suppressed or CP invariant.  This clearly will produce a very rich
phenomenology with many new sources of flavour and CP physics.
  
\subsection{Mass insertion approximation}

To test the impact of a particular SUSY model on the flavour sector we would,
ideally, 
diagonalise the squark mass matrices so we could calculate diagrams 
with the squark mass eigenstates.  Without knowledge of the full mass
matrices however, this is a rather difficult problem, as the matrices are
related to the parameters of the soft SUSY breaking Lagrangian in a
non-trivial manner.  We instead utilise the fact that the flavour changing effects
originate from the off-diagonal elements of the squark (sfermion) mass
matrices.  If the off-diagonal elements are small relative to the diagonal
entries we can use the \textit{mass insertion approximation}
\cite{Hall:1985dx}.  
This enables us to
simply compute ratios of the off-diagonal entries over the diagonal elements
of the squark mass matrices and compare with the bounds we can derive from
experimental data.  

In order to allow the most general method of SUSY breaking and
flavour structure, we can consider a ``generic'' R-parity conserving MSSM.
We parameterise the FCNC and left-right squark mixing by taking the squark
mass matrices as both flavour universal and real.  This mass matrix is
written with the off-diagonal elements denoted $(\Delta_{AB})_{ij}$, where
$AB$ stands for $LL,\,RR,\,LR,\,RL$ and denotes the helicity of the
squarks.  For example, $m_{LL}^2$ can be written
\begin{equation*}
   m_{LL}^2 = \left(\begin{array}{ccc}
                       (m_{LL}^2)_{11} & (\Delta_{LL})_{12} &
                       (\Delta_{LL})_{13} \\
 (\Delta_{LL})_{21} & (m^2_{LL})_{22} &
                       (\Delta_{LL})_{23} \\
    (\Delta_{LL})_{31} & (\Delta_{LL})_{32} &
                       (m^2_{LL})_{33} \\
                   \end{array}\right)
\end{equation*}
   
We can then constrain the 
perturbations from universal scalar masses $m_{\tilde f}^2$ by normalising the
off-diagonal elements by a common mass, normally taken as the average scalar
mass $m_{\tilde f}^2$.  The mass insertions are defined as
\begin{equation} 
   (\delta^d_{AB})_{ij} = \frac{(\Delta_{AB})_{ij}}{m_{\tilde f}^2}
\end{equation}
For the $b\to d$ transition the relevant insertions are
$(\delta^d_{AB})_{13}$ and for the $b\to s$ 
transitions $(\delta^d_{AB})_{23}$.  The universal scalar mass is 
often taken as an approximate mass scale, such as the average squark mass
$m_{\tilde f} \sim 500$ GeV.  

Different realisations of the MSSM will be characterised by different sets of
soft breaking terms which in turn induce 
different sets and combinations of mass insertions.  We classify the effect of
these different MSSM models on $\delta^d_{ij}$ in terms of the helicities AB
= $LL$, $RR$, $LR$, $RL$.  We consider for the most part cases where a single
helicity mass insertion dominates, but also test some cases where two of the
mass insertions are sizable (a double mass insertion).  This allows us to
find the leading 
contributions to various processes by considering diagrams with a single (or
double) mass insertion and express the phenomenological
bounds in a model-independent way.

\section{Gluino contributions in generic MSSM}

We add supersymmetric contributions to the Standard Model expressions using
an effective Hamiltonian induced by gluino exchange as first derived in
\cite{Gabbiani:1996hi} using the mass insertion approximation.  Gluino
exchange is the dominant contribution and we do not consider charged Higgs, chargino or
neutralino exchanges.  This is a good approximation in a preliminary analysis
of this nature as these contributions are proportional
to the weak coupling rather than $\a_s$ as for the gluino
contributions, and so can be safely neglected.

Two classes of diagrams contribute to the $\Delta B=1$ 
processes, namely penguin and box diagrams, which give both additional
contributions to 
the existing Standard Model operators $Q_i$, and induce new operators
$\widetilde{Q}_i$ obtained from the $Q_i$ under the exchange
$\mathrm{L}\leftrightarrow\mathrm{R}$.  The gluino contributions which
interfere with the SM contributions (i.e those from operators $Q_i$) are combined
at the level of the Wilson coefficients.  The contributions from the
remaining new operators $\widetilde{Q}_i$ are added separately since they
do not interfere with the SM amplitude.

The Wilson coefficients for the SUSY operators act to modify the initial
conditions of the SM operators at the scale $\mu=M_W$.  For the radiative
processes, we only need to 
consider the relevant contributions, namely to $C_7^{(0)\,\mathrm{eff}}$ and
$C_8^{(0)\,\mathrm{eff}}$:
\begin{eqnarray}
   \label{eq:SUSYc}
      C_7 &=& \frac{\a_s\pi}{m_{\tilde{q}}^2}
      \left[\left(\delta_{13}^d\right)_{LL}\frac{8}{3}M_3(x) +
      \left(\delta_{13}^d\right)_{LR}\frac{m_{\tilde{g}}}{m_b}\frac{8}{3}M_1(x)\right]\nonumber\\
   C_8 &=& \frac{\a_s\pi}{m_{\tilde{q}}^2}
      \left[\left(\delta_{13}^d\right)_{LL}\left(-\frac{1}{3}M_3(x)-3M_4(x)\right)\right.
      \nonumber\\
     && \qquad \qquad \left.+
      \left(\delta_{13}^d\right)_{LR}\frac{m_{\tilde{g}}}{m_b}\left(-\frac{1}{3}M_1(x)-3M_2(x)\right)\right]  \nonumber\\
\end{eqnarray}
where the $M_i(x)$ are functions of
$x=m_{\tilde{g}}^2/m_{\tilde{q}}^2$ obtained from the calculation of the
gluino penguins \cite{Gabbiani:1996hi}, ($m_{\tilde{q}}$ is the average squark
mass and $m_{\tilde{g}}$ is the gluino mass).  In all of the following analysis we
use equal values for the average squark and gluino masses, i.e. $x=1$.  The
co-efficients for $\widetilde{Q}_7$ and 
$\widetilde{Q}_8$ are obtained from these under the exchange
$\mathrm{L}\leftrightarrow\mathrm{R}$.  

In considering the SUSY contributions to $B\to V\gamma$ the Standard Model contribution
is added to the gluino contribution and we determine the 
possible constraints on the flavour-violating sources in the squark sector.
The bounds we determine on the  $\delta^d_{13}$ and $\delta^d_{23}$
insertions are only indicative as they are extracted
ignoring the error of the theory calculation.  We also indicate the bound on
the parameter space when the $1\sigma$ theory error is included.  

\section{New physics in $b\to d$ transitions}

Although the constraints on the flavour sector are quite stringent there
is still room for new physics to affect the flavour sector.  Using the new
results for the $B\to\rho\gamma$ decays we can constrain the possibility of
new physics in a generic supersymmetric model using the mass insertion
approximation introduced above.  Constraints on  $\delta^d_{13}$
derived from the $\bar{B}_d$---$B_d$ mass difference $\Delta M_d$ and the CP
asymmetry in $B\to J/\psi K_s$ were determined in \cite{Becirevic:2001jj}.
This study found that with NLO evolution, for a single mass insertion and
$m_{\tilde{q}}\sim 
m_{\tilde{g}}$, the limits on $\delta^d_{13}$ are of the order of $10^{-1}$
for the $LL$ or $RR$ insertions,  and $\sim10^{-2}$ for $LR$, $RL$.  

We expect the radiative $b\to d$ decays to show a similar hierarchy, as the
$LR$ and $RL$ insertions are those most effectively constrained by measurements of $B\to
\rho\gamma$ as they enter into the Wilson coefficients with the factor
$(m_{\tilde{g}}/m_b)$.  The $LL$ and $LR$ insertions contribute to the same
operator that is responsible for the $B\to V\gamma$ decay in the SM, and so yield a SUSY contribution which interferes with the SM.  As a
consequence, the rates tend to be larger than the $RR$ (and $RL$) cases since
these insertions do not add to the leading SM amplitude.  The consequences of this in
light of our high central values for the SM amplitudes, are that the resulting
parameter spaces for the $RR$ and $RL$ insertions will be more constrained
than the $LL$ and $LR$ ones.    

With these considerations in mind, we use the gluino
contributions to $\Delta B=1$ 
transitions and the constraints from the newly measured radiative processes
to constrain the $\delta^d_{13}$ insertions.  We can then determine if the
measurement of the $B\to\rho\gamma$ decay can put stringent limits on the
possible SUSY models that may have impact on the $b\to d$ sector. 

We consider single mass insertions for each helicity combination, and double mass
insertions for a left-right symmetric case:
$(\delta^d_{13})_{LL}=(\delta^d_{13})_{RR}$ and
$(\delta^d_{13})_{RL}=(\delta^d_{13})_{LR}$ as motivated in a number of SUSY
models e.g. left-right supersymmetric models.

\subsection{Constraints on $\delta^d_{13}$}

Using the full SM + SUSY contribution to the $B\to\rho\gamma$ processes, we
can determine the allowed parameter space in the $\delta^d_{13}$ plane using
the available experimental results.  We use both the
$B\to(\rho,\omega)\,\gamma$ branching fraction and the ratio 
$R\left[(\rho,\omega)\gamma/K^*\gamma\right]$ as these were fitted separately
from the Belle data set \cite{Abe:2005rj}, so we can apply the constraints
concurrently.  We begin by considering a single mass insertion with an
average squark mass of $m_{\tilde{q}}=500$GeV.  For constraints based on the
ratio $R_{\mathrm{th}}$ we take all insertions 
to $B\to K^*\gamma$ to be zero in the first instance.  

\begin{figure}[h!]
   $$\epsfxsize=0.47\textwidth\epsffile{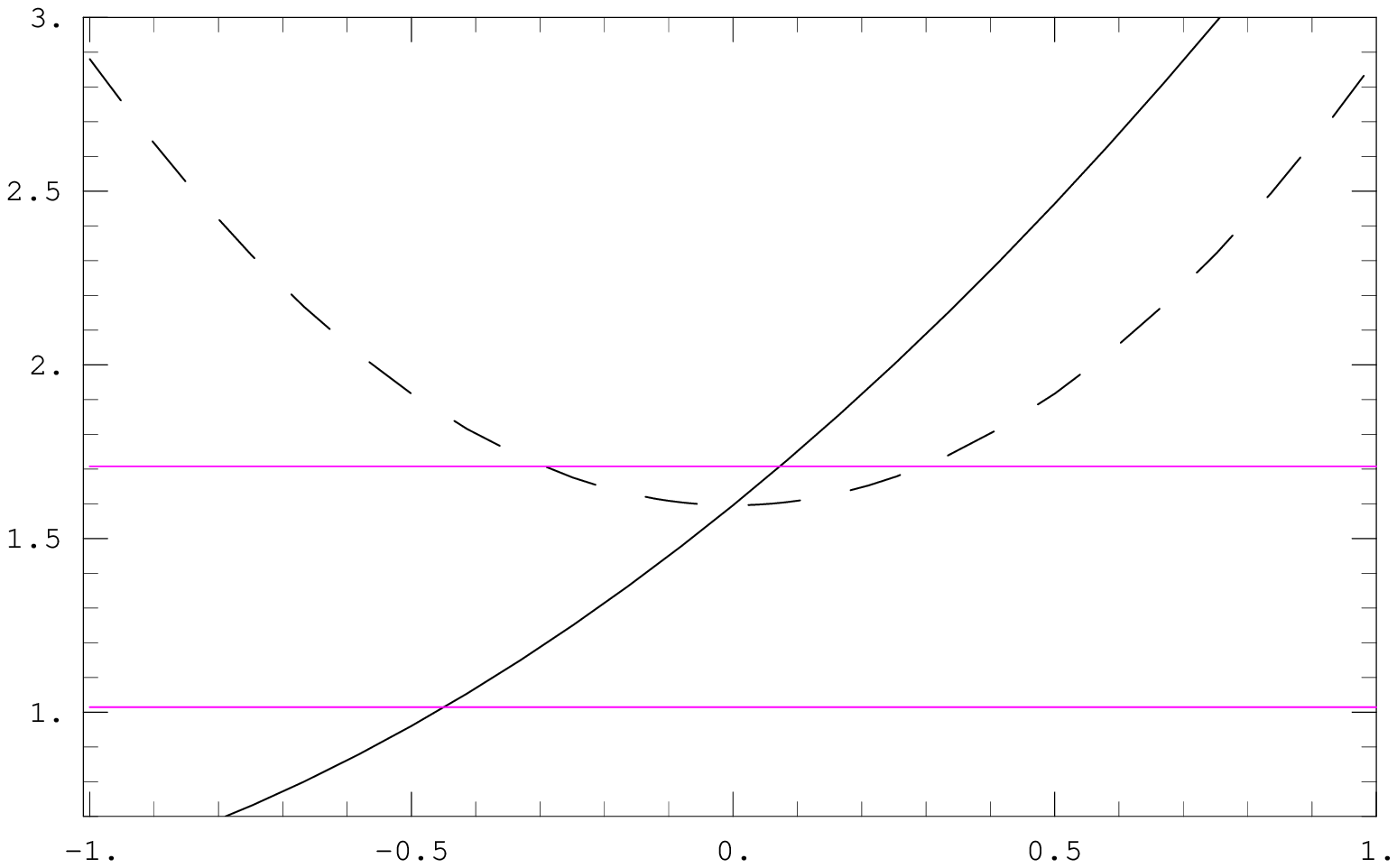}\qquad
   \epsfxsize=0.47\textwidth\epsffile{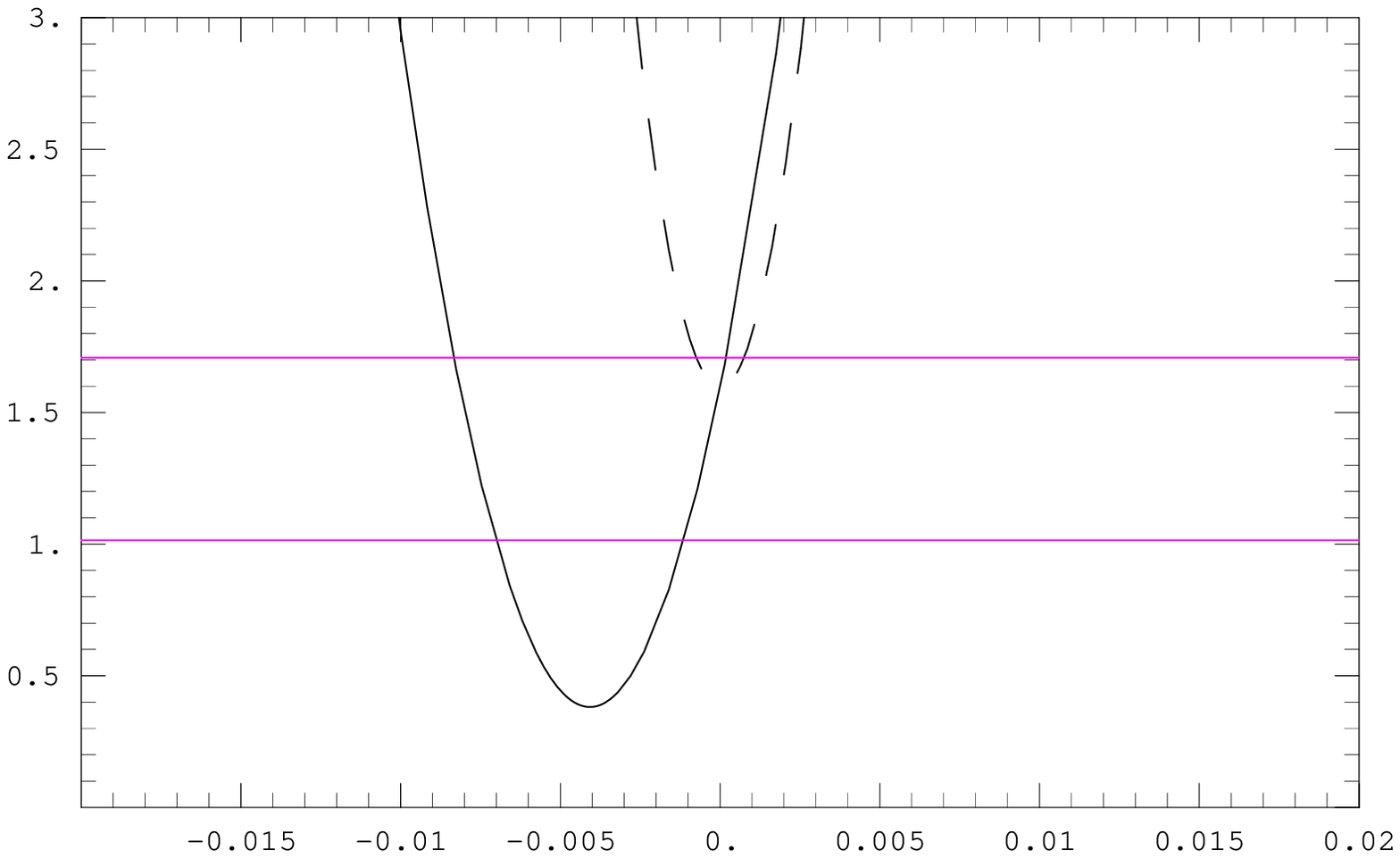}$$
  \vskip-12pt
   \caption[]{Left: Dependence of $\mathrm{BR}(B\to(\rho,\,\omega)\gamma)$ with
   gluino contributions on the single mass insertion $|(\delta_{13}^d)_{LL}|$
   (solid line) and $|(\delta_{13}^d)_{RR}|$ (dashed line) with
   $m_{\tilde{q}}=m_{\tilde{g}}=500$GeV.  Experimental 1$\sigma$ bounds shown
   for comparison.  Right: same for $|(\delta_{13}^d)_{LR}|$ (solid) and
   $|(\delta_{13}^d)_{RL}|$ (dashed).  }  
   \label{fig:massdep}
 \end{figure}

The absolute value of the $B\to(\rho,\,\omega)\gamma$ branching ratio is
plotted against the absolute value of the mass insertion (for each single
mass insertion dominating in turn) in Figure \ref{fig:massdep}. 

The difference between the $LL$ and $RR$ cases is due to the way the gluino
contribution interferes with the Standard Model.  For the case where the
$(\delta_{13}^d)_{LL}$ insertion dominates, the supersymmetric contribution
is to the same operator responsible for the $B\to\rho\gamma$ decay.  This
interference in general causes the decay to proceed at a larger rate than
that of the $RR$ case.  For the $RR$ case where no interference with the
Standard Model is present, the results will be symmetric around
$(\delta_{13}^d)_{RR}=0$, and similarly for $(\delta_{13}^d)_{RL}$. 

The allowed values for the $LR$ and $RL$ insertions are expected to be much
smaller than for the $LL$ or $RR$ insertions as they enter into the Wilson
coefficients with factor $(m_{\tilde{g}}/m_b)$.  Hence $(\delta_{13}^d)_{LR}$ and
$(\delta_{13}^d)_{RL}$ will be the most effectively constrained.  The high
central value of SM amplitude for  $\mathrm{BR}(B\to(\rho,\omega)\gamma)$
will lead to a minimal allowed area of parameter space for the $RR$ and $RL$
insertions.  If the error on the SM amplitude is taken into account the constraints
the parameter space would clearly be less stringent.    

\begin{figure}[hp!]
  $$\epsfxsize=0.47\textwidth\epsffile{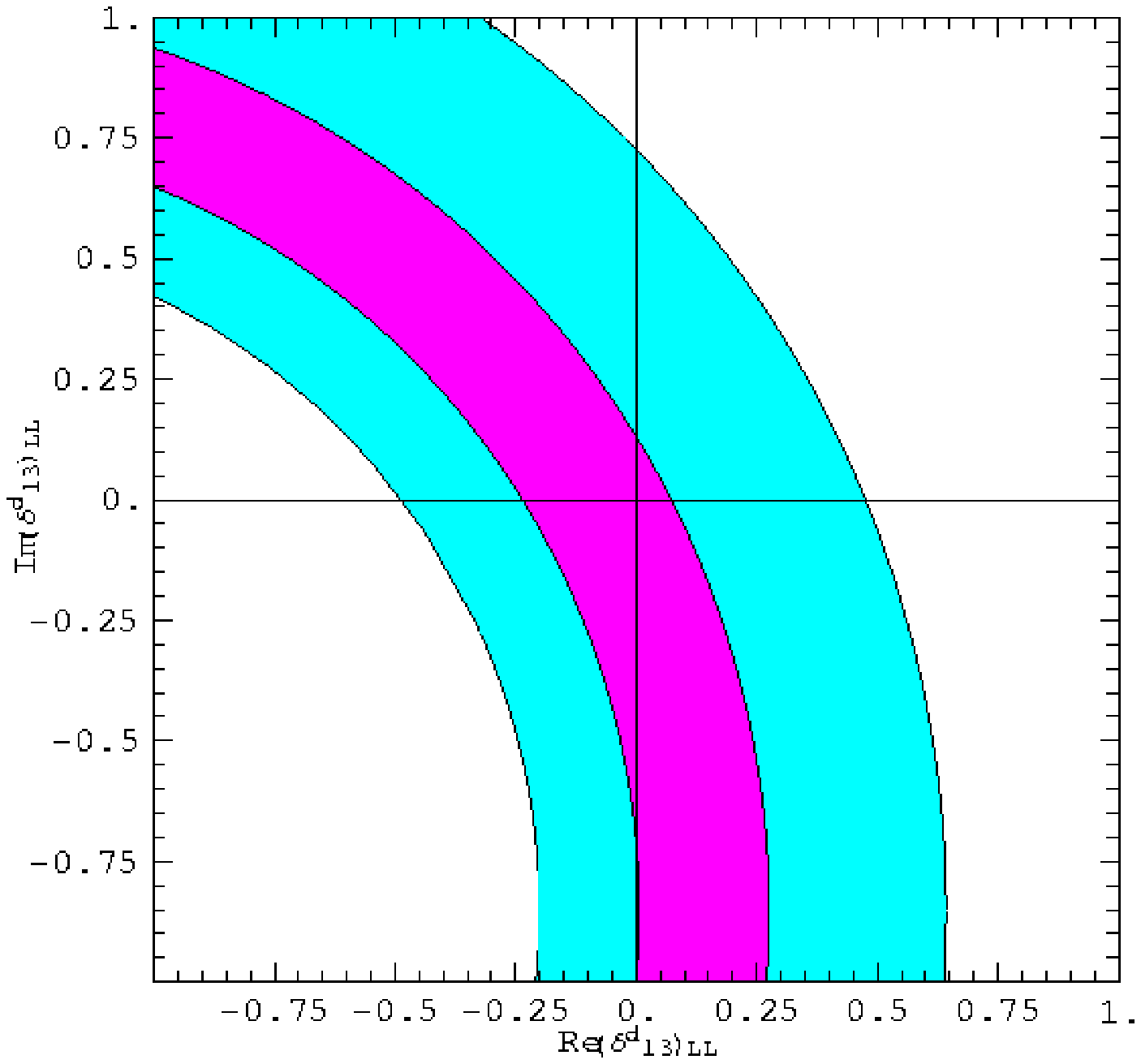}\qquad
  \epsfxsize=0.47\textwidth\epsffile{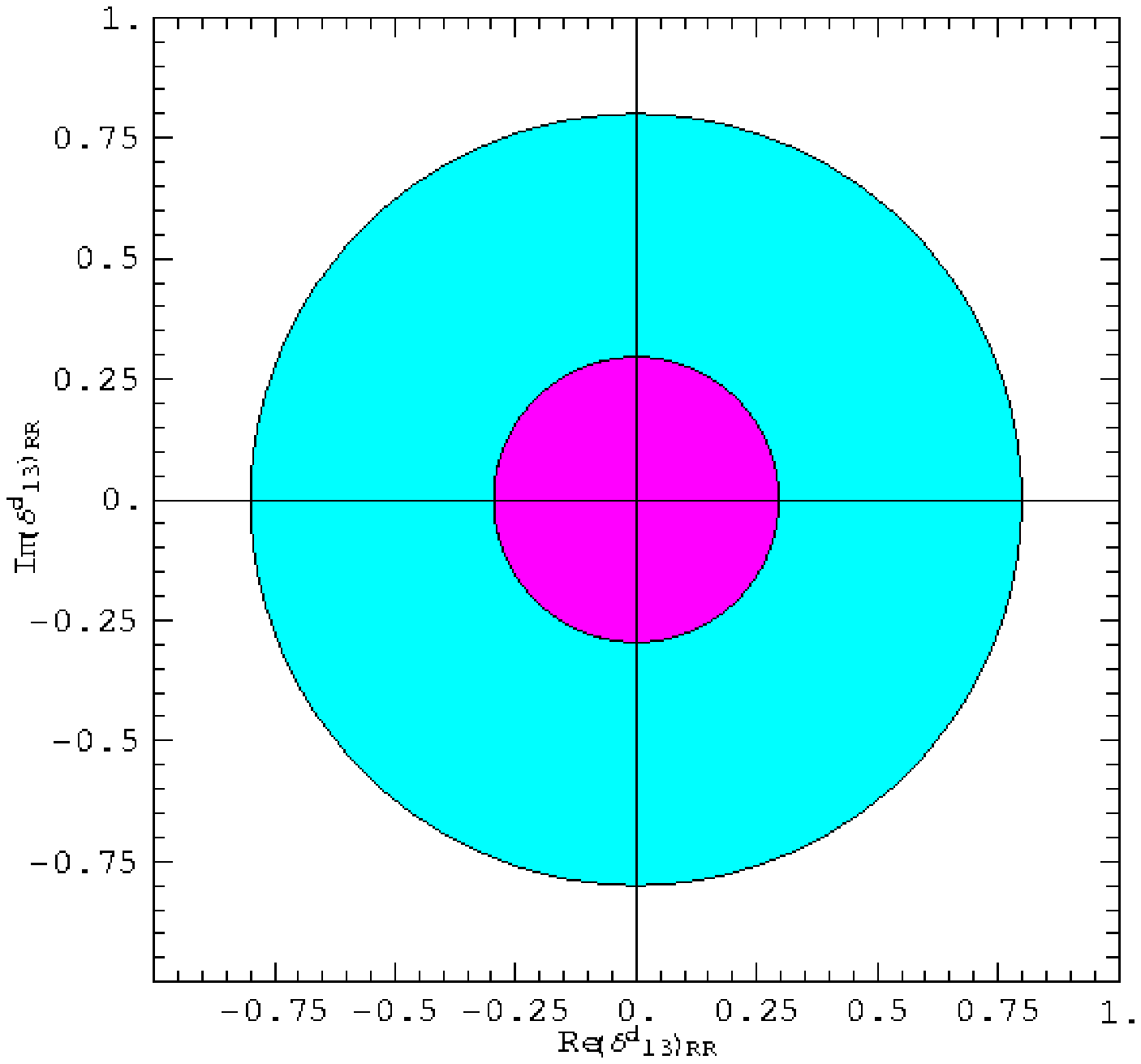}$$ \\
 $$\epsfxsize=0.47\textwidth\epsffile{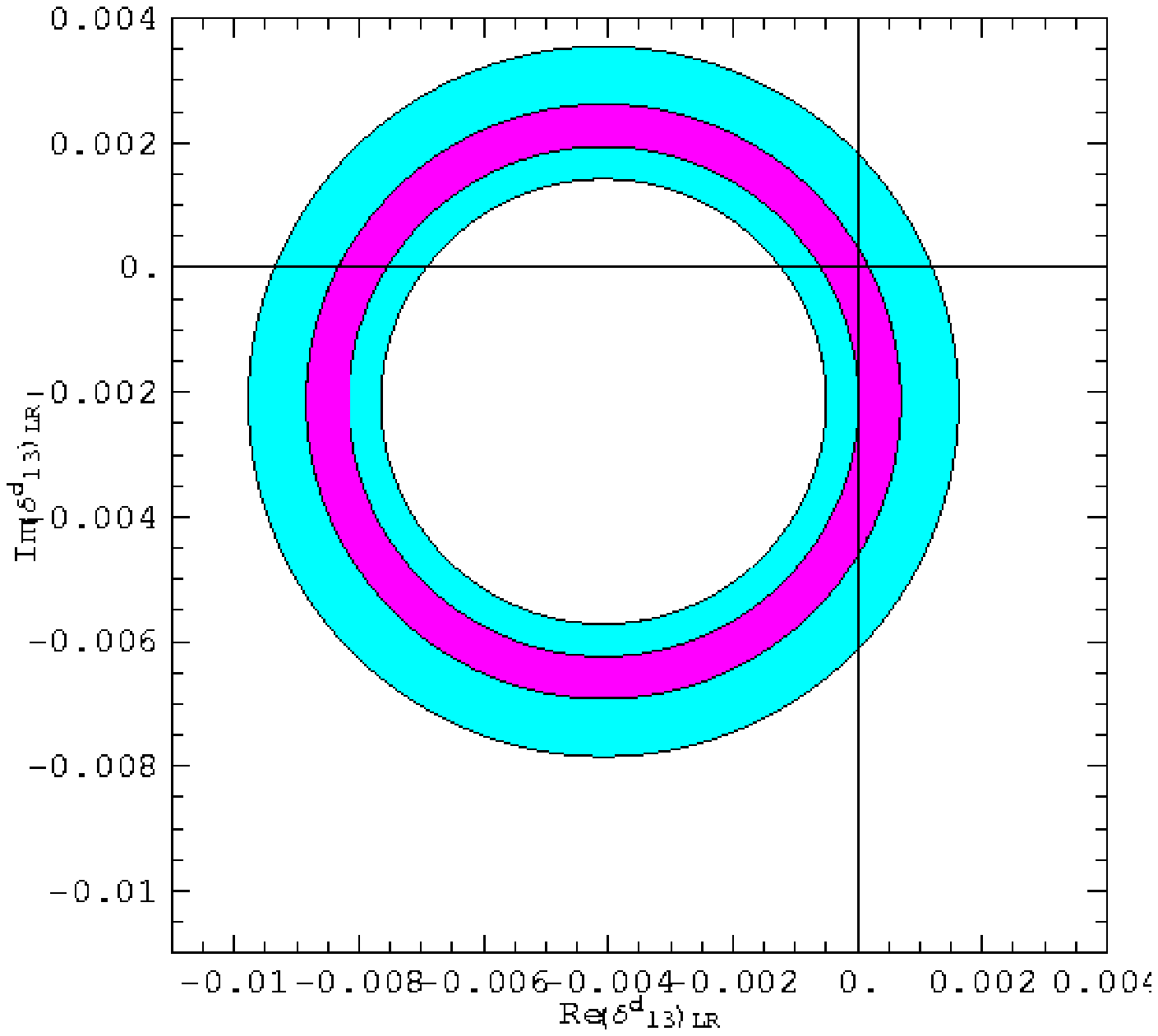}\qquad
  \epsfxsize=0.47\textwidth\epsffile{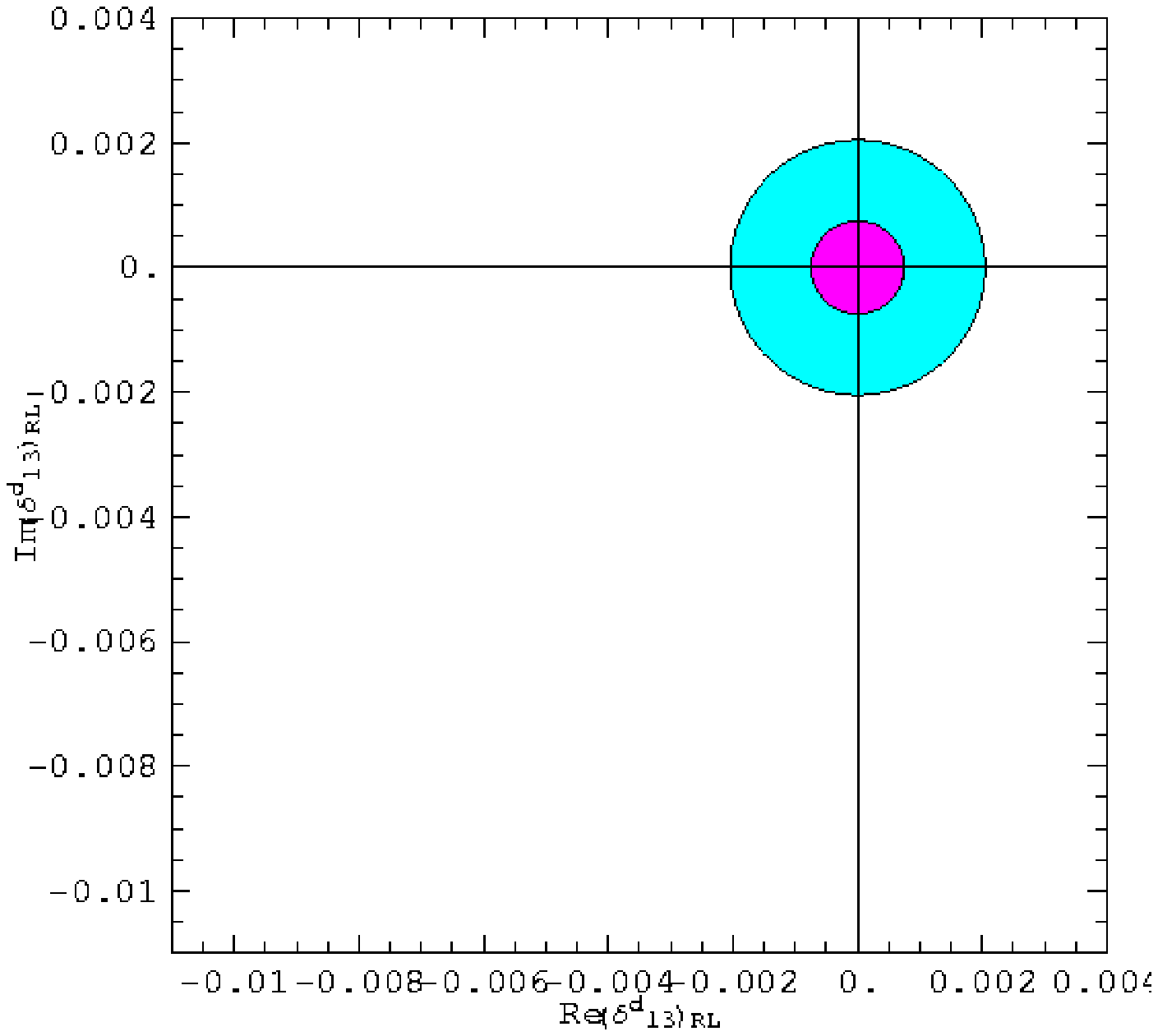}$$ \\
 \vskip-12pt
  \caption[]{Allowed regions in $Re(\delta^d_{13})_{ij}$ --
  $Im(\delta^d_{13})_{ij}$ parameter space for $ij = \mathrm{LL},\mathrm{RR},
  \mathrm{LR},\mathrm{RL}$.  We use $m_{\tilde q} = m_{\tilde g} = 500$ GeV,
  and constraints from Belle measurement of
  $\mathrm{BR}(B\to(\rho,\omega)\,\gamma)$ and additionally
  $R\left[(\rho,\omega)\gamma/K^*\gamma\right]$.  The overlay in magenta
  indicates allowed regions using central value of SM amplitude.}   
  \label{fig:bdyLLRR}
\end{figure}

The full constraints on the parameter space for the four single mass
insertions are shown in Figure \ref{fig:bdyLLRR}, where we extract the
bounds on the insertions using both the central value of the SM amplitude for
$B\to(\rho,\omega)\gamma$  and taking into account the $1\sigma$ error.

The allowed areas in the parameter space indicate regions where
the insertion can produce a supersymmetric contribution which is within
the experimental $1\sigma$ bounds on
$\mathrm{BR}(B\to(\rho,\,\omega)\gamma)$.     
These figures show that using the data imposes quite stringent restrictions
on the parameter space, the most impact coming from the low error on the
$R_{\mathrm{exp}}[(\rho,\omega)\gamma/K^*\gamma]$ measurement.  
 
\begin{figure}[h] 
   $$\epsfxsize=0.47\textwidth\epsffile{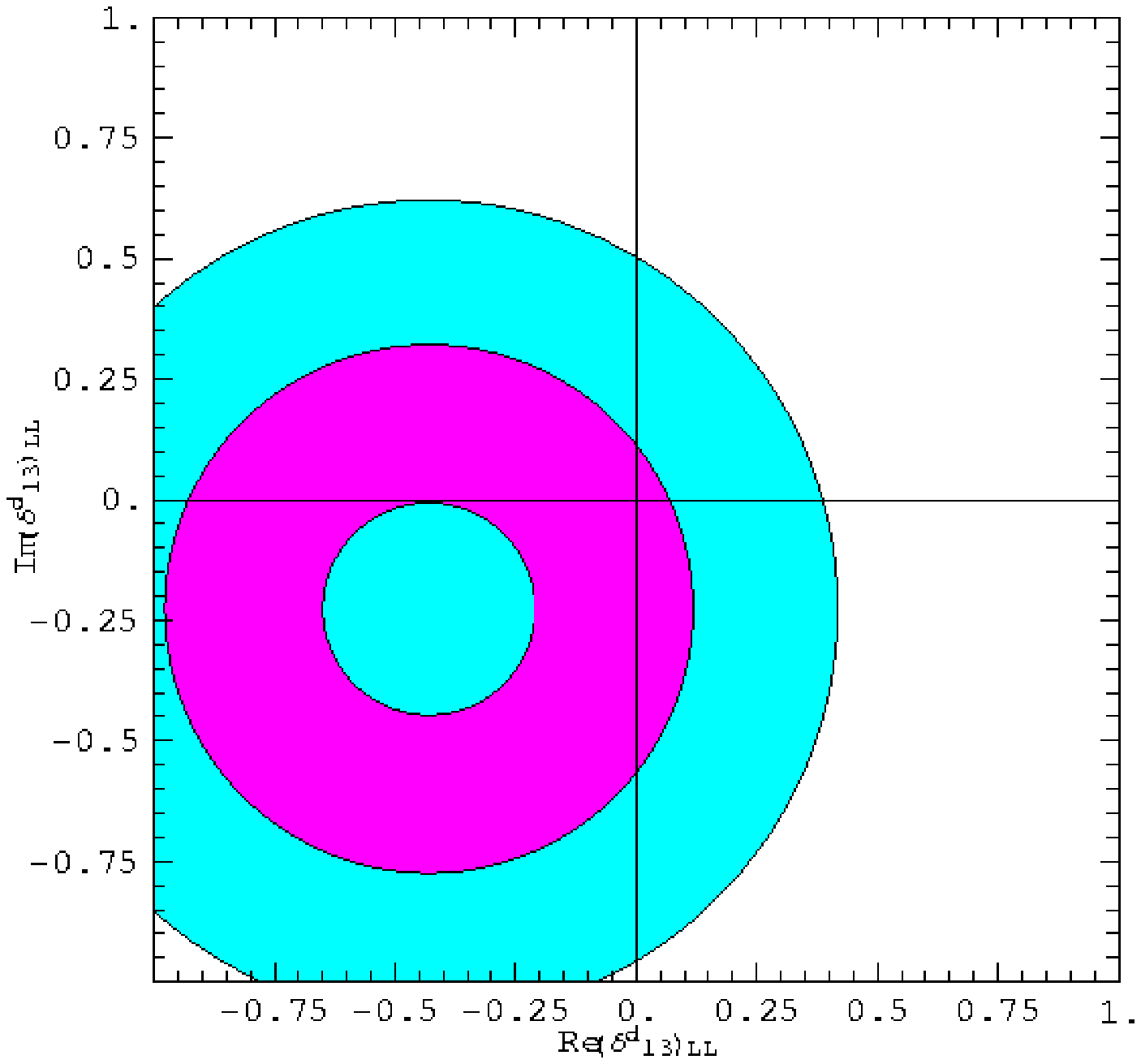}\qquad
   \epsfxsize=0.47\textwidth\epsffile{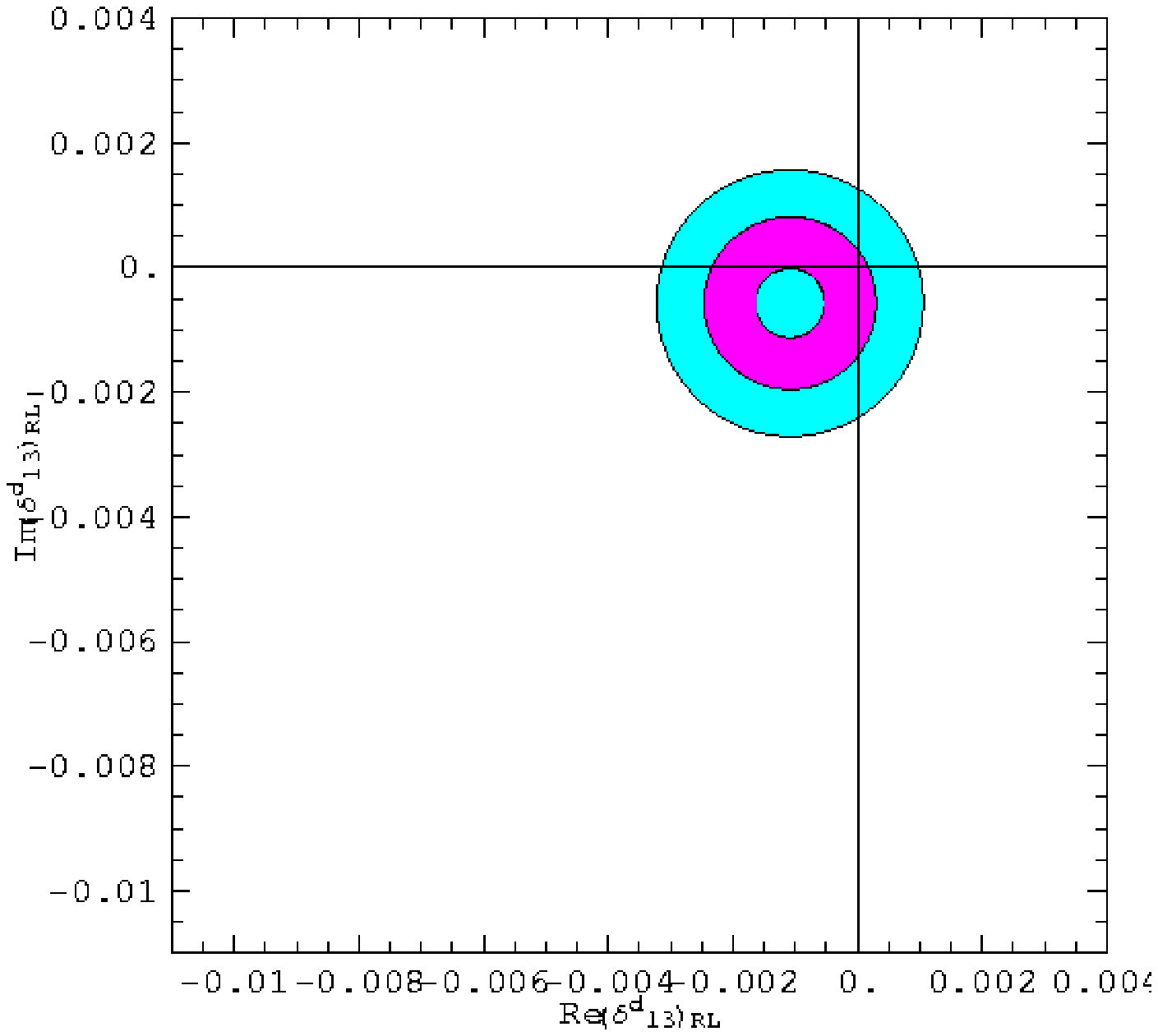}$$
  \vskip-12pt
   \caption[]{Allowed regions in $Re(\delta^d_{13})_{ij}$ --
   $Im(\delta^d_{13})_{ij}$ parameter space for double mass insertions
   $\mathrm{LL}=\mathrm{RR}$ and $\mathrm{LR}=\mathrm{RL}$.  We use
   $m_{\tilde 
   q} = m_{\tilde g} = 500$ GeV, 
   and constraints from Belle measurement of
   $\mathrm{BR}(B\to(\rho,\omega)\,\gamma)$ and $R\left[(\rho,\omega)\gamma/K^*\gamma\right]$.}  
   \label{fig:bdysym}
 \end{figure}

We can also consider the effect of a double mass insertion,
for the cases where  $(\delta_{13}^d)_{LL} = (\delta_{13}^d)_{RR}$ or
$(\delta_{13}^d)_{LR} =  (\delta_{13}^d)_{RL}$.  This sort of scenario is
motivated by a number of different SUSY models, such as the left-right
supersymmetric model or minimal supergravity (mSUGRA) which
demands $\delta_{LL} =  \delta_{RR} = 0$ and $\delta_{LR} = \delta_{RL}$.  The results  for both cases are shown in figure \ref{fig:bdysym}.

We see a similar pattern as with the single mass insertion, of a much greater
constraint on the helicity-changing insertions than the helicity conserving.
The $LL=RR$ case for example, allows ample space for a sizable mass
insertion.  

\section{New physics in $b\to s$ transitions}

\begin{figure}[h]
 $$\epsfxsize=0.5\textwidth\epsffile{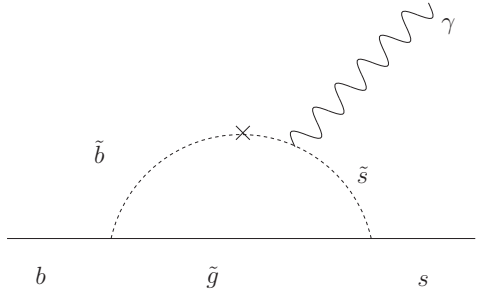} $$  
 \vskip-12pt
  \caption[ ]{Main supersymmetric contribution to $b\to s \gamma$.}
   \label{fig:bsgsusy}
\end{figure}

We now perform a similar analysis with the $B\to K^*\gamma$ decays.  
Although there are strong constraints on new physics contributions to $s\to
d$ and $b\to d $ transitions, this is not the case with $b\to s$
transitions.  The unitarity fit for example strongly constrains the $s\to
d$ and $b\to d $ transitions, but not the $b\to s$ transitions as they do not
affect the fit directly (unless they interfere with the amplitude for $B_s$
mixing and reduce the expected value of $\Delta M_s$ below its experimental
lower bound).   In the radiative processes such as $b\to s\gamma$, there can be 
many new contributions with SUSY particles in the penguin loop.  The main
contribution is shown diagrammatically in Figure \ref{fig:bsgsusy}.  
The cross on the internal squark line represents a flavour changing mass
insertion.  The size of this will be determined by the nature of the
supersymmetric breaking -- our ignorance of this means that the insertion
could in principle be as large as any of the other terms.

\subsection{Constraints on $\delta^d_{23}$}

There have been previous studies on constraining new physics in $b\to
s\gamma$ decays such as \cite{Ciuchini:2002uv}, which concentrates on
using the bounds from the inclusive process $B\to X_s\gamma$.  These studies
have shown that there is still ample room in the parameter space for
SUSY contributions to various different observables.  The constraints that
can be determined from the branching ratio and CP asymmetry in $B\to
X_s\gamma$ mainly affect the helicity-changing contributions i.e. the SUSY
insertions $(\delta_{23}^d)_{LR}$ and $(\delta_{23}^d)_{RL}$.  We can
therefore use the measurement of our exclusive decay $B\to K^*\gamma$, and
see if additional constraints (specifically in the $LL$ and $RR$ sectors) can
be determined.  

\begin{figure}[h] 
   $$\epsfxsize=0.47\textwidth\epsffile{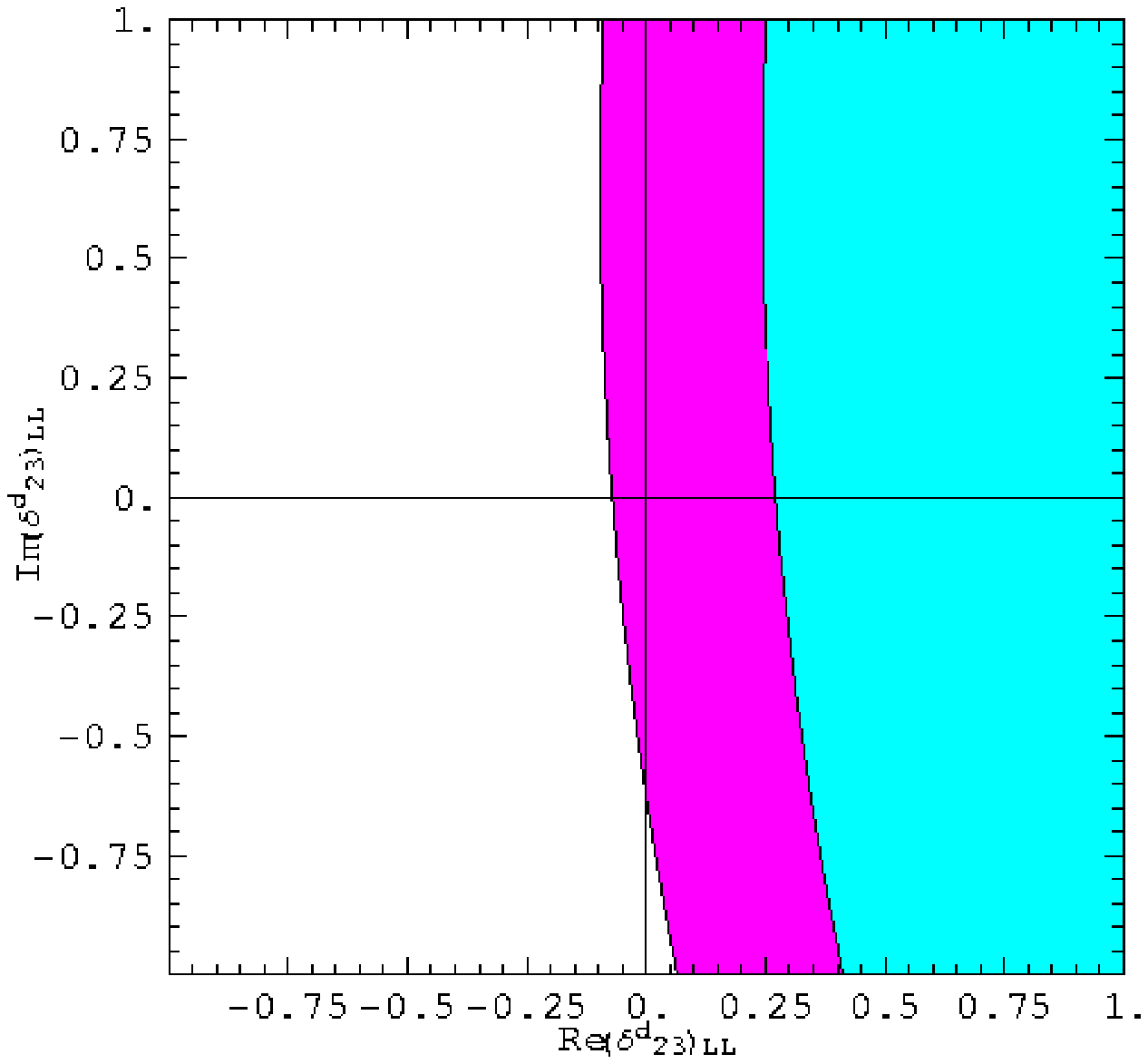}\qquad
   \epsfxsize=0.47\textwidth\epsffile{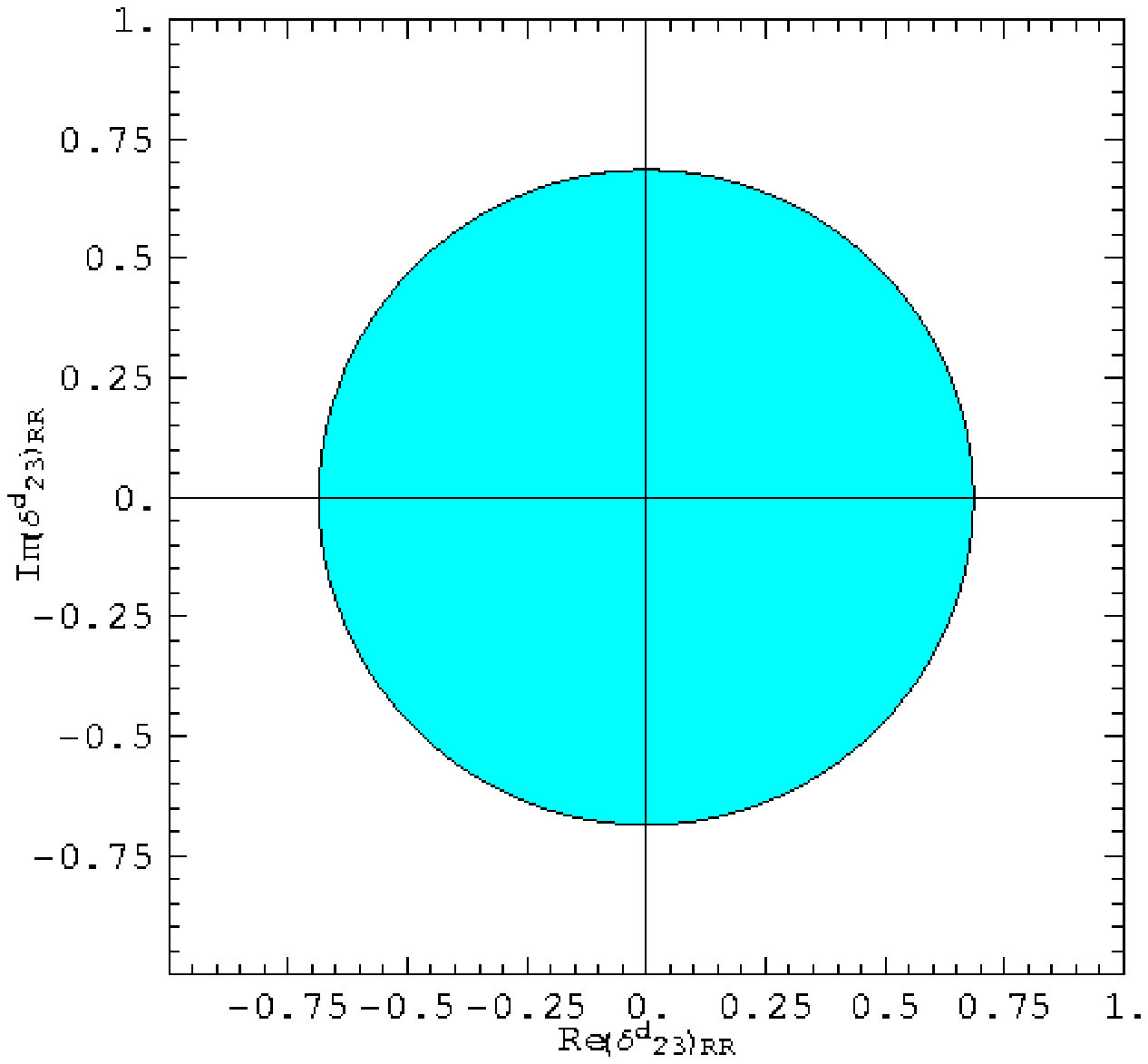}$$ \\
 $$\epsfxsize=0.47\textwidth\epsffile{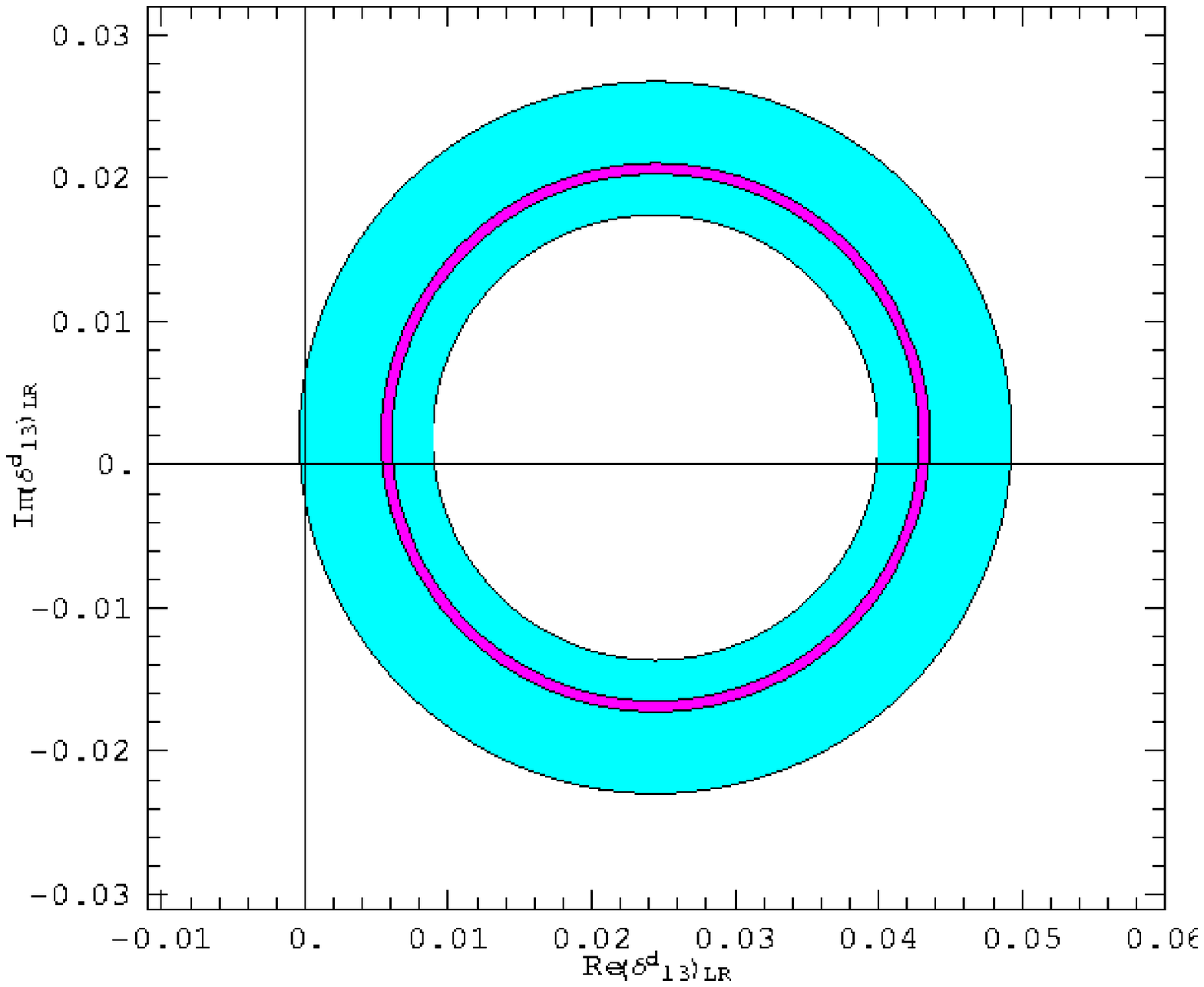}\qquad
   \epsfxsize=0.47\textwidth\epsffile{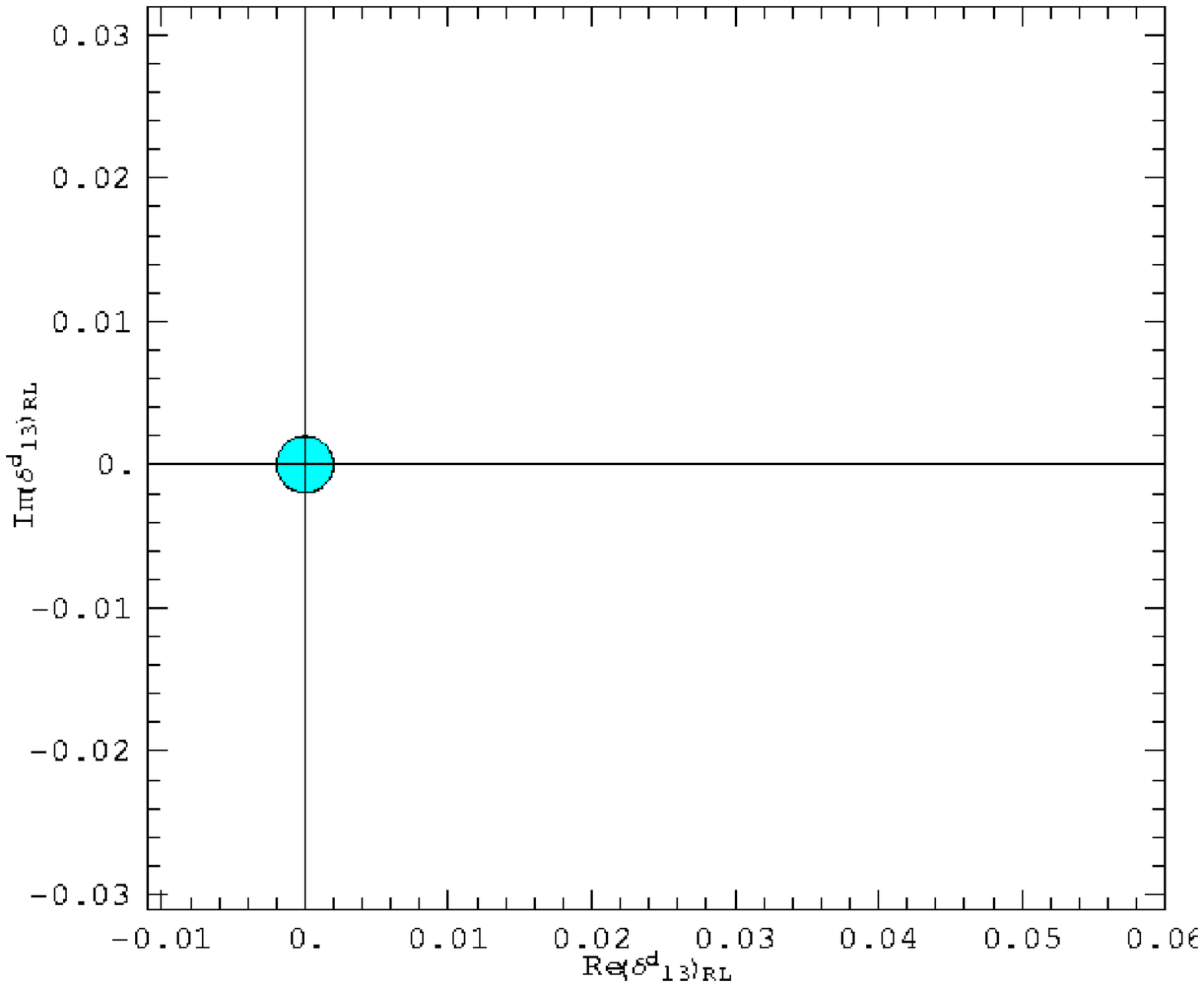}$$
  \vskip-12pt
   \caption[]{Allowed regions in $Re(\delta^d_{23})_{ij}$ --
   $Im(\delta^d_{23})_{ij}$ parameter space for double mass insertions
   $\mathrm{LL}=\mathrm{RR}$ and $\mathrm{LR}=\mathrm{RL}$.  We use
   $m_{\tilde 
   q} = m_{\tilde g} = 500$ GeV, 
   and constraints from Belle measurement of
   $\mathrm{BR}(B\to K^*\gamma)$.}  
   \label{fig:bdysymK}
 \end{figure}

The graphs in Figure \ref{fig:bdysymK} show the constraints on the
   $Re(\delta^d_{13})_{ij}$ --    $Im(\delta^d_{13})_{ij}$ plane once the
   bounds from the $B\to K^*\gamma$ branching ratio have been applied.   

These s show that using the central values for the Standard Model
amplitude for the $K^*\gamma$ decay gives extremely tight constraints on the 
$Re(\delta^d_{23})_{ij}$ -- $Im(\delta^d_{23})_{ij}$ parameter space, leaving
virtually no room for any significant contribution from a single mass
insertion.  Also, since there is no available parameter space for the $RR$
and $RL$ single insertions, no reduction of the constraints is obtained by
considering the double mass insertions.  Including the $1\sigma$ theory error
on the SM contribution opens up the parameter space a little, as we see in
Figure \ref{fig:bdysymK}, specifically with the $LL$ and $RR$ insertions.
Overall however, these results do not support the conclusion that there 
is any new physics in the exclusive $K^*\gamma$ decays.

%%% Local Variables:
%%% mode: latex
%%% TeX-master: t
%%% End:

%% file: chapter6.tex
\chapter{Conclusions and outlook}\label{chp:CONC}
\begin{center}
  \begin{quote}
    \it
    Few can foresee whither their road will lead them, till they come to its end
  \end{quote}
\end{center}
\vspace{-4mm}
\hfill{\small ``The Lord of the Rings'', J.R.R.Tolkien}
%% \begin{center}
%%   \begin{quote}
%%     \it
%%     People do not like to think. If one thinks, one must reach conclusions. Conclusions are not always pleasant.
%%   \end{quote}
%% \end{center}
%% \vspace{-4mm}
%% \hfill{\small Helen Keller}

This thesis has examined the non-leptonic decays both within QCD factorisation
and beyond it, challenging the assumptions and limitations of the method.
Our study was three-fold: we
analysed the treatment of the light-cone distribution amplitudes for the
light mesons and introduced a new model described by simple physical
parameters.  We then scrutinised the charmless non-leptonic decays in the
context of their incalculable non-factorisable contributions -- which we
consider the main limitation of the QCD factorisation method.  Finally, we
analysed the new results on the exclusive $B\to V\gamma$ within QCDF and
generic supersymmetric models.  

Our main new result is the development of a new model for the symmetric part
of the leading-twist light-cone distribution amplitude of light mesons.  This
was done with particular reference to the pion, although the results are
applicable to all light mesons.  The models depend on 
three parameters, two which control the fall-off behaviour of the Gegenbauer
moments $a_n$ in $n$, and a third $\Delta$, which parameterises the maximum
possible impact of the higher-order moments on a factorised physical
amplitude.  This is given by the first inverse moment of the DA.  The model
is defined at a low energy scale ($\mu = 1.2$GeV for the $\pi$) and can be
easily evolved up to higher scales by solving the evolution equation
numerically.  In contrast to the conformal expansion, which in computation
becomes numerically unstable at high orders, our model DAs are very well
behaved numerically and can be easily implemented in computer codes.  We discussed
how the evolution equation can be solved at the leading order, but this could
in principle also be solved to next-to-leading order.

We have introduced a similar model for the antisymmetric part of the
distribution amplitude, relevant for the $K$, and the $K^*$ mesons; these
models are normalised to $a_1$, the first Gegenbauer moment.  For the $\pi$
DA, where experimental data exists, we have formulated constraints on the
model parameters, likely to be valid for other meson DAs as well. The
constraints on the allowed values of $\Delta$ are tighter for the models
$\phi^-$ with an alternating sign fall-off of the Gegenbauer moments than for those with a
equal sign behaviour $\phi^+$.  

We feel that these models are better suited for estimating hadronic
uncertainties of processes calculated within factorisation than using the standard 
truncated conformal expansion.  This is true in particular for
processes that involve convolutions with singular kernels $\sim
1/u$, as the corresponding integral, $\Delta$, is the basic parameter
of the model.

Our models also should prove particularly useful for
describing DAs of mesons other than the $\pi$ which are also symmetric by
virtue of G-parity, but for which no experimental or other reliable
theoretical information is available --- in particular the $\rho$, $\omega$
and $\phi$.  For these particles we argue that existing theoretical
indications from local QCD sum rules \cite{Chernyak:1983ej, Ball:1996tb,
  Ball:2003sc} point to the DAs being narrower 
than the $\pi$, which implies $\Delta <1.2$ at the scale 1.2GeV.  These
results also imply that $a_2$ is positive, hence constraining $\Delta\geq1$.
For the parameter $a$, which controls the fall-off the $a_n$ with $n$, we
have found that the perturbative contributions to QCD sum rules indicate it
to be 3 \cite{Ball:1998sk, Ball:1998ff}, but that smaller values of $a$ are not excluded unless one
can rigorously prove that all leading-twist DAs must behave as $\sim u(1-u)$
near the endpoints $u=0,1$ at all scales.  Any further input, either from
experiment or lattice would be welcome to help constrain the model, however,
even without the constraints being taken literally our models provide a
convenient way to test the impact of non-asymptotic DAs on physical
amplitudes without resorting to the conformal expansion.  

We illustrated the impact of our models on the calculation of the
$B\to\pi\pi$ branching ratios and CP asymmetries and found 
that the branching ratios are more sensitive to the precise
values of the model parameters than the CP-asymmetries. 
The observed discrepancy between experimental data and the predictions
of QCD factorisation cannot however be explained by
nonstandard DAs, which indicates the presence of non-negligible
non-factorisable contributions --- a conclusion that agrees with the
findings of other authors, e.g. \cite{Feldmann:2004mg,Buras:2004ub}.

In order to quantitatively justify this premise, we analysed in detail the
non-factorisable $\Lambda_{\mathrm{QCD}}/m_b$ corrections to $B\to\pi\pi$ decay amplitudes.  We showed
how the experimental information can be used to extract valuable information
on the size and nature of these non-factorisable corrections.  We aimed to move
beyond the model-dependent treatment presented in \cite{Beneke:1999br}, where
non-factorisable contributions are identified from convolution integrals
which have endpoint divergences.  The chirally enhanced power corrections
which originate from hard-spectator scattering diagrams and annihilation topologies,
are each parameterised by an arbitrary complex number, the modulus of which is
estimated from regularising the end-point divergences.  We replaced these
terms with generic non-factorisable corrections via a decomposition of the $B\to\pi\pi$ decays into
isospin amplitudes $A_{1/2}$ and $A_{3/2}$.  Three scenarios were
constructed using differing levels of non-factorisable corrections on top of
the factorisable terms found via QCDF.  

Our analysis included the effect of a non-standard DA, to simulate greater impact from
higher-order moments, as parameterised by the physical parameter $\Delta$.
We found that the maximum contribution allowed within the physical range of
$\Delta$ was required to give the best-fit to the data.  Variation of the
$B\to\pi$ transition form factor also required an extremal (minimum) allowed
value, as found from the $\chi^2$ fitting across the full data set.  Our combined
analysis of all of the most up-to-date experimental data  
available on the $B\to\pi\pi$ system (as of September 2005), showed clearly
that sizable non-factorisable effects are required to bring all of the
$B\to\pi\pi$ results into agreement.  Even for extremal values of our input
parameters, the best-fit scenarios do not give a reasonable $2\sigma$
agreement to the experimental data until at least 30\% contributions are
included.  For the Belle data and combined HFAG results, a minimum level of 40\% is needed to
produce the same level of agreement.
  
Our third scenario showed how the long-distance enhanced charming penguin diagrams can
contribute constructively to the decay amplitudes.  The size of the
enhancement must be tightly constrained in order to keep within the error
bounds of the data (particularly for CP asymmetries) and as taking $|R|>1.5$ will
make the charm penguin the dominant non-factorisable effect.  The branching ratios
and $\pi^+\pi^-$ CP asymmetry can be independently brought into excellent
agreement, however we discussed how the phases of the charm correction are at
odds for these two scenarios.  The best-fit for the scenario with 20\% NF
correction in addition to the charming penguin gives a considerable improvement
above the scenario with only the 20\% correction. However, further NF
corrections are still required to bring the theory into the experimental
$2\sigma$ error bound.  Hence, we can conclude that a sizable NF correction is
still required even in this scenario.   

It will be interesting to watch how any future experimental results on
$B\to\pi\pi$ develop, particularly those on the branching 
ratios for $B\to\pi^+\pi^-$ and $B\to\pi^0\pi^0$.  The current situation makes it
difficult to make conclusive statements, as measurements from
{\textsc{BaBar}} and Belle currently disagree by approximately $2\sigma$ for
both the $\pi^+\pi^-$ and $\pi^0\pi^0$ channels.  A notable rise in the central value
for the $\pi^+\pi^-$ channel (as was the case for {\textsc{BaBar}} and the
figures released in summer 2004 and 2005) would act favourably for a lower
requirement of non-factorisable corrections.  The converse however, could
leave us looking at approaching 50\% non-factorisable corrections in order to obtain a
good $2\sigma$ agreement with the HFAG data.   

The results of our analysis leave us with many more questions than when we
started.  When the experimental situation is resolved and the accuracy improves the
question remains of the source of the disagreement with the $B\to\pi\pi$
data; whether this can be explained by significant non-factorisable effects within
the Standard Model, or if it points to new physics beyond it.  This also
presents the more pertinent question of the validity of the factorisation
scheme itself, if indeed such high levels of power-suppressed corrections are
required to satisfy the data.  The next step in this puzzle will be to
analyse $B\to\pi K$ decays, by fitting a set of generic non factorisable
contributions to the experimental data \cite{BJT05}.  

Finally, we took a step sideways and analysed the recently released data on
$B\to(\rho,\,\omega)\gamma$ decays.  We examined the Standard Model
predictions within QCD factorisation using the technique first presented in
\cite{Bosch:2001gv} using updated non perturbative input,
e.g. form factors \cite{Ball:2004rg}, and testing the effect of our non
standard DA introduced in Chapter \ref{chp:LCDA}.  We found predictions 
for $B\to\rho\gamma$ in agreement with that presented in \cite{Bosch:2001gv}
\begin{equation*}
   \textrm{BR}(B^0\to\rho^0\gamma) = (0.72\pm0.22)\times 10^{-6} \qquad
   \textrm{BR}(B^\pm\to\rho^\pm\gamma) = (1.52\pm0.45)\times 10^{-6}  
\end{equation*}
and estimated the combined branching ratio as
\begin{equation*}
   \textrm{BR}(B\to(\rho,\,\omega)\gamma) = (1.60\pm0.48)\times 10^{-6} 
\end{equation*}
which is within the error bounds of the new Belle measurement.  We found that
the variation of the branching ratio with the non standard DA was minimal,
and that there was less than a 3\% effect across the full physical range of
$\Delta$.  We found the same results for an analogous analysis of $B\to
K^*\gamma$ where we estimated a central value for the branching ratio of 
\begin{equation*}
   \textrm{BR}(B\to K^*\gamma) = (5.49\pm1.64)\times 10^{-5} 
\end{equation*}
This differs from the central value presented in \cite{Bosch:2001gv}
predominantly due to the drop in the value of the $B\to K^*$ form factor.  We
then discussed how these two branching ratios can be combined
into a ratio $R[(\rho,\omega)\gamma/K^*\gamma]$ which can be used to extract
the value of the CKM ratio $|V_{td}/V_{ts}|$.  We obtained a value of
\begin{equation*}
   R[(\rho,\omega)\gamma/K^*\gamma] = 0.030\pm0.008 
\end{equation*}
in excellent agreement with the measured result.  This led to an extracted
value of $|V_{td}/V_{ts}| = 0.195\pm0.051$.  This shows how well the ratio
can reduce the theoretical uncertainty.  

Since previously only an upper bound for the $B\to(\rho,\,\omega)\gamma$
has been available, this was an opportune moment to examine how much
room for new physics remains in the $b\to d\gamma$  decays.  We examined
these decays and $B\to K^*\gamma$ in the context of a generic minimal
supersymmetric model, using the dominant gluino contributions as derived using the mass
insertion approximation.  The contributions from charged Higgs,
neutralino and chargino contributions are sub-leading and can be safely
neglected in an indicative analysis of this type.  
  
We found constraints on the parameter space for the mass insertion
$\delta_{13}^d$ which is relevant for the $b\to d$ transitions.  We showed
how the allowed regions are very stringently constrained, primarily due to the constraint
from  $R[(\rho,\omega)\gamma/K^*\gamma]$.  Taking into account the $1\sigma$
theory error opens up more of the parameter space, but still leaving only the $LL$
and $RR$ insertions as realistic possibilities.  We examined the parameter
space for the $\delta_{23}^d$ insertion using the bounds from the $B\to
K^*\gamma$ decay.  We used the central value of the SM amplitude and found
virtually no parameter space available for a significant mass insertion.
This is to be expected, considering the good agreement of
theory with the measurement for the inclusive decay $B\to X_s\gamma$, which implies that all of the
short-distance physics in the $b\to s\gamma$ transition is under good
control.  

As we rapidly approach the new era of particle physics heralded by the
LHC, precision measurements in the flavour sector will become an essential
component of testing the Standard Model expectations, and to uncover
potential sources of new physics.  We need therefore complete
control over the long distance physics to ensure that the theory errors do
not remain significantly larger than the experimental errors as the
measurements continue to improve.  I believe that this work presents an
unbiased examination of methods to tackle and understand the uncertainties
and limitations of the QCD factorisation formalism and points to a direction in
which further study could go.

%% QCD factorisation is a systematic expansion of long-distance hadronic matrix
%% elements in $\Lambda_{\mathrm{QCD}}/m_b$ and the strong coupling $\a_s$.  

%%% Local Variables:
%%% mode: latex
%%% TeX-master: t
%%% End:

%% file: end.tex
\newpage
\onehalfspacing
\vspace*{5cm}
\begin{center}
    \it
The Road goes ever on and on,\\
Down from the door where it began.\\
Now far ahead the Road has gone\\
And I must follow if I can.\\ 
\ \\
Pursuing it with eager feet,\\
Until it meets some larger way,\\
Where many paths and errands meet.\\
And whither then? I cannot say.\\
\end{center}
\hfill{\small J.R.R. Tolkien}

%%% Local Variables: 
%%% mode: latex
%%% TeX-master: "end"
%%% TeX-master: t
%%% End: 

%% file: appendixD.tex
\chapter{Wilson coefficients for $\mathcal{H}_{\mathrm{eff}}^{\Delta B=1}$}\label{chp:AppD}

\section{$6\times6$ operator basis}
We give the explicit formulae for the Wilson coefficients in the naive
 dimensional regularization scheme.  We consider the $6\times6$ matrix which
 is made up of the operators $Q_1\dots Q_6$.  Recalling that we can expand
 the Wilson coefficients in powers of $\a_s$, we have
\begin{equation}\label{expci}
   C_i(\mu)=C_i^{(0)}(\mu)+\frac{\alpha_s(\mu)}{4\pi}C_i^{(1)}(\mu)+\ldots
\end{equation}
The evolution of the effective Wilson coefficients is governed by the
anomalous dimension matrix (ADM).  The expansion of the ADM including
only gluonic corrections (no electroweak penguins) is, to order $\a_s$,
\be
   \hat\gamma = \frac{\a_s}{4\pi}\hat\gamma^{(0)}_s +
   \left(\frac{\a_s}{4\pi}\right)^2\hat\gamma^{(1)}_s
\ee
 The leading order matrix in this expansion is given (using the operators
 in Section \ref{sec:Ham})
 \begin{equation}\label{adm0}
 \gamma^{(0)}=\left(\begin{array}{cccccc}
         -2  & 6 & 0              & 0               & 0             & 0                \\[0.2cm]
         6 & -2  & -\frac{2}{9}   & \frac{2}{3}     & -\frac{2}{9}  & \frac{2}{3}      \\[0.2cm]
         0  & 0  & -\frac{22}{9}  & \frac{22}{3}    & -\frac{4}{9}  & \frac{4}{3}      \\[0.2cm]
         0  & 0  & 6-\frac{2f}{9} & -2+\frac{2f}{3} & -\frac{2f}{9} & \frac{2f}{3}     \\[0.2cm]
         0  & 0  & 0              & 0               & 2             & -6               \\[0.2cm]
         0  & 0  & -\frac{2f}{9}  & \frac{2f}{3}    & -\frac{2f}{9} & -16+\frac{2f}{3} \\[0.2cm]
 \end{array}\right)
 \end{equation}
The number of quark colours has been set to $N_c = 3$, and the number of
 active flavours is denoted by f.   At next-to-leading order we have the following matrix governing the mixing of current-current and QCD-penguin operators among each other:
 \begin{equation}\label{adm1}
   \gamma^{(1)}=\left(\begin{array}{cccccc}
       -\frac{21}{2}-\frac{2f}{9} & \frac{7}{2}+\frac{2f}{3} & \frac{79}{9}   & -\frac{7}{3}                  & -\frac{65}{9}                    &
   -\frac{7}{3}                  \\[0.2cm]
\frac{7}{2}+\frac{2f}{3}   & -\frac{21}{2}-\frac{2f}{9}  & -\frac{202}{243}                & \frac{1354}{81}               & -\frac{1192}{243}                & \frac{904}{81}                \\[0.2cm]
       0                          & 0                          & -\frac{5911}{486}+\frac{71f}{9} & \frac{5983}{162}+\frac{f}{3}  & -\frac{2384}{243}-\frac{71f}{9}  & \frac{1808}{81}-\frac{f}{3}   \\[0.2cm]
       0                          & 0                          & \frac{379}{18}+\frac{56f}{243}  & -\frac{91}{6}+\frac{808f}{81} & -\frac{130}{9}-\frac{502f}{243}  & -\frac{14}{3}+\frac{646f}{81} \\[0.2cm]
       0                          & 0                          & -\frac{61f}{9}                  & -\frac{11f}{3}                & \frac{71}{3}+\frac{61f}{9}       & -99+\frac{11f}{3}             \\[0.2cm]
       0                          & 0                          & -\frac{682f}{243}               & \frac{106f}{81}               & -\frac{225}{2}+\frac{1676f}{243} & -\frac{1343}{6}+\frac{1348f}{81}
   \end{array}\right)
 \end{equation}

At $\mu_0=M_W$ the matching conditions can be split into the leading order
and NLO corrections, $C^{(0)}_i(M_W)$ and $C^{(1)}_i(M_W)$ respectively.
These are given by
\begin{equation}\label{wcinilo}
   C_i^{(0)}(M_W)=\left\{\begin{array}{cl}
         1 & \mbox{for } i=2\\
         0 & \mbox{ } \textrm{otherwise}\\
         \end{array}\right.
  \end{equation}
and
\begin{equation}\label{wcinilo}
   C_i^{(1)}(M_W)=\left\{\begin{array}{cl}
      \frac{11}{2} & \mbox{for } i=1\\
     -\frac{11}{6} & \mbox{for } i=2\\
     -\frac{1}{6}\widetilde{E_0}(x_t) & \mbox{for } i=3,5\\
     \frac{1}{2}\widetilde{E_0}(x_t) & \mbox{for } i=4,6\\
          \end{array}\right.
  \end{equation}
with $\widetilde{E_0}(x)$ being the Inami-Lim function
 \begin{eqnarray}
      \widetilde{E_0}(x) &=& E_0(x) - \frac{2}{3}\nonumber\\
      E_0(x) &=& \left(-\frac{2}{3}\ln(x) + \frac{x(18-11x-x^2)}{12(1-x)^3} + \frac{x^2(15-16x+4x^2)}{6(1-x)^4}\ln(x)\right) \nonumber\\
 \end{eqnarray}
and 
 \begin{equation}
   x_t = \frac{m_t^2}{M_W^2} 
 \end{equation}

\section{$10\times10$ operator basis, including electroweak penguins}

The expansion of the ADM including gluonic and photonic corrections is
 \be
    \hat\gamma = \frac{\a_s}{4\pi}\hat\gamma^{(0)}_s +
    \frac{\a_s}{4\pi}\hat\gamma^{(0)}_e +
    \left(\frac{\a_s}{4\pi}\right)^2\hat\gamma^{(1)}_s + \frac{\a_s}{4\pi}\frac{\a_e}{4\pi}\hat\gamma^{(1)}_{se}
\ee
where we include the leading QED contributions from $\hat\gamma^{(0)}_e$. but
 do not include the two-loop QCD-QED anomalous dimension matrix
 $\hat\gamma^{(1)}_{se}$.  The other matrices are given below, again in the
 NDR scheme. 

%\hskip 4cm
%\rotate[l]{
\vskip 0.5cm
\hskip -1cm
$ \gs =
\left(
\begin{array}{cccccccccc}
-2 & 6 & 0 & 0 & 0 & 0 & 0 & 0 & 0 & 0 \\ \svs
6 & -2 & {{-2}\over 9} & {2\over 3} & {{-2}\over {9}} &\
  {2\over 3} & 0 & 0 & 0 & 0 \\ \svs
0 & 0 & {{-22}\over {9}} & {{22}\over 3} & {{-4}\over {9}} & {4\over 3} &\
  0 & 0 & 0 & 0 \\ \svs
0 & 0 & 6 - {{2 f}\over {9}} & {-2 + {{2 f}\over 3}} & {{-2\
  f}\over {9}} & {{2 f}\over 3} & 0 & 0 & 0 & 0 \\ \svs
0 & 0 & 0 & 0 & {2} & -6 & 0 & 0 & 0 & 0 \\ \svs
0 & 0 & {{-2 f}\over {9}} & {{2 f}\over 3} & {{-2 f}\over {9}} & {-16 + {{2 f}\over 3}} & 0 & 0 & 0 & 0 \\ \svs
0 & 0 & 0 & 0 & 0 & 0 & 2 & -6 & 0 & 0 \\ \svs
0 & 0 & {{-2 \left( u-d/2 \right) }\over {9}} & {{2 \left(\
  u-d/2 \right) }\over 3} & {{-2 \left( u-d/2 \right)\
  }\over {9}} & {{2 \left( u-d/2 \right) }\over 3} & 0 & -16 & 0 & 0 \\ \svs
0 & 0 & {2\over {9}} & -{2\over 3} & {2\over {9}} & -{2\over 3} & 0 & 0 &\
  -2 & 6 \\ \svs
0 & 0 & {{-2 \left( u-d/2 \right) }\over {9}} & {{2 \left(\
  u-d/2 \right) }\over 3} & {{-2 \left( u-d/2 \right)\
  }\over {9}} & {{2 \left( u-d/2 \right) }\over 3} & 0 & 0 & 6\
  & -2
\end{array}
\right) $\\
%}
\vskip 0.5cm
\hskip -1cm
$ \gem =
\left(
\begin{array}{cccccccccc}
-{8\over 3} & 0 & 0 & 0 & 0 & 0 & {{16\,N}\over {27}} & 0 & {{16\,N}\over\
  {27}} & 0 \\ \svs
0 & -{8\over 3} & 0 & 0 & 0 & 0 & {{16}\over {27}} & 0 & {{16}\over {27}} & 0\
  \\ \svs
0 & 0 & 0 & 0 & 0 & 0 & -{{16}\over {27}} + {{16\,N\,\left( u-d/2\
  \right) }\over {27}} & 0 & -{{88}\over {27}} + {{16\,N\,\left(
  u-d/2 \right) }\over {27}} & 0 \\ \svs
0 & 0 & 0 & 0 & 0 & 0 & {{-16\,N}\over {27}} + {{16\,\left( u-d/2\
  \right) }\over {27}} & 0 & {{-16\,N}\over {27}} + {{16\,\left(
  u-d/2 \right) }\over {27}} & -{8\over 3} \\ \svs
0 & 0 & 0 & 0 & 0 & 0 & {8\over 3} + {{16\,N\,\left( u-d/2\
  \right) }\over {27}} & 0 & {{16\,N\,\left( u-d/2 \right) }\over\
  {27}} & 0 \\ \svs
0 & 0 & 0 & 0 & 0 & 0 & {{16\,\left( u-d/2 \right) }\over {27}} &\
  {8\over 3} & {{16\,\left( u-d/2 \right) }\over {27}} & 0 \\ \svs
0 & 0 & 0 & 0 & {4\over 3} & 0 & {4\over 3} + {{16\,N\,\left( u+d/4\
  \right) }\over {27}} & 0 & {{16\,N\,\left( u+d/4 \right) }\over\
  {27}} & 0 \\ \svs
0 & 0 & 0 & 0 & 0 & {4\over 3} & {{16\,\left( u+d/4 \right) }\over\
  {27}} & {4\over 3} & {{16\,\left( u+d/4 \right) }\over {27}} & 0\
  \\ \svs
0 & 0 & -{4\over 3} & 0 & 0 & 0 & {8\over {27}} + {{16\,N\,\left(
  u+d/4 \right) }\over {27}} & 0 & -{{28}\over {27}} + {{16\,N\,\left(
  u+d/4 \right) }\over {27}} & 0 \\ \svs
0 & 0 & 0 & -{4\over 3} & 0 & 0 & {{8\,N}\over {27}} + {{16\,\left(
  u+d/4 \right) }\over {27}} & 0 & {{8\,N}\over {27}} + {{16\,\left(
  u+d/4 \right) }\over {27}} & -{4\over 3}
\end{array}
\right) $

\begin{displaymath}
\gssndr =
\left(
\begin{array}{cccccccccc}
-{{209}\over 18} & {{41}\over 6} & {79\over 9} & -{{7}\over 3} & -{{65}\over
  9} & -{7\over 3} & {0} & {0} & {0} & {0} \\ \mvs
{41\over 6}  & -{{209}\over 18} &
  -{{202}\over {243}} & {{1354}\over {81}} & -{{1192}\over {243}} &
  {{904}\over 81}  & 0 & 0  & 0 & 0\\ \mvs
0 & 0 & -{{13259}\over {486}}  & {{6253}\over {162}} & -{{11959}\over {243}}
  & -{{1673}\over {81}} & 0 & 0 & 0 & 0\\ \mvs
0 & 0 & {{10793}\over {486}} & -{{5623}\over 162} & -{{6020}\over 243} &
  -{{2852}\over {81}} & 0 & 0  & 0 & 0\\ \mvs
0 & 0 & -{{305}\over 9} & {{-55}\over 3} & {{518}\over 9} & -{{242}\over 3} &
  0 & 0 & 0 & 0\\ \mvs 
0 & 0 & -{{3410}\over {243}} & {{530}\over {81}} & -{{37915}\over 486} &
  -{{22781}\over 162} & 0 & 0 & 0 & 0  \\ \mvs
0 & 0 & -{{61}\over 18} & -{{11}\over 6} & {{83}\over 18} & -{{11}\over 6} &
  {{103}\over 9} & -{187\over 3} & 0 & 0\\ \mvs
0 & 0 & -{{682}\over {729}} & {{53}\over {81}} & {{352}\over {243}} &
  {{368}\over 81} & -{185\over 2} & -{3349\over 18} & 0 & 0\\ \mvs
0 & 0 & {{2375}\over {486}} & -{{2735}\over 162} & {{467}\over {486}} &
  -{{1835}\over {162}} & 0 & 0 &  -{{209}\over 18} & {41\over 6} \\ \mvs
0 & 0 & -{{2186}\over 243} & {602\over 81} & {{1504}\over 243} & {{512}\over
  81} & 0  & 0 & {41\over 6} & -{209\over 18}
\end{array}
\right)
\end{displaymath}
%% \begin{displaymath}
%% \gssndr =
%% \left(
%% \begin{array}{ccccc}
%% -{{21}\over 2} - {{2\,f}\over 9} & {7\over 2} + {{2\,f}\over 3} & {{79}\over\
%%   9} & -{7\over 3} & -{{65}\over 9} \\ \mvs
%% {7\over 2} + {{2\,f}\over 3} & -{{21}\over 2} - {{2\,f}\over 9} &\
%%   -{{202}\over {243}} & {{1354}\over {81}} & -{{1192}\over {243}} \\ \mvs
%% 0 & 0 & -{{5911}\over {486}} + {{71\,f}\over 9} & {{5983}\over {162}} +\
%%   {f\over 3} & -{{2384}\over {243}} - {{71\,f}\over 9} \\ \mvs
%% 0 & 0 & {{379}\over {18}} + {{56\,f}\over {243}} & -{{91}\over 6} +\
%%   {{808\,f}\over {81}} & -{{130}\over 9} - {{502\,f}\over {243}} \\ \mvs
%% 0 & 0 & {{-61\,f}\over 9} & {{-11\,f}\over 3} & {{71}\over 3} + {{61\,f}\over\
%%   9} \\ \mvs
%% 0 & 0 & {{-682\,f}\over {243}} & {{106\,f}\over {81}} & -{{225}\over 2} +\
%%   {{1676\,f}\over {243}} \\ \mvs
%% 0 & 0 & {{-61\,(u-d/2)}\over 9} & {{-11\,(u-d/2)}\over 3} & {{83\,
%% (u-d/2)}\over 9} \\ \mvs
%% 0 & 0 & {{-682\,(u-d/2)}\over {243}} & {{106\,(u-d/2)}\over {81}} &\
%%   {{704\,(u-d/2)}\over {243}} \\ \mvs
%% 0 & 0 & {{202}\over {243}} + {{73\,(u-d/2)}\over 9} & -{{1354}\over {81}} -\
%%   {{(u-d/2)}\over 3} & {{1192}\over {243}} - {{71\,(u-d/2)}\over 9} \\ \mvs
%% 0 & 0 & -{{79}\over 9} - {{106\,(u-d/2)}\over {243}} & {7\over 3} +\
%%   {{826\,(u-d/2)}\over {81}} & {{65}\over 9} - {{502\,(u-d/2)}\over\
%%   {243}}
%% \end{array}
%% \right.
%% \end{displaymath}

% \newsection{$\rsndr$, $\rshv$}

$N$ is the number of colours, $f$, the number of active flavours and
$u$ and $d$ denote the number of up- and down-type flavours respectively
($u+d =f$).  We have also specified $f=5$ for presentation of $\gamma_s^{(1)}$.

The additional initial conditions for the Wilson coefficients of order $\a_e$
are:
\begin{equation}\label{wcinilo}
   C_i^{e}(M_W)=\frac{\a_e}{4\pi}*\left\{\begin{array}{cl}
      0 & \mbox{for } i=1,4-6,8,10\\
      -\frac{35}{18} & \mbox{for } i=2\\
    \frac{2}{3}\frac{1}{s_w^2}\left(2B_0(x_t) + C_0(x_t)\right) & \mbox{for }
      i=3\\ 
     \frac{2}{3}\left(4C_0(x_t)+\widetilde{D}_0(x_t)\right) & \mbox{for }
      i=7\\ 
   \frac{2}{3}\left(4C_0(x_t)+\widetilde{D}_0(x_t) +
      \frac{1}{s_w^2}\left(10B_0(x_t) -4C_0(x_t)\right)\right) & \mbox{for } i=9 \\
          \end{array}\right.
  \end{equation}
with 
\begin{eqnarray}
\begin{split}
B_0 (x_t) &= \frac{x_t}{4 (x_t - 1)} + \frac{x_t}{4 (x_t -
  1)^2} \ln x_t \,  \\ 
C_0 (x_t) &= \frac{x_t ( x_t-6)}{8 (x_t - 1)} + \frac{x_t (2 + 3
  x_t)}{8 (x_t - 1)^2} \ln x_t \,  \\ 
\widetilde{D}_0 (x_t) &= -\frac{4}{9}ln x_t (\frac{-19x_t^3 +
    25x_t^2}{36(x_t=1)^3} + \frac{x_t^2(5x_t^2-2x_t-6)}{18(x_t-1)^4}\ln x_t -
  \frac{4}{9}\\
\end{split}
\end{eqnarray}

%% \section{Magnetic penguins*}

%% The magnetic penguins -- the electromagnetic penguin $Q_{7\gamma}$ and the
%% chromomagnetic penguin $Q_{8g}$ are normally described by the ``effective
%% coefficients''  %REF!
%%  \begin{eqnarray}\label{c78effdef}
%%    C_{7\gamma}^{(0)eff}(\mu) &=& C_{7\gamma}^{(0)} +\sum_{i=1}^6 y_i C_i^{(0)}(\mu)\\
%%    C_{8g}^{(0)eff}(\mu) &=& C_{8g}^{(0)} +\sum_{i=1}^6 z_i C_i^{(0)}(\mu)
%%  \end{eqnarray}
%% with $\vec y=(0,0,0,0,-\frac{1}{3},-\frac{N}{3})$ and $\vec z=(0,0,0,0,1,0)$
%% in the NDR scheme. This ensures the leading order effective coefficients $C^{(0)eff}_i$ with the corresponding anomalous dimension matrix $\gamma^{(0)eff}$ then are regularization and renormalization scheme independent. 

%% %%% Local Variables: 
%% %%% mode: latex
%% %%% TeX-master: t
%% %%% End: 

%% file: appendixA.tex
\chapter{Additional formulae from QCD factorisation}
\label{chp:AppA}

\section{Decay amplitudes for $B\to\pi\pi$}

In terms of the QCD factorisation coefficients defined in Section
(\ref{sec:factcont}), the $B\to\pi\pi$ decay amplitudes are written as
follows:
\begin{eqnarray}
  \label{eq:piampQCDF}
  -\sqrt{2}\mathcal{A}(B^-\to\pi^-\pi^0) &=& \left[\lud a_1 + \tfrac{3}{2}\lpd(-a_7 +
 a_9 + a_{10}^p + r_\chi^\pi a_8^p)\right]A_{\pi\pi}\nonumber\\
 -\mathcal{A}(\bar{B}^0\to\pi^+\pi^-) &=& \left[\lud a_1 +
 \lpd(a_4^p+a_{10}^p) + \lpd r_\chi^\pi(a_6^p + a_8^p)\right]A_{\pi\pi}\nonumber\\
 -\sqrt{2}\mathcal{A}(\bar{B}^0\to\pi^0\pi^0) &=& \left[\lud a_2
 -\lpd(a_4^p-\tfrac{1}{2}a_{10}^p) - \lpd r_\chi^\pi(a_6^p -
 \tfrac{1}{2}a_8^p) + \tfrac{3}{2}\lpd(a_9-a_7)\right]A_{\pi\pi}\nonumber\\
\end{eqnarray}
with $a_i \equiv a_i(\pi\pi)$, $\lpd = V_{pb}V_{pd}^*$ for $p=u,c$ and 
\be
   A_{\pi\pi} = i\frac{G_F}{\sqrt{2}}(m_B^2 -
   m_\pi^2)F_0^{B\to\pi}(m_\pi^2)f_\pi 
\ee
The charge conjugate decays are obtained by replacing
$\lambda_p\to\lambda_p^*$.   

%% \section{$B\to\pi K$}

%% We can similarly write the decays amplitudes for $B\to\pi K$ as follows:
%% \begin{eqnarray}
%%   \label{eq:piKampQCDF}
%%   \mathcal{A}(B^-\to\pi^-\bar{K}^0) &=& \lps\left[\left(a_4^p - \tfrac{1}{2}a_{10}^p\right) +
%%  r_\chi^K\left(a_6^p - \tfrac{1}{2}a_8^p\right)\right]A_{\pi K}\nonumber\\
%%  \sqrt{2}\mathcal{A}(B^-\to\pi^0K^-) &=& \left[\lus a_1 + \lps\left(a_4^p +
%%  a_{10}^p\right) + \lps r_\chi^K\left(a_6^p + a_8^p\right)\right]A_{\pi K}
%%  \nonumber\\
%%      && + \left[\lus a_2 + \tfrac{3}{2}\lps\left(-a_7 +
%%  a_9\right)\right]A_{K\pi} \nonumber\\
%%     \mathcal{A}(\bar{B}^0\to\pi^+K^-) &=& \left[\lus a_1 + \lps\left(a_4^p +a_{10}^p\right) +
%%  \lps r_\chi^K\left(a_6^p + a_8^p\right)\right]A_{\pi K}\nonumber\\
%%   \sqrt{2}(\bar{B}^0\to\pi^0\bar{K}^0) &=& \left[-\lps\left(a_4^p -
%%  \tfrac{1}{2}a_{10}^p\right)-\lps r_\chi^K\left(a_6^p -\tfrac{1}{2}
%%  a_8^p\right)\right]A_{\pi K} \nonumber\\
%%      && + \left[\lus a_2 +
%%  \tfrac{3}{2}\left(-a_7+a_9\right)\right]A_{K\pi}
%% \end{eqnarray}
%% with $a_i \equiv a_i(\pi K)$ or $a_i(K\pi)$ where appropriate, $\lambda_p^{s}
%% = V_{pb}V_{ps}^*$ for $p=u,c$ and 
%% \begin{eqnarray}
%%   A_{\pi K} &=& i\frac{G_F}{\sqrt{2}}(m_B^2 -
%%   m_\pi^2)F_0^{B\to\pi}(m_K^2)f_K \nonumber\\
%%  A_{K\pi} &=& i\frac{G_F}{\sqrt{2}}(m_B^2 -
%%   m_K^2)F_0^{B\to K}(m_\pi^2)f_\pi 
%% \end{eqnarray}
%% The charge conjugate decays are again obtained by replacing
%% $\lambda_p\to\lambda_p^*$. 

\section{Annihilation contributions to $B\to V\gamma$}

The contributions to the annihilation amplitudes contain two different
factors depending if the photon emission if from the light quark in the
$B$-meson, or from one of the constituent quarks of the vector meson.  The
former give the factor $b^V$, and the latter $d^V$, which are expressed as
\cite{Bosch:2002bw}:
\begin{eqnarray*}
   b^V = \frac{2\pi^2}{F_V}\frac{f_Bm_Vf_V}{m_Bm_b\lambda_B} \qquad
   d^V(v) = -\frac{4\pi^2}{F_V}\frac{f_Bf_V^\perp}{m_Bm_b}\int_0^1\frac{\phi_V^\perp(v)}{v}\,dv 
\end{eqnarray*}
The integral over the distribution amplitude in the factor $d^V$ will give
a dependence on $\Delta$ (and $a_1$ where relevant) for all
mesons.  We include these components in the decay amplitudes via 
\begin{eqnarray*}
   a_7^u\longrightarrow a_7^u + a_{ann}^u \\
   a_7^c\longrightarrow a_7^c + a_{ann}^c
\end{eqnarray*}  
where
\begin{eqnarray*}
   a^u_{ann}(\rho^0\gamma) &=& Q_d\left[-a_2b^{\,\rho} + a_4b^{\,\rho} + 2a_6d^\rho(v)\right]\\
   a^c_{ann}(\rho^0\gamma) &=&  Q_d\left[a_4b^{\,\rho} + 2a_6d^\rho(v)\right]\\ \\
   a^u_{ann}(\rho^-\gamma) &=& Q_u\left[a_1b^\rho + a_4b^\rho +
   (Q_s/Q_u+1)\,a_6d^\rho(v)\right] \\
   a^c_{ann}(\rho^-\gamma) &=& Q_u\left[a_4b^\rho + (Q_s/Q_u+1)\,a_6d^\rho(v)\right]  
\end{eqnarray*}
\begin{eqnarray*}
   a^u_{ann}(\bar{K}^{*0}\gamma) &=& Q_d\left[a_4b^{K^*} + a_6(d^{K^*}(v)
   + d^{K^*}(\bar v)) \right]\\
   a^c_{ann}(\bar{K}^{*0}\gamma) &=&  Q_d\left[a_4b^{K^*} + a_6(d^{K^*}(v)
   + d^{K^*}(\bar v)) \right]\\\\
   a^u_{ann}(\bar{K}^{*-}\gamma) &=& Q_u\left[a_1b^{K^*} +
   a_4b^{K^*}+Q_s/Q_u\,a_6d^{K^*}(v)+a_6d^{K^*}(\bar v)\right]\\
   a^c_{ann}(\bar{K}^{*-}\gamma) &=& Q_u\left[a_4b^{K^*}+Q_s/Q_u\,a_6d^{K^*}(v)+a_6d^{K^*}(\bar v)\right]
\end{eqnarray*}  
The $a_i$ are combinations of Wilson coefficients:
\begin{eqnarray*}
   a_{1,2} &=& C_{1,2} + \frac{1}{N_c}C_{1,2} \\
   a_4 &=& C_4 + \frac{1}{N_c}C_{3} \\
   a_6 &=& C_6 + \frac{1}{N_c}C_{5}
\end{eqnarray*}

%%% Local Variables: 
%%% mode: latex
%%% TeX-master: t
%%% End: 

%% file: appendixB.tex
\chapter{Analytic evolution of light-cone distribution
amplitudes}\label{chp:AppB} 

%\section{Full analytic expressions}

The pion distribution amplitude is by nature a non-perturbative quantity, but
has evolution governed by perturbative QCD.  The evolution is described by
the ER-BL equation given in (\ref{eq:ER-BL}); we can recast this as
\begin{eqnarray}
  \label{eq:Uev}
  \phi_{\pi}(u, \mu^2)=\int_0^1dv\,U(u, v, \mu^2, \mu_0^2)\,\phi_{\pi}(v, \mu_0^2),
\end{eqnarray}
where the operator $U(u, v, \mu, \mu_0^2)$ represents the solution to an
evolution equation equivalent to (\ref{eq:ER-BL}) and describes evolution
from the scale $\mu_0^2$ to some scale $\mu^2$.  This presents a general
solution to NLO \cite{Bakulev:2004cu} of
\begin{eqnarray}
  \label{eq:U}
  \hspace{-25pt}U(u, v, \mu^2, \mu_0^2)&=&\sum_{n=0}^\infty E_n(\mu^2,\mu_0^2)\Bigg[C_n^{3/2}(2u-1)+\frac{\a_s(\mu^2)}{4\pi}
         \sum_{k=n+2}^\infty d_{kn}(\mu^2, \mu_0^2)C_k^{3/2}(2u-1) \nonumber\\
         && +\,\mathcal{O}(\a_s^3)\Bigg]\frac{u(1-u)}{N_n}C_n^{3/2}(2v-1).
\end{eqnarray}

At leading order, the mixing of the moments of the distribution ampiltude are
triangular, so that the LO kernal is diagonal with respect to the Gegenbauer
polynomials.  This implies that only the diagonal terms of the (triangular)
anomalous dimension matrix $\gamma_{nn}^{(0)}\equiv\gamma_n^{(0)}$ will
appear and are completely encoded in the function $E_{n}^{LO}(\mu^2, \mu_0^2)$
\be
\label{eq:ELO} E_n^{LO}(\mu^2,\mu_0^2) =
\exp\left[-\int_{\a_s(\mu_0^2)}^{\a_s(\mu^2)}d\a_s\frac{\gamma_n(\a_s)}{2\beta(\a_s)}\right]
= \left[\frac{\a_s(\mu^2)}{\a_s(\mu_0^2)}\right]^{\gamma_n^{(0)}/2b_0}, 
   \ee
with
\be
  \label{eq:anom0}
  \gamma_n^{(0)}=2C_F\left[-3-\frac{2}{(n+1)(n+2)}+4\sum_{i=1}^{n+1}\frac{1}{i}\right]
\ee
The anomalous dimensions are identical to those known from deep inelastic
scattering \cite{Bakulev:2004cu,Gonzalez-Arroyo:1979df}. 

At next-to-leading order, the conformal operators mix under renormalisation,
so that both the diagonal and non-diagonal terms will contribute to  the
evolution.  The equivalent function to (\ref{eq:ELO}), $E_n^{NLO}$, is
written 
\begin{eqnarray}
  \label{eq:ENLO}
  E_n^{NLO}(\mu^2,\mu_0^2) =
  \left[\frac{\a_s(\mu^2)}{\a_s(\mu_0^2)}\right]^{\gamma_n^{(0)}/2b0}\left[\frac
  {1+\frac{b_1} {4\pi 
  b_0}\a_s(\mu^2)}{1+\frac{b_1}{4\pi   
  b_0}\a_s(\mu_0^2)}\right]^{w(n)},
\end{eqnarray}
with
\begin{eqnarray}
  \label{eq:wn}
  w(n)\equiv\frac{\gamma_n^{(1)}b_0-\gamma_n^{(0)}b_1}{2b_0b_1}.
\end{eqnarray}
The off-diagonal contributions to $U(u,v,\mu^2,\mu_0^2)$ denoted $d_{nk}$ in the full solution (\ref{eq:U}) given by
\begin{eqnarray}
  d_{kn}(\mu^2, \mu_0^2)=2\frac{N_n}{N_k}S_{kn}(\mu^2, \mu_0^2)\,c_{kn},
\end{eqnarray}
where
\bea
 && c_{kn} = (2n+3)\left\{\frac{-\gamma_n^{(0)}-2b_0+8C_FA_{kn}}{2(k-n)(k+n+3)} + \frac{2C_F[A_{kn}-\psi(k+2)+\psi(1)]}{(n+1)(n+2)}\right\} \nonumber\\
  && S_{kn}(\mu^2, \mu_0^2) = \frac{\gamma_k^{(0)}-\gamma_n^{(0)}}{\gamma_k^{(0)}-\gamma_n^{(0)}-2b_0}\left\{1-\left[\frac{\a_s(\mu^2)}{\a_s(\mu_0^2)}\right]^{-1+(\gamma_k^{(0)}-\gamma_n^{(0)})/2b_0}\right\},\nonumber\\
  && A_{kn}=\psi\left(\frac{k+n+4}{2}\right)-\psi\left(\frac{k-n}{2}\right)+2\psi(k-n)-\psi(k+2)-\psi(1) \;\nonumber\\
  \eea
and with the function $\psi(z)$ defined as $\psi(z)\equiv\frac{d\ln{\Gamma(z)}}{dz}$.  The NLO anomalous dimensions are given \cite{Floratos:1977au,Gonzalez-Arroyo:1979df} as
\begin{eqnarray} \gamma_n^{(1)}&=&-(C_F^2-\tfrac{1}{2}C_FC_A)\left\{16S_1(n)\frac{2n+1}{n^2(n+1)^2}+16\left[2S_1(n)-\frac{1}{n(n+1)}\right][S_2(n)-S_2^\prime(\tfrac{1}{2}n)]\right.  \nonumber\\
      && +
      \left.64\tilde{S}(n)+24S_2(n)-3-8S_3^\prime(\tfrac{1}{2}n)-8\frac{3n^3+n^2-1}{n^3(n+1)^3}
      -16(-1)^n\frac{2n^2+2n+1}{n^3(n+1)^3}\right\} \nonumber\\
      && -
      C_FC_A\left\{S_1(n)\left[\tfrac{536}{9}+8\frac{2n+1}{n^2(n+1)^2}\right]-16S_1(n)S_2(n)+S_2(n)\left[-\tfrac{52}{3}+\frac{8}{n(n+1)}\right]\right.
      \nonumber\\
     &&
      \left.-\tfrac{43}{6}-4\frac{151n^4+263n^3+97n^2+3n+9}{9n^3(n+1)^3}\right\}-\frac{C_Fn_f}{2}\left\{-\tfrac{160}{9}S_1(n)+\tfrac{32}{3}S_2(n)+\tfrac{4}{3}\right.
      \nonumber\\
   && + \left.16\frac{11n^2+5n-3}{9n^2(n+1)^2}\right\}
\end{eqnarray}
where the $n$-dependent functions are
\begin{eqnarray}
  \label{eq:ratfun}
  S_i(n)&=&\sum_{j=1}^n\frac{1}{j^i} \nonumber\\
  S_i^\prime(\tfrac{1}{2}n)&=&\frac{1+(-1)^n}{2}S_i(\tfrac{1}{2}n)+\frac{1-(-1)^n}{2}S_i(\tfrac{n-1}{2})
  \nonumber\\
  \tilde{S}(n) &=& \sum_{j=1}^n\frac{(-1)^j}{j^2}S_1(j)
\end{eqnarray}
and $C_F=\tfrac{4}{3}$, $C_A=3$.

The distribution amplitude to NLO will then be of the form
\begin{eqnarray}
  \label{eq:DA}
  \phi_\pi(u,
  \mu^2) &=& 6u(1-u)\sum_{n=0}^\infty\left\{\vphantom{\frac{a_s(\mu^2)}{4}}a_n(\mu_0^2)E_n^{NLO}(\mu^2, \mu_0^2)C_n^{3/2}(2u-1)\right. \nonumber\\
  && +\left.\frac{\a_s(\mu^2)}{4\pi}\sum_{k=n+2}^\infty a_n(\mu_0^2)E_n^{NLO}(\mu^2, \mu_0^2)d_{kn}(\mu^2,\mu_0^2)C_k^{3/2}(2u-1)\right\} \nonumber\\
\end{eqnarray}

The Gegenbauer moments $a_n$ contain all of the important non-perturbative information.  They can be independently evolved from their input values at some scale $\mu_0^2$ to any scale $\mu^2$ using the (NLO) evolution functions discussed above.  Generalising to include the dependence on all higher-order Gegegnbauer moments we can write the following as an alternative to equation (\ref{eq:DA}), using the shorthand $a_n(\mu_0^2)\equiv a_n^0$:
\be
 \phi_\pi(u,\mu^2) = 6u(1-u)\sum_{n=0}^\infty a_n(\mu^2) C_n^{3/2}(2u-1)
\ee
where
\be
  \label{eq:anNLO}
a_n(\mu^2)\equiv a_n^0\,E_n^{NLO}(\mu^2, \mu_0^2)+\frac{\a_s(\mu^2)}{4\pi}\sum_{k=0}^{n-2}d_{nk}(\mu^2,
  \mu_0^2)E_k(\mu^2, \mu_0^2)a_k^0
\ee
for example:
\begin{eqnarray}
  && a_2(\mu^2)=a_2^0E_2+\frac{\a_s}{4\pi}d_{20} \nonumber\\
  && a_4(\mu^2)=a_4^0E_4+\frac{\a_s}{4\pi}(d_{40}+d_{42}E_2a_2^0) \nonumber\\
  && a_6(\mu^2)=a_6^0E_6+\frac{\a_s}{4\pi}(d_{60}+d_{62}E_2a_2^0+d_{64}E_4a_4^0) \nonumber
\end{eqnarray}

When the evolution from $\mu_0^2\to\mu^2$ crosses any heavy
flavour thresholds these expressions will be modified.  Taking $\mu_0^2$ at
a low scale $\sim1\text{GeV}^2$ where the number of active flavours $n_f$ is three, the
above expressions apply for all $\mu^2\leq m_c^2$.  Above this value
there are two distinct regions: (i) with $n_f=4$ when $m_c^2<\mu^2\leq m_b^2$
and (ii) with $n_f=5$ for $\mu^2>m_b^2$. These can be summarised in a similar
expression to (\ref{eq:anNLO})
\bea
  \label{eq:anlonf}
   a_n^{NLO}(\mu^2)=a_n^0E_n^{(n_f)}(\mu^2, \mu_0^2)+\frac{\a_s(\mu^2)}{4\pi}\sum_{k=0}^{n-2}d_{nk}^{(n_f)}(\mu^2,
  \mu_0^2)E_k^{(n_f)}(\mu^2, \mu_0^2)a_k^0 \nonumber\\
\eea
where the evolution function $E_n^{(n_f)}$ takes on a different form in each region above the charm and beauty thresholds.

Region (i)
\begin{eqnarray}
  \label{eq:E4}
  E_n^{(n_f)}(\mu^2, \mu_0^2)|_{R1}&=&E_n^{(4)}(\mu^2, m_c^2)E_n^{(3)}(m_c^2,
    \mu_0^2)\nonumber\\
  d_{nk}^{(n_f)}(\mu^2, \mu_0^2)&=&d_{nk}^{(4)}(\mu^2, m_c^2) E_n^{(n_f)}(\mu^2, \mu_0^2)|_{R1}+d_{nk}^{(3)}(m_c^2,
    \mu_0^2)E_n^{(4)}(\mu^2, m_c^2)\nonumber\\
\end{eqnarray}
Region (ii)
\bea
  \label{eq:E5}
  E_n^{(n_f)}(\mu^2, \mu_0^2)|_{R2}&=&E_n^{(5)}(\mu^2, m_b^2)E_n^{(4)}(m_b^2, m_c^2)E_n^{(3)}(m_c^2,
    \mu_0^2)\nonumber\\
 d_{nk}^{(n_f)}(\mu^2, \mu_0^2)&=&\left\{d_{nk}^{(5)}(\mu^2,
    m_b^2)+d_{nk}^{(4)}(m_b^2,\mu_0^2)\right\}E_n^{(n_f)}(\mu^2,
    \mu_0^2)|_{R2} \nonumber\\
     && \qquad + \,d_{nk}^{(3)}(m_c^2, \mu_0^2)\,E_n^{(5)}(\mu^2, m_b^2)\,E_n^{(4)}(m_b^2, m_c^2)\nonumber\\
\eea

%% file: appendixC.tex
\chapter{Summary of input parameters}\label{chp:AppC}

 %%%%%%%%%%%%%%%%%%%%%%%%%%%%%%%%%%%%%%%%%%%%%%%%%%%%
   \begin{center}
 \begin{tabular*}{115mm}{@{\extracolsep\fill}|c|c|c|c|}\hline
      %\begin{tabular}{|c|c|c|c|}\hline
          \multicolumn{4}{|c|}{Masses}\\  \hline
           $m_c$ & $m_{b,pole}$ &  $m_{t,pole}$ & $M_W$ \\ \hline
           1.3GeV & 4.8GeV & 174.3GeV & 80.4GeV  \\        \hline
           $m_\pi$ & $m_\rho$ & $m_\omega$ & $m_{K^*}$\\   \hline
           140MeV & 770MeV & 782MeV & 894MeV \\            \hline
      \end{tabular*}

\begin{tabular*}{115mm}{@{\extracolsep\fill}|c|c|c|c|}\hline
      %\begin{tabular}{|c|c|c|c|}\hline
          \multicolumn{4}{|c|}{CKM Parameters and couplings}\\  \hline
           $\bar\eta$ & $\bar\rho$ & $V_{cb}$ & $V_{cd}$ \\\hline
            0.347 & 0.196 & 0.0415 & -0.2258\\ \hline
            $\gamma$ & $\beta$ & $\a$ & $\a_s(M_Z)$\\   \hline
            $60.3^\circ\pm6.8^\circ$ & $23.4^\circ$ & 1/137 & 0.1187\\ \hline
      \end{tabular*}

\begin{tabular*}{115mm}{@{\extracolsep\fill}|c|c|c|c|}\hline
      %\begin{tabular}{|c|c|c|c|}\hline
          \multicolumn{4}{|c|}{Decay constants \cite{Ball:2004rg}}\\  \hline
           $f_B$ & $f_\pi$ & $f_{K^*}$ & $f_{K^*}^\perp$ \\\hline
            $200\pm30$MeV & 133MeV & 217MeV & 170MeV\\ \hline
            $f_\rho$ & $f_\rho^\perp$ & $f_\omega$ & $f_\omega^\perp$\\   \hline
            205MeV & 160MeV & 195MeV & 145MeV\\ \hline
      \end{tabular*}

\begin{tabular*}{115mm}{@{\extracolsep\fill}|c|c|c|c|}\hline
      %\begin{tabular}{|c|c|c|c|}\hline
          \multicolumn{4}{|c|}{Form factors \cite{Ball:2004ye,Ball:2004rg}}\\  \hline
           $F_\pi$ & $F_{K^*}$ & $F_\rho$ & $F_\omega$ \\\hline
            $0.258\pm0.031$ & $0.333\pm0.028$ & $0.267\pm0.021$ &
           $0.242\pm0.022$\\ \hline 
      \end{tabular*}

\begin{tabular*}{115mm}{@{\extracolsep\fill}|c|c|c|c|}\hline
      %\begin{tabular}{|c|c|c|c|}\hline
          \multicolumn{4}{|c|}{B-meson parameters}\\  \hline
           $m_B$ & $\lambda_B$ & $\tau_{B^+}$ & $\tau_{B^0}$ \\\hline
            5.28GeV & $350\pm150$MeV & 1.65ps & 1.53ps\\ \hline
      \end{tabular*}

   \end{center}

%% \begin{table}[h]
%%  \begin{center}
%%  \begin{tabular*}{140mm}{@{\extracolsep\fill}|c|c|c|c|c|}
%%  \hline
%%  $V_{us}$ & $V_{cb}$ & $\left|V_{ub}/V_{cb}\right|$ & 
%%  $\rho$ & $\eta$ \\
%%  \hline
%%  0.22 & 0.041 & $0.085 \pm 0.025$ & X & X \\
%% %%  $\Lambda_{\overline{MS}}^{(5)}$ & $\alpha$ & $G_F$\\
%% %%  \hline
%% %%  $0.085 \pm 0.025$ & $(225 \pm 25)$ MeV & X &  
%% %%  \mbox{GeV}^{-2}$\\
%%   \hline
%%  \end{tabular*}
%% \end{center}
%% CKM parameters

%%  \begin{center}
%%  \begin{tabular}{|c|c|c|c|}
%%  \hline
%%  $m_c$ & $m_{b,pole}$ &  $m_{t,pole}$ & $M_W$ \\
%%  \hline
%%  1.3GeV & 4.8GeV & 174.3GeV & 80.4GeV  \\
%%  \hline
%%  $m_\pi$ & $m_\rho$ & $m_\omega$ & $m_{K^*}$\\
%%  \hline
%%  140MeV & 770MeV & 782MeV & 894MeV \\
%%   \hline
%%  \end{tabular}
%% \end{center}
%% Masses

 %\caption[]{Summary of input parameters.\label{tab:input}}
 %\end{table}
 %%%%%%%%%%%%%%%%
%%%%%%%%%%%%%%%%%%%%%%%%%%%%%%%%%%%%%%%%%%%%%%%%%

%%% Local Variables: 
%%% mode: latex
%%% TeX-master: t
%%% TeX-master: t
%%% End: 